%% file: tesis_final.tex
\documentclass[12pt, twoside, openright]{report}
\usepackage{fancyhdr}
\fancypagestyle{plain}{%
\fancyhead{} % get rid of headers on plain pages
}

\usepackage[activeacute, spanish]{babel}
\usepackage[latin1]{inputenc}

\usepackage[spanish]{babel}

\usepackage{a4wide,natbib}
\usepackage{epsfig}
\usepackage{graphicx}
\usepackage{psfrag}

\usepackage{latexsym}
\usepackage{amsfonts}
\usepackage{amsmath}
\usepackage{amssymb}

\decimalpoint

\begin{document}
\def\tablename{Tabla}

%\pagenumbering{roman}

\include{tapa}
\include{dedicatoria}

\pagenumbering{roman}
\include{prologo_final}

\pagenumbering{arabic}
\tableofcontents

\addcontentsline{toc}{chapter}{Resumen}
\include{resumen_final}

\addcontentsline{toc}{chapter}{Abstract}
\include{abstract_final}

\pagestyle{fancy}
%\pagenumbering{arabic}

\include{Intro_final}

\part{Procesos no t'ermicos}
\include{Choques_final}

\include{Proc_rad_final}

\part{Estudios de fuentes a diferentes escalas}
\include{YSO_final}

\include{MQs_final}
\include{AGNs_final}

\include{clusters_final}

\include{Conclusiones_final}

\addcontentsline{toc}{chapter}{Referencias}
\pagestyle{empty}
\include{referencias_final}

\pagestyle{fancy}
\appendix
\include{acronimos}

\include{PublicationList}

\end{document}

%% file: tapa.tex
\begin{titlepage}

\vspace{1cm}
\begin{center}
  \LARGE Tesis de Doctorado \\[0.75cm]
  \LARGE \textsc{\huge Radiaci'on no-t'ermica asociada a ondas de 
choque astrof'isicas} \\[1cm]
  \Large {Lic. Anabella T. Araudo} \\
  \Large Director: Prof. Dr. Gustavo E. Romero \\
\end{center}

\vspace{3cm}

\begin{figure}[h!]
\centering
\includegraphics[width=3cm,height=4cm]{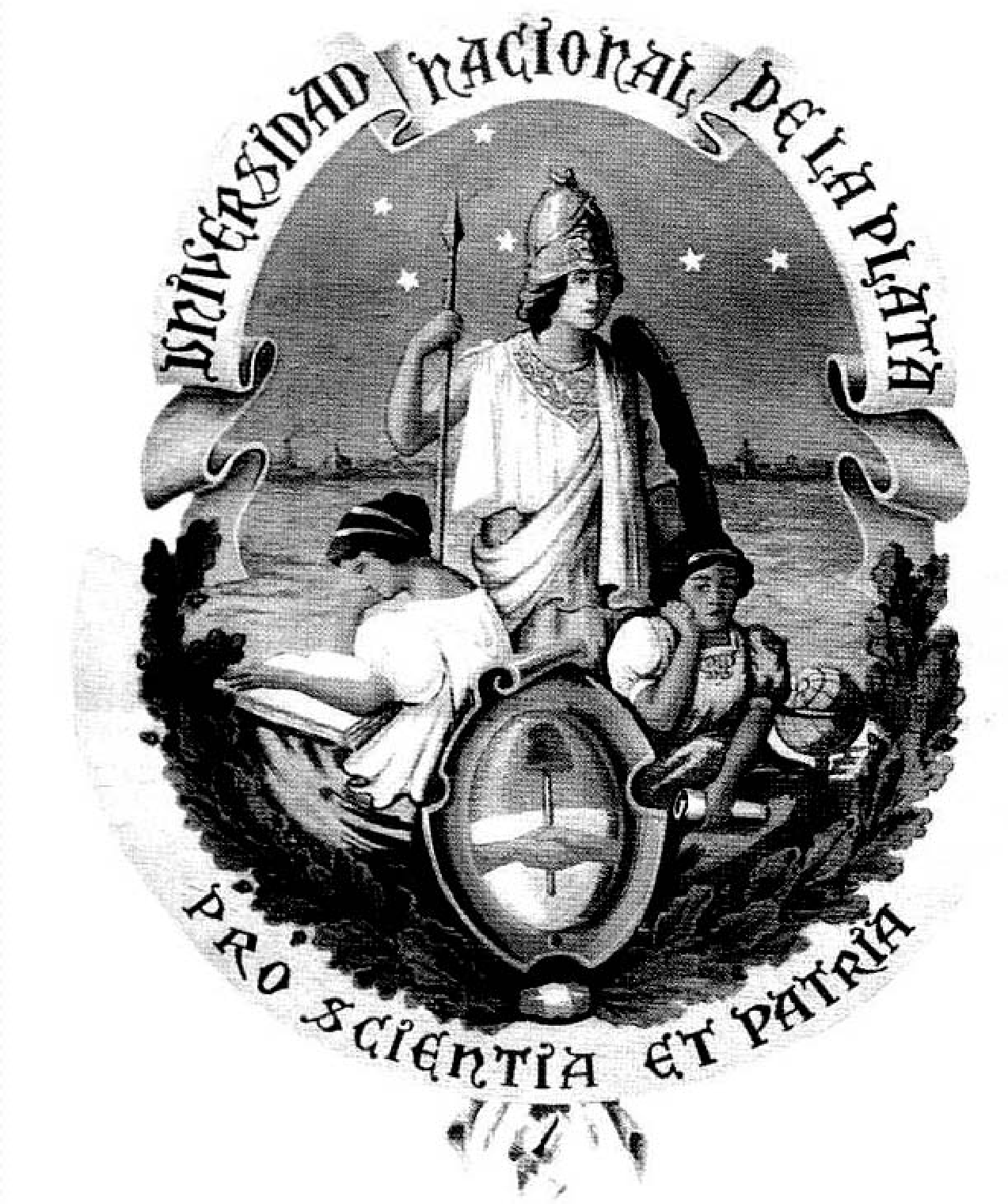}
\end{figure}
\vspace{0.01cm}
\begin{center}
\large Universidad Nacional de La Plata \\
  Facultad de Ciencias Astron\'omicas y Geof\'{\i}sicas
\end{center}

\begin{center}
\normalsize \textsc{20 de septiembre de 2010}
\end{center}

\end{titlepage}

%% file: dedicatoria.tex
\thispagestyle{empty}

\pagestyle{empty}

\vspace{20cm}

\begin{flushleft}
\vspace{20cm}
\hspace{9cm} Dedico esta tesis a Valent'i\\
\hspace{9cm} y a mis padres, \\
\hspace{9cm} Susana y Arsenio.
\end{flushleft}

\clearpage
\thispagestyle{empty}

%% file: prologo_final.tex
\chapter*{Prefacio}

Los trabajos que, junto con mi director, presentaremos a continuaci'on 
se engloban dentro de la
tesis de doctorado titulada ``Emisi'on no-t'ermica asociada a ondas de choque
astrof'isicas''. La misma no podr'ia haberse llevado a cabo
sin la ayuda de muchas personas y el apoyo de algunas instituciones.
Es por esto que, a modo personal, quisiera hacerles a ellas un sencillo, 
pero muy sincero, agradecimiento a trav'es de las siguientes l'ineas.\\ 

En primer lugar quisiera agradecer a mi director, el Prof. Gustavo E. 
Romero, por 
haberme guiado en cada trabajo que hemos realizado. Le  agradezco el 
haber compartido sus ideas conmigo, las discusiones que hemos tenido para poder
implementarlas y por sobre todo sus consejos. Me enseñ'o 
a trabajar procurando un entendimiento
global de los temas e inculcandom'e un profundo interes 'etico en cada 
problema que hemos encarado.  
La siguiente frase ilustra como sus consejos me han servido para 
realizar la tesis, aprendiendo a salvar los inconvenientes del camino: 
\emph{``La casa queda lejos de aqu'i, pero usted no se perder'a si toma 
ese camino a la izquierda y en cada encrucijada del camino dobla a la 
izquierda.''}\footnote{J.L. Borges, 
Extracto del cuento \emph{``El jard'in de los senderos que se bifurcan''}, 
Ficciones.}\\

A los profesores Ana Mar'ia Platzeck y Osvaldo Civitarese y al Dr. 
Felix Mirabel,
todos ellos integrantes del jurado examinador de esta tesis, les agradezco los 
fruct'iferos comentarios realizados. En particular, quisiera agradecer
a la Prof. Platzeck por la lectura
minusciosa de cada cap'itulo (y por sus excelentes clases).\\

Al Consejo Nacional de Investigaciones Cient'ificas y T'ecnicas (CONICET)
le agradezco el haberme otorgado dos becas para realizar mis estudios e
investigaciones doctorales. \\
 
Al Instituto Argentino de Radioastronom'ia (IAR) y a la Facultad de Ciencias
Astron'omicas y Geof'isicas (FCAG) de la UNLP quisiera agradecerles el 
haberme facilitado las condiciones y los insumos necesarios para realizar los 
trabajos que forman parte de esta tesis. \\

I want to acknowledge the hospitality and suport of Prof. Felix Aharonian.
He hosted me during three stays in the Max Plank Institut für Kernphysik
in Heidelberg. \\

Al Prof. Josep Mar'ia Paredes, le agradezco haberme invitado a exponer mis
trabajos en la Facultat de F'isica de la Universitat de Barcelona.\\

A los integrantes del grupo de Astrof'isica Relativista y Radioastronom'ia 
(GARRA) quisiera agradecerles el haberme permitido desarrollar la tesis
en un ambiente cordial de trabajo. En particular, 
quisiera agradecer el apoyo brindado por Paula Benaglia e Ileana Andruchow. \\

A Federico Bareilles le agradezco  la ayuda brindada para salvar  
diferentes (y muchos) problemas inform'aticos. \\

A todos mis amigos. En particular, voy a destacar a algunas personas 
muy importantes: Ceci Fariña, Vero Firpo, Andrea Fortier, Anah'i Granada, 
Ielca Martinic, Antonella Monachesi y Claudia Sc'occola. 
Con ellas he compartido gran parte de la carrera en la FCAG,  
adem'as de charlas hermosas y discusiones de toda 'indole.
A Javier Vasquez, a Nicol'as Duronea y a Gonzalo De El'ia, un agradecimiento 
especial por el compañerismo. A Pol Bordas, la alegr'ia.\\ 

El agradecimiento m'as profundo y eterno es para mi familia. 
Ellos han sido (y ser'an siempre) el sustento emocional y la raz'on por 
la cual he podido realizar muchas acciones. A mis sobrinos (Bianca, 
Ernestina, Rom'an e Iv'an) les 
quiero agradecer que desde que est'an en este mundo han llenado mi vida de 
ternura. \\

L$^{\prime}$agraïment més especial de tots és per a una persona 
extraordinària, en Valentí Bosch-Ramon. Com a col·laborador, el Dr. 
Bosch-Ramon m$^{\prime}$ha ajudat 
en cadascun dels treballs que hem realitzat plegats (i també en aquells en 
què ell no hi participava directament!). Amb ell he après molta astrofísica, 
m$^{\prime}$ha assistit fins al més mínim detall de cada article, i 
m$^{\prime}$ha aconsellat 
sobre com desenvolupar-me en diverses situacions. Però més enllà de tot això, 
gràcies Valentí per fer-me part de la teva vida. Gràcies per donar-me tant 
d$^{\prime}$amor i fer de la vida quotidiana una bellíssima obra 
d$^{\prime}$art. Les següents 
línies són d$^{\prime}$una cançó que coneixes bé, i m$^{\prime}$ajuden per 
a dir-te per que et necessito al meu costat:\\

\noindent\emph{Para decidir si sigo poniendo esta sangre en tierra, \\
Este corazón que va de su parte, sol y tinieblas\\
Para continuar caminando al sol por estos desiertos\\
Para recalcar que estoy viva en medio de tantos muertos.\\
Para decidir, para continuar, para recalcar y considerar\\
Sólo me hace falta que estés aquí...}\footnote{Raz'on de vivir, Victor 
Heredia.}   

\newpage

Finalmente, con la ayuda de todas las personas que mencion'e antes, 
he encontrado la casa que quedaba tan lejos. Aludiendo una vez m'as 
a J.L. Borges,  los invito a leer la tesis. \\

\bigskip

\emph{``Alguna vez, los senderos de ese laberinto convergen: por ejemplo, usted
llega a esta casa, pero en uno de los pasados posibles usted es mi enemigo, 
en otro mi amigo. Si se resigna usted a mi pronunciaci'on incurable, leeremos
unas p'aginas.''}\footnote{J.L. Borges, 
Extracto del cuento \emph{``El jard'in de los senderos que se bifurcan''}, 
Ficciones.}

\bigskip
\bigskip
\hspace{9cm} Anabella Araudo

%% file: resumen_final.tex
\chapter*{Resumen}

El objetivo principal de esta tesis es investigar
los procesos físicos que dan lugar a emisión no-t'ermica a
altas energías en objetos astrof'isicos capaces de acelerar partículas
hasta velocidades relativistas. 
En particular, se ha estudiado la emisi'on de rayos gamma
producida en fuentes c'osmicas con diferentes escalas espaciales, desde 
objetos estelares j'ovenes hasta c'umulos de galaxias, pasando por 
microcuasares y n'ucleos de galaxias activas. 
En los dos primeros tipos de objetos se 
ha modelado la emisi'on de rayos gamma a partir de los 
datos obtenidos en frecuencias radio de las fuentes
IRAS~16547-4247 y Abell~3376. En los dos 'ultimos,
se ha desarrollado un modelo espec'ifico de emisi'on basado en la 
interacci'on de inhomogeneidades del medio externo con los \emph{jets} 
producidos por el objeto compacto. Espec'ificamente, se han considerado
\emph{clumps} o grumos del viento de la estrella compañera en los 
microcuasares y nubes de la regi'on de formaci'on de l'ineas anchas en las 
galaxias activas, interactuando con los \emph{jets} de las fuentes.  
En todos los casos, los modelos desarrollados permiten realizar predicciones 
contrastables
por la nueva generación de instrumentos que operan en altas energías, tales
como los satélites \emph{Fermi} y \emph{AGILE} y los telescopios 
Cherenkov HESS, MAGIC y el planeado CTA.

\vspace{1.5cm}
Palabras claves: 
\begin{itemize}
\item Rayos gamma: general
\item Ondas de choque: general
\item Ondas de choque: aceleraci'on de part'iculas
\item Procesos radiativos: no t'ermicos
\item Estrellas de gran masa: formaci'on: emisi'on no t'ermica
\item Estrellas de gran masa: vientos
\item Sistemas binarios: microcuasares: general
\item Galaxias activas: general
\item C'umulos de galaxias: Abell~3376

\end{itemize}

%% file: abstract_final.tex
\chapter*{Abstract}

The main goal of this thesis is to study the physical processes
that can produce non-thermal emission  at high energies in astrophysical
objects capable to accelerate particles up to relativistic velocities.
In particular, we have studied the gamma-ray emission  produced in 
cosmic sources with different spatial scales, from young stellar 
objects  to clusters of galaxies,
going through microquasars and active galactic nuclei.  
In the former cases, we have modeled the gamma-ray emission using the 
radio data  from the sources  IRAS~16547-4247 and Abell~3376. 
In the latter, we have developed a specific radiation model based on the 
interaction of the inhomogeneities of the external medium 
with the jets generated by the compact object. Specifically,
we have considered clumps of the 
massive stellar wind in microquasars, and clouds of the broad 
line region in active galactic nuclei, interacting 
with the jets of the sources.   
In all  cases, the developed models allow us to make predictions 
testables with the new generation of instruments operating at high energies, 
such as the satellites \emph{Fermi} and \emph{AGILE}, and the Cherenkov
telescopes HESS, MAGIC, and the forthcoming CTA.

\vspace{1.5cm}
Key words: 
\begin{itemize}
\item Gamma-rays: general
\item Shock waves: general
\item Shock waves: particle acceleration
\item Radiative processes: non-thermal
\item Massive stars: formation: non-thermal emission
\item Massive stars: winds
\item Binary systems: microquasars: general
\item Active galaxies: general
\item Clusters of galaxies: Abell~3376

\end{itemize}

%% file: Intro_final.tex
\chapter{Introducci\'on  general}
\label{intro-general}

La astronom'ia es la ciencia que  estudia los fen'omenos
que ocurren fuera de nuestro planeta. La tecnolog'ia
actual ha permitido estudiar in situ los planetas y sat'elites naturales 
cercanos, accediendo a ellos a trav'es de sondas y naves tripuladas. 
Sin embargo, para los procesos que ocurren fuera del sistema solar, 
la astronom'ia a'un sigue investigando casi como en la antiguedad,
es decir, a trav'es de la luz que llega a nuestros detectores
desde los objetos celestes. 
Esta radiaci'on, en cada banda de frecuencia, nos ofrece una 
fenomenolog'ia  distinta, ya que los procesos
f'isicos subyacentes pueden ser muy variados. 
En particular, la emisi'on de rayos gamma da cuenta de los procesos
no t'ermicos, es decir, fuera del equilibrio termodin'amico, que puedan 
tener lugar en la fuente. 
El Universo en rayos gamma es puramente no t'ermico, ya que la
temperatura requerida para emitir fotones gamma t'ermicamente
es extremadamente alta ($\sim 10^{13}$~K) y dif'icilmente puedan 
tenerla los sistemas f'isicos conocidos\footnote{Temperaturas $\sim 10^{13}$~K
podr'ian alcanzarse por per'iodos de tiempo cortos en eventos explosivos 
muy energ'eticos como el \emph{Big Bang}
o los eruptores de rayos gamma (GRBs, por \emph{Gamma Ray Bursts}).}.    

Los rayos gamma forman la 'ultima banda del espectro electromagn'etico,
abarcando m'as de 14 'ordenes de magnitud en energ'ia:
\begin{equation}
5\times10^5\,{\rm eV} \lesssim E_{\gamma} \lesssim 10^{20}\,{\rm eV}.
\end{equation}
El l'imite inferior, $E_{\gamma} \sim m_e c^2 \sim 5\times10^5$~eV, 
corresponde a la emisi'on de 
l'ineas, como la de aniquilaci'on de los pares 
electr'on-positr'on $e^{\pm}$, mientras que el valor superior, 
$E_{\gamma} \sim 10^{20}$~eV,
corresponde a los rayos c'osmicos m'as energ'eticos que han sido detectados.
Dada la gran amplitud de esta banda de energ'ia, 
resulta conveniente subdividirla en las siguientes 
regiones\footnote{1 TeV $= 10^{12}$~eV, 1 PeV $= 10^{15}$~eV y
1 EeV $= 10^{18}$~eV.}:
\begin{itemize}
\item $E_{\gamma} < 30$~MeV: Baja energ'ia (LE, por \emph{Low Energy}).
\item $30\,{\rm MeV} < E_{\gamma} < 30$~GeV: Alta energ'ia (HE, por \emph{High 
Energy}).
\item $30\,{\rm GeV} < E_{\gamma} < 30$~TeV: Muy alta energ'ia (VHE, por 
\emph{Very High Energy}).
\item $30\,{\rm TeV} < E_{\gamma} < 30$~PeV: Ultra alta energ'ia (UHE, por 
\emph{Ultra High Energy}).
\item $E_{\gamma} > 30$~PeV: Extremadamente alta energ'ia (EHE, por 
\emph{Extremely High Energy}).
\end{itemize}
La observaci'on en las bandas LE y HE debe realizarse desde sat'elites
espaciales, ya que
la atm'osfera terrestre es opaca para las frecuencias correspondientes.
Por otro lado, los
rayos gamma de energ'ias $\gtrsim 30$~GeV pueden detectarse indirectamente
desde la superficie de la Tierra mediante telescopios Cherenkov. 
Los rayos gamma, al penetrar en la atm'osfera interact'uan con los
campos all'i presentes (magn'eticos, de materia y de fotones) produciendo 
pares $e^{\pm}$. Estos leptones, interactuando con los mismos campos, 
pueden crear m'as
fotones y pares menos energ'eticos, desarrollando as'i una cascada 
electromagn'etica. La luz (visible) Cherenkov producida por los leptones
relativistas en la atm'osfera puede detectarse desde la superficie terrestre.
Reconstruyendo la cascada se puede determinar la energ'ia
del fot'on gamma original y la direcci'on de arribo, que nos indica la
localizaci'on de la fuente. 
Sin embargo, s'olo hay detecci'on de fuentes hasta la regi'on de las UHE  
y por lo tanto solo para  $E_{\gamma} <$~PeV puede hablarse de una 
astronom'ia de rayos gamma. 

\begin{figure}
\begin{center}
\includegraphics[angle=0, width=0.8\textwidth]{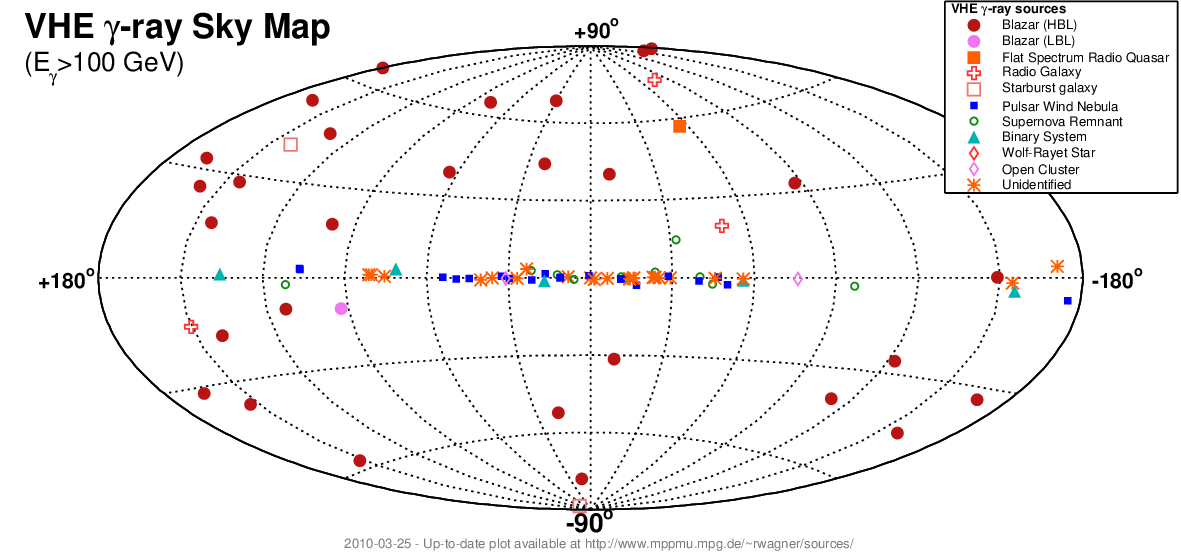}
\caption{Mapa del cielo en rayos gamma de VHE. Se muestran las fuentes 
puntuales detectadas hasta la actualidad por diferentes telescopios. 
Entre las fuentes identificadas, las m'as comunes son los AGN,
en particular, los blazares 
(http://www.mppmu.mpg.de/~rwagner/sources/).}
\label{sky-map}
\end{center}
\end{figure}

Los telescopios de rayos gamma que actualmente est'an en funcionamiento
son los siguientes:
\begin{itemize}
\item Telescopios espaciales ($E_{\gamma} \lesssim 30$~GeV)
\begin{itemize}
\item \emph{Fermi}: Es el instrumento m'as nuevo que se ha puesto en 
funcionamiento.
Fue lanzado por la NASA en el año 2008 y puede detectar fotones con energ'ias
$20$~MeV $\lesssim E_{\gamma} \lesssim  100$~GeV.
\item \emph{AGILE}: Es un sat'elite italiano que funciona en el rango de
energ'ias 0.1~GeV $\lesssim  E_{\gamma} \lesssim  30$~GeV. 
\item \emph{Swift}: Lanzado por la NASA en el año 2004, puede detectar fotones 
gamma con energ'ias de $\sim 150$~keV.
Adem'as del instrumento BAT de rayos X duros (pero que llega a detectar algunos
fotones gamma blandos), a bordo del 
sat'elite hay un telescopio  ultravioleta (UV) y otro 'optico, con el
fin de poder detectar GRBs en diferentes longitudes de onda.  
\end{itemize}
\item Telescopios Cherenkov ($E_{\gamma} \gtrsim 30$~GeV)
\begin{itemize}
\item HESS: Es un arreglo de 4 telescopios franco-alemanes de 12~m 
de di'ametro cada uno  ubicados en Gamsberg, Namibia. 
Opera en el rango de energ'ias
50~GeV $\lesssim  E_{\gamma} \lesssim  10$~TeV. 
\item MAGIC: Es un 'unico receptor germano-español de 17~m de di'ametro 
ubicado en La Palma y detecta fotones gamma de energ'ias
50~GeV $\lesssim  E_{\gamma} \lesssim  10$~TeV.
\item VERITAS:  Es un arreglo de 4 telescopios estadounidenses de 12~m
de di'ametro cada uno, ubicados 
en Arizona, y que opera en la banda de energ'ias 
100~GeV $\lesssim  E_{\gamma} \lesssim  50$~TeV.
\item Cangaroo III: Es una colaboraci'on japonesa-australiana que opera
un arreglo de 4 telescopios de 10~m cada uno ubicados en Woomera, 
en el sur de Australia. Detecta rayos gamma de $E_{\gamma} \sim 100$~GeV.
\end{itemize}
\end{itemize}
Todos estos instrumentos detectan fuentes, y de ah'i que los llamemos 
telescopios. Estas fuentes pueden ser puntuales o extendidas, 
como por ejemplo la regi'on central de la Galaxia.
Entre las fuentes detectadas por los telescopios listados
anteriormente, 
hay n'ucleos de galaxias activas (AGN, por \emph{Active Galactic Nuclei}),
 pulsares, remanentes de supernovas, binarias de rayos X,
GRBs y algunas regiones de formaci'on estelar. 
Hay otras fuentes que han sido observadas, pero cuya detecci'on 
\footnote{En frecuencias gamma, para que  una detecci'on sea 
confiable se requiere una relaci'on se~nal (S) ruido (N) tal que 
$S/N \geq 6$.}
no ha sido clamada a'un como ocurre con los c'umulos de galaxias y
algunas regiones de formaci'on estelar. 
Por otro lado, hay muchas fuentes que a'un no han sido identificadas. 
En la Figura~\ref{sky-map}
se muestra un mapa del cielo con las fuentes detectadas 
en el rango de las VHE por diferentes telescopios. 

La detecci'on de fuentes gamma es importante porque 
nos provee informaci'on sobre procesos f'isicos extremos.
La astrof'isica de rayos gamma es una disciplina 'unica para estudiar los
fen'omenos  fuera del equilibrio 
que involucran las enormes cantidades de energ'ia requeridas para  
emitir los fotones  
m'as energ'eticos del espectro electromagn'etico. As'i,
los emisores de rayos gamma deben ser regiones del espacio en las cuales 
las part'iculas relativistas all'i presentes pueden enfriarse 
eficientemente debido a interacciones con
grandes densidades de materia y/o radiaci'on.

En esta tesis nos hemos propuesto estudiar los procesos no t'ermicos que 
puedan desarrollarse en una variedad de fuentes astrof'isicas 
con diferentes escalas
espaciales, desde objetos estelares j'ovenes (YSOs, por 
\emph{Young Stellar Objects}) hasta c'umulos de galaxias,
pasando por microcuasares (MQs, por \emph{Microquasars}) y AGNs. 
En particular, nos hemos concentrado en estos cuatro tipos de objetos 
para analizar como las ondas 
de choque pueden acelerar part'iculas cargadas
 hasta velocidades relativistas y como luego estas part'iculas 
pueden interactuar con el medio y producir fotones mediante procesos
radiativos no t'ermicos.
Hemos calculado el espectro de fotones producido, haciendo 'enfasis en el
an'alisis de la emisi'on de rayos gamma. 

\begin{itemize}
\item 
En los YSOs  masivos, estudiamos la
aceleraci'on de part'iculas en los choques terminales de los \emph{jets}, 
producidos 
cuando 'estos son frenados por el medio circundante. 
Hemos calculado el espectro de emisi'on de rayos gamma de la fuente 
IRAS 16547-4247 obteniendo niveles de luminosidad detectables con los
telescopios que actualmente est'an funcionando.  

\item
En los microcuasares de alta masa (HMMQ, por \emph{High Mass Microquasar}), 
el viento de la estrella compañera 
puede tener estructura (inhomogeneidades en la densidad). Estas
inhomogeneidades al llegar al \emph{jet} 
pueden penetrar en 'el y producir  ondas de choque. Las part'iculas all'i
aceleradas pueden producir emisi'on variable y, en rayos gamma  detectable 
por los 
telescopios actuales. Este mecanismo podr'ia explicar cierta 
 variabilidad observada
en fuentes como Cygnus~X-1, LS~5039 y LSI~+31~603.

\item
En los AGNs estudiamos una
situaci'on similar a la analizada en los HMMQs. Sin embargo, en los primeros,
ser'ia la interacci'on de las nubes que se encuentran orbitando al agujero 
negro central con el \emph{jet} la que dar'ia como resultado la emisi'on de 
rayos gamma.
Este mecanismo podr'ia explicar la emisi'on observada en algunas galaxias.  

\item 
Finalmente, en el c'umulo de galaxias Abell 3376 estudiamos el espectro
en rayos gamma producido por la  poblaci'on de part'iculas relativistas
generada en  choques detectados en el borde de la fuente. 
\end{itemize}

Los resultados de nuestra investigaci'on sobre
 ``\emph{Radiaci'on no t'ermica asociada a las ondas de choque
astrof'isicas}'', tem'atica que da nombre a esta tesis, ser'an presentados 
de la 
siguiente manera: en los Cap'itulos~2 y 3 haremos una 
somera descripci'on de los procesos no t'ermicos: ondas de choque y 
radiaci'on. Los cuatro cap'itulos siguientes est'an dedicados a cada una de
las fuentes estudiadas en esta tesis: YSOs (Cap'itulo~4), HMMQs 
(Cap'itulo~5), AGNs (Cap'itulo~6) y c'umulos de galaxias
(Cap'itulo~7). Finalmente, en el Cap'itulo~8,
discutiremos los resultados y las conclusiones. A lo largo de toda la tesis,
los valores de las magnitudes f'isicas consideradas estar'an dados en 
unidades cgs 
y las unidades electromagn'eticas seran Gauss (G) y esu.

%% file: Choques_final.tex
\chapter{Ondas de choque y aceleraci'on de part'iculas}
\label{cap2}

\section{Introducci'on}

En un plasma que se mueve a una velocidad $\vec v$,  tiene un campo 
magn'etico $\vec B$,  una presi'on
$P$ y una densidad $\rho$, en el cual pueda aplicarse la 
aproximaci'on magnetohidrodin'amica\footnote{La aproximaci'on 
magnetohidrodin'amica considera que el campo magn'etico es relevante en la
din'amica del plasma. En las ecuaciones de Maxwell, esto se introduce
despreciando la corriente de desplazamiento.} (MHD), 
las ecuaciones que describen 
el estado del sistema pueden separarse en 3 grupos: mec'anicas,
electromagn'eticas y termodin'amicas.
\begin{itemize}
\item Las {\bf mec'anicas} incluyen la conservaci'on de la masa
\begin{equation}
\label{masa} 
\frac{\rm{D}\rho}{\rm{D} t} + \rho \vec\nabla \cdot \vec v = 0,
\end{equation}
y la ecuaci'on de movimiento 
\begin{equation}
\label{mov}  
\rho \frac{{\rm D}\vec v}{{\rm D} t} = -\vec\nabla P + \vec I \times \vec B + \vec F',
\end{equation}
donde $\vec F'$ incluye a todas las dem'as fuerzas, adem'as del gradiente de 
$P$ y de la fuerza de Lorentz ($\vec I \times \vec B$), que act'uan sobre el 
sistema.
La fuerza $\vec F'$ podr'ia ser la gravedad, la fuerza viscosa, etc.  
Por otro lado, la derivada convectiva D/D$t$ se define como
D/D$t \equiv \partial /\partial t + \vec v \cdot \vec\nabla$. En la 
aproximaci'on MHD, la relaci'on entre $\vec I$ y $\vec B$ est'a dada por la
ecuaci'on de Ampere-Maxwell, $(4\pi/c^2)\,\vec I = \vec\nabla \times \vec B$. 
\item Dentro de las ecuaciones {\bf electromagn'eticas} incluimos a la 
 de divergencia nula:
\begin{equation} 
\vec\nabla \cdot \vec B = 0,
\end{equation}
y a la ecuaci'on de inducci'on
\begin{equation}
\label{induccion} 
\frac{\partial \vec B}{\partial t} = \vec\nabla \times(\vec v \times \vec B)
+ \frac{c^2}{4\pi\sigma_{\rm B}} \,\nabla^2 \vec B,
\end{equation}
donde $\sigma_{\rm B}$ es la conductividad magn'etica.
\item Finalmente las ecuaciones {\bf termodin'amicas} son la ecuaci'on de 
estado, por ejemplo la de gas ideal,
\begin{equation}
\label{gasideal} 
P = n\, K_{\rm B}\, T,
\end{equation}
y la ecuaci'on de conservaci'on de la energ'ia:
\begin{equation}
\label{energia} 
\frac{\rho^{\gamma_{\rm ad}}}{\gamma_{\rm ad} -1} \frac{\rm{D}}{\rm{D} t} \left(
\frac{P}{\rho^{\gamma_{\rm ad}}}\right) = - \mathcal{L},
\end{equation}
donde $\gamma_{\rm ad}$ es el 'indice adiab'atico y toma el valor $5/3$ para 
gases monoat'omicos no relativistas y $4/3$ para los relativistas.
Las p'erdidas de energ'ia $\mathcal{L}$ pueden estar debidas a la viscosidad,
a la difusi'on t'ermica o a la radiaci'on. En la ecuaci'on~(\ref{gasideal}) 
$n = \rho/m$ es la densidad num'erica de part'iculas de masa $m$, $T$ es la
temperatura y $K_{\rm B}$ la constante de Boltzmann.
\end{itemize}

Si provocamos una pequeña perturbaci'on en el movimiento del plasma, 
las propiedades del sistema se ver'an modificadas de la siguiente manera:
\begin{equation} 
\xi = \xi_0 + \xi_1(\vec r, t),
\end{equation}
donde $\xi_1(\vec r, t)$ es la perturbaci'on respecto del estado de 
equilibrio $\xi_0$. La magnitud $\xi$ puede ser la presi'on, 
la densidad, la velocidad o el campo magn'etico.
Si la perturbaci'on es chica, es decir, $|\xi_1| \ll |\xi_0|$, 
entonces
las ecuaciones de la MHD perturbadas se pueden linealizar considerando
s'olo los t'erminos de primer orden en las perturbaciones $\xi_1$.
Si no hay campos magn'eticos, la perturbaci'on  
satisface la ecuaci'on de onda:
\begin{equation} 
\nabla^2\xi_1 - \frac{1}{C^2}\frac{\partial^2\xi_1}{\partial t^2}=0,
\end{equation}
donde $C$ es (el m'odulo de) la velocidad de propagaci'on de la 
perturbaci'on\footnote{A los m'odulos de las magnitudes vectoriales los
representamos a trav'es de la letra usada para representar a la magnitud pero
sin la flecha que indica la naturaleza vectorial de la misma. Por
ejemplo, $|\vec A| = A$.}. 

La ecuaci'on 
de movimiento (\ref{mov}) depende de las fuerzas que act'uen sobre 
el sistema: el 
gradiente de presi'on, la fuerza 
de Lorentz u otro tipos de fuerzas  $\vec F'$.
Dependiendo de la importancia relativa de unas respecto de las otras, 
tendremos diferentes tipos de ondas:
\begin{itemize}
\item Si $|\vec\nabla P| \gg |\vec I \land \vec B|$ y 
$|\vec\nabla P| \gg |\vec F'|$, y consideramos un fluido 
inicialmente en reposo, entonces tendremos variaciones 
en $P$, $\rho$ y $v$. Estas son las llamadas ondas de sonido y se propagan 
a la velocidad del sonido $C_{\rm s} = \sqrt{\gamma_{\rm ad}\, P/\rho}$ en el
medio no perturbado.

\item Si $|\vec I \land \vec B| \gg |\vec\nabla P|$ y 
$|\vec I \land \vec B| \gg |\vec F'|$,
entonces tendremos las llamadas ondas magnetohidrodin'amicas o de Alfv'en.
Estas tienen diferentes modos dependiendo de que la variaci'on sea en la 
direcci'on o en la intensidad de $\vec B$. 
Las ondas de Alfv'en  de corte son transversales (se propagan 
perpendicularmente a $\vec B_0$) y var'ian s'olo la direcci'on
de $\vec B$. La velocidad de fase de estas ondas es 
$\vec v_{\rm f} = \pm v_{\rm A}\,\cos(\theta_{\rm B})\,\vec k/k$, donde 
$\vec k$ es el vector de onda y $\theta_{\rm B}$ es el 'angulo entre $\vec k$
y $\vec B_0$, y la velocidad de grupo es $\vec v_{\rm g} = \vec v_{\rm A}$. 
La velocidad de Alfv'en se define de la siguiente manera:
$v_{\rm A} = \sqrt{\gamma_{\rm ad}\, B_0/\rho_0}$.
Por otro lado, las ondas de Alfv'en de compresi'on son longitudinales; 
modifican tanto la direcci'on como la intensidad de $\vec B$ y se propagan
con una velocidad  $\vec v_{\rm f} = \vec v_{\rm g} = \pm  v_{\rm A}\,\vec k/k$. 

\item Si  $|\vec I \land \vec B| \sim |\vec\nabla P|$ y $|\vec F'| \sim 0$,
entonces tenemos las llamadas ondas magnetoac'usticas. La velocidad 
$C_{\rm{sA}}$ de propagaci'on (de fase) de estas ondas es  
$0 \le C_{\rm{sA}} \le \sqrt{V_{\rm A}^2 + C_{\rm s}^2}$. Cuando 
$\max\{v_{\rm A}, C_{\rm s}\} \leq C_{\rm{sA}} \leq \sqrt{V_{\rm A}^2 + C_{\rm s}^2}$ se dice que la onda est'a en el modo r'apido, mientras que si 
$0 \le C_{\rm{sA}} \le \min\{v_{\rm A}, C_{\rm s}\}$ 
se dice que est'a en el modo lento. 
\end{itemize}

Notemos que en todos estos casos, la velocidad de fase, $\vec v_{\rm f}$, 
y por lo tanto la  de grupo, son  constantes. Esto implica que no hay
dispersi'on, es decir, que el perfil 
de las ondas no cambia, contrariamente a lo que ocurre, como veremos
luego, en las ondas de choque. 
Hasta aqu'i hemos realizado un breve resumen de lo que sucede cuando 
las perturbaciones 
son pequeñas. ¿Qu'e ocurre cuando las perturbaciones ya no son tan chicas, 
comenzando a hacerse importantes los t'erminos no lineales?

\section{Definici'on y estructura de las ondas de choque}

\begin{figure}
\begin{center}
\includegraphics[angle=0, width=0.4\textwidth]{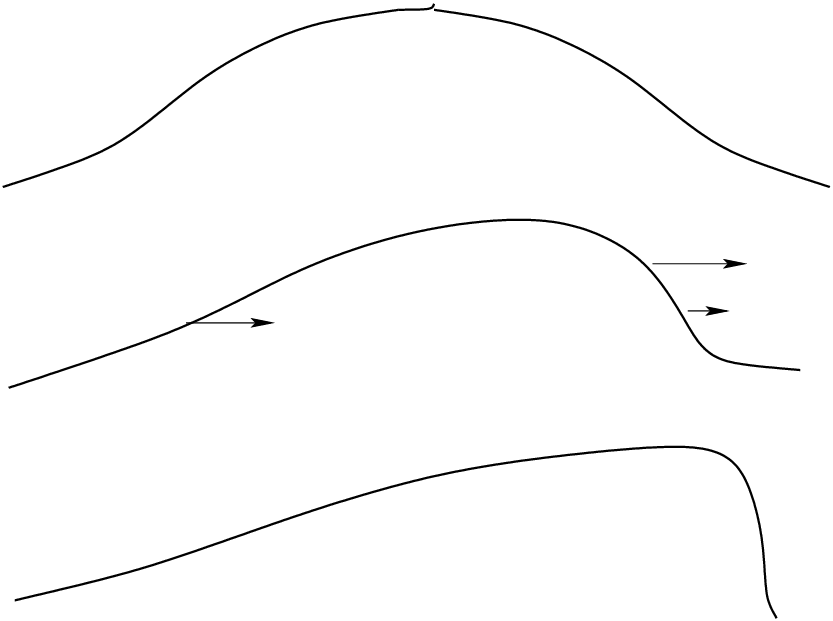}
\caption{Empinamiento del frente de la onda a medida que los t'erminos
no lineales de las ecuaciones perturbadas se hacen m'as importantes.}
\label{empinado}
\end{center}
\end{figure}

Dada una onda sonora o una onda magnetoac'ustica (de compresi'on), 
cuando la amplitud es grande, es decir cuando los t'erminos no lineales 
de las ecuaciones son 
importantes, el perfil de la onda se modifica con el tiempo. 
Esto se debe a que las perturbaciones de mayor amplitud viajan m'as
r'apidamente que las de menor amplitud, es decir $\vec v_{\rm g}$ 
no es constante,
y el perfil de la onda  se modifica empin'andose cada vez m'as 
a medida que la onda avanza,
como se muestra en la Figura~\ref{empinado} (Priest 1982, Platzeck 2009). 

Este comportamiento es similar tanto en $\vec v$ como en $P$ y en $\rho$.
Los gradientes de estas cantidades se hacen cada vez m'as grandes de tal manera
que efectos disipativos como la viscosidad o el t'ermino disipativo de la 
ecuaci'on de inducci'on no pueden despreciarse. Son estos efectos los
que impiden que esta situaci'on contin'ue indefinidamente. Cuando la 
fuerza viscosa
(no consideraremos disipaci'on magn'etica) se equilibra con los gradientes de 
$\vec v$, $P$ y $\rho$ se alcanza el estado estacionario y se frena el 
empinamiento del perfil de la onda. Luego contin'ua propag'andose con este 
perfil constante y empinado. Esto es lo que llamamos una onda de choque.
Cuando se llega a esta situaci'on, la velocidad de propagaci'on, que 
llamaremos $v_{\rm ch}$, es mayor
que la velocidad de la onda linealizada correspondiente, $C_{\rm s}$ o
$C_{\rm sA}$.

\begin{figure}
\begin{center}
\includegraphics[angle=0, width=0.6\textwidth]{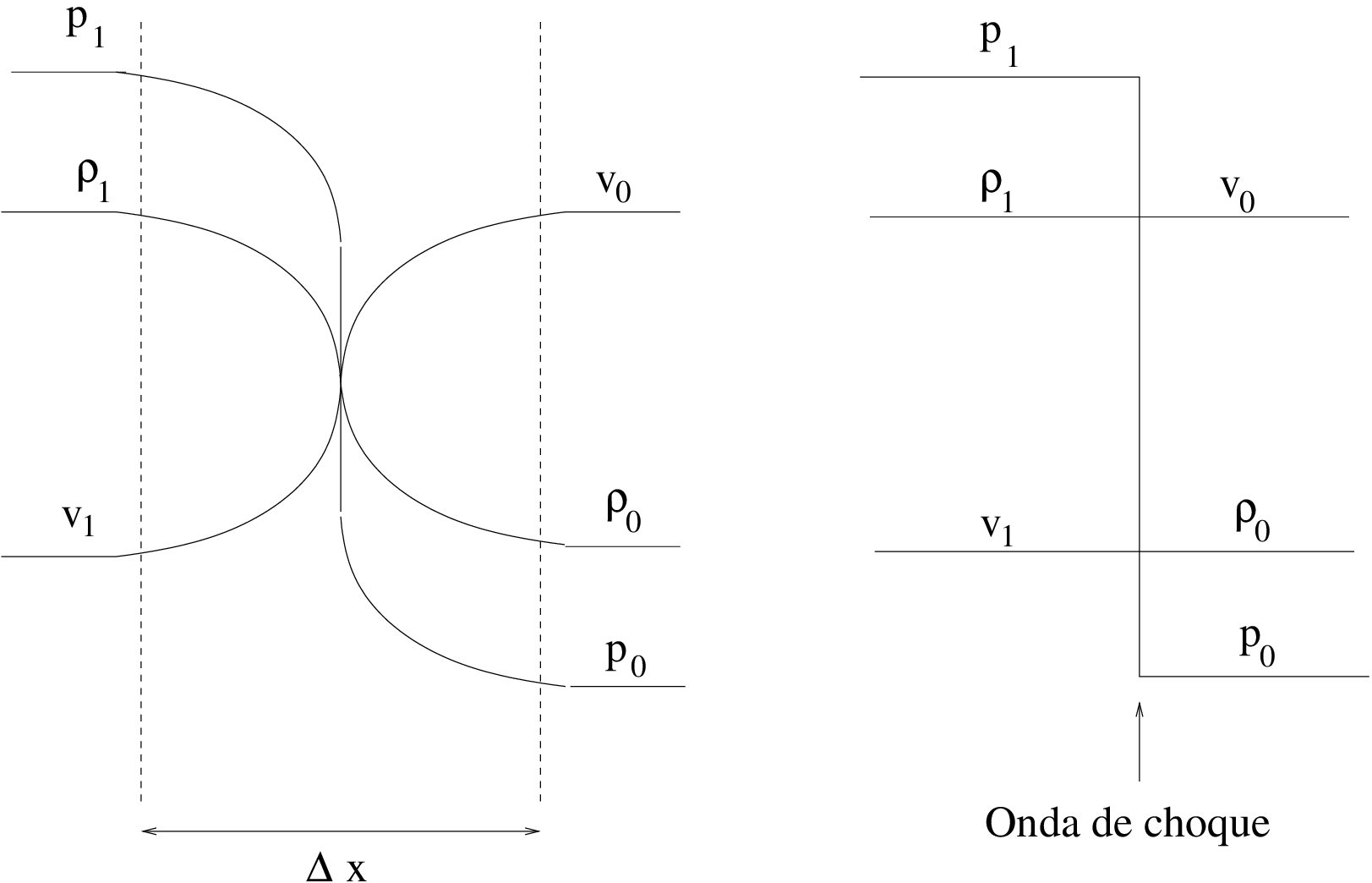}
\caption{Esquema de las variaciones en la presi'on ($P$), la densidad ($\rho$),
y la velocidad ($\vec v$) del fluido. Con los sub'indices 0 y 1 indicamos las 
magnitudes en la regi'on no perturbada y perturbada, respectivamente.
En la figura de la izquierda se muestra la region de transici'on $\Delta x$ 
ampliada con respecto a la figura de la derecha. En HD y MHD,  
$\Delta x \rightarrow 0$,
es decir, la regi'on de transici'on puede 
considerarse como superficie de discontinuidad de las magnitudes graficadas.}
\label{jump}
\end{center}
\end{figure}

El equilibrio entre la fuerza viscosa y los gradientes ocurre en una regi'on 
muy peque~na comparada con el tamaño del sistema. El ancho $\Delta x$ de 
esta regi'on de transici'on se estima que es del orden de algunos 
caminos libres medios $\lambda$ (longitud que recorre una part'icula antes de
desviarse significativamente\footnote{Si el mecanismo de interacci'on dominante
entre las part'iculas es colisional, entonces $\lambda$ es la distancia 
recorrida por la part'icula antes de interactuar con otra. En el pr'oximo
cap'itulo (ver ecuaci'on~(\ref{camino-libre-medio})) definiremos $\lambda$
de una manera m'as apropiada.}) y decrece 
a medida que $\vec v_{\rm ch}$ crece. 
De esta  manera, si $v_{\rm ch} \gg C_{\rm s}$ el choque se dice
fuerte y la regi'on de transici'on  puede considerarse
como una ``superficie'' de discontinuidad. 
El tiempo que tarda el fluido en 
atravesar la regi'on de transici'on es muy chico comparado con el 
tiempo de los cambios
en las regiones fuera de ella. De esta manera podemos considerar que 
el material que se encuentra fuera de la regi'on de transici'on est'a en 
un estado estacionario, como se muestra en la Figura~\ref{jump}. En la
aproximaci'on hidrodin'amica (HD) y MHD $\Delta x \sim 0$. 
Podemos definir a las ondas de choque de la siguiente manera: 

\begin{itemize}
\item
{\bf Onda de choque:} Onda de gran amplitud que produce una 
discontinuidad en los valores de las funciones que representan a las 
propiedades del plasma. El frente de choque divide al medio
en dos regiones: chocada y no chocada, en las cuales los valores de 
$P$, $\rho$, $T$, $\vec v$ y $\vec B$ son muy distintos.
\end{itemize}

\begin{figure}
\begin{center}
\includegraphics[angle=0, width=0.7\textwidth]{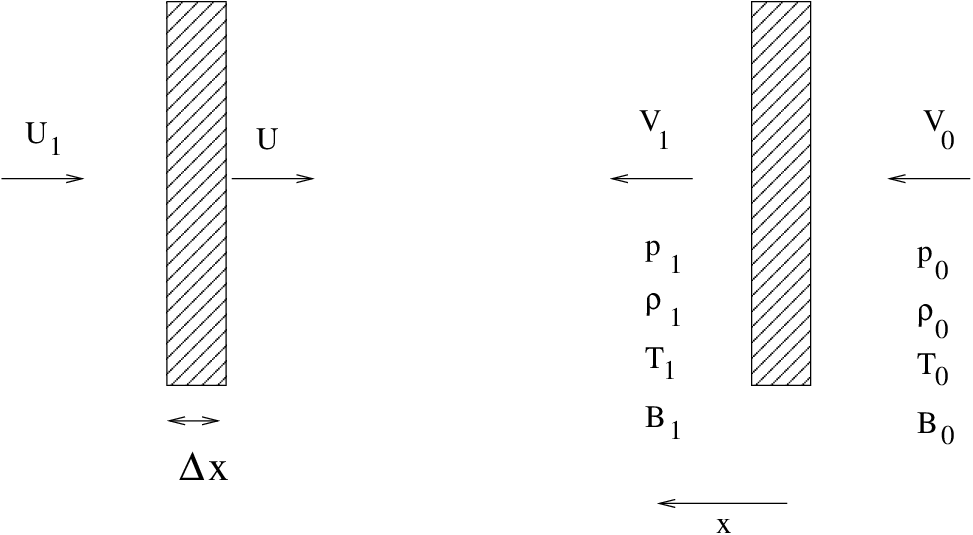}
\caption{Velocidades en el SR con el medio no perturbado en reposo (izquierda)
y en el SR solidario al choque (derecha).}\label{choque}
\end{center}
\end{figure}

Denominamos con el sub'indice 0 a las propiedades en el medio sin perturbar 
(delante del frente de onda) y 
con el sub'indice 1 a aquellas en el medio chocado 
(detr'as del frente de onda). 
Suponemos que ambos medios son uniformes. En el sistema de referencia (SR)
con el medio 0 en reposo, el choque se mueve con una velocidad $\vec U$ y el
medio perturbado con una velocidad $\vec U_1$. En el SR
fijo al frente de choque, el medio no perturbado se acerca al choque 
con una velocidad $\vec v_0$ mientras que el material perturbado se aleja del
choque con una velocidad $\vec v_1$. En la Figura~\ref{choque} se esquematizan 
ambas situaciones. Notar que 
\begin{equation}
\vec v_0 = -\vec U \qquad {\rm y} \qquad \vec v_1 = \vec U_1 - \vec v_0.
\end{equation}
Los m'odulos de las velocidades se relacionan de la siguiente manera: 
$v_0 = U$ y $v_1 = v_0 - U_1 = U - U_1$.

\section{Ondas de choque no relativistas}
\label{NR}

Las ondas de choque no relativistas (NR) son aquellas tales que 
$v_{\rm ch} \lesssim 0.1 c$, siendo $c$ la velocidad de la luz en el vac'io. 
Veremos a continuaci'on que en el caso de choques no relativistas 
los valores de la velocidad, la presi'on, 
la densidad y el campo magn'etico 
en los medios chocado y no chocado pueden relacionarse teniendo en cuenta 
las ecuaciones (\ref{masa})-(\ref{energia}). 

\subsection{Ondas de choque hidrodin'amicas}

Las ondas de choque hidrodin'amicas son aquellas que se producen en un medio en 
el cual la fuerza din'amicamente relevante es  $\vec\nabla P$. Esto es
$|\vec\nabla P| \gg |\vec I \land \vec B|$ y 
$|\vec\nabla P| \gg |\vec F'|$. As'i, las ondas de choque 
hidrodin'amicas se propagan a una velocidad mayor que las
ondas de sonido, es decir, $v_{\rm ch} > C_{\rm s}$. 

Consideremos un medio no conductor y en estado estacionario, con densidad
$\rho_0$, presi'on $P_0$,  campo magn'etico $\vec B_0 \sim 0$
y un choque propagandose con una velocidad  $\vec v_{\rm ch}$.
En el SR solidario al frente
de choque, la velocidad del medio no chocado es $\vec v_0$ y 
debido a que hemos tomado convenientemente el eje $x$ como la direcci'on de 
propagaci'on del choque, s'olo habr'a variaciones espaciales en tal direcci'on.
En este caso de ondas de choque propag'andose en un medio con $\vec B \sim 0$,
las ecuaciones que describen el estado del sistema son la ecuaci'on (\ref{masa})
de conservaci'on de la masa, la ecuaci'on (\ref{mov}) de movimiento  
y la de conservaci'on de la energ'ia (\ref{energia}).

Debido a que el sistema est'a en estado estacionario, la 
ecuaci'on~(\ref{masa}) se simplifica a
\begin{equation} 
\frac{{\rm d}(\rho \,\vec v)}{{\rm d} x}=0.
\end{equation}
Esta condici'on debe cumplirse a lo largo de cualquier l'inea de flu'ido, 
por lo cual, si la integramos entre un punto en la regi'on 1  ($x_1$) y 
otro en la regi'on no perturbada ($x_0$) obtenemos: 
\begin{equation} 
\label{RH_prim_1}
\rho_1 \,\vec v_1 = \rho_0 \,\vec v_0.
\end{equation}
Esta ecuaci'on describe que el flujo de masa $\rho \,\vec v$ que pasa a trav'es 
de la onda de choque se conserva. 
Por otro lado, la ecuaci'on de movimiento en un medio no conductor 
($\vec I \land \vec B =0$) y estacionario se reduce a  
\begin{equation} 
\label{mov_2}
\rho \,\vec v \cdot\frac{{\rm d}\vec v}{{\rm d}x} = 
-\frac{{\rm d}p}{{\rm d}x} + 
\frac{4}{3}\frac{{\rm d}}{{\rm d}x}\left(\lambda_{\rm v}.
\frac{{\rm d}\vec v}{{\rm d}x}\right).
\end{equation}
La fuerza viscosa 
$\vec F' = -(4/3)\,{\rm d}(\lambda_v{\rm d}v/{\rm d}x)/{\rm d}x\,\check i$,
donde $\lambda_v$ es una constante, 
est'a presente en la regi'on de transici'on $\Delta x$ pero es nula 
tanto en el medio chocado como en el no chocado. 
Por lo tanto, si integramos la ecuaci'on~(\ref{mov_2}) entre $x_1$ y $x_0$
y usamos que $\rho_1 v_1 = \rho_0 v_0$, obtenemos
\begin{equation}
\label{RH_prim_2} 
\rho_0\, v_0^2 + P_0 = \rho_1 \,v_1^2 + P_1.
\end{equation}  
La magnitud $(\rho \vec v)\cdot\vec v$, conocida como ``presi'on cin'etica'', 
es la 
cantidad de impulso transportado por unidad de tiempo y por unidad de 
'area transversal al movimiento. Por otro lado, $P$ es la fuerza que act'ua    
sobre dicha 'area. 

Finalmente, en la ecuaci'on de conservaci'on de la energ'ia 
consideraremos las p'erdidas por conducci'on t'ermica 
$\mathcal{L}_{\rm c} = \vec\nabla(-\lambda_{\rm c}\vec\nabla T)$, donde 
$\lambda_{\rm c}$ es una constante, y por viscosidad
$\mathcal{L}_{\rm v} = (-2/3)\rho\lambda_{\rm v}(\vec\nabla \cdot\vec v)^2$, 
ambas presentes solo en la regi'on
de transici'on. La ecuaci'on (\ref{energia}) escrita en funci'on de la energ'ia
interna por unidad de masa $\epsilon$ y teniendo en cuenta la viscosidad y la 
conducci'on de calor toma la forma
\begin{equation} 
\rho v\left(\frac{{\rm d}\epsilon}{{\rm d}x} - \frac{p}{\rho^2}
\frac{{\rm d}\rho}{{\rm d}x}\right)=  
\frac{{\rm d}}{{\rm d}x}\left(\lambda_{\rm c}\frac{{\rm d}T}{{\rm d}x}\right) + 
\frac{2}{3}\rho\lambda_{\rm v}
\left(\frac{{\rm d} v}{{\rm d}x}\right)^2.
\end{equation}  
Integrando como antes entre $x_1$ y $x_0$, donde ${\rm d}T/{\rm d}x=0$ y 
${\rm d}\vec v/{\rm d}x=0$, y usando las relaciones (\ref{RH_prim_1}) y 
(\ref{RH_prim_2}), obtenemos:
\begin{equation} 
\label{}
\left(\frac{1}{2}\rho_1 v_1^2 + \epsilon_1\rho_1\right)\vec v_1 + P_1 \vec v_1 =
\left(\frac{1}{2}\rho_0 v_0^2 + \epsilon_0\rho_0\right)\vec v_0 + P_0 \vec v_0,
\end{equation}
donde $((1/2)\rho v^2 + \epsilon\rho)$ es la cantidad densidad de 
energ'ia cin'etica
e interna transportada por unidad de superficie y de tiempo; $p\,v$ es el
trabajo por unidad de tiempo y de 'area realizado por la presi'on del gas.   
Si consideramos que $\rho_1\vec  v_1 = \rho_0 \vec v_0$ tenemos
\begin{equation} 
\epsilon_1 + \frac{P_1}{\rho_1}+ \frac{v_1^2}{2} =
\epsilon_0 + \frac{P_0}{\rho_0}+ \frac{v_0^2}{2}.
\end{equation}
Teniendo en cuenta que para un gas ideal $\epsilon = p/\rho(\gamma - 1)$, la 
ecuaci'on de conservaci'on de la energ'ia nos da la relaci'on
\begin{equation} 
\frac{\gamma_{\rm ad}}{\gamma_{\rm ad} -1}\frac{P_1}{\rho_1}+ \frac{v_1^2}{2} =
\frac{\gamma_{\rm ad}}{\gamma_{\rm ad} -1}\frac{P_0}{\rho_0}+ \frac{v_0^2}{2}.
\end{equation}

As'i hemos obtenido las {\bf relaciones de Rankine-Hugoniot}, o relaciones de
salto:
\begin{equation} 
\label{RH_1}
\rho_0 \vec v_0 = \rho_1 \vec v_1
\end{equation}        
\begin{equation} 
\label{RH_2}
\rho_0 v_0^2 + P_0 = \rho_1 v_1^2 + P_1
\end{equation}
\begin{equation}
\label{RH_3}
\frac{\gamma_{\rm ad}\, P_0}{(\gamma_{\rm ad}-1)\rho_0} + \frac{1}{2}v_0^2 =
\frac{\gamma_{\rm ad}\, P_1}{(\gamma_{\rm ad}-1)\rho_1} + \frac{1}{2}v_1^2.
\end{equation}
Resolviendo estas ecuaciones podemos obtener informaci'on  sobre 
como la onda de choque modifica el medio. De la ecuaci'on (\ref{RH_1}) 
obtenemos que 
\begin{equation} 
\frac{v_1}{v_0} = \frac{\rho_0}{\rho_1}.
\end{equation} 
Por otro lado, de las ecuaciones (\ref{RH_2}) y (\ref{RH_3}) podemos hallar 
expresiones para $\rho_0/\rho_1$ y para $P_0/P_1$ en funci'on del 
n'umero de Mach $M_0$ en el medio no perturbado:
\begin{equation} 
M_0 = \frac{v_0}{C_{\rm s_0}} = \frac{v_0}{\sqrt{\gamma_{\rm ad}\, P_0/\rho_0}}.
\end{equation} 
Se obtienen las siquientes relaciones (Landau \& Lifshitz, 1959):
\begin{equation} 
\label{v1v0}
\frac{\rho_1}{\rho_0} = \frac{v_0}{v_1} = \frac{(\gamma_{\rm ad}+1)M_0^2}
{(\gamma_{\rm ad}-1)M_0^2 + 2}
\end{equation} 
\begin{equation}
\label{p1p0} 
\frac{P_1}{P_0} = \frac{2\gamma _{\rm ad}M_0^2 - (\gamma_{\rm ad}-1)}
{\gamma_{\rm ad}+1}.
\end{equation}
Si adem'as consideramos la ecuaci'on de estado (\ref{gasideal}), entonces la
temperatura del medio chocado es $T_1 = P_1/(K_{\rm B} n_1)$ y en funci'on de
par'ametros conocidos el cociente $T_1/T_0$ resulta: 
\begin{equation}
\label{T1T0} 
\frac{T_1}{T_0} = \frac{[2 \gamma_{\rm ad} M_0^2 - (\gamma_{\rm ad}-1)]
[(\gamma_{\rm ad}-1) M_0^2 + 2]}{(\gamma_{\rm ad} + 1)^2 M_0^2}.
\end{equation}

Recordemos que todas estas expresiones (\ref{v1v0})-(\ref{T1T0}) han sido 
deducidas en el SR que se mueve
con el frente de choque. Teniendo en cuenta que $v_1 \le v_0$, 
podemos decir:
\begin{itemize}
\item La onda de choque {\bf comprime el gas}: $\rho_1 > \rho_0$. 
\item  La {\bf velocidad de la onda de choque es supers'onica}:
$U = v_0 > C_{\rm s_0}$.
\item El {\bf material chocado es subs'onico}: $M_1 < 1$.
\item La onda de choque {\bf aumenta la presi'on  del medio}: $P_1 > P_0$.
\item La {\bf entrop'ia aumenta}: $S_1 > S_0$.
\item La onda de choque {\bf calienta el plasma}: $T_1 > T_0$. 
\end{itemize}

Si bien la presi'on crece si la velocidad del choque aumenta, el incremento
de la densidad est'a acotado:
\begin{equation}
\label{}
\lim_{M_0 \rightarrow \infty}\frac{\rho_1}{\rho_0} = 
\frac{\gamma_{\rm ad} + 1}{\gamma_{\rm ad} - 1}. 
\end{equation}
Para el caso particular de un gas monoat'omico, $\gamma_{\rm ad} = 5/3$ 
y se obtiene $\rho_1/\rho_0 = 4$. 
Cuando $M_0 \gg 1$ se dice que el choque es fuerte y en este caso las
relaciones de salto se expresan de una manera muy sencilla:
\begin{equation}
\label{chf_nv}
\rho_1 = 4 \rho_0 \qquad {\rm y} \qquad v_1 = \frac{v_0}{4},
\end{equation}
\begin{equation}
\label{chf_pT}
P_1 = \frac{3}{4}\rho_0 v_0^2 \qquad {\rm y} \qquad 
T_1 \sim 2\times10^{-9} v_0^2\,{\rm K},
\end{equation}
y estas relaciones valen en toda la regi'on chocada. 

Los choques que hemos estudiado en esta secci'on se conocen como
{\bf choques adiab'aticos}, ya que hemos considerado que no hay p'erdidas de
energ'ia ni en la regi'on chocada ni en la no chocada, es decir, 
$\mathcal{L}_1 = \mathcal{L}_0 = 0$.
Sin embargo, hay situaciones en las cuales  las condiciones en la regi'on 
1 (originalemente 
adiab'atica)  son tales que las p'erdidas radiativas comienzan a hacerse 
relevantes. En este caso, la propagaci'on del choque se modifica
disminuyendo $vec v_{\rm ch}$ y el estado del medio 1 cambia,
haci'endose mucho m'as denso y fr'io, como veremos a continuaci'on.

\subsubsection{Choques  radiativos}

Dado un choque que se propaga a una velocidad $\vec v_{\rm ch}$, y cuyas
perturbaciones en el medio est'an descriptas a trav'es de las ecuaciones
(\ref{v1v0}), (\ref{p1p0}) y (\ref{T1T0}), estas 'ultimas pueden no valer en 
toda la regi'on chocada si las p'erdidas radiativas en dicho medio son 
importantes. 
Si bien en la regi'on m'as cercana al choque el estado del medio es aquel 
descripto por las condiciones de un choque adiab'atico, a mayores distancias
la densidad crece y la temperatura disminuye. La regi'on 1 es entonces
muy densa y fr'ia.

Para estimar la importancia de las p'erdidas radiativas en el medio chocado
debemos comparar el tiempo de enfriamiento $t_{\rm ter}$ con otros tiempos
caracter'isticos de cada sistema. Si consideramos las p'erdidas por emisi'on
t'ermica (de l'ineas y continuo) $\mathcal{L_{\rm ter}} \sim n^2 \Lambda(T)$,
donde $\Lambda(T) = A_{\rm t} T^{\alpha_{\rm t}}$ (Myasnikov et al. 1998):
\begin{equation}
\label{Temp-rad}
\Lambda(T) = \left\{
\begin{array}{ll}
7\times10^{-19}\, T^{-0.6} \qquad& {\rm si} \, \, \, 10^4 < T < 4\times10^7 
\,{\rm K}\\
3\times10^{-27}\, T^{0.5} \qquad&{\rm si} \, \, \, T > 4\times10^7 \,{\rm K}, 
\end{array} \right.
\end{equation}
entonces $t_{\rm ter} = (5/2) P_1 / \mathcal{L_{\rm ter}}$.
Por otro lado, si consideramos como tiempo caracter'istico aquel en el cual 
el material chocado recorre una distacia caracter'istica
$X$, entonces $t_{\rm d} \sim X/U_1$ y el choque ser'a radiativo si
$t_{\rm ter} < t_{\rm d}$.

\begin{figure}
\begin{center}
\includegraphics[angle=0, width=0.7\textwidth]{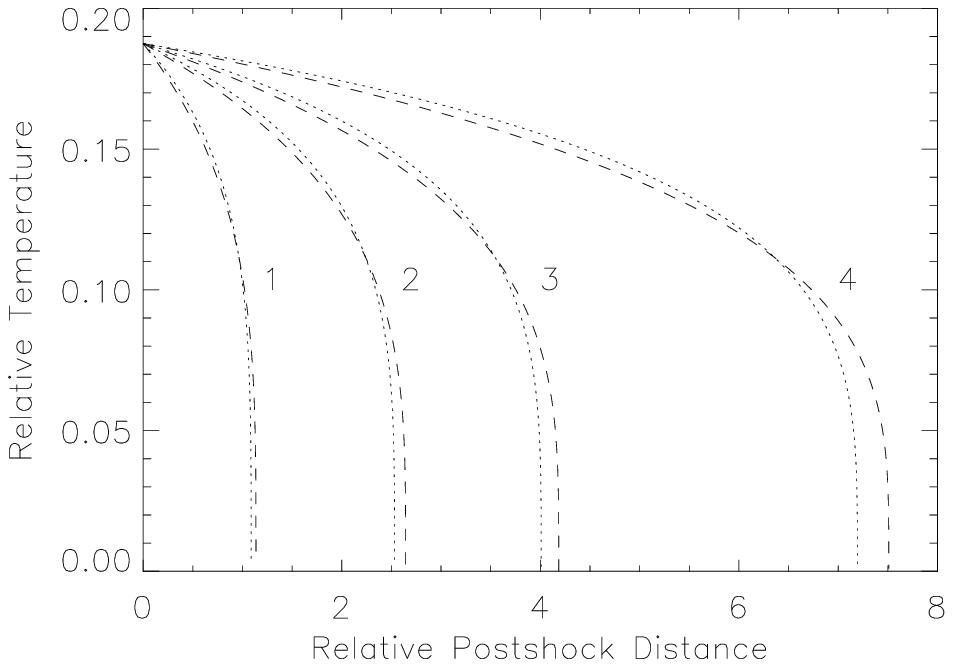}
\caption{Perfil de temperatura en el medio chocado. Se 
muestran las soluciones exactas (l'inea punteada) y las calculadas con 
la aproximaci'on de presi'on constante (l'inea de rayas). Los n'umeros 
1, 2, 3 y 4 indican que las velocidades de los choques consideradas han sido 
$v_{\rm ch} = 500$, 600, 750 y 900~km~s$^{-1}$ (Zhekov \& Palla, 2007).}
\label{Temp-radiativo}
\end{center}
\end{figure}

En este caso, debido a las p'erdidas radiativas, la energ'ia no se conserva 
a lo largo de una l'inea de flu'ido en la regi'on chocada. 
La ecuaci'on~(\ref{energia}) se escribe ahora como (Zhekov \& Palla, 2007)
\begin{equation}
\frac{{\rm d}}{{\rm d}x}\left(\frac{\gamma_{\rm ad}}
{\gamma_{\rm ad} -1} P_1 v_1 +
\frac{\rho_1 v_1^3}{2} \right) = -\mathcal{L_{\rm ter}},
\end{equation}
y si consideramos que $P_1$ se mantiene constante,
entonces la ecuaci'on anterior puede reescribirse como
\begin{equation}
\frac{5}{2} \,n_1\, v_1\, K_{\rm B}\, \frac{{\rm d}T}{{\rm d}x} \sim
-\mathcal{L_{\rm ter}} = - n_1^2 \,\Lambda(T_1).
\end{equation}
Resolviendo esta ecuaci'on obtenemos el siguiente perfil de temperatura 
en la regi'on chocada:
\begin{equation}
\label{T-rad}
T_1(x) = \left(- \frac{2\, P_1^2}{5 \, n_0 \, v_0 \, K_{\rm B}^3} 
A_{\rm t} (3 - \alpha_{\rm t}) \, x + 
\left(T_1^{\rm ad}\right)^{3 - \alpha_{\rm t}}\right)^{1/(3 - \alpha_{\rm t})},
\end{equation} 
donde $P_1$ y $T_1^{\rm ad}$ son la presi'on y la temperatura obtenidas con 
las relaciones de Rankine-Hugoniot. 
%La presi'on $P_1$ se asume constante en toda la regi'on 1.
En la Figura~\ref{Temp-radiativo} se muestra el perfil de temperatura en 
la regi'on chocada de un choque radiativo obtenido bajo la aproximaci'on 
de presi'on constante en dicha regi'on.
Conociendo $T_1(x)$, la densidad puede hallarse a partir
de la ecuaci'on de estado. Para fluidos ideales la densidad en la regi'on
1 var'ia de la siguiente manera: 
\begin{equation}
\label{n-rad}
n_1(x) = \frac{P_1}{K_{\rm B}\,T_1(x)}.
\end{equation}

De las ecuaciones (\ref{T-rad}) y (\ref{n-rad}) vemos que a medida que el 
material se aleja del frente de choque, $T_1$ disminuye dr'asticamente 
(ver la Figura~\ref{Temp-radiativo}) y $n_1$ crece. 
De esta manera, el medio chocado se enfr'ia en un tiempo menor que el
tiempo en el cual el choque recorre una distancia del orden del tamaño del 
sistema. La regi'on adiab'atica del medio 1 es pequeña
y luego el material se enfr'ia hasta que se llega a un estado de
alta densidad; muy compacto.

\subsection{Ondas de choque magnetohidrodin'amicas}

En el caso en que $\vec B \neq 0$ y relevante din'amicamente, entonces tendremos
ondas de choque MHD. Las hay de diferentes tipos dependiendo de la geometr'ia
de $\vec B$. 

\subsubsection{Choques transversales}

Los choques transversales son aquellos en los cuales el campo 
magn'etico en la regi'on no perturbada, $\vec B_0$, es perpendicular a 
la velocidad de propagaci'on  del choque. Es decir, 
$\vec B_0 \perp \vec v_{\rm ch}$. 
En este caso, adem'as de considerar las ecuaciones de movimiento, de 
conservaci'on de la masa y de conservaci'on de la energ'ia, se considera
tambi'en la ecuaci'on de inducci'on. As'i, aparece en las relaciones de 
Rankine-Hugoniot (\ref{RH_2}) y (\ref{RH_3})  el t'ermino $B^2/8 \pi$ 
correspondiente a la presi'on y a la densidad de energ'ia magn'etica.
Se tiene la relaci'on adicional
\begin{equation}   
\frac{B_1}{B_0} = \frac{\rho_1}{\rho_0}.
\end{equation}  

Definiendo el par'ametro  $\zeta \equiv \rho_1/\rho_0$, se tiene
que $\zeta$ satisface una ecuaci'on c'ubica. 
La soluci'on $\zeta = 1$ corresponde a una onda de Alfv'en de corte 
si $v_0^2 = v_{\rm A_1}^2$, que no son ondas de choque. (Las ondas de Alfv'en 
de corte son ondas MHD que no se deforman aunque la amplitud de la pertubaci'on 
sea grande.) La otra soluci'on 
es $\zeta < 0$, pero se descarta por no tener sentido f'isico. 
Finalmente, la soluci'on $\zeta > 0$ corresponde a una onda de choque.
En esta situaci'on se tiene un choque propag'andose perpendicularmente 
a $\vec B$ y
el efecto de este 'ultimo es reducir el valor de $\zeta$ respecto del caso 
hidrodin'amico. Es decir, en presencia de un campo magn'etico, el 
aumento de la densidad de la regi'on chocada es menor que en el caso
hidrodin'amico.   

\begin{figure}
\begin{center}
\includegraphics[angle=0, width=0.7\textwidth]{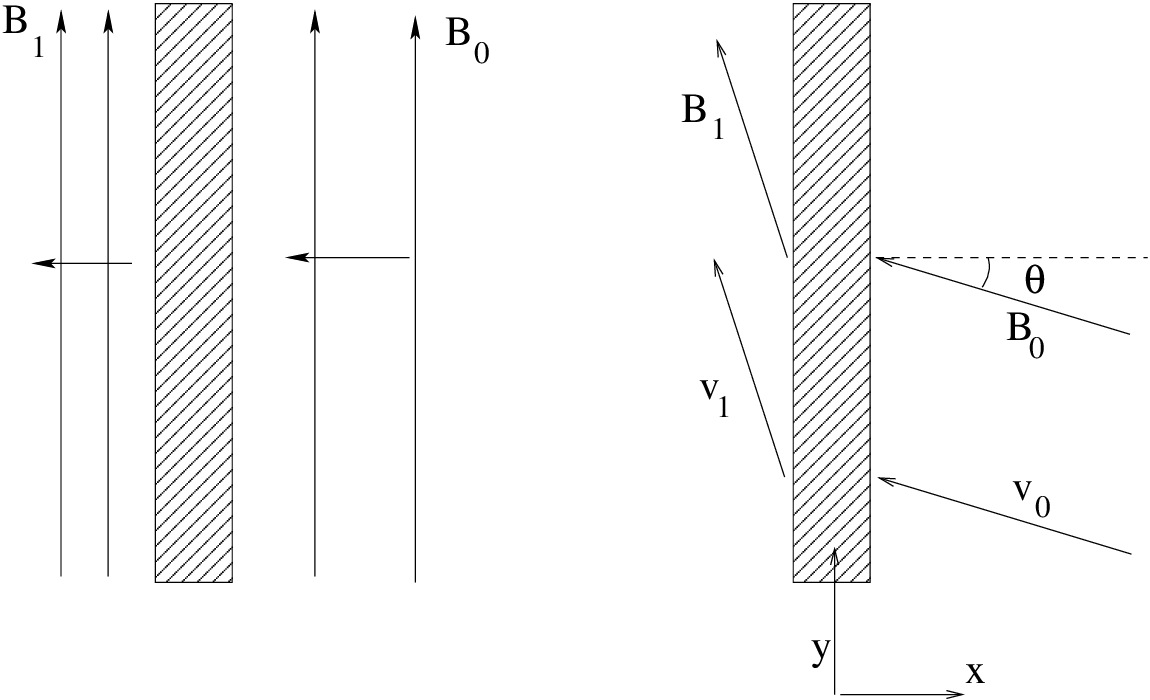}
\caption{Ondas de choque transversales (izquierda) y oblicuas (derecha).
El ancho del choque es $\Delta x \rightarrow 0$ y $\vec B_0$ y $\vec B_1$ 
son el campo magn'etico en la regi'on no perturbada y perturbada, 
respectivamente.}
\label{SR_B}
\end{center}
\end{figure}

\subsubsection{Choques oblicuos}

Este caso corresponde a aquel en el cual $\vec B_0$ es oblicuo a 
$\vec v_{\rm ch}$.
Aqu'i, para obtener las relaciones de salto, adem'as de tener en cuenta 
las ecuaciones consideradas en el caso de los choques transversales,
usamos tambi'en la ecuaci'on 
$\vec \nabla \cdot \vec B = 0$.

Resolviendo como antes estas ecuaciones  se llega
nuevamente a una relaci'on c'ubica para $\zeta$. 
Una de las soluciones posibles
es la trivial, $\zeta = 1$, en la cual no hay onda de choque.
La soluci'on 
$\zeta = 1$ corresponde a una onda de Alfv'en si $v_0^2 = v_{\rm A_1}^2$ 
que, como mencionamos antes, no son ondas de choque. 
Se las llama ondas intemedias porque se propagan a la velocidad de Alfv'en
en la direcci'on $x$, es decir, $v_{\rm ch} = v_{{\rm A}_{0x}}$. 
Las otras dos soluciones, dadas por $\zeta > 1$, corresponden a las ondas 
de choque  asociadas a las ondas magnetoac'usticas r'apida 
($v_{0 x} > v_{{\rm A}_{0  x}}$) y lenta ($v_{0 x} < v_{{\rm A}_{0  x}}$).

\section{Ondas de choque relativistas}

Un fluido que se mueve con una velocidad $U \gtrsim 0.1 c$ 
%(en el SR de un observador en el infinito) 
es relativista. La informaci'on sobre su estado 
esta contenida en el tensor de energ'ia-impulso $T_{\mu\nu}$. Para un fluido
ideal: 
\begin{equation}
T_{\mu\nu} = \frac{1}{c^2}\,\omega \, u_{\mu}u_{\nu} + P \,g_{\mu\nu},
\end{equation}
donde $u_{\mu} = \Gamma (c, \vec{u})$ es la tetra-velocidad, $\beta = u/c$ y 
$\Gamma = 1/\sqrt{1 - \beta^2}$ es el factor de Lorentz y 
$\omega = e_{\rm t} + P$ es la entalp'ia. 
La densidad de energ'ia total $e_{\rm t}$ de un fluido relativista se escribe 
$e_{\rm t} = n mc^2 + e$, donde $n mc^2$ es la densidad de energ'ia en reposo
y $e$ es la cin'etica. Finalmente, $g_{\mu\nu}$ es la m'etrica del 
espacio-tiempo. En esta tesis consideraremos la m'etrica de
Minkowski, con $g_{tt} = -1$ y
$g_{xx} = g_{yy} = g_{zz}= 1$. En el SR propio, es decir, que se
mueve con el fluido, $T_{\mu\nu}$ es una matr'iz diagonal:
\begin{equation}
T_{\mu\nu} = \left(
\begin{array}{cccc}
e_{\rm t} & 0 & 0& 0\\
0 & P & 0& 0\\
0 & 0 & P& 0\\
0 & 0 & 0& P
\end{array}
\right).
\end{equation}

El 'indice adiab'atico para gases monoat'omicos var'ia de la siguiente 
manera: $4/3 < \gamma_{\rm ad} < 5/3$.
En el caso no relativista (NR) $\gamma_{\rm ad} = 5/3$ mientras que en el caso 
ultra-relativista (UR) $\gamma_{\rm ad} = 4/3$. En esta 'ultima situaci'on, 
la ecuaci'on de estado es $P = e/3$ y la velocidad del sonido
resulta $C_{\rm s} = c/\sqrt{3}$.
Las ecuaciones de movimiento y de conservaci'on de la energ'ia est'an 
contenidas en la derivada $\partial T_{\mu}^{\nu}/\partial x^{\mu} = 0$
mientras que la ecuaci'on de continuidad de la masa, o de conservaci'on de la
densidad de part'iculas $n$, se obtiene a partir de  
$\partial (n u^{\nu})/\partial x^{\nu} = 0$.

La teor'ia de las ondas de choque en fluidos relativistas se construye de la 
misma manera que para las NR presentadas en la Secci'on~\ref{NR}
(Taub 1948).
Si consideramos un choque que se propaga en la 
direcci'on $x$, 
las condiciones de salto se obtienen de considerar la conservaci'on del 
n'umero de part'iculas ($n \Gamma \beta$), y las densidades
de flujo de momento ($\omega \Gamma^2 \beta^2 + P$) y de energ'ia 
($\omega \Gamma^2 \beta$). 
En el SR del choque, estas ecuaciones se escriben de la
siguiente manera:
\begin{equation} 
\label{RHrel_1}
n_0 \,\Gamma_0 \,\beta_0 = n_1 \,\Gamma_1 \,\beta_1,
\end{equation}        
\begin{equation} 
\label{RHrel_2}
\omega_0 \,\Gamma_0^2 \,\beta_0^2 + P_0 = \omega_1 \,\Gamma_1^2 \,\beta_1^2+P_1,
\end{equation}
\begin{equation}
\label{RHrel_3}
\omega_0 \,\Gamma_0^2 \,\beta_0 = \omega_1 \,\Gamma_1^2\, \beta_1.
\end{equation}

Para resolverlas consideraremos el caso UR, en el cual la ecuaci'on de estado 
es $P = e/3$ y la entalp'ia se reduce a $\omega = 4 P$.
Adem'as consideraremos que $P_0 \ll \omega_0 \,\Gamma_0^2 \,\beta_0^2$,
es decir, un choque fuerte.
En el SR del choque, el
medio no chocado llega  al frente de choque con una velocidad $v_0 \sim c$, con
lo cual $\beta_0 \sim 1$ y $\Gamma_0 \rightarrow \infty$. El medio chocado
se aleja del choque a una velocidad menor que la del medio no chocado, pero 
que a'un es moderadamente relativista:  
\begin{equation} 
\beta_1 = \frac{1}{3} \qquad \longrightarrow  \qquad
\Gamma_1 = \frac{3}{2 \sqrt{2}} \sim 1.06.
\end{equation}        
Por otro lado, la densidad chocada resulta $n_1 = 2\sqrt{2} \,n_0 \,\Gamma_0$.
Pero recordemos que $n_0$ est'a medida en el SR del flu'ido. Si transformamos
esta magnitud al SR del observador a trav'es de la relaci'on 
$n' = \Gamma n$, donde las magnitudes primadas son las medidas en este SR, 
obtenemos  
\begin{equation} 
n_1' \sim 3 \,n_0'.
\end{equation}      
Finalmente, la presi'on  en el medio chocado resulta
\begin{equation} 
P_1 = \frac{2}{3}\, m\, n_0'\, \Gamma_0 \,c^2.
\end{equation}  
Al ser la presi'on un invariante relativista, $P_1 = P_1'$.

Si consideramos campos magn'eticos, entonces, al igual que en el caso
no relativista, s'olo debemos hacer los siguientes reemplazos en las
ecuaciones (\ref{RHrel_2}) y (\ref{RHrel_3}): $P \rightarrow P + B^2/(8 \pi)$ 
y $e \rightarrow e + B^2/(8 \pi)$.

\section{Producci'on de ondas de choque}

Hay muchas maneras de producir ondas de choque en sistemas astrof'isicos. 
En esta tesis, hemos estudiado aquellas que se forman por perturbar un fluido 
que se mueve con una  velocidad mayor que la del sonido. 
Esto es, dado un choque que se propaga por un medio, estudiamos de que manera
la interacci'on con el medio externo  puede producir otros choques.

\subsection{Ondas de choque por perturbaciones locales}
\label{pert-locales}

Una de las formas m'as comunes de perturbar un fluido en movimiento es 
interponiendol'e un obst'aculo. Si el flu'ido es supers'onico y el
obst'aculo es r'igido, entonces se produce 
una 'unica onda de choque que se propaga en el sentido opuesto al movimento
del fluido. Si el obst'aculo no es r'igido, 
se puede producir adem'as un choque que se propaga a trav'es del obst'aculo,
como se muestra en la Figura~\ref{jet-obstaculo}. 
Un escenario posible para esta interacci'on es un \emph{jet} al cual 
se le interpone una inhomogeneidad del medio circundante. En esta tesis
hemos estudiado dos situaciones similares: en un HMMQ y en un 
AGN. En el primer caso, estudiamos la
interacci'on de grumos presentes en el viento de la estrella compañera
con el \emph{jet} del objeto compacto y en el segundo hemos estudiado la
interacci'on de nubes que circundan el agujero negro del centro de la
galaxia con
el \emph{jet} del AGN. 

A fin de saber cuales son los procesos f'isicos m'as relevantes en este 
tipo de interacciones, haremos un an'alisis de las escalas de 
tiempo de los procesos que tienen lugar debido a  la 
interacci'on de  un obst'aculo con un \emph{jet}.

\subsubsection{Escalas de tiempo}
 
Supongamos un obst'aculo esf'erico, de radio $R_{\rm o}$, con una densidad 
homog'enea $n_{\rm o}$ y que se mueve con una
velocidad $\vec v_{\rm o}$ en la direcci'on perpendicular a la direcci'on de 
propagaci'on del \emph{jet}, que designaremos $z$. 
Por otro lado supondremos un \emph{jet} que se propaga con una velocidad 
$v_{\rm j}$ y con un factor de Lorentz $\Gamma_{\rm j}$. 
Considerando un 'angulo
 de abertura del \emph{jet} $\phi \sim 6^{\circ}$ (valor t'ipico de \emph{jets}
colimados de MQs y AGNs), la relaci'on entre 
el radio y la altura es $R_{\rm j} \sim 0.1 z_{\rm j}$. 
Considerando  que la luminosidad cin'etica del \emph{jet}, $L_{\rm j}$, 
es constante,
la densidad, $n_{\rm j}$ (en el SR del laboratorio), de 'este var'ia con 
$z_{\rm j}$ a trav'es de la relaci'on
\begin{equation} 
\label{L_jet}
L_{\rm j} = (\Gamma_{\rm j} -1)\, m_p c^2 \,\sigma_{\rm j}\,n_{\rm j}\,v_{\rm j},
\end{equation} 
donde $\sigma_{\rm j} = \pi r_{\rm j}^2$ es la secci'on del \emph{jet} a 
la altura $z_{\rm j} = 10\, r_{\rm j}$.
Un par'ametro importante en nuestro estudio ser'a el cociente entre la
densidad del obst'aculo y la del \emph{jet}: $\chi \equiv n_{\rm o}/n_{\rm j}$.

El tiempo de penetraci'on, $t_{\rm o}$, del obst'aculo en el \emph{jet}  
puede estimarse a trav'es de la ecuaci'on:
\begin{equation} 
t_{\rm o}\sim \frac{2 \,R_{\rm o}}{v_{\rm o}}.
\end{equation} 
Por otro lado, si suponemos que dentro del \emph{jet} el obst'aculo se mueve
siguiendo su trayectoria original, el
tiempo que tardar'ia en cruzar el \emph{jet} (recorriendo su
ancho igual a $2 R_{\rm j}$) est'a dado por
\begin{equation} 
\label{t_j}
t_{\rm j}\sim \frac{2\; R_{\rm j}}{v_{\rm o}} \sim 
\frac{2\;(0.1 z_{\rm j})}{v_{\rm o}}.
\end{equation}

\begin{figure}
\begin{center}
\includegraphics[angle=0, width=0.7\textwidth]{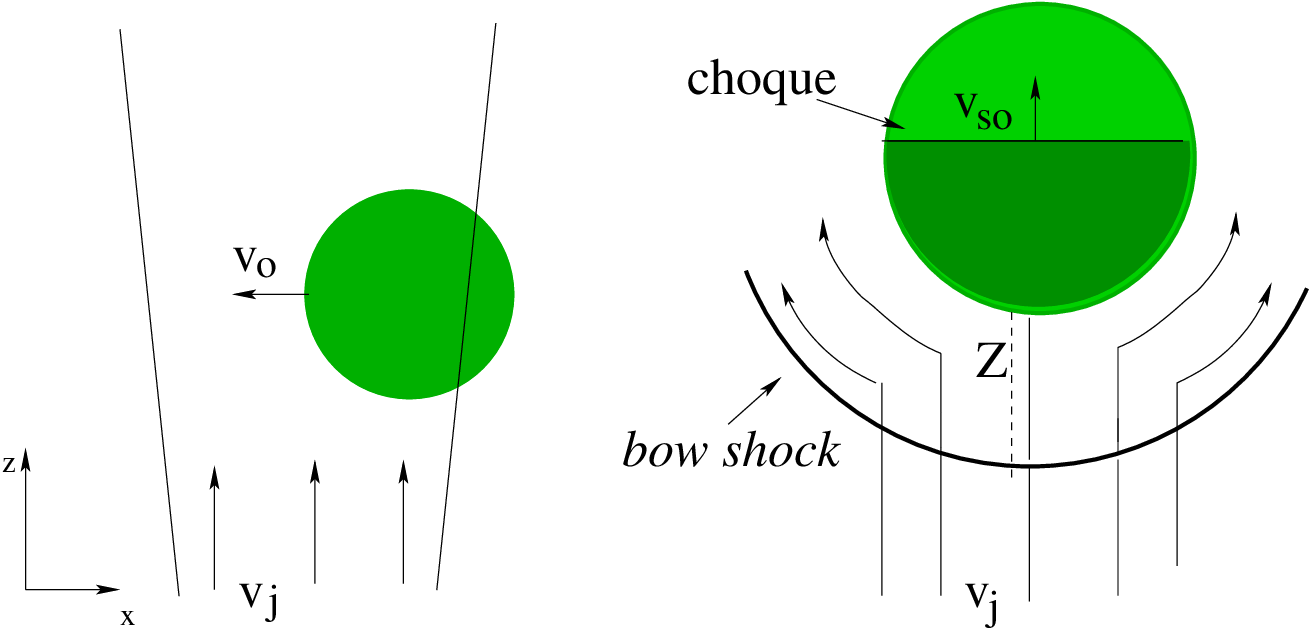}
\caption{Esquema de la interacci'on de un \emph{jet} con un obst'aculo
(verde).
Izquierda: el obst'aculo penetra en el \emph{jet} moviendos'e a una velocidad
$\vec v_{\rm o}$. Derecha: Una vez que el obst'aculo est'a completamente adentro
del \emph{jet}, se forma un \emph{bow shock} sim'etrico en el \emph{jet}.
Mientras, un choque se propaga en el
obst'aculo a una velocidad $\vec v_{\rm co}$. El \emph{bow shock} est'a
a una distancia $Z$ del obst'aculo. (El gr'afico no est'a a escala.)}
\label{jet-obstaculo}
\end{center}
\end{figure}

Una vez que el obst'aculo ha entrado en el \emph{jet}, la 
interacci'on con el material de este 'ultimo  producir'a dos choques:
uno en el \emph{jet} y otro (o una onda s'onica) en el obst'aculo, como se 
muestra en la Figura~\ref{jet-obstaculo}. 
El choque en el primero se mueve a una velocidad $\sim v_{\rm j}$ y lo 
llamaremos \emph{bow shock} (``choque de proa''). Al cabo de
recorrer una distancia $Z$ en el \emph{jet}, este choque alcanza el estado 
estacionario, es decir,
la separaci'on entre el \emph{bow shock} y el obst'aculo es fija e igual a $Z$. 
Para estimar esta distancia consideramos que el n'umero de part'iculas  del 
\emph{jet} que por unidad de tiempo atraviesan el \emph{bow shock} 
($\sim n_{\rm j} v_{\rm j} \sigma_{\rm o}$, donde 
$\sigma_{\rm o} = \pi\,R_{\rm o}^2$ es la secci'on del obst'aculo) y luego 
escapan siendo  advectadas de la regi'on chocada a la velocidad $v_{\rm j1}$ 
del medio chocado del jet ($v_{\rm j1} \sim v_{\rm j}/4$ 'o $v_{\rm j}/3$, 
para los casos NR y UR, respectivamente) se conserva.
De esta manera, de la igualdad 
$n_{\rm j} v_{\rm j} \sigma_{\rm o} = n_{\rm j1}\,v_{\rm j1}\, 2 \pi\, 
R_{\rm o}\, Z$,
obtenemos que $Z \sim 0.2 R_{\rm o}$ y $\sim 0.3 R_{\rm o}$ en los casos 
de choques UR y NR, respectivamente. Luego, el tiempo en el cual el
\emph{bow shock} alcanza el estado estacionario es
\begin{equation} 
\label{t_bs}
t_{\rm bs}\sim\frac{Z}{v_{\rm j1}}.
\end{equation}

Respecto del choque en el obst'aculo, la velocidad de propagaci'on 
puede determinarse asumiendo que se establece el
equilibrio de presiones en la superficie de separaci'on entre la
regi'on chocada del \emph{jet} y del obst'aculo y usando las relaciones de 
salto.
Se obtiene la siguiente expresi'on para la velocidad:
\begin{equation}
\label{v-choque-obstaculo} 
v_{\rm co}\sim \frac{v_{\rm j} (\Gamma_{\rm j} -1)}{\sqrt{\chi}}.
\end{equation} 
Luego el tiempo en el cual el choque recorre todo el obst'aculo es
\begin{equation} 
t_{\rm co}\sim \frac{2\; R_{\rm o}}{v_{\rm co}}\sim 
\frac{2\; R_{\rm o}\sqrt{\chi}}{v_{\rm j}(\Gamma_{\rm j} -1)}.
\end{equation} 
Esta magnitud $t_{\rm co}$ puede considerarse un tiempo caracter'istico
(o de vida) del sistema, ya que como veremos a continuaci'on, las 
inestabilidades
pueden destruir el obst'aculo en un tiempo $\sim t_{\rm co}$ 
o 'este podr'ia
ser acelerado y comenzar a moverse con el \emph{jet} antes de escapar del 
mismo (en un tiempo $t_{\rm j}$).

\begin{figure}
\begin{center}
\includegraphics[angle=0, width=0.7\textwidth]{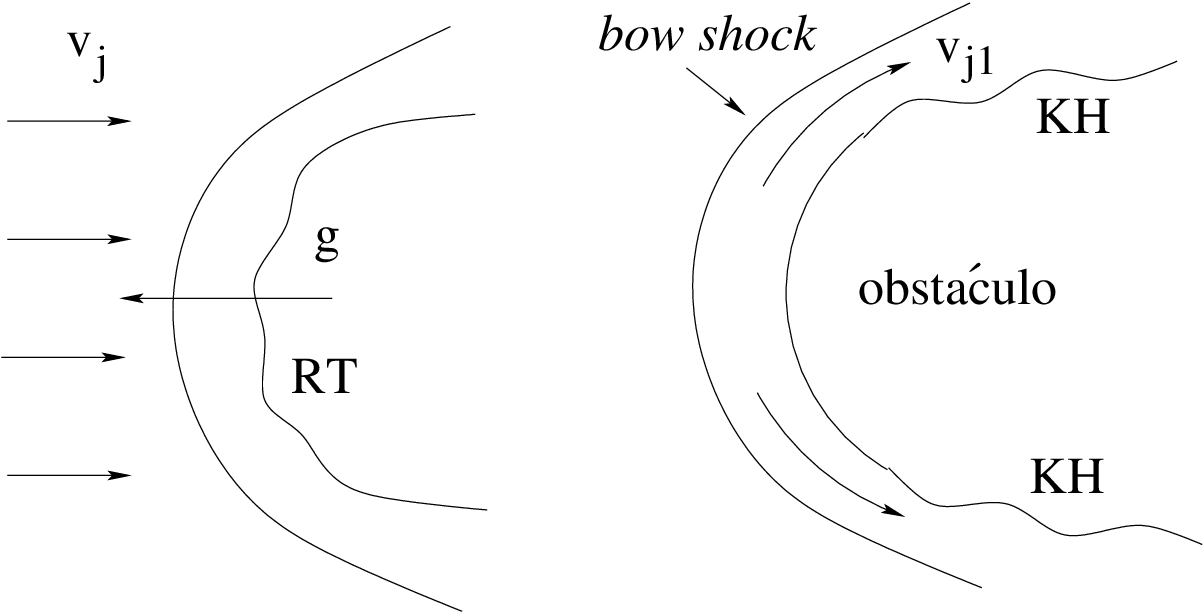}
\caption{Diagrama del obst'aculo y del \emph{bow shock}. Las inestabilidades 
de RT se desarrollan en la superficie frontal del obst'aculo (figura izquierda),
mientras que las de KH lo hacen en las superficies laterales (figura derecha).}
\label{inestabilidades}
\end{center}
\end{figure}

La aceleraci'on $\vec g$ que el material del \emph{jet} ejerce contra el 
obst'aculo puede
determinarse considerando conservaci'on de la entalp'ia y suponiendo que 
toda la presi'on cin'etica del \emph{jet} se 
convierte en presi'on t'ermica del material chocado en la superficie
de contacto con el obst'aculo. As'i obtenemos que  
\begin{equation} 
\label{g_general}
g \sim \frac{(\Gamma_{\rm j} -1)\,v_{\rm j}^2}{\chi R_{\rm o}},
\end{equation}
con lo cual el tiempo de aceleraci'on del obst'aculo hasta la velocidad del
\emph{jet} es 
\begin{equation} 
t_{\rm g} \sim \frac{v_{\rm j}}{g} \sim \sqrt\chi\; t_{\rm co}.
\end{equation}
Sin embargo, en un tiempo $\sim t_{\rm co}$ el obst'aculo se acelera hasta la 
velocidad $v_{\rm co}$ del choque que se propaga en su interior. 
La aceleraci'on $\vec g$ adem'as de acelerar el obst'aculo, 
hace que se desarrollen inestabilidades de Rayleigh-Taylor (RT) en la 
parte de 'este que se encuentra m'as cerca de la base del \emph{jet}, 
como se muestra en la Figura~\ref{inestabilidades}.
El tiempo en el cual estas inestabilidades crecen hasta una longitud de escala 
$l \sim R_{\rm o}$ est'a dado por
\begin{equation}
\label{t_RT} 
t_{\rm RT}\sim \sqrt\frac{l}{g} \sim \sqrt\frac{R_{\rm o}}{g} \sim t_{\rm co}.
\end{equation} 
Por otro lado, la diferencia relativa $v_{\rm rel}$ entre la velocidad 
del material chocado del
jet que rodea al obst'aculo, $v_{\rm j1}$, y la velocidad del material
chocado del obst'aculo,  $v_{\rm o1}$, desarrolla inestabilidades de
Kelvin-Helmholtz (KH), con una escala de tiempo dada por:
\begin{equation} 
t_{\rm KH}\sim\frac{l\;\sqrt{\chi}}{v_{\rm rel}} \sim 
\frac{R_{\rm o}\;\sqrt{\chi}}{v_{\rm rel}},
\end{equation}  
donde nuevamente hemos considerado $l \sim R_{\rm o}$. Notar que 
como $v_{\rm rel} \sim v_{\rm j1}$, $t_{\rm KH}\sim  t_{\rm co}$. 
En las estimaciones de $t_{\rm RT}$ y de  $t_{\rm KH}$
no hemos tenido en cuenta al campo magn'etico, que 
podr'ia estabilizar el sistema (Blake 1972, Romero 1995).

De las diferentes escalas de tiempo estimadas anteriormente, podemos 
resumir 
que una vez que el obst'aculo penetra en el \emph{jet}, el 
\emph{bow shock} alcanza la
distacia $Z$ r'apidamente, en un tiempo $t_{\rm bs} \ll t_{\rm co}$ y luego 
el obst'aculo podr'ia ser destruido por las inestabilidades, acelerarse
y comenzar a moverse conjuntamente con el \emph{jet} o escapar de 'este.
Sin embargo, para hacer un an'alisis m'as exaustivo de cuales ser'an los 
procesos m'as relevantes necesitamos especificar la localizaci'on en el 
\emph{jet}
a la cual se produce la interacci'on. La determinaci'on de este par'ametro,
conjuntamente con otras magnitudes 
del sistema (\emph{jet} y obst'aculo), ser'an expuestas en los Cap'itulos 
\ref{MQs} y \ref{AGNs}, donde los escenarios considerados ser'an HMMQs y AGNs, 
respectivamente.

\subsection{Ondas de choque por interacciones globales}
\label{pert-globales}

Cuando un fluido supers'onico choca con  un  medio externo,
la onda de choque se disipa a medida
que se propaga si la inyecci'on de energ'ia es puntual en el tiempo,
como ocurre en las explosiones de supernovas.
Por otro lado, si la inyecci'on de energ'ia es continua
el choque se frena, pero m'as lentamente que en el caso anterior. 
Este choque llega a un estado estacionario en el caso radiativo. 
En las dos situaciones descriptas anteriormente se produce un choque 
reverso que se propaga en el sentido
opuesto al movimiento del fluido. En el caso del evento puntual, 
el choque reverso al cabo de 
un cierto tiempo se disipar'a. Si la inyecci'on inicial
de energ'ia es continua, se formar'a un choque estacionario.

Situaciones en las cuales pueden producirse choques de las maneras
descriptas anteriormente pueden darse  cuando 
los \emph{jets} son frenados por el medio en el cual se propagan y se producen
los llamados choques terminales, como ocurre por ejemplo en los \emph{jets} 
de YSOs masivos y en los \emph{hot-spots} de las radio-galaxias.
Otra situaci'on en la cual se desarrollan choques es
cuando dos c'umulos de galaxias colisionan. Estos choques a gran escala
se propagan hasta una cierta distancia del centro de gravedad de la fusi'on
de ambos c'umulos y luego se disipan.

A diferencia de lo expuesto en la secci'on~\ref{pert-locales}, aqu'i
estudiaremos dos sistemas con propiedades muy diferentes: el \emph{jet} 
de un YSO y un c'umulo de 
galaxias. Es por esto que la descripci'on de las escalas 
de tiempo que haremos a continuaci'on ser'a muy general y luego, en los 
cap'itulos correspondientes a cada fuente particular: YSOs 
(Cap'itulo~\ref{yso})
y c'umulos de galaxias  (Cap'itulo~\ref{clusters}), haremos un estudio 
m'as detallado de
las escalas de tiempo con f'ormulas pertinentes a cada fuente.

\subsubsection{Escalas de tiempo}

Supongamos un choque que se propaga a una velocidad $\vec v_{\rm ch}$
en un medio externo.
El choque, a medida que se propaga empuja y apila material delante de 'el, 
y as'i el frente de choque es ahora la ``cabeza'' de
ese material que se mueve a la velocidad de la onda.
Definimos el tiempo de propagaci'on $t_{\rm p}$ como aqu'el en el cual el choque
se frena sustancialmente, es decir, podemos suponer que durante 
$t_{\rm p}$ 'este se propaga ``libremente'' por el medio una distancia
$Z_{\rm p}$ en un tiempo: 
\begin{equation}
t_{\rm p} \sim \frac{Z_{\rm p}}{v_{\rm ch}}.
\end{equation}
%
%La distancia $Z_{\rm p}$ es un par'ametro que puede determinarse 
%observacionalmente.  

\begin{figure}
\begin{center}
\includegraphics[angle=0, width=0.5\textwidth]{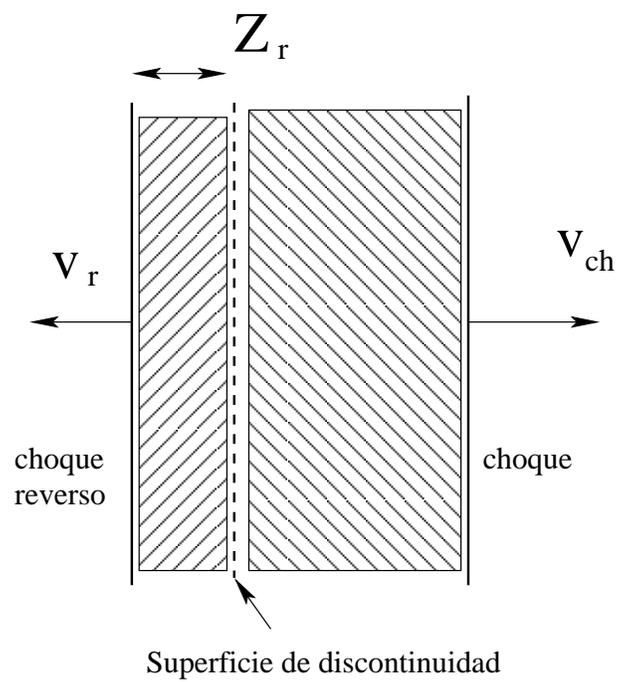}
\caption{Esquema de un choque que se propaga en un medio externo con una
velocidad $\vec v_{\rm ch}$. Se forma un choque reverso que se propaga con una
velocidad $\vec v_{\rm r}$ en sentido opuesto a $\vec v_{\rm ch}$, y entre
ambos medios chocados se forma una superficie de discontinuidad en la cual 
pueden desarrollarse inestabilidades de RT.}
\label{reverse_shock}
\end{center}
\end{figure}

Sin embargo, mientras m'as material se haya acumulado, m'as dificil 
ser'a moverlo y as'i la velocidad del choque va disminuyendo.  
Si la velocidad decrece, entonces la luminosidad tambi'en (de acuerdo a la
ecuaci'on~(\ref{L_jet})) y es as'i como el choque se va debilitando. 
Si $\vec v_{\rm ch}$ disminuye, entonces 
hay fuertes variaciones de presi'on en la regi'on chocada y se produce 
una onda. Si 'esta se propaga a una velocidad mayor que  $C_{\rm s}$ 
('o $C_{\rm sA}$) 
entonces ser'a una onda de choque y la llamaremos {\bf choque reverso}. 
Este choque reverso se disipa o llega al estado estacionario 
si la inyecci'on de energ'ia es puntual o
continua, respectivamente. En ambos casos, el choque reverso recorre
una distancia $Z_{\rm r}$ a la velocidad $\vec v_{\rm r}$ en un tiempo
\begin{equation}
t_{\rm r} \sim \frac{Z_{\rm r}}{v_{\rm r}}.
\end{equation}
La velocidad  $\vec v_{\rm r}$ puede estimarse a trav'es del contraste de 
densidades entre ambos medios chocados. 
Entre las regiones de ambos medios perturbados (el externo y el chocado 
por el choque reverso)
se genera una superficie de discontinuidad, 
como se muestra en la Figura~\ref{reverse_shock}. 
En esta superficie, las inestabilidades de
RT pueden ser importantes y mezclarse material de 
ambos medios chocados. La escala temporal de esta inestabilidad est'a
determinada por la ecuaci'on~(\ref{t_RT}): $t_{\rm RT} \sim (l/g)^{1/2}$,
pero considerando $l \sim Z_{\rm r}$ y que la aceleraci'on es ejercida
o bien por el material chocado del \emph{jet} o bien por el centro del c'umulo
de galaxias (en los casos particulares que estudiamos en esta tesis). 

Finalmente, el choque reverso puede ser radiativo si $l_{\rm ter} < Z_{\rm r}$,
siendo $l_{\rm ter} \sim t_{\rm ter}\,v_{\rm r}$ y el tiempo de 
enfriamiento por emisi'on t'ermica
\begin{equation}
\label{t_ter}
t_{\rm ter} \sim \frac{5}{2} \frac{P_1}{\mathcal{L}_{\rm ter}}.
\end{equation}
La funci'on $\mathcal{L}_{\rm ter}$ depende de la temperatura del medio 
chocado a trav'es de la funci'on $\Lambda(T)$, como se indica en la 
ecuaci'on~(\ref{Temp-rad}).

\section{Aceleraci'on de part'iculas en choques}
\label{particle_accel}

Las ondas de choque, adem'as de modificar las propiedades
termodin'amicas  del medio por el cual se propagan, pueden
tambi'en acelerar part'iculas hasta energ'ias relativistas.  
Estas part'iculas aceleradas son llamadas {\bf part'iculas no t'ermicas},
ya que su distribuci'on en energ'ia no es Maxweliana.
El n'umero de estas part'iculas no t'ermicas es una fracci'on pequeña del
n'umero de part'iculas termalizadas que 
forman el medio en el cual se propaga el choque.
Es decir, el choque acelera s'olo una fracci'on peque~na de las part'iculas 
t'ermicas del medio sac'andolas del estado de equilibrio termodin'amico y
convirti'endolas en no t'ermicas.
La presencia de estas part'iculas en fuentes astrof'isicas es usualmente
detectada a trav'es de la emisi'on no t'ermica en radio 
y de la radiaci'on gamma (la cual no puede ser t'ermica).

El mecanismo propuesto originalmente por Fermi (1949) describe como las
part'iculas podr'ian ser aceleradas por rebotes sucesivos entre centros
dispersores con una distribuci'on  aleatoria de velocidades.
Este proceso resulta poco eficiente ya que la ganancia de energ'ia $E$
cada vez que la part'icula rebota es
$\langle\Delta E\rangle/E \propto \beta^2$, y como $\beta  \ll 1$ 
la ganancia es muy poca ($\beta c$ es la velocidad de la part'icula).  
Por otro lado, adem'as de ser poco eficiente, la situaci'on en la cual tiene
lugar el mecanismo es poco  
frecuente en la naturaleza. Sin embargo, es f'acilmente adaptable 
a un escenario astrof'isico si reemplazamos los centros dispersores
por un choque e inhomogeneidades magn'eticas en el medio.
La teor'ia original de Fermi fue modificada en los años 70 por
diversos autores, entre ellos Axford, Lear \& Skadron (1977) y Bell (1978)
para describir un mecanismo m'as eficiente 
($\langle\Delta E\rangle/E \propto \beta$) y que tuviese 
lugar en sistemas astrof'isicos con choques. 
Para que este proceso tenga lugar, es necesario que las part'iculas
en el medio perturbado puedan difundir y alcanzar el choque. 
A continuaci'on describiremos someramente el mecanismo de aceleraci'on de Fermi 
modificado y que se conoce como mecanismo de Fermi de tipo I. 
Este proceso tambi'en se conoce como
mecanismo de aceleraci'on difusiva ya que la difusi'on juega un rol importante
en el desarrollo del mismo. 

En un plasma en el cual el campo magn'etico es uniforme, las part'iculas 
cargadas se mueven siguiendo trayectorias helicoidales alrededor 
de las l'ineas de $\vec B$. El radio de giro ($r_{\rm g}$) est'a determinado 
por la energ'ia $E$ de las part'iculas y por la intensidad del campo magn'etico:
\begin{equation}
r_{\rm g} = \frac{E}{Z_e \,q_e \,B},
\end{equation}  
donde $Z_e$ es el n'umero at'omico\footnote{En todas las aplicaciones 
realizadas durante esta tesis $Z_e = 1$.} y $q_e$ es la carga el'ectrica del
electr'on.
Sin embargo, en los sistemas astrof'isicos que nos interesan el campo $\vec B$ 
presenta irregularidades que perturban el movimiento helicoidal de las
part'iculas produciendo una reorientaci'on de las mismas, y por lo tanto 
un movimiento desordenado.  
Las part'iculas con velocidad $\vec v$ difunden en el medio con un camino 
libre medio $\lambda = 3 D /v$, donde $D$ es el coeficiente de 
difusi'on\footnote{En el Cap'itulo~\ref{proc-rad} se dar'a la definici'on del 
camino libre medio.}. 
Si el movimiento de la part'icula es en una direcci'on que forma un 'angulo
$\theta$ con $\vec B$, entonces $D = D_{\parallel} \cos^2(\theta) + 
D_{\perp} \sen^2(\theta)$, siendo $D_{\parallel}$ y $D_{\perp}$
los coeficientes de difusi'on en la direcci'on paralela y perpendicular a 
$\vec B$, respectivamente. Debido a que $D$ es
desconocido en la mayor'ia de los sistemas astrof'isicos, se suele considerar
que $D_{\parallel}$ es un n'umero $\eta$ veces el coeficiente de 
difusi'on m'inimo o de Bohm, $D_{\rm B} = r_{\rm g} c/3$, y que significa 
considerar que el camino libre medio en la direcci'on paralela a $\vec B$ es
$\lambda_{\parallel} \sim r_{\rm g}$. Luego resulta
$D_{\perp} \approx D_{\parallel}/(1 + \eta^2)$ (Jokipii 1987).

Con cada ciclo en el cual la part'icula va del medio no chocado al chocado y
vuelve al medio original, la ganancia de energ'ia es 
$\langle\Delta E\rangle/E \sim (4/3)\beta$. Al cabo de $k_{\rm c}$ ciclos, 
la energ'ia de la part'icula, que inicialmente era $E_i$, resulta 
\begin{equation}
E = E_i \left(1 + \frac{\Delta E}{E}\right)^{k_{\rm c}}.
\end{equation}  
%
%donde $E_i$ es la energ'ia inicial de la part'icula.
El tiempo en el cual se realiza cada ciclo est'a determinado por como las
part'iculas difunden en cada medio, es decir, de los coeficientes de
difusi'on $D_0$ y $D_1$, a trav'es de la  expresi'on 
$t_{\rm ciclo} \sim (4/c)(D_0/v_0 + D_1/v_1)$ (Protheroe 1999).
El ``tiempo de aceleraci'on'' es el tiempo requerido para que
las part'iculas alcancen una energ'ia $E$ a trav'es de un mecanismo de 
aceleraci'on dado; para el mecanismo de Fermi de tipo I resulta:
\begin{equation}
t_{\rm ac} \equiv \left(\left. \frac{1}{E}\frac{{\rm d}E}{{\rm d}t}
\right|_{\rm ac}\right)^{-1}
= \frac{E}{\langle\Delta E \rangle} t_{\rm ciclo}
\sim \frac{3 \zeta}{(\zeta -1) v_0}\left(\frac{D_0}{v_0} + \frac{D_1}{v_1}
\right).
\end{equation}
As'i, el tiempo de aceleraci'on depende  de la geometr'ia
de $\vec B$. A continuaci'on damos expresiones sencillas para $t_{\rm ac}$
correspondientes a configuraciones particulares de  $\vec B$.
\begin{itemize}
\item Choques paralelos ($\vec B \parallel \vec v_{\rm ch}$):
suponiendo que  $D_0 = D_1 = D_{\parallel}$ y que $\vec B_0 = \vec B_1$
 se tiene que el tiempo de aceleraci'on es
\begin{equation}
\label{t_acc_par}
t_{\rm ac}^{\parallel} \sim \frac{20}{3}\frac{\eta E}{q_e\, B_0 \,v_{\rm ch}^2}.
\end{equation}

\item Choques transversales ($\vec B \perp \vec v_{\rm ch}$):
suponiendo que  $D_0 = D_1 = D_{\perp}$ y que $\vec B_1 \sim 4\vec B_0$ 
se tiene que el tiempo de aceleraci'on es
\begin{equation}
\label{t_acc_perp}
t_{\rm ac}^{\perp} \sim \frac{8}{3}\frac{E}{\eta\, q_e\, B_0\, v_{\rm ch}^2}.
\end{equation}
\end{itemize}
Por otro lado, debido a que en los choques relativistas el c'alculo de $D$ es 
a'un m'as complicado que en los NR, se puede considerar que en los primeros
$t_{\rm ac} \sim E/(0.1 \,q_e \,B_0 \,c)$ (como en el caso de la supernova del
Cangrejo -\emph{Crab}-), sin especificar la geometr'ia de $\vec B$.

El n'umero de ciclos que realiza la part'icula depende b'asicamente 
del tamaño $L_{\rm ac}$ del acelerador. Las part'iculas pueden cruzar
el choque sucesivas veces antes de que $r_{\rm g}$ ($\propto E$) crezca lo 
suficiente, es decir, $r_{\rm g} \sim L_{\rm ac}$, como para escapar del 
acelerador (Hillas 1984). 
Al cabo de $k_{\rm c}$ ciclos el espectro de las part'iculas aceleradas 
e inyectadas en el medio chocado ser'a $Q(E) = K_0 E^{-p}$, donde  
\begin{equation}
p = \frac{\ln (E/E_i)}{\ln(1 + \Delta E/E)}
\end{equation}
y  $[Q] =$~erg$^{-1}$~s$^{-1}$. En la ecuaci'on anterior
la dependencia con $k_{\rm c}$ est'a impl'icita en $E/E_i$.
Esta inyecci'on $Q(E)$ de part'iculas por unidad de tiempo en una determinada 
regi'on de la fuente dar'a lugar, al cabo de un tiempo $t$, a una 
distribuci'on $N(E,t)$ de part'iculas no t'ermicas.

\subsection{Poblaci'on de part'iculas no t'ermicas}

Dada una distribuci'on  $Q(E)$ de part'iculas aceleradas e 
inyectadas en una regi'on, nos interesa conocer cual ser'a la
distribuci'on de part'iculas $N(E,t)$ en un tiempo $t$.

A lo largo de esta tesis consideraremos que tanto la regi'on donde se
aceleran las part'iculas, el acelerador, como aquella en la cual rad'ian, 
el emisor, son homog'eneas\footnote{Notamos que el campo magn'etico debe 
tener una componente inhomog'enea para que se el mecanismo de Fermi pueda 
desarrollarse.}. Es decir, las magnitudes f'isicas no depender'an de las
coordenadas espaciales. Bajo esta hip'otesis, la ecuaci'on que describe 
la evoluci'on temporal de $N(E,t)$ es
la siguiente (Ginzburg \& Syrovatskii, 1964):
\begin{equation}
\label{kinetic}
\frac{\partial N(E,t)}{\partial t} = \frac{\partial}{\partial E}
\left(\left.\frac{{\rm d}E}{{\rm d}t}\right|_{\rm rad} N(E,t)\right) - 
\frac{N(E,t)}{t_{\rm esc}} + Q(E)\,,
\end{equation}
donde $[N(E,t)] =$~erg$^{-1}$. Esta ecuaci'on tiene en cuenta 
la inyecci'on de part'iculas $Q(E)$ que en un tiempo $t$ nos da la distribuci'on
$N(E,t)$. Las part'iculas desde que se inyectaron en el tiempo $t_0 = 0$ hasta
el tiempo $t$ sufren p'erdidas radiativas que est'an contempladas en el 
t'ermino ${\rm d}E/{\rm d}t|_{\rm rad} \equiv \dot E_{\rm rad}$. 
Adem'as, puede haber p'erdidas 
de energ'ia en el sistema
debido a que las part'iculas pueden escapar del acelerador. El escape de 
las part'iculas est'a considerado en el t'ermino $N(E,t)/t_{\rm esc}$, donde 
\begin{equation}
\label{t_esc}
t_{\rm esc}^{-1} =  \left( \frac{1}{t_{\rm conv}} + 
\frac{1}{t_{\rm dif}} \right).
\end{equation} 
Esta escala de tiempo tiene en cuenta el escape por convecci'on 
($t_{\rm conv}$) y por difusi'on ($t_{\rm dif}$). Mientras 
que el primero contempla el arrastre de las part'iculas por el movimiento
del medio chocado (que se mueve a una velocidad $\sim v_{\rm ch}/4$),
el segundo considera las p'erdidas de energ'ia por difusi'on de las 
part'iculas. 

La soluci'on m'as general de la ecuaci'on~(\ref{kinetic}) considerando que
$t_{\rm esc} = t_{\rm conv}$, es decir, $t_{\rm conv} \ll t_{\rm diff}$, es 
(Khangulyan et al. 2007)
\begin{equation}
\label{N_solution}
N(E,t) = \frac{1}{|\dot E|} \int_E^{E_{\rm eff}}\, Q(t,E') \,
\exp\left(-\frac{\tau(E,E')}{t_{\rm esc}}\right)\, {\rm d}E',
\end{equation} 
donde $E_{\rm eff}$ es la energ'ia m'axima de las part'iculas que en el tiempo
$t$ pueden enfriarse hasta una energ'ia $E$, es decir,
\begin{equation}
t(E) = \int_E^{E_{\rm eff}}  \frac{{\rm d}E'}{|\dot E'|} \qquad {\rm y} \qquad
\tau(E,E') = \int_E^{E'}  \frac{{\rm d}E''}{|\dot E''|}.
\end{equation} 
Dada una energ'ia $E'$, $\tau$ es el tiempo en el cual una part'icula
de energ'ia $E'$ se enfr'ia hasta obtener una energ'ia $E$.
Notemos que  $\tau \leq t$.  
La ecuaci'on~(\ref{N_solution}) toma una expresi'on muy sencilla en dos 
casos l'imites que veremos a continuaci'on, si $p > 1$: 
\begin{itemize}
\item Si $t(E) \ll t_{\rm esc}$ entonces $\tau (E, E') \ll t_{\rm esc}$ y
 entonces el t'ermino $\exp(-\tau/t_{\rm esc}) \sim 1$, con lo cual
$N(E,t) \sim (-Q(E) + Q(E_{\rm eff})) (E_{\rm eff} -E)/|\dot E|\sim Q(E)\,t(E)$ y
las part'iculas se acumulan a medida que $t$ crece. Este proceso de 
acumulaci'on de part'iculas de energ'ia $E$ contin'ua hasta que $E_{\rm eff}$
coincide con la energ'ia m'axima de la distribuci'on $Q(E)$ o el tiempo
de escape se hace menor que el tiempo de enfriamiento. 
\item Si $t_{\rm esc} \ll t(E) $: 
$N(E) \sim Q(E)\, t_{\rm esc}$ y el sistema llega a un estado estacionario.
\end{itemize}

Para conocer la distribuci'on $N(E)$ necesitamos conocer las p'erdidas 
radiativas
que sufren las part'iculas. Para esto, en el pr'oximo cap'itulo
repasaremos los procesos radiativos que hemos estudiado durante esta tesis.

%% file: Proc_rad_final.tex
\chapter{Procesos radiativos}
\label{proc-rad}

Todo lo que conocemos de la mayor'ia de las fuentes 
astrof'isicas es gracias a que ellas rad'ian y a que es posible 
detectar parte de los fotones emitidos con instrumentos adecuados. 
Por esta raz'on,  conocer los procesos a trav'es de los
cuales se genera la emisi'on es indispensable
para modelizar correctamente las fuentes que nos interesa estudiar.

Como  mencionamos en el Cap'itulo~\ref{intro-general}, la emisi'on de
rayos gamma no es posible mediante procesos en equilibrio termodin'amico.
La producci'on de estos fotones tan energ'eticos ocurre a trav'es de
 procesos no t'ermicos,
esto es, por la interacci'on de part'iculas relativistas
con campos de materia, de fotones y  magn'eticos. 
Todas estas interacciones pueden describirse como dispersiones de las
part'iculas relativistas por la interacci'on con
otras part'iculas (en ocaciones virtuales) y que como resultado  producen 
fotones. Por esto,
para estudiar la producci'on de radiaci'on (gamma, en particular), 
es conveniente definir algunos conceptos b'asicos que son comunes a todos los
procesos radiativos no t'ermicos y que usaremos a lo largo de toda la tesis. 

\section{Conceptos b'asicos}

Dada una poblaci'on de part'iculas de tipo $i$ contenidas en un volumen 
$V$, si todas
ellas tienen energ'ias $E_i$ diferentes tales que 
$E_i^{\rm min} \leq E_i \leq E_i^{\rm max}$, la densidad total de estas
particulas ser'a
\begin{equation}  
\label{n_i}
n_i = \int_{E_i^{\rm min}}^{E_i^{\rm max}} 
\frac{{\rm d}n_i}{{\rm d}E_i}\,{\rm d}E_{i},
\end{equation} 
donde $[n_{i}] =$~cm$^{-3}$. La densidad de energ'ia de estas part'iculas 
es 
\begin{equation}  
%\label{n_int}
u_i = \int_{E_i^{\rm min}}^{E_i^{\rm max}} 
\frac{{\rm d}n_i}{{\rm d}E_i}\,E_{i}\,{\rm d}E_{i},
\end{equation} 
siendo $[u_{i}] =$~erg~cm$^{-3}$. 
Si la distribuci'on fuese monoenerg'etica, es decir, $E_{i}$
constante, entonces $u_{i} =n_{i} \,E_{i}$. 
Si la velocidad de las part'iculas es $v_{i}$,  
la luminosidad de 'estas ser'a 
\begin{equation}  
\label{L_ii}
%\mathcal{L}_{i} = u_{i} \,v_{i} S,
L_{i} = u_{i} \,v_{i} S,
\end{equation} 
donde la superficie $S = \partial V$ y $[L_i] =$~erg~s$^{-1}$.
%$[\mathcal{L}_{i}] =$~erg~s$^{-1}$.
Si las part'iculas son relativistas, con una energ'ia cin'etica
$(\Gamma_i -1)\,m_i\,c^2$, la ecuaci'on~(\ref{L_ii}) se escribe como 
(\ref{L_jet}). 

Supongamos que inyectamos una part'icula  relativista, con energ'ia 
$E_i$, en un medio en el 
cual existe una densidad $n_{\rm b}$ de part'iculas que llamaremos blanco
y cuya energ'ia es $E_{\rm b}$.
En el SR de la part'icula relativista, las part'iculas del medio se mueven
a una velocidad $\sim c$ y  el n'umero de interacciones que 
ocurren por unidad de tiempo es 
\begin{equation}  
\label{n_int}
N_{\rm int} = \sigma(E_i)\,n_{\rm b}\, c,
%\int\frac{{\rm d}\sigma(E_{\rm b}', E)}{{\rm d}E_{\rm b}} 
%\,n_{\rm b}\, c\,{\rm d}E_{\rm b}
\end{equation} 
donde $\sigma(E_i)$ es una medida de la superficie efectiva 
([$\sigma$] = cm$^2$)
de interacci'on del proceso de dispersi'on 
y conocida como {\bf secci'on eficaz}\footnote{Debido
a que los valores de $\sigma$ son muy  chicos, la unidad de medida que
se suele usar es el Barn: 1b $= 10^{-24}$~cm$^2$.}. 
Mientras mayor sea $\sigma$, mayor es la probabilidad de dispersi'on.
A la energ'ia de los fotones creados en la interacci'on 
la llamaremos $E_{\rm ph}$ y depende
de la energ'ia inicial $E_{\rm b}$, de la energ'ia de la part'icula
relativista, $E_i$, y del 'angulo de interacci'on. Sin embargo, $E_{\rm ph}$ no
est'a completamente determinada ya que depende de las
caracter'isticas de la colisi'on. Por lo tanto, el estado final,
caracterizado por $E_{\rm ph}$ y por el 'angulo de dispersi'on $\Omega$, tendr'a
asociada una funci'on de probabilidad. Esta informaci'on est'a contenida
en la secci'on eficaz diferencial 
${\rm d}\sigma(E_{\rm ph}, E_i, \Omega)/{\rm d}E_{\rm ph}{\rm d}\Omega$, 
de tal manera que la secci'on eficaz total es
\begin{equation}  
\sigma(E_i) = \int \frac{{\rm d}^2 \sigma(E_{\rm ph}, E_i, \Omega)}
{{\rm d}E_{\rm ph} \,{\rm d}\Omega}{\rm d}E_{\rm ph}\,{\rm d}\Omega.
\end{equation} 
Debido a la naturaleza del proceso, 
la distribuci'on energ'etica de los fotones creados o dispersados
 depender'a fuertemente del 'angulo de interacci'on.
Sin embargo, ya que en la mayor'ia de las situaciones astrof'isicas este 
'angulo no se conoce, se suelen tomar las magnitudes isotropizadas, 
es decir, integradas en todos los posibles 'angulos de interacci'on (o de
dispersi'on). 
En esta tesis no consideraremos las dependencias angulares
(ni espaciales: emisor homog'eneo) de las magnitudes. 
De la ecuaci'on~(\ref{n_int}), si $x = c\,t = (\sigma\,n_{\rm b})^{-1}$, 
entonces se tiene que ocurre solo una interacci'on en el tiempo $t$. 
Se define el {\bf camino libre medio}
como la distancia $\lambda$ que las part'iculas recorren entre dos 
interacciones sucesivas:
\begin{equation} 
\label{camino-libre-medio} 
\lambda \equiv \frac{1}{\sigma(E_i)\, n_{\rm b}}.
\end{equation} 

El {\bf tiempo de enfriamiento} $t_{\rm rad}$ es una medida de la eficiencia
de un proceso de interacci'on de una part'icula relativista con un blanco de 
densidad $n_{\rm b}$.
Si en este proceso la part'icula relativista pierde una
cantidad de energ'ia $E_i - E_f$, donde $E_i$ y $E_f$ son las energ'ias inicial 
y final, respectivamente, y definimos la inelasticidad $f$ del proceso
como $f = (E_i - E_f)/E_i$, entonces 
\begin{equation}  
\label{t_cool}
t_{\rm rad}=\frac{1}{f \, \sigma(E_i) \, n_{\rm b} \,c}.
\end{equation}
Las p'erdidas de energ'ia que mencionamos en el cap'itulo anterior, en la
ecuaci'on~(\ref{kinetic}), se definen ahora como
\begin{equation} 
\label{t_cool-1} 
\left.\frac{{\rm d}E_i}{{\rm d}t}\right|_{\rm rad} \equiv  -\,
\frac{E_i}{t_{\rm rad}}.
\end{equation}

Supongamos que  inyectamos una distribuci'on $Q_i(E_i)$ de part'iculas 
relativistas en el medio. Al cabo de un tiempo $\geq t_{\rm rad/esc}$, la 
distribuci'on de 'estas ser'a $N_i(E_i) \sim Q_i(E_i)\,t_{\rm rad/esc}$.
%(si no hay otro mecanismo de enfriamiento). 
El n'umero de 'estas part'iculas acumuladas
en el volumen $V$ ser'a\footnote{A las magnitudes intensivas las llamamos con
letras min'usculas y a las extensivas con may'usculas. Por ejemplo, 
las distribuciones de energ'ia de las part'iculas ser'an $n_i(E_i)$ si son por
unidad de volumen y $N_i(E_i)$ en caso contrario. De esta manera, bajo la 
suposici'on de que $n_i(E_i)$ no depende de las coordenadas espaciales, 
$N_i(E_i) = n_i(E_i)\,V$.}:
\begin{equation}  
N_i^{\rm rel} = \int_{E_i^{\rm min}}^{E_i^{\rm max}} N_i(E_i)\, {\rm d}E_i.
\end{equation} 

Para conocer el espectro de fotones producidos en la interacci'on de
las $N_i^{\rm rel}$ part'iculas relativistas y cuya distribuci'on en energ'ias 
es $N_i(E_i)$,
necesitamos conocer la emisividad $J_{\rm ph}$ de la fuente. 
Si las part'iculas son relativistas podemos suponer que se mueven a una 
velocidad $\sim c$ y entonces 
\begin{equation}  
\label{emisividad}
J_{\rm ph}(E_{\rm ph}) \sim \frac{c}{4 \pi} \int_{E_i =E_{\rm ph}}^{E_i^{\rm max}} \,n_{\rm b} 
%n_{\rm b} \,
\frac{{\rm d}\sigma(E_{\rm ph}, E_i)}{{\rm d}E_{\rm ph}}\, N_i(E_i)\, {\rm d}E_i,
\end{equation}
donde $[J_{\rm ph}] =$~erg$^{-1}$~s$^{-1}$.
Si conocemos $J_{\rm ph}$ luego la luminosidad espec'ifica en el caso 
de un emisor homog'eneo e isotr'opico es
\begin{equation}  
L_{E_{\rm ph}} \sim E_{\rm ph}\,J_{\rm ph}
\end{equation}
y $[L_{E_{\rm ph}}] = $s$^{-1}$.

A continuaci'on haremos una somera descripci'on de los cuatro procesos 
radiativos que hemos considerado en esta tesis: radiaci'on sincrotr'on, 
dispersi'on Compton inversa (IC),
Bremsstrahlung relativista e interacciones prot'on-prot'on ($pp$). 
 
\section{Procesos radiativos}

Existen varias maneras de producir fotones. Una de ellas es a trav'es de
la aceleraci'on de part'iculas cargadas a trav'es de la fuerza ejercida sobre 
ellas por alg'un campo. Otra manera es a trav'es del decaimiento de 
part'iculas.   

\subsection{Radiaci'on sincrotr'on}

Las part'iculas cargadas sienten la fuerza de Lorentz que ejerce sobre ellas
el campo electromagn'etico. Esta fuerza hace que las
part'iculas describan un movimiento helicoidal alrededor de las l'ineas 
de campo magn'etico $\vec B$. Las part'iculas relativistas que son 
aceleradas por
la fuerza de Lorentz producen la llamada radiaci'on sincrotr'on. 
Como veremos luego, este proceso radiativo es mucho m'as eficiente para 
leptones que para hadrones, por lo cual en esta tesis s'olo consideraremos 
la radiaci'on sincrotr'on producida por los primeros. Otra
caracter'istica de este proceso es que los fotones son emitidos en una 
direcci'on preferencial, aquella correspondiente al movimiento de la 
part'icula. 

\begin{figure}[]
\begin{center}  
\includegraphics[angle=0, width=0.5\textwidth]{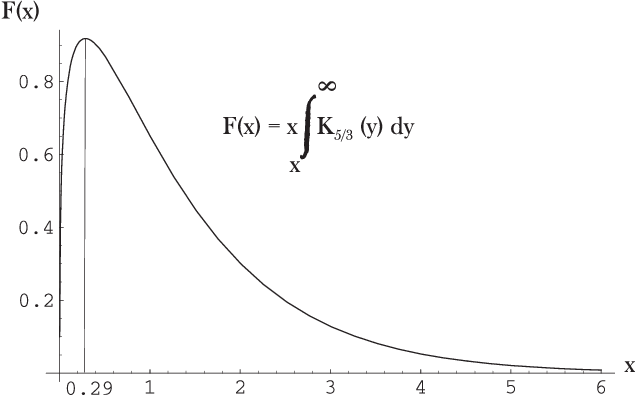}
\caption{Gr'afico de la funci'on $F(x)$, donde $x = E_{\rm ph}/E_{\rm c}$.
Como puede apreciarse, el m'aximo corresponde a $x = 0.3$, esto es, 
$E_{\rm ph} = 0.3\,E_{\rm c}$. Gracias a esta forma tipo funci'on $\delta$ que
muestra $P_{\rm sin}(E_{\rm ph}, E_e)$ es posible adoptar la aproximaci'on 
de que los electrones de energ'ia $E_e$
emiten fotones de energ'ia $\sim 0.3 E_{\rm c}(E_e)$.}
\label{sinc_aprox}
\end{center}
\end{figure}

La radiaci'on es emitida en un cono cuyo 'angulo de apertura 
llamaremos $\theta \sim 1/\Gamma_e$, donde
$\Gamma_e = E_e/(m_ec^2)$ es el factor de Lorentz de un electr'on con energ'ia
$E_e$. La distribuci'on en energ'ia de la potencia sincrotr'on por
part'icula es
\begin{equation}
\label{P_1e_exac}
P_{\rm sin}(E_e,E_{\rm ph})=\frac{1}{h}\,\frac{\sqrt{3}e^3}{m_e c^2}\,B_{\perp}\,
\frac{E_{\rm ph}}{E_{\rm c}}\int_{E_{\rm ph}/E_{\rm c}}^{\infty} K_{5/3}(\zeta) \,
{\rm d}\zeta,
\end{equation}
donde $B_{\perp}$ es la componente de $\vec B$ perpendicular a la direcci'on
del movimiento de la part'icula. Si asumimos que $\vec B$ es
isotr'opico, entonces $B_{\perp} = \sqrt{2/3} B$. La energ'ia caracter'istica
de los fotones producidos es $E_{\rm c}(E_e)= 5.1\times10^{-8}\,B\,E_e^2$~erg.
La funci'on de Bessel de
segunda especie y de orden $5/3$, $K_{5/3}$, tiene una forma tal que 
la integral de ella por el cociente $E_e/E_c$ puede aproximarse de la 
siguiente manera
\begin{equation}
\frac{E_{\rm ph}}{E_{\rm c}}\int_{E_{\rm ph}/E_{\rm c}}^{\infty} K_{5/3}(\zeta) \,
{\rm d}\zeta \sim
1.85\,\left(\frac{E_{\rm ph}}{E_{\rm c}}\right)^{1/3}
\exp\left(\frac{-E_{\rm ph}}{E_{\rm c}}\right).
\end{equation}
La funci'on $P_{\rm sin}(E_e,E_{\rm ph})$ tiene un m'aximo muy pronunciado
en la energ'ia $E_{\rm ph} \sim 0.3 E_{\rm c}$, como se muestra en la 
Figura~\ref{sinc_aprox}.

La p'erdida total de energ'ia radiada por un electr'on de
energ'ia $E_e$ se obtiene integrando la distribuci'on~(\ref{P_1e_exac}) en todas
las energ'ias radiadas $E_{\rm ph}$:
\begin{equation}
\label{P_1e_total} 
\left.\frac{{\rm d}E_e}{{\rm d}t}\right|_{\rm sin} \equiv -P_{\rm sin}(E_e)  = -
\int P_{\rm sin}(E_e, E_{\rm ph})\, {\rm d}E_{\rm ph}.
\end{equation} 
Luego, el tiempo de enfriamiento resulta 
\begin{equation}
\label{t_sin} 
t_{\rm sin} = \frac{2\,\pi}{3\,c \,\sigma_{\rm T}} \frac{(m_e c^2)^2}{E_e \,B^2}
\sim \frac{4.1\times10^2}{B^2\,E_e}~\rm{s}\,.
\end{equation} 
Siendo $t_{\rm sin} \propto m^4$, donde $m$ es la masa de la part'icula 
que est'a radiando, el tiempo de enfriamiento por radiaci'on sincrotr'on 
ser'a $\sim 10^{12}$ veces  m'as corto para 
electrones que para protones, ya que $m_e \sim 10^{-3}\, m_p$. 

Si ahora consideramos que inyectamos una poblaci'on 
$Q_e(E_e) \propto E_e^{-p}$ de electrones relativistas, 
las p'erdidas que sufrir'an por radiaci'on sincrotr'on modificar'an $Q_e(E_e)$ 
dando lugar a la distribuci'on de electrones
\begin{equation}
N_e(E_e) \sim Q_e(E_e)\,t_{\rm sin} \propto E_e^{-p -1}.
\end{equation}
Dada la distribuci'on $N_e(E_e)$, para obtener la potencia radiada por 
todos estos 
electrones lo que debemos hacer es integrar el siguiente producto:
\begin{equation}
\label{P_sin} 
P_{\rm sin}(E_{\rm{ph}}) =
\int_{E_e^{\rm{min}}}^{E_e^{\rm{max}}}P(E_e,E_{\rm{ph}})\;N_e(E_e)\,{\rm d}E_e,
\end{equation}
donde $E_e^{\rm{min}}$ y $E_e^{\rm{max}}$ son las energ'ias m'inima y m'axima,
respectivamente, de la distribuci'on de electrones relativistas. 
Sin embargo, debido a que cada electr'on radiar'a la mayor parte
de su energ'ia $E_e$ en fotones de energ'ia 
$E_{\rm{ph}}\sim 0.3 E_{\rm c}(E_e)$, podemos
calcular la integral~(\ref{P_sin}) considerando una aproximaci'on tipo 
$\delta$. De esta manera, utilizando una aproximaci'on 
$\delta(E_{\rm ph} - 0.3E_{\rm c})$ podemos hallar una soluci'on anal'itica 
aproximada de $P_{\rm sin}(E_{\rm{ph}})$. 

Para obtener la luminosidad a la energ'ia $E_{\rm ph}$ calculamos 
$E_{\rm ph}L_{\rm ph} = E_{\rm ph}P_{\rm sin}(E_{\rm{ph}})$.
Es interesante notar que el espectro de fotones emitidos tiene una 
distribuci'on en energ'ia tipo ley de potencias de 'indice $\alpha$
($E_{\rm ph} L_{\rm sin} \propto E_{\rm ph}^{-\alpha}$), si el 
espectro de electrones
tambi'en es una ley de potencias $N_e \propto E_e^{-p'}$. Las relaciones entre
ambos 'indices es $\alpha = (p^{\prime} -1)/2$.

\subsection{Radiaci'on Compton inversa}

Consideremos  un gas de fotones de energ'ia $E_{\rm phm}$ cuya densidad de 
energ'ia es $u_{\rm phm}$ y un electr'on relativista que
atraviesa dicho gas. Si en el SR del laboratorio
los fotones son menos energ'eticos que el electr'on,
entonces los primeros ser'an dispersados por el segundo. Como
producto de esta interacci'on los fotones ganan energ'ia, a diferencia de lo
que ocure en la interacciones Compton (directas), y de aqu'i que se las  
llame dispersiones Compton inversas.   

\begin{figure}[]
\begin{center}  
\includegraphics[angle=270, width=0.5\textwidth]{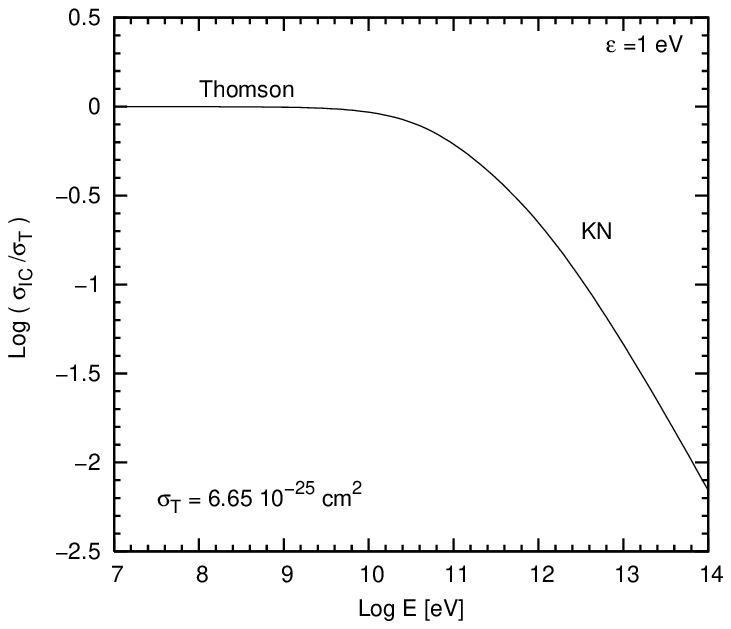}
\caption{Secci'on eficaz de la interacci'on IC en los reg'imenes Th
y KN para fotones semilla con energ'ia $\tilde E_{\rm phm} = 1$~eV.}
\label{fig_sigma_IC}
\end{center}
\end{figure}

En el SR del electr'on, si se acerca un fot'on poco energ'etico, 
'este sufrir'a un cambio mayor en su energ'ia y momento que uno con 
m'as energ'ia. 
Esto se refleja en la secci'on 
eficaz de la interacci'on, $\sigma_{\rm IC}$.
Si integramos en todos los posibles 'angulos de incidencia, $\sigma_{\rm IC}$
depende de la energ'ia de los fotones semilla y del electr'on 
a trav'es de la expresi'on (Vila \& Aharonian 2009)
\begin{equation}
\sigma_{\rm IC} = \frac{3 \sigma_{\rm T}}{8 y} \left[\left(
1-\frac{2}{y}-\frac{2}{y^2}\right)\ln(1 + 2y) + \frac{1}{2} + \frac{8}{y}
-\frac{1}{2(1 + 2y)^2}\right],
\end{equation}
donde $y \equiv E_{\rm phm} E_e/(m_e^2 c^4)$ y $\sigma_{\rm T} \sim 0.67$~b es la
secci'on eficaz de Thomson. Notamos que para $y \ll 1$, 
$\sigma_{\rm IC} \sim \sigma_{\rm T}$ y se dice que 
la interacci'on ocurre en el r'egimen de Thomson (Th), mientras que si 
$y \gg 1$
entonces $\sigma_{\rm IC} \sim (3\sigma_{\rm T}/8)\ln(4y)/y$ y se dice que
 interacci'on se desarrolla en el r'egimen de Klein-Nishina (KN).
Como se muestra
en la Figura~\ref{fig_sigma_IC}, en el r'egimen de KN $\sigma_{\rm IC}$ cae 
abruptamente. Sin embargo, las p'erdidas de energ'ia por interacci'on 
son mayores en el
r'egimen KN que en el Th, siendo en el primero catastr'oficas, esto es,
$E_{\rm ph} \sim E_e$. En el r'egimen Th, la energ'ia m'axima que pueden
alcanzar los fotones es 
$E_{\rm ph}^{\rm max} \sim (4/3)\, \Gamma_e^2 \,E_{\rm phm}$. 

Dado un electr'on relativista inmerso en un gas de fotones, el n'umero de 
interacciones por unidad de tiempo
puede calcularse a trav'es de la ecuaci'on~(\ref{n_int}), considerando que
$n_{\rm b} = n_{\rm phm}$, con lo cual
obtenemos que el n'umero de interacciones por unidad de tiempo es 
$N_{\rm int} = n_{\rm phm} \,\sigma_{\rm IC} \,c$. 
El n'umero de interacciones es igual al 
n'umero de fotones dispersados. 
Si ahora queremos conocer la distribuci'on energ'etica de los fotones 
dispersados por un electr'on, debemos tener en cuenta que la energ'ia 
$E_{\rm ph}$ de estos 
fotones no est'a fija sino que es una distribuci'on y esto se muestra
en la secci'on eficaz diferencial (Blumenthal \& Gould, 1970)
\begin{equation}
\label{sigma-ci}
\frac{{\rm d}\sigma_{\rm{IC}}}{{\rm d}E_{\rm{ph}}}(E_{\rm{ph}},E_e) = 
\frac{3\sigma_{\rm T}}{4 E_{\rm{ph}} \Gamma_e^2} f(x_{\rm ic})
\end{equation}
con
\begin{equation}
f(x_{\rm ic}) = \left[ 2x_{\rm ic}\ln{x_{\rm ic}} + x_{\rm ic} +1-2x_{\rm ic}^{2} +
\frac{(4\epsilon_{\rm{ph}}\Gamma_e x_{\rm ic})^{2}(1-x_{\rm ic})}{2(1+4\epsilon_{\rm{ph}}\Gamma_e x_{\rm ic})} \right] P_{\rm ic}\left(
\frac{1}{4\Gamma_e^{2}},1,x_{\rm ic} \right),
\end{equation}
donde $\epsilon_{\rm{ph}} = E_{\rm{ph}}/m_ec^2$. La
funci'on $P_{\rm ic}$ es igual a 1 cuando $1/4\Gamma_e^2 \leq x_{\rm ic} \leq 1$
 y $P_{\rm ic} = 0$ en cualquier otro caso. Siendo 
$\epsilon_{\rm{phm}} = E_{\rm{phm}}/m_ec^2$, el par'ametro adimensional 
$x_{\rm ic}$ se define de la siguiente manera
\begin{equation}
x_{\rm ic} =\frac{\epsilon_{\rm ph}}{4\epsilon_{\rm{phm}} 
\Gamma_e^{2}(1-\epsilon_{\rm ph}/\Gamma_e)}.
\end{equation}

La interacci'on de un electr'on con un campo de fotones dispersa a 'estos
dando un espectro de fotones dispersados con energ'ias 
$E_{\rm ph}^{\rm min} \leq E_{\rm ph} \leq E_{\rm ph}^{\rm max}$.
Para calcular
la potencia radiada por un electr'on por interacciones IC debemos integrar
sobre todas las energ'ias $E_{\rm phm}$ a trav'es de la expresi'on 
\begin{equation} 
\label{P_IC}
P_{\rm IC}(E_e) = -\left.\frac{{\rm d}E}{{\rm d}t}\right|_{\rm IC} 
= c\,\int_{E_{\rm phm}^{\rm min}}^{E_{\rm phm}^{\rm max}}\,\int_{E_{\rm ph}^{\rm min}}^{E_{\rm ph}^{\rm max}} 
\frac{{\rm d}\sigma_{\rm IC}(E_{\rm ph}, E_e)}{{\rm d}E_{\rm ph}}\, 
\frac{{\rm d}n_{\rm phm}(E_{\rm phm})}{{\rm d}E_{\rm phm}} \,E_{\rm ph}\,{\rm d}E_{\rm phm}\,
{\rm d}E_{\rm ph}. 
\end{equation}
Para el caso en el cual los fotones semilla siguen una distribuci'on 
(quasi)monoenerg'etica con energ'ia $\tilde E_{\rm phm}$, las p'erdidas 
de un electr'on pueden obtenerse a trav'es de una aproximaci'on   
$\delta(E_{\rm phm} - \tilde E_{\rm phm})$. El tiempo
de enfriamiento tanto en el r'egimen de Th como en el de KN queda determinado 
por la expresi'on (Bosch-Ramon \& Khangulyan, 2009)
\begin{equation} 
\label{t_ci}
t_{\rm IC} =  \frac{6.1\times10^{12}\,\tilde E_{\rm phm}}{u_{\rm phm}}
\frac{(1 + 8.3\,y)}{\ln(1+0.2\;y)}\frac{(1 + 1.3\,y^2)}{(1 + 0.5\,y + 1.3\,y^2)}
~\rm{s}\,,
\end{equation}
donde ahora $y = \tilde E_{\rm phm} E_e/(m_e^2 c^4)$ y $u_{\rm phm}$ es la 
densidad de energ'ia de los fotones ambientales.
Considerando los l'imites a bajas  energ'ias de (\ref{t_ci}), es
posible hallar $t_{\rm IC}$ en el r'egimen de Th, resultando
\begin{equation} 
t_{\rm IC}^{\rm Th} \sim  \frac{24}{E_e\,u_{\rm phm}} \,\,{\rm s}. 
\end{equation}

Si ahora inyectamos una distribuci'on $Q_e(E_e) \propto E_e^{-p}$ de electrones
relativistas en el gas de 
fotones, las p'erdidas por IC en el r'egimen Th modifican el espectro de 
la misma manera
que las p'erdidas por radiaci'on sincrotr'on, esto es,
\begin{equation}
N_e(E_e) \sim Q_e(E_e)\,t_{\rm IC}^{\rm Th} \propto E_e^{-p -1}.
\end{equation}
Dada la distribuci'on $N_e(E_e)$, para obtener la distribuci'on de fotones 
dispersados calculamos la emisividad de los mismos a trav'es de
\begin{equation}
\label{q_ci}
J_{\rm IC}(E_{\rm ph}) = \frac{c}{4 \pi} \; \int_{E_e^{\rm min}}^{E_e^{\rm max}} 
\int_{E_{\rm phm}^{\rm min}}^{E_{\rm phm}^{\rm max}} \frac{{\rm d}\sigma_{\rm IC}(E_{\rm ph},E_e)}{{\rm d}E_{\rm ph}}\frac{{\rm d}n_{\rm phm}(E_{\rm phm})}{{\rm d}E_{\rm phm}} 
\,N_e(E_e)\,  {\rm d}E_{\rm phm}\,{\rm d}E_e.
\end{equation}

\subsubsection{Auto Compton}

Si los fotones semilla son creados externamente a la fuente (es decir, 
fuera del volumen V) entonces el proceso se dice Compton externo (EC, por
\emph{External Compton}).
Por otro lado, en el caso particular de que los fotones semilla sean 
producidos por 
radiaci'on sincrotr'on de la misma poblaci'on de electrones $N_e(E_e)$ que 
interact'uan por IC, entonces el proceso se llama auto Compton sincrotr'on 
(SSC, por 
\emph{Synchrotron Self Compton}). Bajo esta situaci'on, la distribuci'on
de fotones semilla es una ley de potencias y para calcular el espectro de los
fotones producidos debemos usar la ecuaci'on~(\ref{q_ci}), considerando que
${\rm d}n_{\rm phm}/{\rm d}E_{\rm phm} = {\rm d}n_{\rm sin}/{\rm d}E_{\rm sin}$ 
es la densidad de fotones producidos por radiaci'on sincrotr'on.

\subsection{Bremsstrahlung relativista}

Un electr'on relativista inmerso
en un campo de materia (no relativista) ser'a acelerado por el
el campo coulombiano producido por los n'ucleos de los 'atomos que forman
'este 'ultimo. La materia puede estar constitu'ida
por n'ucleos desnudos (es decir, ionizada) o por 'atomos (n'ucleos 
apantallados por los electrones).
En ambos casos, la interacci'on puede esquematizarse de la forma 
\begin{equation} 
e + (e,N) \longrightarrow e + (e,N) + \gamma,
\end{equation}
aunque la secci'on eficaz no es la misma. Esta viene dada por la expresi'on
(Bosch-Ramon 2006)
\begin{equation}
\frac{{\rm d}\sigma_{\rm Brem}(E_e,E_{\rm ph})}{{\rm d}E_{\rm ph}}  = 
\frac{4\,\alpha_{\rm ef}\, r_e^2\,Z_e^2}{E_{\rm ph}} \phi(E_e,E_{\rm ph}), 
\end{equation}
donde $\alpha_{\rm ef} \sim 1/137$ es la constante de estructura fina y
$r_e = q_e^2/m_ec^2$ es el radio cl'asico del electr'on. Para el caso de
un n'ucleo desnudo, la funci'on $\phi$ esta dada por
\begin{equation}
\phi (E_e,E_{\rm ph})= \left[1+\left(1-\frac{E_{\rm ph}}{E_e}\right)^2
  - \frac{2}{3}\left(1-\frac{E_{\rm ph}}{E_e}\right)\right]
\left\{\ln\left[\frac{2 E_e (E_e - E_{\rm ph})}{m_e c^2
    E_{\rm ph}}\right] -\frac{1}{2} \right\},
\end{equation}
mientras que para el caso en que el n'ucleo est'a
completamente apantallado por todos los electrones
\begin{equation}
\phi (E_e,E_{\rm ph})= \left[1+\left(1-\frac{E_{\rm ph}}{E_e}\right)^2
  - \frac{2}{3}\left(1-\frac{E_{\rm ph}}{E_e}\right)\right]
\ln\left(\frac{191}{Z_e^{1/3}}\right)
+\frac{1}{9}\left(1-\frac{E_{\rm ph}}{E_e}\right).
\end{equation}

El electr'on entrega casi toda su energ'ia a los fotones en estas 
interacciones, es decir, $E_{\rm ph} \sim E_e$, y por esto se dice que las
p'erdidas que sufre son catastr'oficas. Sin embargo, es posible hallar 
una expresi'on continua para el tiempo de enfriamiento. 
Para calcularlo, hacemos el mismo 
an'alisis que en el caso de las interacciones IC, sin embargo, debido a que 
hay dos parametrizaciones de $\sigma_{\rm Brem}$, para el caso en el que
la materia est'a apantallada o aquel donde el
medio est'a completamente ionizado, obtendremos dos expresiones diferentes para
el tiempo de enfriamiento. Siendo $n_{\rm m}$ la densidad del campo de materia,
en el primer caso tenemos 
\begin{equation} 
%\label{t_Brem}
t_{\rm Brem}^{\rm neu} \sim \frac{1.4\times10^{16}}{n_{\rm m}\,Z_e^2
\left(\ln\left(\frac{183}{Z_e^{1/3}}\right) -\frac{1}{18}\right)}~\rm{s}\,,
\end{equation} 
mientras que en el segundo
\begin{equation} 
\label{t_Brem}
t_{\rm Brem}^{\rm ion} \sim \frac{1.4\times10^{16}}{n_{\rm m}\,Z_e^2
\left(\ln\left(\frac{E_e}{m_ec^2}\right) + 0.36\right)
}~\rm{s}\,.
\end{equation}

Si ahora inyectamos una distribuci'on $Q_e(E_e) \propto E_e^{-p}$ 
de electrones relativistas,
las p'erdidas por las interacciones con la materia casi no  modifican  
$Q_e(E_e)$ ya que la dependencia con $E_e$ de $t_{\rm Brem}$ es despreciable. 
Por esto,
\begin{equation} 
\label{}
N_e(E_e) \sim Q_e(E_e)\, t_{\rm Brem} \propto E_e^{-p}.
\end{equation}

La emisividad producida por una distribuci'on de electrones $N_e(E_e)$  est'a 
dada por la ecuaci'on~(\ref{emisividad}), donde $n_{\rm b}$ es la densidad
de materia y $\sigma = \sigma_{\rm Brem}$, con lo cual hallamos
%y luego la luminosidad espec'ifica nos da...
%
\begin{equation}  
\label{emisividad_Bremm}
J_{\rm Brem}(E_{\rm ph}) \sim \frac{c}{4 \pi} \,n_{\rm m} \int_{E_{\rm ph}}^{\infty} 
\frac{{\rm d}\sigma_{\rm Brem}(E_e, E_{\rm ph})}{{\rm d}E_{\rm ph}}\, 
N_e(E_e)\, {\rm d}E_e.
\end{equation}

\subsection{Interacciones prot'on-prot'on}

Adem'as de procesos asociados a cargas aceleradas, los fotones tambi'en 
pueden producirse por 
decaimientos de part'iculas. Por ejemplo, los piones neutros $\pi^0$ decaen
en dos rayos gamma. Una manera de  producir $\pi^0$ es a trav'es de 
colisiones inel'asticas 
 prot'on-prot'on, en las cuales un prot'on relativista interact'ua con 
un prot'on no relativista. El canal de interacciones $pp$ m'as importante para
generar $\pi^0$ es el siguiente (Romero 2010):
\begin{equation}
\label{int_pp}
p + p  \longrightarrow p + \Delta^+ + \tilde n_0\,\pi^0 + 
\tilde n_{\pm}\,(\pi^+ + \pi^-),
\end{equation}
donde $\pi^+$ y $\pi^-$ son piones cargados positiva y negativamente, 
respectivamente, y $\tilde n_0$ y $\tilde n_{\pm}$ son las multiplicidades 
(n'umeros enteros positivos). 
En este proceso, el prot'on relativista pierde $\sim 50$\% de su energ'ia,
con lo cual la inelasticidad de la interacci'on es $f_{pp}\sim 0.5$. 
El decaimiento $\Delta^+ \longrightarrow p + \pi^0$ entrega al $\pi^0$ 
aproximadamente un 17\% de la energ'ia $E_p$ del prot'on relativista.
% y se lo
%llama ``pi'on l'ider'', ya que el resto de los piones creados
%tienen mucha menos energ'ia.  
El $33\%$ restante de $E_p$ es entregada al
resto de los piones creados. Un 11\% va a los piones neutros y un 22\% a los 
cargados. Sin embargo, debido a que las multiplicidades $\tilde n_{0}$ y
$\tilde n_{\pm}$ son grandes, la energ'ia de cada una de estas part'iculas es 
poca. Es por esto que al $\pi^0$  con energ'ia $\sim 0.17 E_p$ se lo llama
``pi'on lider''.
%Si los 
%protones son muy energ'eticos, la probabilidad de crear piones de cada 
%tipo ($\pi^0$ y $\pi^{\pm}$) es la misma y la fracci'on de la energ'ia del 
%proton que va a cada tipo de piones es $\kappa_{\pi}\sim f_{pp}/3 \sim 0.17$. 

\begin{figure}[]
\begin{center}  
\includegraphics[angle=0, width=0.5\textwidth]{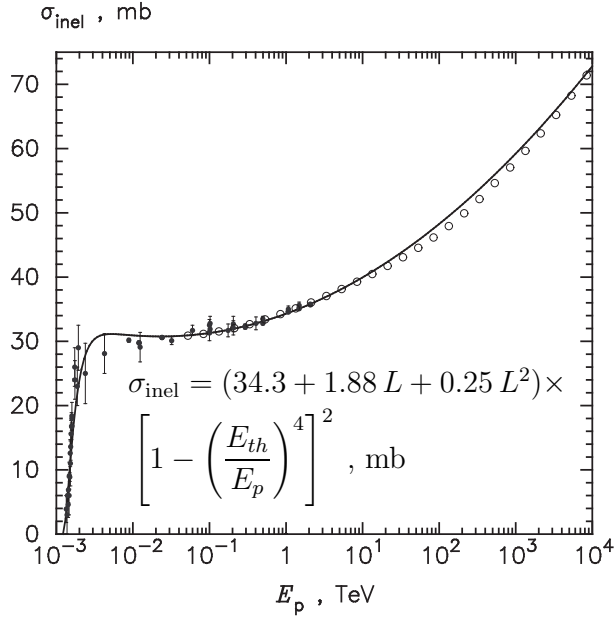}
\caption{Secci'on eficaz de la interacci'on $pp$ (Kelner et al. 2006).}
\label{Kelner_sigma}
\end{center}
\end{figure}

Para que la interacci'on $pp$ ocurra,
la energ'ia del prot'on relativista debe ser mayor que un valor umbral
$E_{\rm u} \sim m_pc^2 + 2 m_{\pi} c^2 (1 + m_{\pi}/4m_p) \sim 1.22$~GeV.   
La parametrizaci'on  m'as reciente de la secci'on eficaz total de la 
interacci'on $pp$ es la dada por Kelner y colaboradores (2006):
\begin{equation}
\sigma_{pp}(E_p) = (34.3 + 1.88\;L + 0.25\;L^2)\left[1 -
\left(\frac{E_{\rm u}}{E_p}\right)^4\right]^2 \qquad \rm{mb},
\end{equation} 
donde $L = \ln(E_p/ 1 \; \rm{TeV})$ y su forma se muestra en la 
Figura~\ref{Kelner_sigma}.

Como mostramos en la ecuaci'on~(\ref{t_cool}) el tiempo de enfriamiento 
depende de la
secci'on eficaz de la interacci'on, con lo cual para diferentes 
parametrizaciones de 'esta tendremos diferentes valores del tiempo de 
enfriamiento por $pp$.
Sin embargo, debido a que todas las parametrizaciones halladas son parecidas
y que la dependencia con $E_p$ es chica, 
usaremos una expresi'on sencilla pero 'util para los fines de esta tesis.
Considerando que $\sigma_{pp} \sim 30$~mb y que la inelasticidad de la 
interacci'on es $f_{pp} \sim 0.5$, el tiempo de enfriamiento para un medio
de densidad $n_{\rm m}$ resulta
\begin{equation} 
\label{t_pp}
t_{pp} \sim  \frac{2\times10^{15}}{n_{\rm m}}~\rm{s}\,.
\end{equation}

Como ocurre con el proceso Bremsstrahlung relativista, las p'erdidas por $pp$
no modifican la forma del espectro de inyecci'on. Si inyectamos protones
relativistas con una distribuci'on $Q_p(E_p) \propto E_p^{-p}$, 
al cabo de un tiempo $\gtrsim t_{pp}$ tendremos una distribuci'on estacionaria
de protones tal que
\begin{equation}
%\label{q_pi}
N_p (E_p) \sim Q_p(E_p)\,t_{pp} \propto E_p^{-p}.
\end{equation}
Estos  protones relativistas, al interactuar con el medio
con densidad $n_{\rm m}$, producir'an $\pi^0$ cuya emisividad est'a dada por:
\begin{equation}
\label{q_pi}
J_{\pi^0}(E_{\pi^0}) = \frac{c}{4\pi}\; n_{\rm m}\; \int_{E_{p}^{\rm min}}^{E_{p}^{\rm max}}
\frac{{\rm d}\sigma_{pp}(E_{\pi^0},E_p)}{{\rm d}E_{\pi^0}}\; N_p(E_p)\; {\rm d}E_p\,. 
\end{equation}
Para $E_p \lesssim 0.1$~TeV, los piones creados tendr'an una energ'ia promedio
$E_{\pi^0} \sim \kappa_{\pi} E_p$, donde $\kappa_{\pi} \sim 0.17$ (Gaisser 1990),
 y por lo tanto es  posible considerar la aproximaci'on 
$\delta(E_{\pi^0} - \kappa_{\pi}E_p^{\rm cin})$ para calcular $J_{\pi^0}$. 
Siendo $E_p^{\rm cin} = E_p - m_pc^2$ la energ'ia cin'etica de los protones, 
se llega al siguiente resultado:
\begin{equation}
J_{\pi^0}(E_{\pi^0}) \sim \frac{c}{4\pi}\frac{\tilde n_0}{\kappa_{\pi}}\,n_{\rm m}\, 
N_p\left(m_pc^2 + \frac{E_{\pi^0}}{\kappa_{\pi}}\right)\,
\sigma_{pp}\left(m_pc^2 + \frac{E_{\pi^0}}{\kappa_{\pi}}\right).  
\end{equation}
Sin embargo, la aproximaci'on   $\delta$ no es muy buena si 
$E_p \gtrsim 0.1$~TeV ya que  en este caso la distribuci'on de
piones se ensancha en energ'ia y la parte de m'as baja energ'ia se hace m'as
``blanda'' para $E_p$ mayores, ya que se crean muchos piones de poca energ'ia
(es decir, la multiplicidad crece). En este rango de energ'ia de los protones,
las f'ormulas adecuadas para calcular $J_{\pi^0}$ son las dadas por Kelner y
colaboradores (2006). De esta manera, usando la aproximaci'on $\delta$ 
(modificada)
para $E_p \lesssim 0.1$~TeV y las parametrizaciones de Kelner para 
$E_p \gtrsim 0.1$~TeV,
la multiplicidad   $\tilde n_0$ puede considerarse un par'ametro libre y 
ajustarlo de tal manera que $J_{\pi^0}$ sea una funci'on continua. 
Finalmente, calculada $J_{\pi^0}$, podemos obtener la emisividad de los rayos
gamma a trav'es de 
\begin{equation}
\label{q_pp}
J_{\rm ph}(E_{\rm ph}) = 2  \int_{E_{\pi^0}^{\rm min}}^{E_{\pi^0}^{\rm max}}  
\frac{J_{\pi^0}(E_{\pi^0})}{\sqrt{E_{\pi^0}^2 - m_{\pi}^2c^4}}\, {\rm d}E_{\pi^0}\,,
\end{equation}
donde $E_{\pi^0}^{\rm min} = E_{\rm ph}  + m_{\pi}^2c^4/(4E_{\rm ph})$.
El factor 2 en la expresi'on (\ref{q_pp}) 
tiene en cuenta que por cada $\pi^0$ se producen dos rayos gamma.

\subsubsection{Creaci'on de pares electr'on-positr'on}

De acuerdo a (\ref{int_pp}), adem'as de $\pi^0$ tambi'en se crean $\pi^{\pm}$
en las interacciones $pp$. Estos 'ultimos luego decaen en muones que a su vez 
producen neutrinos y pares electr'on-positr'on ($e^{\pm}$). 
La mayor'ia de los pares producidos tienen energ'ias $E_{e_2} \gg m_ec^2$ 
 y por esto nos interesa conocer la distribuci'on $N_{e_2}(E_{e_2})$
de estas part'iculas  ya que son una poblaci'on adicional de leptones 
relativistas que se enfriar'an de la misma manera que los electrones 
primarios acelerados, por ejemplo, en frentes de ondas de choque.

Si $N_p(E_p)$ es una ley de potencias, entonces la distribuci'on 
de pares $e^{\pm}$ creados tambi'en ser'a  una ley de potencias 
$N_{e_2}(E_{e_2}) = K_{e_2}' E_{e_2}^{-p_2'}$, donde tanto $p_2'$ como $K_{e_2}'$ 
dependen de
los par'ametros de $N_p(E_p)$. Calculando $N_{e_2}(E_{e_2})$  trav'es de las 
f'ormulas dadas por Kelner y colaboradores (2006) y considerando  
$K_{e_2}' = 1$ podemos hallar
$p_2'$ de hacer un ejuste gr'afico del espectro calculado. Luego, conociendo 
el 'indice  del espectro de los pares 
podemos hallar  $K_{e_2}'$  a trav'es de la igualdad $u_{e_2} = f_{e^{\pm}} u_p$, 
donde
$f_{e^{\pm}}$ depende de $p_2$. Las cantidades $u_{e_2}$ y $u_p$ son las 
densidades de energ'ia de los pares $e^{\pm}$ y de los protones relativistas,
respectivamente.

\section{Absorci'on}  

Anteriormente hemos visto diferentes procesos no t'ermicos 
que producen fotones. Sin embargo, 'estos pueden a su vez ser absorbidos
mediante interacciones con otras part'iculas.
Estas interacciones pueden ocurrir con campos de materia,
%(ref),
con campos magn'eticos (Sturrock 1971) o con campos de fotones 
(Coppi \& Blandford 1990).  

Supongamos que un fot'on de energ'ia $E_{\rm ph}$ se propaga en un medio de
densidad $n_{\rm b}$ y contenido en una regi'on de tamaño $L_{\rm m}$.
Si la emisividad de los fotones es $J_{\rm ph}^0$,  luego
de atravesar la regi'on de tamaño $L_{\rm m}$, la emisividad ser'a reducida a 
$J_{\rm ph}$ de manera tal que
\begin{equation}
J_{\rm ph}(E_{\rm ph}) = J^0_{\rm ph}(E_{\rm ph})\, e^{-\tau}.
\end{equation}
La profundidad 'optica del medio, $\tau$, se 
define de la siguiente manera: 
\begin{equation}
\tau(E_{\rm ph}) = \int_0^{L_{\rm m}} \int_{E_{\rm b}^{\rm{min}}}^{\infty} 
n_{\rm b}(E_{\rm b},r)\,
\sigma_{\rm abs}(E_{\rm ph},E_{\rm b})\, {\rm d}E_{\rm b}\, {\rm d}r,
\end{equation}
donde  $\sigma_{\rm abs}$ es la secci'on eficaz del proceso y 
$E_{\rm b}^{\rm{min}}$ es la energ'ia m'inima de las part'iculas del medio
para que ocurra la interacci'on
entre una de ellas  y un fot'on. Si $\tau \ll 1$ entonces 
no hay absorci'on y el medio se dice 'opticamente delgado. Por otro lado,
si $\tau \geq 1$ la absorci'on puede ser significativa y el medio se dice
 'opticamente grueso.

\subsection{Absorci'on por creaci'on de pares electr'on-positr'on}

\begin{figure}[]
\begin{center}  
\includegraphics[angle=270, width=0.6\textwidth]{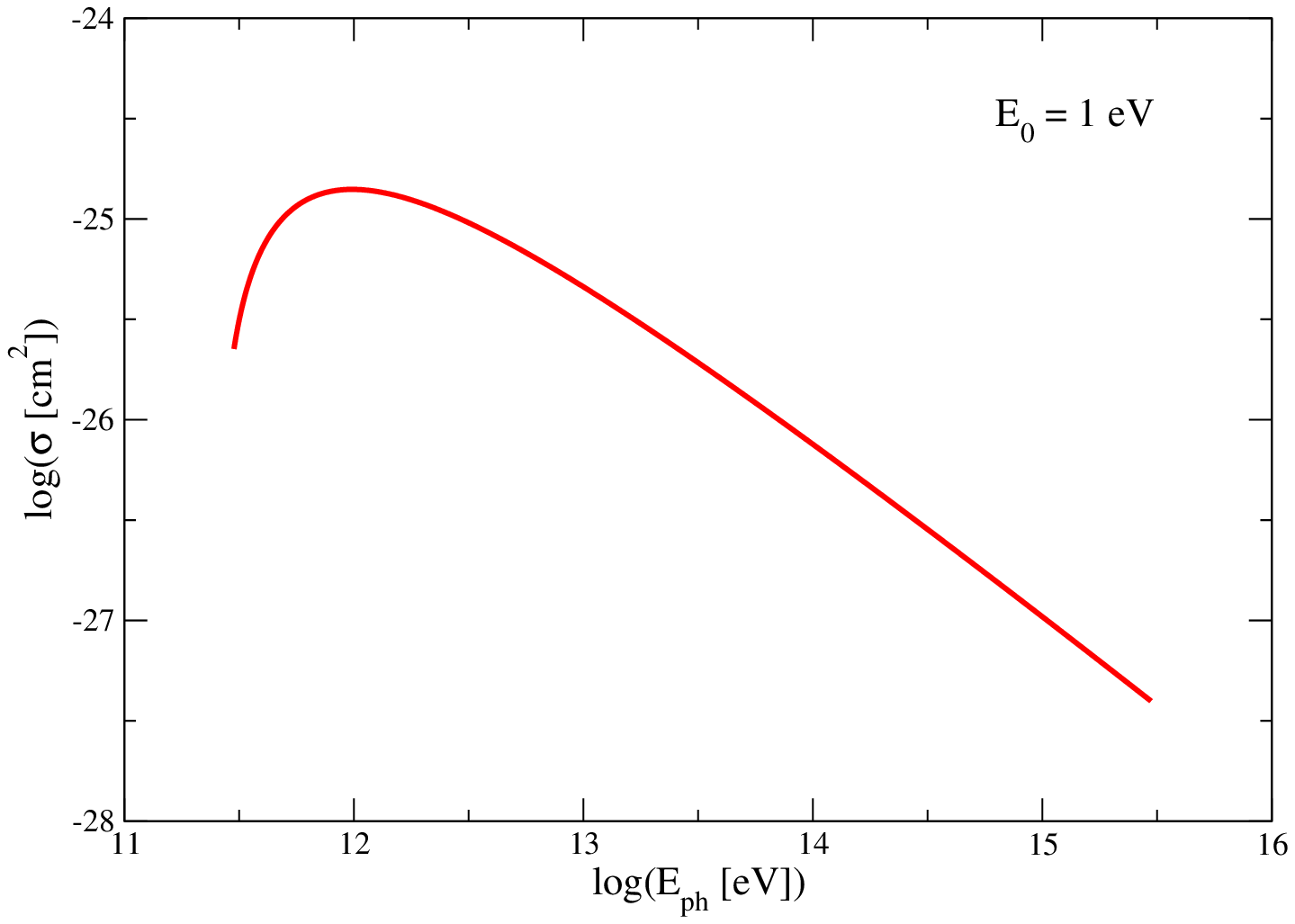}
\caption{Secci'on eficaz isotr'opica ($\sigma_{\gamma\gamma}$) del proceso 
de creaci'on de pares 
$e^{\pm}$ por aniquilaci'on de fotones. El grafico corresponde a 
$E_{\rm ph}^0 = 1$~eV.}
\label{sigma-abs}
\end{center}
\end{figure}

 En esta tesis s'olo consideraremos un mecanismo de absorci'on; aquel
en el cual dos fotones se aniquilan produciendo un par $e^{\pm}$ de la
siguiente manera
\begin{equation}
\gamma + \gamma \longrightarrow e^+ + e^-.
\end{equation}
Si consideramos que el fot'on de energ'ia $E_{\rm ph}$ producido en la 
fuente se propaga en un gas isotr'opico y (quasi)monoenerg'etico 
de fotones de energ'ia $E_{\rm ph}^0$, 
para que la interacci'on ocurra, debido a
que  se crean dos part'iculas con masa $m_e$, la energ'ia umbral de los
fotones  debe ser tal que $E_{\rm ph} E_{\rm ph}^0 \geq 2 m_e^2c^4$. 
La secci'on eficaz de esta interacci'on fot'on-fot'on es (Coppi \& 
Blandford 1990)
\begin{equation}
\sigma_{\gamma\gamma}(E_{\rm ph},E_{0}) \sim \sigma_{\rm T}
\frac{(x_{\gamma\gamma}-1)^{1/3}}{x_{\gamma\gamma}^{5/2}} \left(
\frac{1}{2\,x_{\gamma\gamma}^{1/2}} + \frac{3}{4}\ln(x_{\gamma\gamma})\right)
H(x_{\gamma\gamma} -1)
\end{equation}
donde $x_{\gamma\gamma} \equiv E_{\rm ph} E_{\rm ph}^0/m_e^2c^4$ y 
$H(x_{\gamma\gamma} -1)$ es la funci'on de Heaviside.
Como se muestra en la Figura~\ref{sigma-abs}, $\sigma_{\gamma\gamma}$ tiene 
un m'aximo cuando $E_{\rm ph}\,E_{\rm ph}^0 \sim 3.7\, m_e^2c^4$.

Si la densidad del gas de fotones en el cual se propaga el rayo gamma es
$n_{\rm{phm}}$, la
profundidad 'optica ser'a:
\begin{equation}
\tau_{\gamma\gamma}(E_{\rm ph}, E_{\rm ph}^0) = 
n_{\rm ph}\,\sigma_{\gamma\gamma}(E_{\rm phm},E_{\rm ph}^0)\, L_{\rm m}.
\end{equation}

%% file: YSO_final.tex
\chapter{Objetos estelares j'ovenes}
\label{yso}

\section{Introducci'on}

Las estrellas se clasifican de acuerdo a su masa $M_{\star}$ en de 
gran masa 
($M_{\star} \geq 8 M_{\odot}$) y de baja masa ($M_{\star} \leq 8 M_{\odot}$).
Numerosas de 'estas 'ultimas est'an cerca y  son fáciles de detectar.
Por esto se 
conoce bastante de su formación y evolución. Sin embargo, no ocurre lo
mismo con las estrellas de gran masa, las cuales se encuentran embebidas
en grandes condensaciones de gas y polvo con lo cual la extinción de la luz 
que emiten es significativa  y poco llega de ella a nuestros detectores. 
Si adem'as tenemos en cuenta que el tiempo de vida de estas estrellas es
muy corto ($\sim 1 - 10$~Myr), la observaci'on y detecci'on de las 
estrellas tempranas (tipos espectrales O y B) en cada estado evolutivo 
es extremadamente dif'icil.  
Es por esto que el estudio de 
la formaci'on de las estrellas de gran masa es uno de los grandes t'opicos 
 de la astrof'isica actual. 

 La formaci'on estelar comienza cuando una nube de gas en el espacio se 
vuelve inestable y colapsa bajo la acci'on de su propia gravedad. Durante 
el colapso, la nube se fragmenta. Esto es todo lo que puede decirse 
respecto de la etapa inicial de la formaci'on de una estrella, ya que  
se sabe muy poco de como ocurre la fragmentaci'on de la nube.  
Luego, la nube queda dividida en varias partes, cada una de ellas con una 
distribuci'on de densidad inhomog'enea la cual induce el proceso de acreci'on
y as'i se forman n'ucleos m'as densos. Dependiendo de la masa de estos n'ucleos
y de la tasa de acreci'on de materia, se formar'an estrellas de diferentes 
masas. 

 Para los n'ucleos poco densos que luego dar'an lugar a las estrellas de baja
masa, la secuencia de eventos hasta llegar a la formaci'on de la estrella
pareciera ser clara y la resumimos de la siguiente manera (Shu et al. 1987): 
\begin{enumerate}
\item Un estado inicial en el cual la regi'on central de los n'ucleos 
densos se contrae, intensificando as'i su campo gravitacional y formando una 
protoestrella en cada n'ucleo.
\item Un estado de acreci'on caracterizado por la formaci'on de un disco 
alrededor de cada protoestrella y a trav'es del cual la misma acreta 
materia del medio circundante. 
\item La fase en la cual se producen los flujos bipolares (\emph{outflows}
o \emph{jets} dependiendo del 'angulo de colimaci'on de los mismos), por 
los cuales 
la protoestrella deposita materia con momento angular y energ'ia cin'etica 
en sus alrededores. Estos flujos bipolares, al chocar violentamente con 
el medio circundante, producen los llamados objetos Herbig-Haro (HH).  
\item Finalmente, la etapa en la cual la protoestrella se sit'ua en la 
secuencia principal de edad cero (ZAMS, por \emph{Zero Age Main Sequence}).
Es en este momento cuando comienza la combusti'on eficiente de hidr'ogeno 
en el n'ucleo de la estrella.
\end{enumerate}

\begin{figure}
\begin{center}
\includegraphics[angle=0, width=0.6\textwidth]{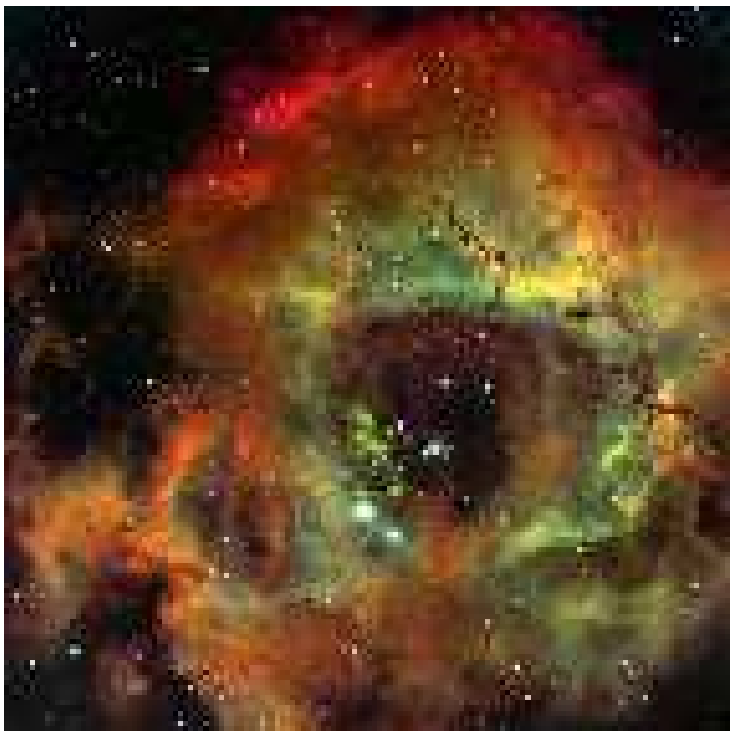}
\caption{Regi'on de formaci'on estelar masiva conocida como Messier~20. 
Esta imagen ha sido tomada en el IR por el sat'elite Spitzer.}\label{msfr}
\end{center}
\end{figure}

Por otro lado, el mecanismo de formaci'on de las estrellas de gran masa 
no es tan claro hoy en d'ia. 
Si bien no hay dudas de que las estrellas tempranas 
se formar'ian en nubes moleculares
gigantes, el proceso de fragmentaci'on de dicha nube para formar 
grumos autogravitantes de gran masa es a'un desconocido, como as'i tambi'en 
los subsecuentes
procesos din'amicos que dar'ian origen a estas estrellas.
Se han planteado dos modelos para explicar la formaci'on de las estrellas
de gran masa: 
uno es aquel en el cual el mecanismo de formaci'on es similar al que 
opera en las 
estrellas de baja masa (Shu et al. 1987), mientras que el
otro 
es un modelo tipo jer'arquico donde la coalescencia de estrellas menos
masivas dar'ia origen a estrellas con m'as masa (Bonnell et al. 1998). 
Observaciones recientes 
de regiones de formaci'on estelar gigantes y masivas 
(ver la Figura~\ref{msfr}) han detectado la presencia
de discos de acreci'on  y de flujos bipolares emanando de protoestrellas 
de gran masa, como describiremos a continuaci'on.

{\bf C'umulos:}
 Observaciones en radio de regiones de formaci'on estelar masivas de
la Galaxia y de galaxias cercanas muestran que las estrellas de gran 
masa se forman en grupos. 
 Las fuentes IRAS\footnote{Se conocen como fuentes IRAS a aquellas detectadas 
por el sat'elite IR que lleva dicho nombre.} m'as luminosas asociadas 
a regiones compactas de gas ionizado (y por ende a objetos estelares 
j'ovenes), muestran una morfolog'ia compleja cuando se las observa
en frecuencias radio. Garay y colaboradores ($1993$) sugieren que la
presencia de estrellas embebidas en la regi'on (un c'umulo de estrellas 
O y B) excitar'ian el
gas de la misma y esto producir'ia la compleja estructura
observada. 
Algunas de las regiones m'as estudiadas, que
contienen estrellas masivas, son W$3$ Orion-Trapezium y NGC $2024$.
La densidad
t'ipica de estos c'umulos de estrellas j'ovenes es de 
$\sim 10^4$ estrellas por $\rm{pc^{-3}}$ y el tama~no es de $\sim 0.2 -
0.4$ pc.

{\bf Discos de acreci'on:} Los discos detectados en regiones de
formaci'on de estrellas de gran masa tienen un di'ametro entre $0.1$ y $1$
pc, y sus masas pueden ir desde las $10$ a las 2000
$\rm{M_{\odot}}$. 
Una de las evidencias observacionales m'as importantes de la
existencia de discos circumestelares en regiones de formaci'on estelar
masiva es la detecci'on de la emisi'on de gas y polvo en el infrarrojo (IR).

\begin{figure}
\begin{center}
\includegraphics[angle=0, width=0.6\textwidth]{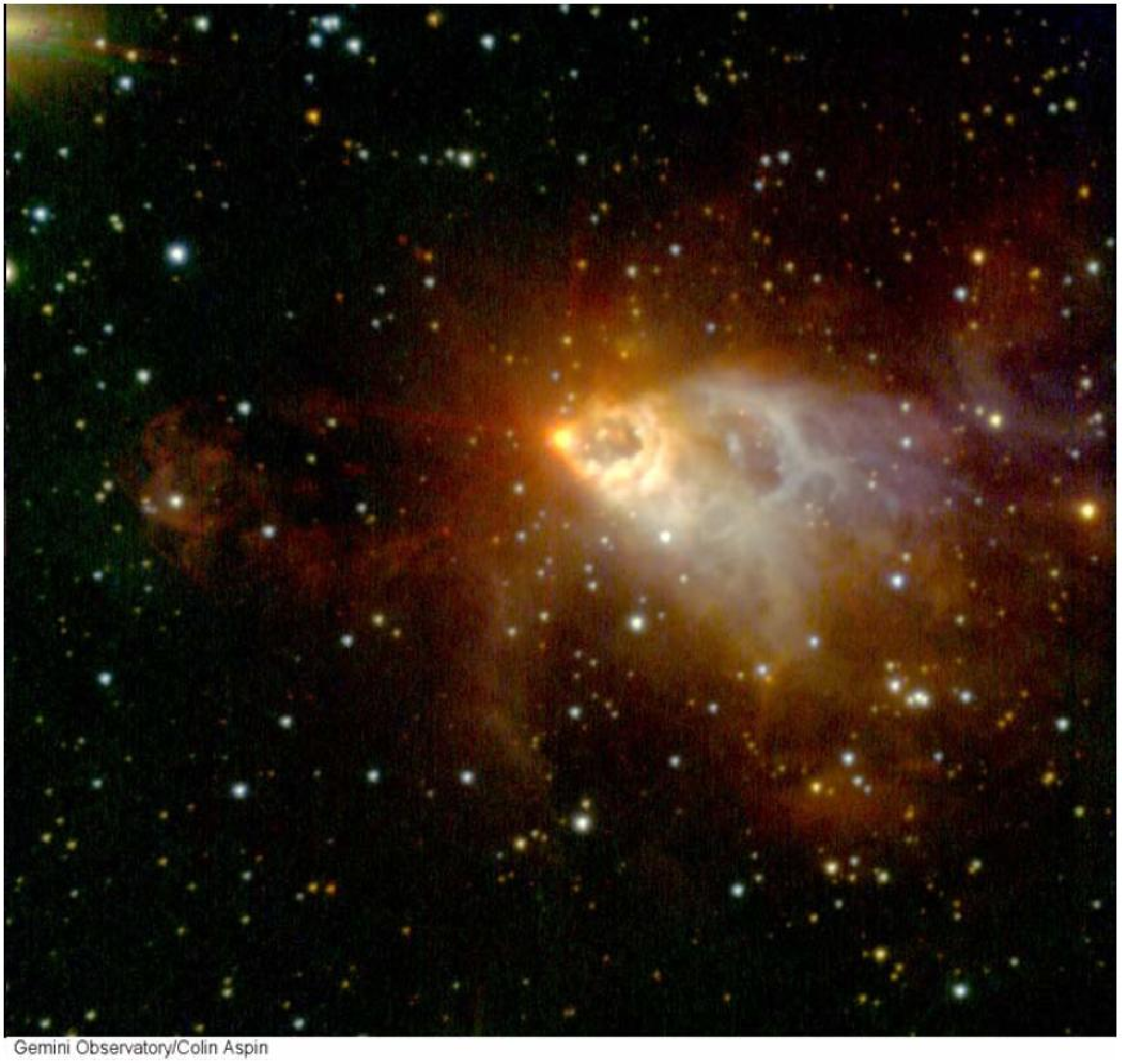}
\caption{\emph{Outflow} masivo asociado a la fuente AFGL~2591. La imagen
ha sido tomada en el IR por el telescopio
Gemini Norte (Zinnecker \& Yorke, 2007).}\label{outflow}
\end{center}
\end{figure}

{\bf \emph{Jets} y \emph{outflows}:} 
La presencia de flujos bipolares  en la formaci'on  de
estrellas de gran masa fue determinada recientemente mediante observaciones
en radio. 
 Sobre una muestra de aproximadamente 120 regiones de formaci'on de estrellas
tempranas, en el 90~\% de los casos  se observ'o gas  moviendos'e a
velocidades altas ($> 100$~km~s$^{-1}$). Luego, si las  
velocidades elevadas del gas est'an
asociadas a la presencia de \emph{outflows}, puede concluirse que
'estos son comunes tambi'en en la formaci'on de las estrellas de gran masa.
Algunas de las propiedades de los \emph{outflows} detectados en
regiones de formaci'on de estrellas tempranas son las siguientes:
\begin{itemize}
\item Tienen una masa promedio de $130 \;\rm{M_{\odot}}$, aunque en
algunos caso puede alcanzar valores de hasta $\sim 4800 \;\rm{M_{\odot}}$.
\item La tasa de p'erdida de masa abarca desde $\sim 3\times 10^{-5}
\rm{M_{\odot}}$ $\rm{yr^{-1}}$ hasta  $\sim 3\times 10^{-2}
\rm{M_{\odot}}$ $\rm{yr^{-1}}$.
\item La energ'ia cin'etica toma valores desde  $\sim 10^{46}$
hasta $\sim 6\times 10^{48}$~erg.
\end{itemize}
Estas cantidades son $\sim 100$ veces mayores que las correspondientes a los
\emph{outflows} presentes durante la formaci'on de estrellas de baja
masa. Luego puede concluirse que los \emph{outflows} de alta masa
inyectan mayor cantidad de energ'ia al medio circundante que los de
baja masa. En la Figura~\ref{outflow} se muestra el \emph{outflow} detectado
en la fuente AFGL~2591.

Los resultados observacionales descriptos anteriormente sugieren que 
las estrellas de gran masa se forman a trav'es de un mecanismo similar 
al que opera 
para formar las estrellas de baja masa, pero con una tasa de acreci'on
de materia a trav'es de los discos y de eyecci'on a trav'es de los 
\emph{outflows}
$\sim 100$ veces mayor a la observada en estrellas de baja masa.   
Sin embargo, no se han detectado discos de acreci'on en estrellas
de $M_{\star} > 50\,M_{\odot}$, las cuales al ser tan luminosas
($L_{\star}  > 10^5\,L_{\odot}$) podr'ian evaporar los discos a trav'es de 
un mecanismo conocido como ``fotoerosi'on''. 
Zinnecker \& Yorke (2007) sugieren que el mecanismo para formar a las estrellas 
m'as masivas ($M_{\star} > 50\,M_{\odot}$) no ser'ia un 
simple escaleo del proceso de formaci'on de las de baja masa. 
Los fuertes vientos de las estrellas m'as masivas (junto a la fuerte
radiaci'on) disipar'ian el disco y favorecer'ian las 
interacciones entre protoestrellas cercanas, siendo entonces la 
coalescencia de estrellas o protoestrellas el mecanismo de formaci'on.  

Por todo lo expuesto anteriormente, es claro que todav'ia hay mucho para 
decir respecto de como se forman las estrellas de gran masa. 
Todos los estudios y observaciones de regiones de formaci'on estelar masivas
son en general en las bandas de radio, IR y rayos~X, mientras que 
a m'as altas energ'ias  s'olo se han observado algunos c'umulos y asociaciones 
de estrellas O y B. Hasta el momento, no se han detectado estrellas en
formaci'on en el rango de los rayos gamma, aunque posiblemente se han
detectado rayos gamma de estrellas de gran masa evolucionadas, 
en general Wolf-Rayet (WR) o estrellas luminosas variables azules (LBV, por
\emph{Luminous Blue Variable}). Las regiones en las que se ha detectado 
emisi'on  significativa en rayos gamma son  Westerlund~2 (Aharonian et al. 
2007), Cygnus~OB2
y Carina (Tavani et al. 2009a).  
Las part'iculas relativistas que emiten esta radiaci'on gamma
detectada podr'ian acelerarse en las ondas de choque producidas por las 
explosiones de supernovas dentro de la regi'on o bien en choques de vientos 
de estrellas de gran masa. 
Sin embargo, estos c'umulos o asociaciones de estrellas tempranas tambi'en  
pueden ser regiones de formaci'on estelar, aunque 
nada se hab'ia dicho de la posible 
emisi'on a altas energ'ias de las protoestrellas masivas. 
Los rayos gamma, de ser producidos
en regiones de  formaci'on estelar, aportar'ian informaci'on adicional a la 
que se obtiene de otras frecuencias sobre como es el mecanismo de formaci'on 
de las estrellas de gran masa, motivando esto el primer estudio realizado 
en esta tesis. 

Sabemos que para emitir rayos gamma es necesaria la presencia de part'iculas 
no t'ermicas. Una de las evidencias de la existencia de electrones relativistas
es la detecci'on en frecuencias radio de emisi'on no t'ermica. 
Se han detectado en radio   
muchos YSOs de gran masa, pero s'olo algunos con emisi'on no t'ermica. 
Para nuestro estudio hemos elegido una fuente muy particular detectada en 
radio (no t'ermico) por Garay y colaboradores (2003). 
Estos autores han detectado en frecuencias
radio un sistema triple compuesto por una protoestrella de gran masa y dos
l'obulos. Este sistema se asocia con la fuente infrarroja
IRAS 16547-4247.

\section{La fuente IRAS 16547-4247}

La fuente IRAS~16547-4247 es una regi'on de formaci'on estelar de gran masa
ubicada a una distancia $d_{\rm nm} \sim 2.9$~kpc y tiene un di'ametro 
angular de $\sim 27''$ (Garay et al. 2003), que a la distancia $d_{\rm nm}$ 
corresponde a un tamaño lineal 
$D_{\rm nm} \sim 0.38$~pc ($\sim 1.1\times10^{18}$~cm). 
La luminosidad bolom'etrica es
$L_{\star} \sim 6.2\times10^4 L_{\odot} \sim 2.4\times10^{38}$~erg~s$^{-1}$
donde $L_{\odot} = 4\times10^{33}$~erg~s$^{-1}$ es la luminosidad bolom'etrica 
del Sol. 
De la densidad columnar de hidr'ogeno estimada, resulta que la 
masa de la nube es $M_{\rm nm} \sim 9\times10^2\,M_{\odot} \sim 10^{36}$~gr,
siendo $M_{\odot} \sim 2\times10^{33}$~gr una masa solar, y as'i la densidad
num'erica promedio de part'iculas resulta 
$n_{\rm nm} \sim 5.2\times10^5$~cm$^{-3}$. 
Los par'ametros conocidos y estimados de esta fuente est'an listados en
la Tabla~\ref{Param_YSO}.

\begin{table}[h]
\begin{center}
\begin{tabular}{ll}
\hline
\hline
Par'ametro & Valor \\
\hline
Distancia & $d_{\rm nm} = 2.9$~kpc\\
Tamaño & $D_{\rm nm} = 0.38$~pc\\
Masa & $M_{\rm nm} = 9\times 10^2\;M_{\odot}$\\
Densidad & $n_{\rm nm} = 5.2\times 10^5\;\rm{cm}^{-3}$\\
Luminosidad & $L_{\star} = 6.2\times 10^4 L_{\odot}$ \\
\hline
\end{tabular}
\caption{Par'ametros de la fuente IRAS~16547-4247.}\label{Param_YSO}
\end{center}
\end{table}

\begin{figure}
\begin{center}
\includegraphics[angle=0, width=0.45\textwidth]{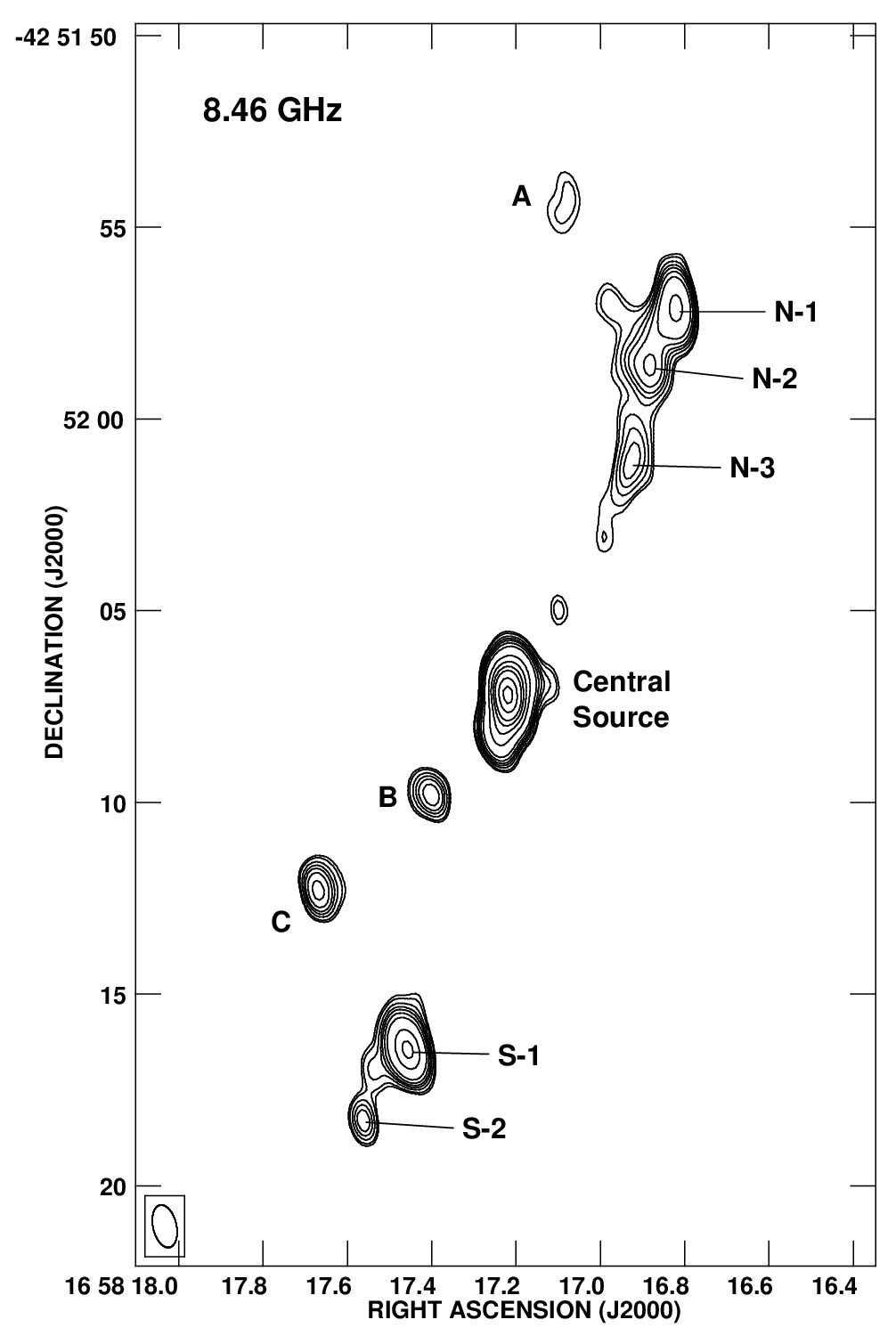}
\includegraphics[angle=0, width=0.45\textwidth]{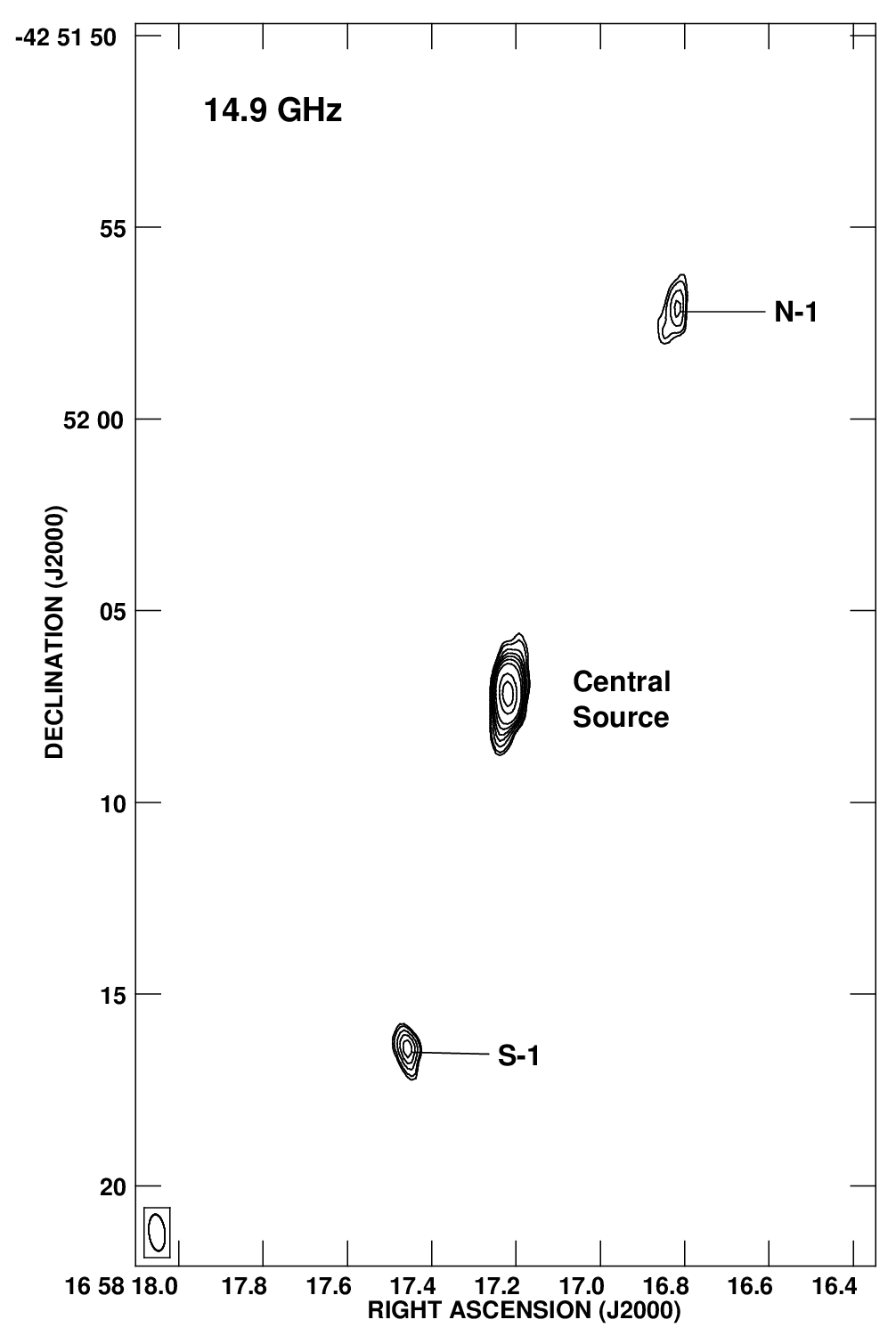}
\caption{Im'agenes tomadas con VLA de la fuente IRAS~16547-4247 en
$\nu =$ 8.46 
(izquierda) y 14.9~GHz (derecha). A 'estas frecuencias se ve claramente el
sistema triple compuesto por la fuente central (\emph{central source}) y 
los l'obulos norte (N) y sur (S) (Rodr'iguez et al. 2005).}
\label{Garay_figs}
\end{center}
\end{figure}

Mediante observaciones llevadas a cabo con el interfer'ometro ATCA
(\emph{Australia Telescope Compact Array}),
Garay y colaboradores (2003) detectaron un sistema triple embebido en la
nube molecular asociada a la fuente IRAS 16547-4247. Este sistema
est'a formado por una protoestrella de gran masa y dos l'obulos
alineados con 'esta y ubicados sim'etricamente a una distancia $\sim 0.14$~pc
($\sim 10^{''}$) de la protoestrella central, como se muestra 
en la Figura~\ref{Garay_figs}. 
Con observaciones posteriores realizadas con el radiotelescopio 
VLA (\emph{Very Large Array})
por Rodr'iguez y colaboradores (2005) a las frecuencias $\nu = 8.46$ y
14.9~GHz, se calcul'o el 'indice espectral $\alpha$ del flujo observado 
$S_{\nu}\,(\propto \nu^{-\alpha}$) de cada componente del sistema triple. 
Para la fuente central se estim'o $\alpha = -0.33 \pm 0.05$ mientras que
para los l'obulos norte y sur, $\alpha = 0.17 \pm 0.39$ y  $0.59 \pm 0.15$,
respectivamente. Estos valores son muy similares a los calculados 
por Garay y colaboradores (2003) con los datos de ATCA. 

Si la emisi'on continua detectada en radio es producida 'unicamente a trav'es
de interacciones libre-libre, entonces el espectro observado debe 
corresponder a un valor 
$\alpha < 0.1$, independientemente de las caracter'isticas del emisor.
Por otro lado, valores $\alpha \gg 0.1$ son debidos a radiaci'on sincrotr'on 
'opticamente delgada (Rodr'iguez et al. 1993). 
Con este criterio,
la emisi'on radio de la fuente central ser'ia t'ermica, asociada a la 
radiaci'on libre-libre de la base del jet y que se conoce como \emph{jet} 
t'ermico, mientras que la emisi'on proveniente de los l'obulos, en
particular del sur, ser'ia no t'ermica.  
Sin embargo, para confirmar que la emisi'on de los
l'obulos es no t'ermica, deben hacerse estudios polarim'etricos.

\begin{figure}
\begin{center}
\includegraphics[angle=0, width=0.5\textwidth]{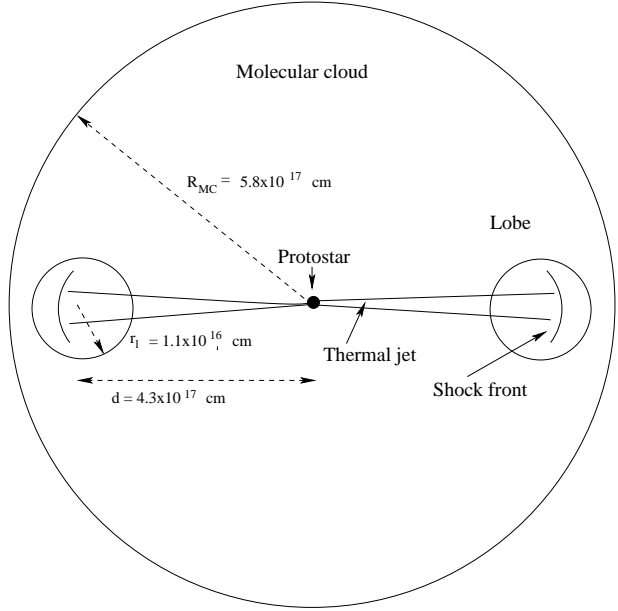}
\caption{Esquema del modelo propuesto para la fuente IRAS~16547-4247
(Araudo et al. 2007).}
\label{yso_model}
\end{center}
\end{figure}

La detecci'on de emisi'on no t'ermica proveniente de los l'obulos
es una evidencia de la
presencia de electrones relativistas en tal lugar (donde esta radiaci'on
es producida). 
Estas part'iculas ser'ian aceleradas hasta energ'ias relativistas en los 
choques producidos cuando el \emph{jet} es frenado  por el medio en el 
cual se est'a propagando, es decir, la nube molecular. 
En esta tesis modelamos a la fuente como se muestra en la 
Figura~\ref{yso_model}. 

Adem'as de radiaci'on sincrotr'on, los electrones relativistas podr'ian
producir fotones a m'as altas energ'ias por otros mecanismos radiativos como 
el Bremsstrahlung relativista y la dispersi'on Compton inversa. 
Por otro lado, debido a que el mecanismo de Fermi puede acelerar
tambi'en protones, las interacciones $pp$ entre protones
acelerados y protones fr'ios de la nube son tambi'en posibles. Como
consecuencia de esto se producen, adem'as de los fotones, pares $e^{\pm}$ 
secundarios los cuales se enfr'ian de la misma manera que los 
electrones primarios. De esta menera, podr'iamos tener tres poblaciones
diferentes de part'iculas no t'ermicas: electrones y protones primarios
 y pares secundarios  $e^{\pm}$. A continuaci'on estudiamos como se 
producen estas poblaciones de part'iculas  relativistas.

\section{Poblaci'on de part'iculas relativistas}

La emisi'on detectada en frecuencias radio proveniente de los l'obulos
de la fuente IRAS~16547-4247 es no t'ermica (de acuerdo al criterio 
determinado por Rodr'iguez et al. 1993) y consideramos que dicha 
radiaci'on es producida por electrones relativistas acelerados 
en los choques teminales de los \emph{jets} del YSO o por pares $e^{\pm}$
producidos por protones  relativistas acelerados en los mismos choques. 
Mientras los \emph{jets}  se propagan (libremente) por la nube
molecular van chocando el material de 'esta 
produciendo as'i un \emph{bow shock} en el medio.
Cuando los \emph{jets} son frenados significativamente por este medio 
externo chocado, se forma un choque reverso en los  \emph{jets},
como se coment'o  en la secci'on~\ref{pert-globales}.
A continuaci'on haremos una descripci'on m'as detellada que la expuesta
en el Cap'itulo~\ref{cap2}, estudiando espec'ificamente el caso
de los YSOs, para luego estimar la poblaci'on de part'iculas
relativistas en los l'obulos de la fuente IRAS~16547-4247.
En particular, nos concentramos en el estudio del l'obulo sur, ya que la 
emisi'on
radio proveniente de 'este tiene un indice espectral puramente no t'ermico,
$\alpha \sim 0.6$, mientras que la emisi'on del l'obulo norte podr'ia estar
contaminada de radiaci'on t'ermica, ya que el $\alpha$ ($\sim 0.2$) 
es m'as cercano al valor cr'itico $\alpha_{\rm c} = 0.1$.

\subsection{Choques terminales}

Para determinar las caracter'isticas de los choques terminales, un 
par'ametro importante es el contraste entre la densidad del \emph{jet} y 
de la nube, $\chi_{\rm yso} \equiv n_{\rm j}/n_{\rm nm}$.
Adem'as de $\chi_{\rm yso}$ es necesario conocer la velocidad del \emph{jet}, 
$v_{\rm j}$, para poder
estimar las velocidades del \emph{bow shock} ($v_{\rm bs}$) y del 
choque reverso ($v_{\rm r}$), de la siguiente manera (Bosch-Ramon et al. 2010):
\begin{equation}
\label{v_bs-v_rs}
v_{\rm bs} \sim \frac{v_{\rm j}}{(1 + \chi_{\rm yso}^{-1/2})} \qquad
{\rm y} \qquad
v_{\rm r} \sim  v_{\rm j} - \frac{3}{4}\, v_{\rm bs},  
\end{equation}
donde se ha supuesto que el  \emph{bow shock} es adiab'atico.
Debido a que la velocidad de los \emph{jets} de la fuente IRAS~16547-4247 no se 
conoce, adoptamos para esta magnitud un valor que sea concordante con 
aquellos medidos en otros YSOs de gran masa.
Mart'i y colaboradores (1995) han determinado para el \emph{jet} t'ermico de 
la fuente HH~80-81 una velocidad que ronda los $600$ y $1400$~km~s$^{-1}$.
Luego, para la fuente que estamos estudiando en este cap'itulo consideramos 
$v_{\rm j} = 1000$~km~s$^{-1}$.

Si bien no conocemos el valor de $\chi_{\rm yso}$, porque no conocemos
$n_{\rm j}$ en la localizaci'on de los l'obulos 
($Z_{\rm j} \sim 4\times10^{17}$~cm), podemos determinar 
$v_{\rm bs}$ a trav'es del tiempo de vida del \emph{jet}, $\tau_{\rm j}$. 
Esta escala temporal se define como aquella en la cual el \emph{jet}
recorre la distancia $Z_{\rm j}$ a la que se detectan los l'obulos. Esto es,
$\tau_{\rm j} \sim Z_{\rm j}/v_{\rm bs}$ y en el caso de la fuente
IRAS~16547-4247, $\tau_{\rm j} \sim 10^{11}$~s (Garay et al. 2007)
con lo cual, si $Z_{\rm j} \sim 4.3\times10^{17}$~cm obtenemos 
$v_{\rm bs} \sim 5\times10^6$~cm~s$^{-1}$.
Con este valor de $v_{\rm bs}$ y las ecuaciones~(\ref{v_bs-v_rs}) hallamos 
$\chi_{\rm yso} \sim 2.5\times10^{-3}$ y luego 
$n_{\rm j}(Z_{\rm j}) \sim 1.25\times10^3$~cm$^{-3}$.   
Siendo $v_{\rm bs} \ll v_{\rm j}$, resulta que $v_{\rm r} \sim v_{\rm j}$.

La regi'on entre ambos choques puede tener un estructura muy compleja ya 
que estos choques
pueden ser radiativos y en la superficie de discontinuidad entre
ambos medios chocados (conocida como ``superficie de trabajo'') se pueden 
desarrollar inestabilidades de RT que mezclan material de ambos medios
(Blondin et al. 1989). 
Para estimar si los choques son radiativos o no, comparamos el 
tiempo de enfriamiento por emisi'on t'ermica, $t_{\rm ter}$, con 
el tiempo caracter'istico $t_{\rm carac}$ en el cual el medio chocado 
recorre una distancia $\sim R_{\rm lob}$.
Este 'ultimo resulta $t_{\rm carac} \sim 4\,R_{\rm lob}/v_{\rm ch}$.
Para estimar $t_{\rm ter}$ consideramos la ecuaci'on~(\ref{t_ter}) y 
obtenemos (Bosch-Ramon et al. 2010):
\begin{equation}
t_{\rm rad} \sim 2\times10^{-25} \frac{v_{\rm ch}^2}{n_0 \,\Lambda(T)} 
\sim 7\times10^{-18} \,\frac{v_{\rm ch}^{3.2}}{n_0} \,\,{\rm s},
\end{equation}
donde hemos considerado la funci'on de enfriamiento
$\Lambda(T) = 7\times10^{-19}\, T^{-0.6}$ (Bosch-Ramon et al. 2010) y 
las condiciones de salto 
(\ref{chf_nv}) y (\ref{chf_pT}). Luego, comparando $t_{\rm rad}$
con $t_{\rm carac}$ tenemos que los choques son radiativos si 
$t_{\rm rad} < t_{\rm carac}$, es decir, si
\begin{equation}
v_{\rm ch} < \left(\frac{4\,R_{\rm lob}\, n_0}{7\times10^{-18}}\right)^{1/4.2}
%\sim 2.4\times10^4 (R_{\rm lob}\,n_0)^{1/4.2}\,\,{\rm cm\,s^{-1}}.
\sim 1.6\times10^8\,n_0^{1/4.2}\,\,{\rm cm\,s^{-1}}.
\end{equation}
Considerando $v_{\rm ch} = v_{\rm bs}$ y $n_0 = n_{\rm nm}$ para el 
\emph{bow shock} y $v_{\rm ch} = v_{\rm r}$ y $n_0 = n_{\rm j}(Z_{\rm j})$ 
para el choque reverso,
obtenemos que mientras que el primero resulta radiativo, el 'ultimo es
adiab'atico. Luego la densidad del material chocado de la nube aumenta m'as
de un factor 4, pudiendo llegar hasta valores mucho m'as grandes que
el valor adiab'atico.
Con esto, si el factor de mezcla de ambos medios chocados es grande,
podemos considerar que las part'iculas aceleradas en el choque reverso
pueden interactuar, adem'as de con el material del \emph{jet} chocado, 
con material 
chocado de la nube, cuya densidad es mucho m'as alta y as'i las p'erdidas
por Bremsstrahlung relativista e interacciones $pp$ mucho m'as eficientes.  

En resumen, tenemos que los l'obulos son regiones en donde hay choques 
fuertes que 
pueden acelerar part'iculas hasta energ'ias relativistas. Estas part'iculas,
adem'as de producir la radiaci'on sincrotr'on observada, pueden tambi'en radiar 
por otros mecanismos no t'ermicos, como veremos a continuaci'on.

\subsection{Aceleraci'on de part'iculas y p'erdidas radiativas}
\label{Accel_part_yso}

Como vimos en el Cap'itulo~\ref{cap2}, la eficiencia para acelerar 
part'iculas
mediante el mecanismo de Fermi de tipo I es $\propto v_{\rm ch}^2$.
Por esto, nos concentramos en la aceleraci'on de part'iculas en el
choque reverso, ya que $v_{\rm r} \gg v_{\rm bs}$.

El tiempo de aceleraci'on depende, adem'as de $v_{\rm r}^2$,  
del campo magn'etico en la regi'on de aceleraci'on, que 
coincide con el l'obulo sur, y denotaremos $B_{\rm lob}$. 
Para estimar $B_{\rm lob}$ consideramos que la densidad de energ'ia
magn'etica, $B_{\rm lob}^2/(8\pi)$, es igual a la densidad de energ'ia 
$u_{\rm nt}$ de las part'iculas no t'ermicas. 
Sabemos que hay leptones relativistas en el l'obulo que producen la emisi'on
sincrotr'on detectada en frecuencias radio. Estos leptones pueden ser 
electrones primarios ($e_1$) acelerados en el choque reverso o bien 
pares $e^{\pm}$ ($e_2$) producidos por interacciones inel'asticas $pp$ 
de protones ($p$) 
acelerados en el mismo choque con el material de la nube molecular. 
El caso m'as general que podemos plantear es 
\begin{equation}  
\label{equip_yso}
\frac{B_{\rm lob}^2}{8\pi} = u_{\rm nt} = u_{e_1} + u_p + u_{e_2},
\end{equation}
donde $u_{e_1}$ y $u_p$ son las densidades de energ'ia de los 
electrones y protones primarios, respectivamente, y
$u_{e_2}$ corresponde a los  pares $e^{\pm}$.
En cada caso  ($i = e_1, p, e_2$):
\begin{equation}  
u_{i}=\int_{E_i^{\rm min}}^{E_i^{\rm max}} E_i \,n(E_i) \,{\rm d}E_i,
\end{equation}
siendo $n(E_i)=k_i^{\prime} E_i^{-p_i'}\exp(-E_i/E_i^{\rm{max}})$ la
distribuci'on de energ'ia de las part'iculas $i$ por unidad de volumen,
es decir, $[n(E_i)] =$ cm$^{-3}$~erg$^{-1}$ y $E_i^{\rm min}$ y $E_i^{\rm max}$
las energ'ias m'inima y m'axima de la distribuci'on $n(E_i)$.
Como no tenemos ning'un indicio observacional de la presencia de protones 
relativistas en el l'obulo sur, la 'unica manera de estimar $u_p$ es a 
trav'es de un par'ametro
fenomenol'ogico $a$, imponiendo que $u_p = a u_{e_1}$. En esta tesis
consideramos 3 casos: $a = 0, 1$ y $100$.
El valor $a = 100$ se toma porque el espectro de los rayos c'osmicos
gal'acticos pareciera indicar que la fracci'on de protones a electrones es 
$\sim 100$. Por otro lado $u_{e_2} = f_{e^{\pm}} u_p$, donde $f_{e^{\pm}} < 1$ 
(Kelner et al. 2006). 

El espectro observado en radio del l'obulo sur puede ajustarse con una 
ley de potencias 
($\propto \nu^{-\alpha}$) con 'indice $\alpha = 0.59$. Luego, el 'indice 
espectral de los
leptones relativistas que producen esta radiaci'on sincrotr'on resulta
$p_e' = 2 \alpha + 1 = 2.18$.    
En los casos $a = 1$ y $a = 100$ consideramos adem'as
una distribuci'on $n_p (E_p)$  de protones relativistas cuya poblaci'on
de pares $e^{\pm}$ producidos en las interacciones $pp$ puede estimarse 
como se explic'o en el Cap'itulo~\ref{proc-rad}.
En el caso con $a = 1$, $u_{e_2} = f_{e^{\pm}}\,u_{e_1}$ con lo cual 
la emisi'on de los primarios es tambi'en mayor que la de los secundarios y 
ajustamos el espectro observado con la emisi'on de los electrones
acelerados en el choque reverso. Esto es, $p_{e_1}' = 2.18$ y fijando 
$p_p' = p_{e_1}'$ resulta $p_{e_2}' = 2.13$. 
Por otro lado, en el caso
con $a = 100$, resulta $u_{e_2} \sim 50\, u_{e_1}$ y as'i el espectro observado 
se ajusta con la emisi'on sincrotr'on producida por los pares. De esta manera, 
si $p_{e_2}' = 2.18$, entonces $p_p' = 2.27$ y por lo tanto $p_{e_1}' = 2.27$ 
tambi'en.

Fijados los valores de $p_i'$ para cada caso y las relaciones entre 
las densidades de energ'ia $u_i$, podemos determinar las constantes
$k_i^{\prime}$.  Para esto, nos valemos del flujo sincrotr'on observado en radio, 
cuya expresi'on   es la siguiente (Ginzburg \& Syrovatskii 1964): 
\begin{equation}
\label{flujo_obs}
S_{\nu} = 1.35\times10^{-22}\,a_{\rm sin}(p_i')\, \frac{k_i^{\prime} \,V_{\rm lob}
\,B_{\rm lob}^{(p_i'+1)/2}}{d_{\rm nm}^2} 
\left(\frac{6.26\times10^{18}}{\nu}\right)^{(p_i' - 1)/2}  ,  
\end{equation}  
donde $a_{\rm sin}(p_i')$ es una funci'on complicada tabulada 
para diferentes valores de $p_i'$. Para $p_i' = 2.18$ resulta 
$a_{\rm sin} \sim 0.1$, interpolando 
linealmente entre $a_{\rm sin}(2)$ y $a_{\rm sin}(2.5)$. El volumen
del l'obulo sur es $V_{\rm lob} \sim 5.6\times10^{48}$~cm$^{3}$ y $d_{\rm nm}$
es la distancia a la fuente.
Considerando que $S (\nu = 8.46\,{\rm GHz}) = 2.8$~mJy (Garay et al. 2003)
y la ecuaci'on~(\ref{equip_yso}) de equipartici'on de
la energ'ia es posible obtener $B_{\rm lob}$ y $k_i^{\prime}$ 
($i = e_1$ 'o $i = e_2$) 
para cada valor del par'ametro $a$. Luego, teniendo en cuenta las relaciones
entre las densidades de energ'ia ($u_p = a u_{e_1}$ y 
$u_{e_2} \sim f_{e^{\pm}} u_p$),
hallamos $k_p^{\prime}$ y la constante lept'onica que falta. 
En la Tabla~\ref{k_B} se listan los valores de $B_{\rm lob}$, 
$k_{e_1}^{\prime}$, $k_p^{\prime}$ y $k_{e_2}^{\prime}$ calculados para los 
3 valores de $a$ considerados. 

\begin{table*}[]
\begin{center}
\begin{tabular}{ccccc}
\hline
\hline
a  & $B_{\rm lob}$   & $k_{e_1}^{\prime}$ & $k_p^{\prime}$ & $k_{e_2}^{\prime}$ \\
{} & [G] & [erg$^{p_{e_1}'-1}~$cm$^{-3}$] & 
[erg$^{p_p'-1}$~cm$^{-3}$] & [erg$^{p_{\rm{e_2}}'-1}$~cm$^{-3}$] \\
\hline
0 & $2.0\times10^{-3}$ & $2.6\times10^{-9}$ & - & - \\
1   & $2.5\times10^{-3}$ & $2.0\times10^{-9}$ & $6.7\times10^{-9}$ & 
$6.5\times10^{-10}$  \\
100  & $3.0\times10^{-3}$ & $2.7\times10^{-11}$ & $1.7\times10^{-8}$ & 
$1.5\times10^{-9}$\\ 
\hline
\end{tabular}
\caption{Campo magn'etico y constantes de normalizaci'on ($k_i'$)
para los diferentes casos considerados en esta tesis.}\label{k_B}
\end{center}
\end{table*}

Conociendo el valor de $B_{\rm lob}$ podemos estimar el tiempo de 
aceleraci'on tanto para protones como para electrones primarios. 
Aunque no tenemos
informaci'on sobre la geometr'ia de $\vec B_{\rm lob}$ en los l'obulos, 
aqu'i adoptamos el valor m'as conservativo para la eficiencia acelerativa
($\vec B_{\rm lob} \parallel \vec v_{\rm ch}$)
y es por esto que 
consideramos la f'ormula~(\ref{t_acc_par}) para calcular el tiempo de 
aceleraci'on, resultando
\begin{equation}
t_{\rm ac} = 1.4\times10^{10}\;\frac{E_{e_1,p}}{B_{\rm lob}\, v_{\rm r}^2} \sim
1.4\times10^{-6}\;\frac{E_{e_1,p}}{B_{\rm lob}}\,{\rm s},
\end{equation}
si consideramos que $v_{\rm r} \sim v_{\rm j} \sim 1000$~km~s$^{-1}$. 

Como vimos en el Cap'itulo~\ref{cap2}, la aceleraci'on contin'ua hasta
que las part'iculas o se escapan del acelerador o bien se enfr'ian por 
p'erdidas radiativas. En el primer caso, el tiempo
de escape  $t_{\rm esc}$
est'a dado por la ecuaci'on~(\ref{t_esc}), siendo el tiempo de convecci'on
\begin{equation}
t_{\rm conv} \sim \frac{R_{\rm lob}}{v_{\rm bs}/4} \sim 4\times10^8\,\,{\rm s},
\end{equation}
y el tiempo de difusi'on en el r'egimen de Bohm resulta
\begin{equation}
t_{\rm dif} \sim 4.8\times10^{-20}\frac{B_{\rm lob}\,R_{\rm lob}^2}{E_i} \sim 
5.8\times10^{6}\frac{B_{\rm lob}}{E_i}\,\,{\rm s}.
\end{equation}
Adem'as de acelerarse y/o escaparse, las part'iculas primarias y tambi'en 
los pares secundarios pueden perder energ'ia a trav'es 
de interacciones sucesivas con otras part'iculas o campos.  

\subsubsection{P'erdidas lept'onicas}

Como vimos en el Cap'itulo~\ref{proc-rad}, las interacciones de leptones
relativistas con campos magn'eticos producen radiaci'on sincrotr'on,
cuyas p'erdidas resultan (ver ecuaci'on~(\ref{t_sin})):  
\begin{equation}
t_{\rm{sin}} \sim 4\times10^{2}\frac{1}{B_{\rm lob}^2 \,E_{e_1, e_2}}\;\;\rm{s}.
\end{equation}
Debido a que $B_{\rm lob}$ cambia con $a$, para cada valor de este par'ametro 
tenemos un valor diferente de $t_{\rm{sin}}$, al igual que de $t_{\rm ac}$.

Los leptones tambi'en pueden interactuar con los fotones IR producidos
por la protoestrella masiva. La densidad de energ'ia de estos fotones
es $u_{\rm{ph}} = L_{\star}/(4\pi\, R_{\rm nm}^2c)$, donde $L_{\star}$ 
es la luminosidad de
la protoestrella y $R_{\rm nm} = D_{\rm nm}/2$ es el radio de la nube. Resulta
$u_{\rm{ph}} \sim 1.84\times10^{-9}\; \rm{erg\;cm^{-3}}$ y los
fotones correspondientes a esta densidad tienen una energ'ia 
$E_{\rm ph\star} \sim 3 K_{\rm B} T_{\rm nm} \sim 6.6\times10^{-3}$~eV al estar reprocesados
por el polvo. Con fotones semilla
de tan baja energ'ia, las interacciones IC ocurrir'an en el r'egimen de
KN para energ'ias tan altas como 
$E_{e_1,e_2} > (m_e c^2)^2/E_{\rm ph\star} \sim 3.8\times10^{13}$~eV. 
Tanto en el r'egimen de Th como en KN la f'omula valida para el tiempo
de enfriamiento es la ecuaci'on~(\ref{t_ci}), que en el caso del l'obulo 
sur nos queda
\begin{equation} 
\label{t_ci_yso}
t_{\rm IC}  \sim 3.3\times10^{21}\,E_{\rm ph\star}
\frac{(1 + 8.3\,y)}{\ln(1+0.2\;y)}\frac{(1 + 1.3\,y^2)}{(1 + 0.5\,y + 1.3\,y^2)}
~\rm{s}\,,
\end{equation}
donde $y \equiv E_{\rm ph\star} E_{e_1,e_2}/(5.1\times10^5\,\rm{eV})^2$.

Por otro lado, debido a que la nube tiene una densidad alta,
$n_{\rm nm} \sim 5\times10^5$~cm$^{-3}$, las p'erdidas por Bremsstrahlung 
relativista ser'an importantes, con un tiempo caracter'istico 
(ver ecuaci'on~(\ref{t_Brem}))
\begin{equation}
t_{\rm Brem} \sim \frac{2.8\times10^{10}}{\ln\left(\frac{E_{e_1,e_2}}
{m_ec^2}\right) + 0.36}~\rm{s}\,.
\end{equation} 

\begin{figure}
\begin{center}
\includegraphics[angle=270, width=0.49\textwidth]{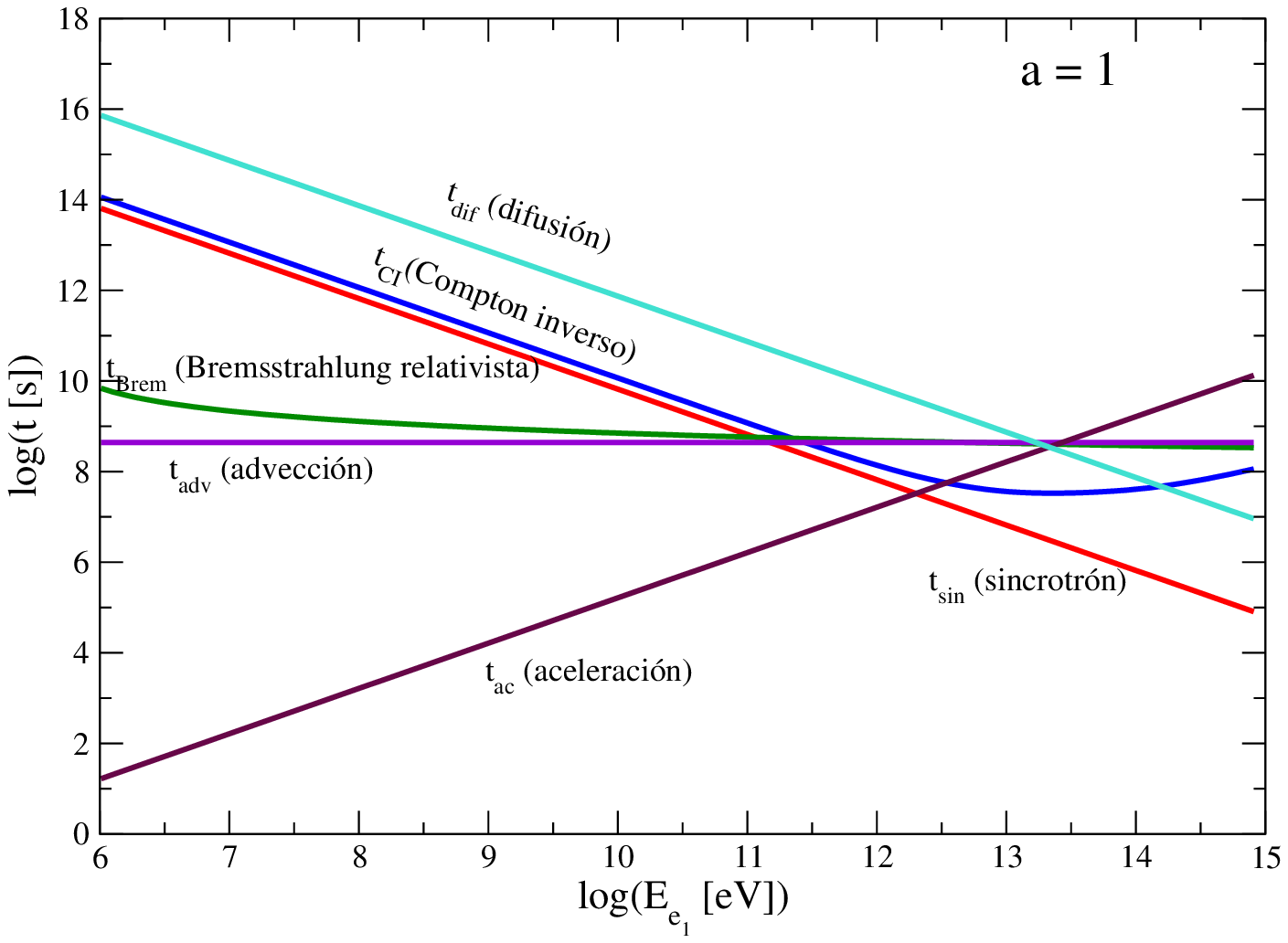}
\includegraphics[angle=270, width=0.49\textwidth]{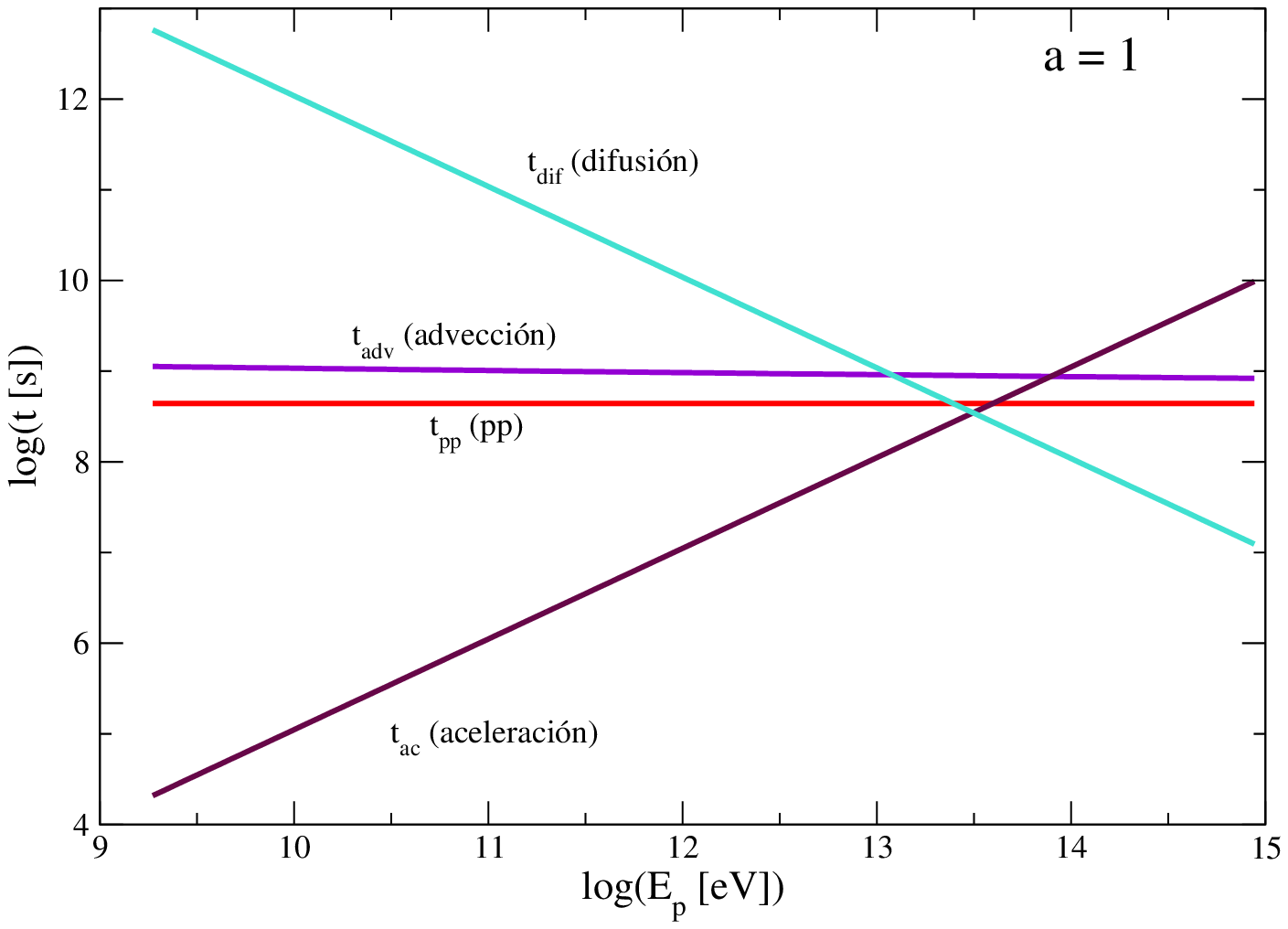}
\caption{Tiempos de aceleraci'on y de enfriamiento 
para los electrones primarios (izquierda) y para los protones (derecha).
Estos gr'aficos corresponden al caso con $a =1$, pero al resto de los casos
les corresponden figuras similares, ya que la 'unica diferencia est'a en 
el valor de $B_{\rm lob}$, que afecta a $t_{\rm ac}, t_{\rm sin}$ y $t_{\rm dif}$,
y 'este no cambia demasiado, como se muestra en la Tabla~\ref{k_B}.}  
\label{losses_YSO}
\end{center}
\end{figure}

Para obtener la energ'ia m'axima que pueden alcanzar los electrones 
acelerados en el l'obulo sur, comparamos las ganancias por aceleraci'on 
con las p'erdidas radiativas. Para esto hemos graficado los tiempos 
definidos antes como se muestra en la Figura~\ref{losses_YSO} (izquierda).
Como vemos en el gr'afico de la izquierda, los electrones alcanzan 
energ'ias m'aximas
$\sim 1$~TeV, para el caso $a=1$, como consecuencia del balance 
entre la aceleraci'on y las
p'erdidas por radiaci'on sincrotr'on. 
Como puede apreciarse tambi'en en la Figura~\ref{losses_YSO} (izquierda), para
$E_e \lesssim 3\times10^{10}$~eV, las p'erdidas por sincrotr'on dejan 
de dominar para hacerse m'as
importantes aquellas por Bremsstrahlung relativista. A esta energ'ia de 
quiebre, que llamamos $E_{\rm q}$, el espectro de electrones primarios 
sufre un cambio en su dependencia 
con $E_e$, ya que las p'erdidas dominantes cambian. El valor de $E_{\rm q}$
puede determinarse igualando $t_{\rm sin} = t_{\rm Brem}$, con lo cual
obtenemos $E_{\rm q} \sim 1.8\times10^5/B_{\rm lob}^2$~eV.
En la Tabla~\ref{E_max_YSO} listamos los
valores de $E_{e_1}^{\rm max}$ y $E_{\rm q}$ para los tres valores de $a$ 
considerados.

\begin{figure}
\begin{center}
\includegraphics[angle=270, width=0.45\textwidth]{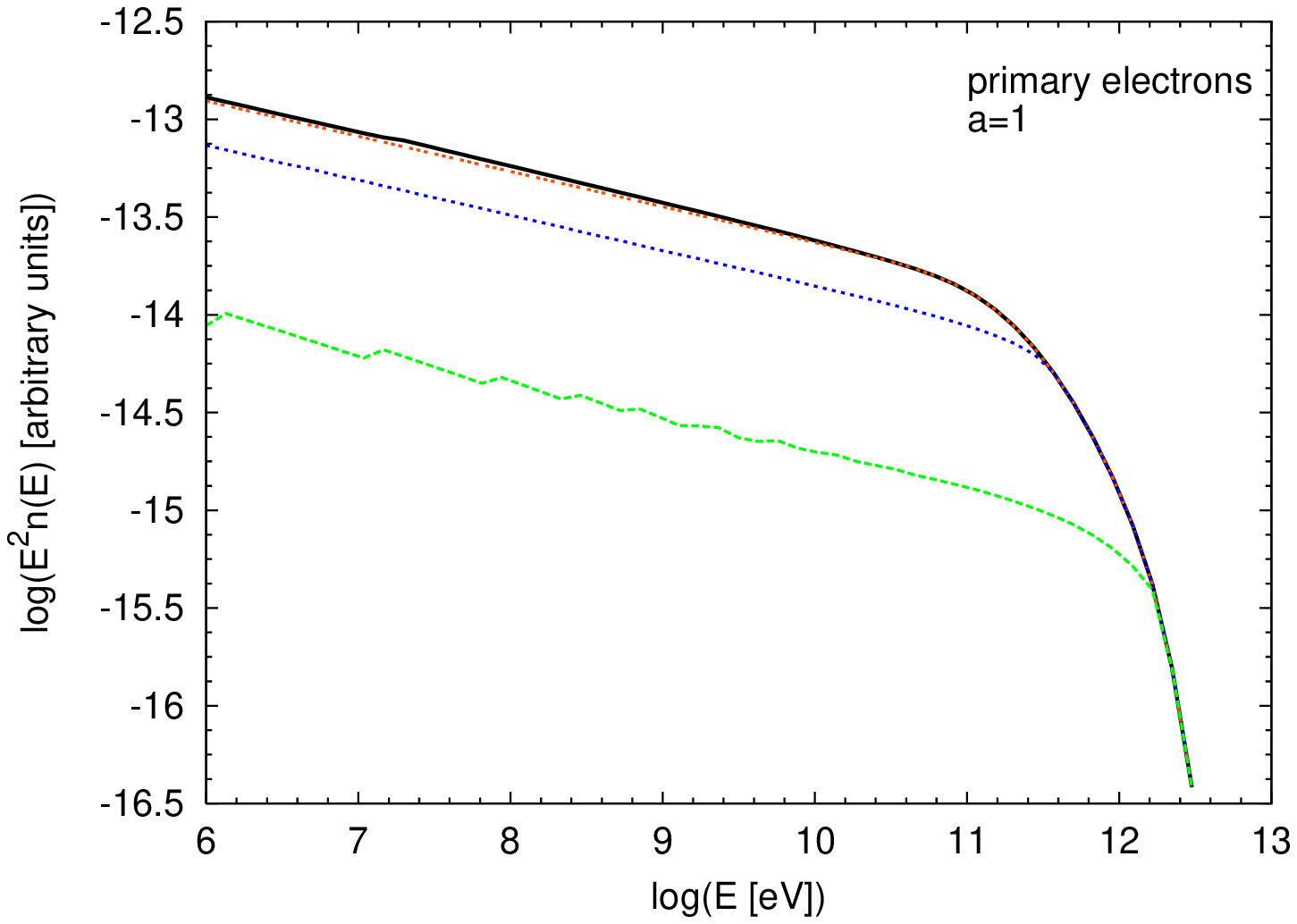}
\includegraphics[angle=270, width=0.45\textwidth]{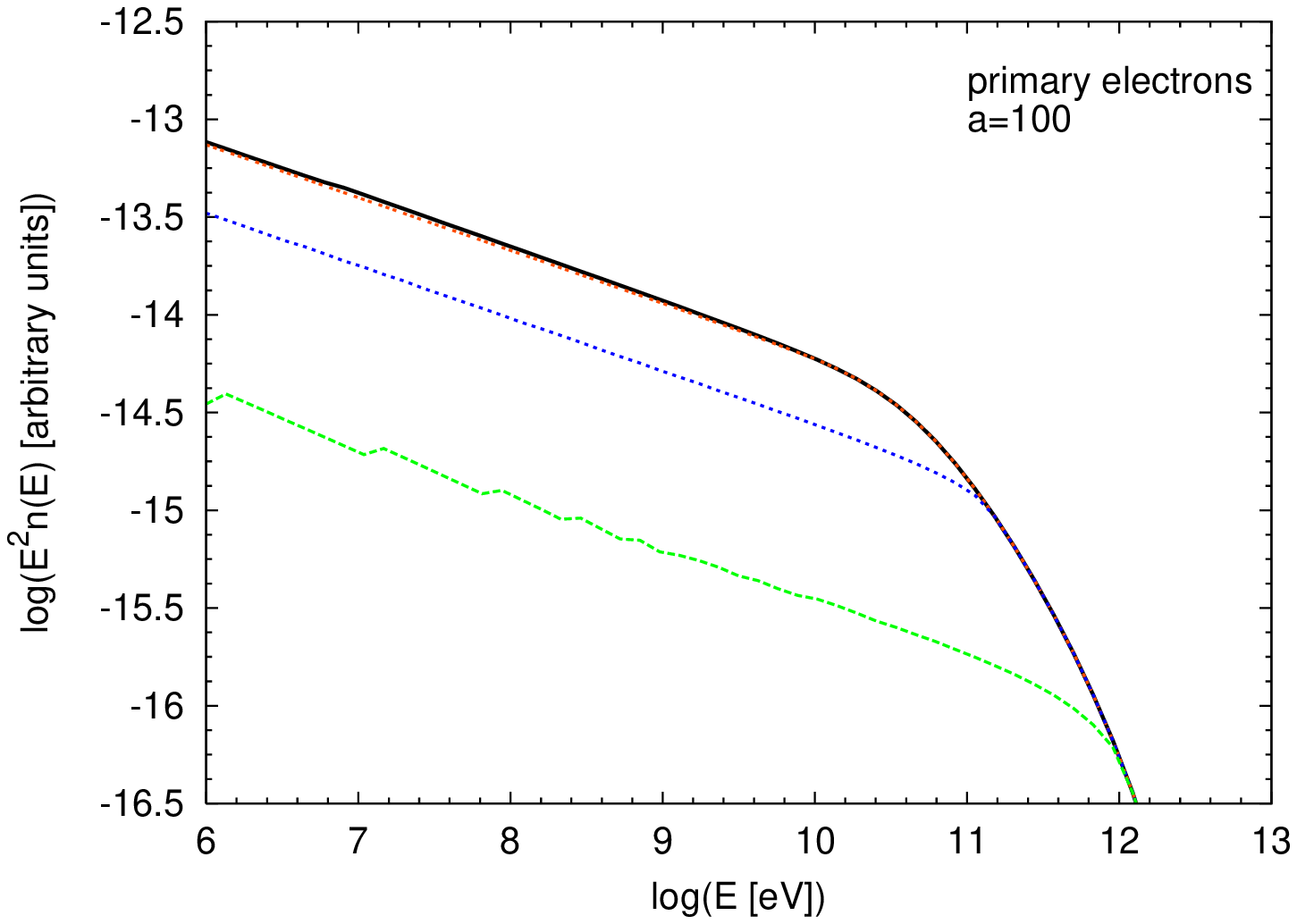}
\caption{Distribuciones de energ'ia de los electrones primarios.
En cada panel mostramos las distribuciones para diferentes tiempos de
inyecci'on de part'iculas ($\tau_{\rm inj}$). 
Las diferentes curvas corresponden a los siguientes valores  de
$\tau_{\rm inj}$: $10^7$ s (l'inea de rayas largas, verde), $10^8$ s (
l'inea de rayas y puntos, azul), $10^9$ s
(l'inea punteada, rojo) and $10^{10}$~s (l'inea s'olida, negro). 
Notar que para edades $> 10^9$~s el estado estacionario
es alcanzado (Araudo et al. 2007).}
\label{evol_YSO_eprimarios}
\end{center}
\end{figure}

Dado que el tiempo de vida de la fuente $\tau_{\rm vida}$ 
($\sim t_{\rm j} \sim 10^{11}$~s) es
mucho mayor que los tiempos de enfriamiento dominantes (sincrotr'on
y Bremsstrahlung relativista), el espectro de energ'ia de los 
electrones relativistas
inyectados en el l'obulo se encuentra en el estado estacionario. 
Es decir, toda la 
energ'ia de estas part'iculas relativistas se rad'ia, b'asicamente, por emisi'on
sincrotr'on y Bremsstrahlung relativista dentro de la fuente (el l'obulo sur), 
ya que los tiempos radiativos son similares al tiempo de escape
($t_{\rm esc} \sim 4\times10^8$~s). 
Para mostrar esto hemos calculado el espectro de part'iculas 
resultante de la inyecci'on $q_{e_1}(E_{e_1})$, a diferentes tiempos de 
inyecci'on $\tau_{\rm iny}$. 
Considerando $\tau_{\rm iny} = 10^7, 10^8, 10^9$ y $10^{10}$~s obtuvimos
los resultados que se muestran en la Figura~\ref{evol_YSO_eprimarios}. 
Finalmente, el espectro de electrones relativistas en el l'obulo sur
resulta 
\begin{equation}
\label{N_e}
n_e(E_e) = \left\{
\begin{array}{ll}
q_e(E_e)  \,t_{\rm Brem} \qquad& {\rm si} \, \, \, E_e \leq E_{\rm q} \\
q_e(E_e)  \,t_{\rm sin}  \qquad& {\rm si} \, \, \, E_e \geq E_{\rm q}.
\end{array} \right.
\end{equation}

\subsubsection{P'erdidas hadr'onicas}

En el caso de los protones, 'estos pierden energ'ia principalmente por
interacciones con los protones fr'ios de la nube molecular, cuya densidad es
$n_{\rm nm}$ y el tiempo de enfriamiento por $pp$ resulta: 
\begin{equation}
t_{pp} \sim 4.3\times10^9 \;\;\rm{s}.
\end{equation}
Como se muestra en la Figura~\ref{losses_YSO} (derecha), las p'erdidas 
por difusi'on son importantes solo a energ'ias muy altas, cercanas a 
$E_p^{\rm max}$. Esta 'ultima queda determinada entonces comparando el
tiempo de aceleraci'on con el de difusi'on, resultando 
$E_p^{\rm max} \sim 1.3\times10^{18}\,B_{\rm lob}^2$~eV, como se muestra en la 
Tabla~\ref{E_max_YSO}.

Considerando que el tiempo m'as relevante es $t_{pp}$, cuya dependencia con 
$E_p$ es despreciable (ver la Figura~\ref{losses_YSO}),  
el espectro de los protones relativistas resulta estacionario 
y con una dependencia con $E_p$ similar a la de la inyecci'on, es decir
\begin{equation}
n_p(E_p) \sim q_p(E_p) \,t_{pp} \propto E_p^{-p_p}.
\end{equation}

\begin{figure}
\begin{center}
\includegraphics[angle=270, width=0.45\textwidth]{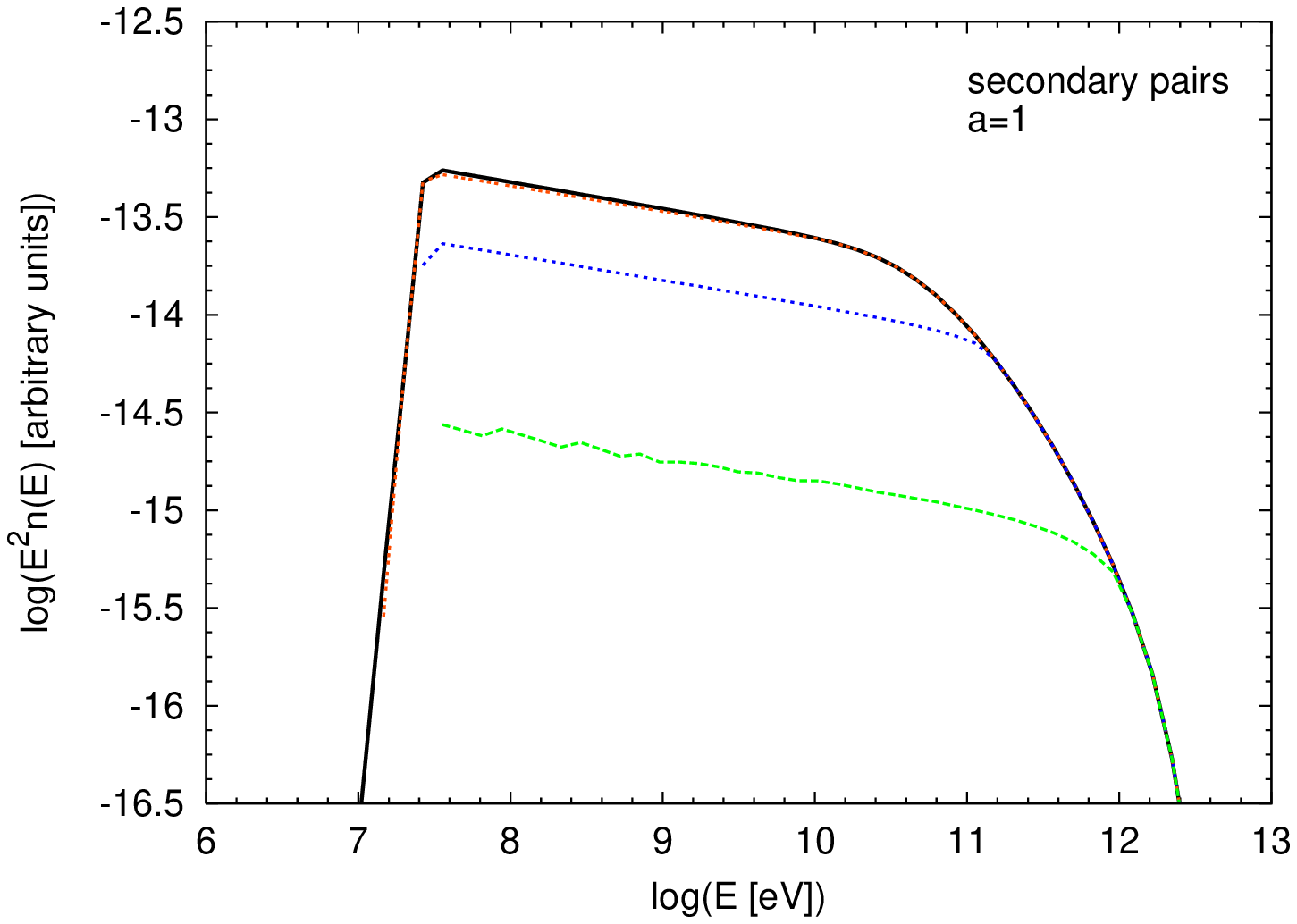}
\includegraphics[angle=270, width=0.45\textwidth]{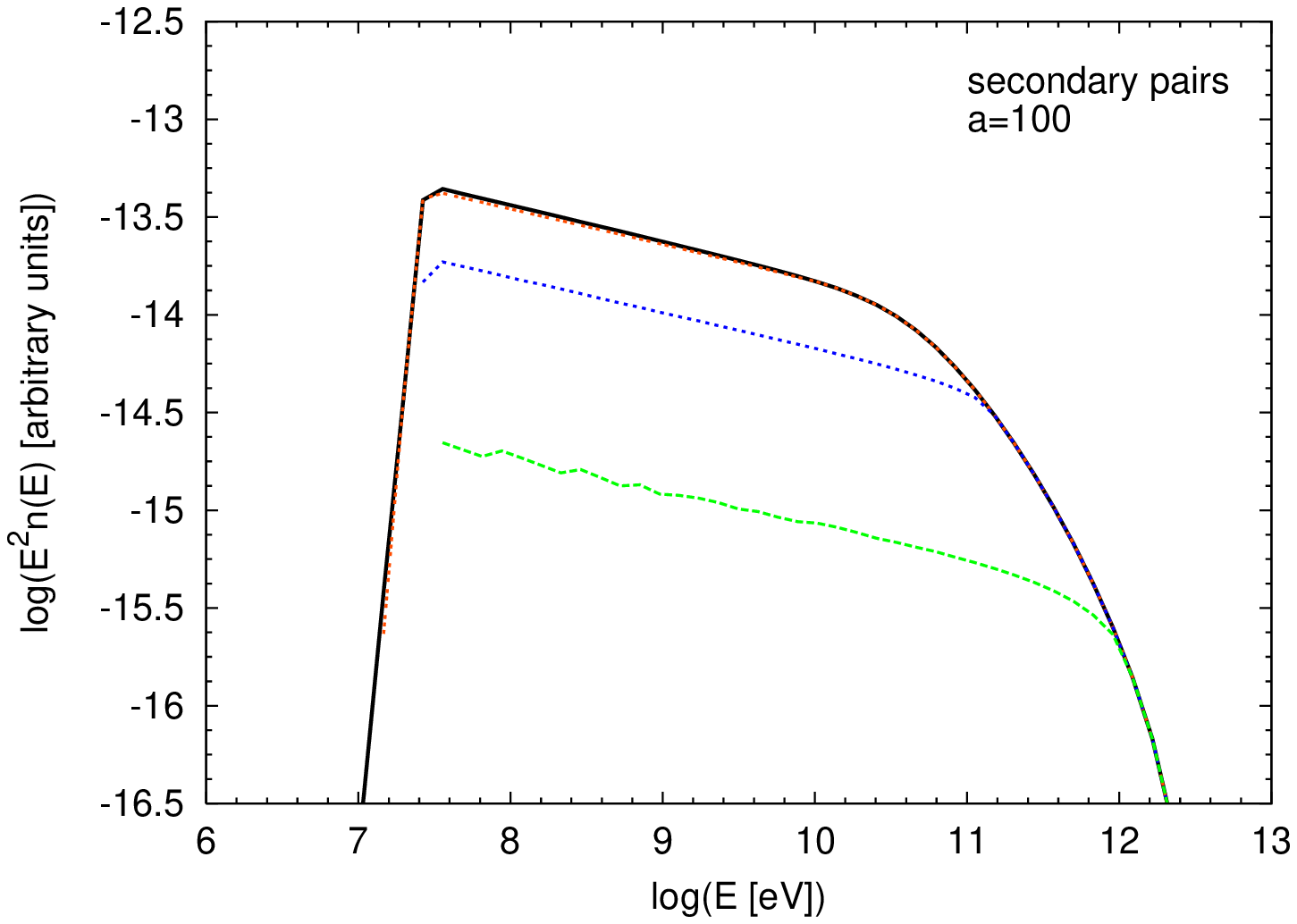}
\caption{Distribuciones de energ'ia de los pares secundarios.
En cada panel mostramos las distribuciones para diferentes tiempos de
inyecci'on de part'iculas ($\tau_{\rm iny}$). 
Las diferentes curvas corresponden a los siguientes valores  de
$\tau_{\rm iny}$: $10^7$ s (l'inea de rayas largas, verde), $10^8$ s (
l'inea de rayas y puntos, azul), $10^9$ s
(l'inea punteada, rojo) y $10^{10}$~s (l'inea s'olida, negro). 
Notar que para edades $> 10^9$~s el estado estacionario
es alcanzado.
La energ'ia m'inima de los pares $e^{\pm}$ inyectados en el l'obulo sur
es $\sim 5\times10^7$~eV, como se aprecia en las figuras (Araudo et al. 2007).}
\label{evol_YSO_esecundarios}
\end{center}
\end{figure}

Las interacciones $pp$ dan lugar a la producci'on de pares $e^{\pm}$,
los cuales rad'ian luego por los mismos procesos lept'onicos descriptos en la
secci'on anterior. Como en el caso de los electrones primarios, hemos 
calculado  la evoluci'on temporal del espectro de los pares obteniendo
que para $\tau_{\rm vida} \sim 10^{11}$~s el espectro se encuentra en el
estado estacionario. En la Figura~\ref{evol_YSO_esecundarios}
se muestran los resultados hallados. 

\begin{table}[]
\begin{center}
\begin{tabular}{cccc}
\hline 
\hline
a    & $E_{\rm q}$ & $k_{e_1,{\rm q}}'$ & $k_{e_2,{\rm q}}'$ \\ 
{} & [eV] & [erg$^{p_{e_1}'}$~cm$^{-3}$] & [erg$^{p_{e_2}'}$~cm$^{-3}$]\\ 
\hline 
$0$   & $9.4\times10^{9}$  & $3.7\times10^{-11}$ & - \\  
$1$   & $3.1\times10^{10}$ & $9.8\times10^{-11}$ & $3.2\times10^{-11}$\\
$100$ & $2.2\times10^{10}$ & $9.5\times10^{-13}$ & $5.4\times10^{-11}$ \\
\hline
\end{tabular}
\caption{Energ'ia de quiebre del espectro de electrones para los diferentes
casos discutidos en esta tesis. Las constantes de normalizaci'on $k_i^{\prime}$ 
($i = e_1, e_2$) de las distribuciones de energ'ia de los leptones para 
$E_{e_1, e_2} > E_{\rm q}$ est'an tambi'en listadas.}\label{break}
\end{center}
\end{table}

\begin{table}[]
\begin{center}
\begin{tabular}{cccc}
\hline
\hline
a  & $E_{\rm{e_1}}^{\rm{max}}$ & $E_{\rm p}^{\rm{max}}$ & $E_{\rm{e_2}}^{\rm{max}}$ \\
{} & [eV]                 & [eV]             & [eV]                 \\
%\hline
\hline
$0$   & $3.1\times10^{12}$ & -                  & -                   \\
$1$   & $3.1\times10^{12}$ & $4.7\times10^{13}$ & $1.1\times10^{12}$ \\ 
$100$ & $2.8\times10^{12}$ & $5.7\times10^{13}$ & $1.8\times10^{12}$ \\
\hline
\end{tabular}
\caption{Energ'ias m'aximas obtenidas para los electrones y protones
acelerados en el l'obulo sur.
% de la radio-fuente asociada con la protoestrella masiva IRAS~16547$-$4247. 
Adem'as, las energ'ias m'aximas de los pares secundarios se
muestran en la 'ultima columna.}\label{E_max_YSO}
\end{center}
\end{table}

\section{Distribuciones espectrales de energ'ia}

En las secciones previas vimos que en los choques terminales (en particular
el choque reverso, en el caso de la fuente que estamos estudiando) se 
pueden acelerar part'iculas hasta energ'ias 
$\gtrsim 10^{13}$~eV y luego 'estas se enfr'ian eficientemente por diversos
procesos radiativos no t'ermicos.  
En esta secci'on calculamos las SEDs producidas por estos procesos 
radiativos, teniendo en cuenta las distribuciones $n_i(E_i)$  
estimadas anteriormente y los campos magn'eticos, de materia y de fotones 
presentes en el l'obulo sur de la fuente IRAS~16547-4247.

Asumiendo que el emisor es homog'eneo y
esf'erico con un radio $R_{\rm lob}$ y volumen $V_{\rm lob}$,
consideraremos en todos los casos que la luminosidad espec'ifica
es 
\begin{equation}
E_{\rm ph}\,L_{\rm ph} (E_{\rm ph}) = V_{\rm lob}\,E_{\rm ph}\int 
j_{\rm ph}(E_{\rm ph}, E_i) \, {\rm d}E_i, 
\end{equation}
donde $j_{\rm ph}(E_{\rm ph}, E_i)$ es la emisividad de cada proceso radiativo.

\subsection{Interacciones lept'onicas}

\begin{figure}
\begin{center}
\includegraphics[angle=270, width=0.5\textwidth]{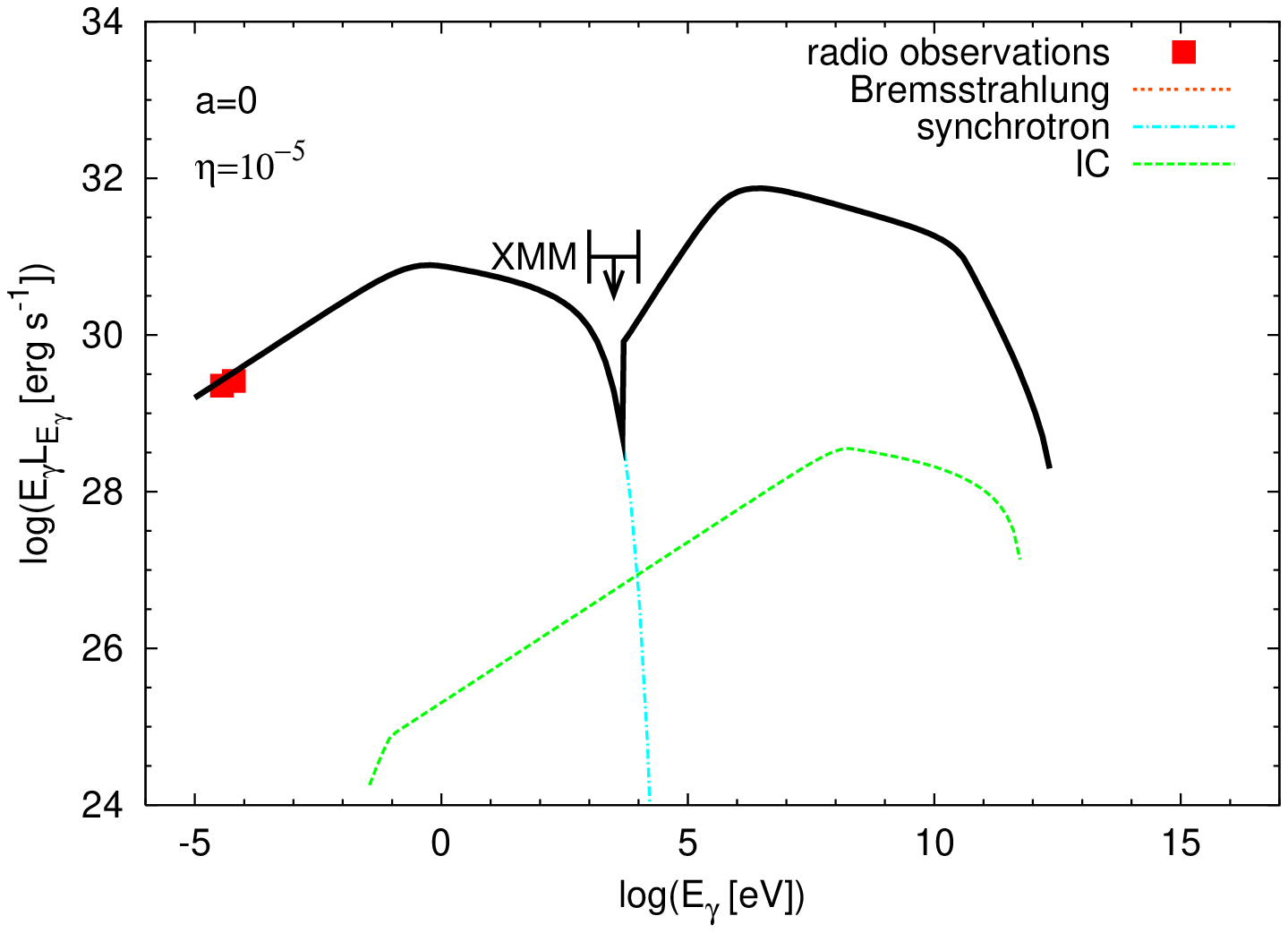}
\caption{Distribuci'on espectral de energ'ia para el caso puramente 
lept'onico, $a = 0$.
La l'inea negra representa la suma de las tres contribuciones (sincrotr'on,
IC y Bremsstrahlung relativista) a la SED total. En las frecuencias 
$\nu = 8.4$ y $14.9$~GHz ($E_{\rm ph} \sim 3.5\times10^{-5}$ y 
$6.2\times10^{-5}$~eV, respectivamente) se representa con cuadrados rojos 
la luminosidad observada 
con VLA a estas frecuencias. En el rango de los rayos~X se muestra la
sensibilidad del sat'elite \emph{XMM}, mostrando que nuestra predicci'on
te'orica est'a de acuerdo con el hecho de que este instrumento no haya 
detectado a la fuente (Araudo et al. 2007).}\label{SED1}
\end{center}
\end{figure}

Considerando los valores de $B_{\rm lob}$ mostrados en la Tabla~\ref{k_B}
calculamos la luminosidad espec'ifica emitida en el l'obulo sur
por radiaci'on sincrotr'on. 
Por otro lado, con los valores de $u_{\rm ph}$ y $n_{\rm nm}$  calculamos
las emisividades de los procesos IC y Bremsstrahlung relativista
usando las f'ormulas descriptas en el Cap'itulo~3. En la 
Figura~\ref{SED1} se muestra la SED correspondiente al caso puramente 
lept'onico, $a = 0$.

Como se muestra en la figura antes mencionada, la radiaci'on a energ'ias altas
est'a dominada por la emisi'on producida por el mecanismo Bremsstrahlung 
relativista, con un pico 
de $\sim 10^{32}$~erg~s$^{-1}$ a $E_{\rm ph} \sim 1$~MeV. A energ'ias 
$E_{\rm ph} \gtrsim 1$~GeV, la fuente presenta luminosidades de
$\sim 10^{31}$~erg~s$^{-1}$, con un \emph{cut-off} a $E_{\rm ph} \sim 10$~GeV.
En rayos X, en el rango 1-10 keV, las luminosidades esperadas son de 
$\sim 10^{30}$~erg~s$^{-1}$, presentando un ablandamiento del espectro  
debido al  \emph{cut-off}  exponencial $\exp(-E_{e_1}/E_{e_1}^{\rm max})$ de 
la distribuci'on de los electrones primarios.
En las energ'ias m'as bajas, el espectro calculado (sincrotr'on) ajusta muy
bien los puntos observados a las frecuencias $\nu = 8.4$ y $14.9$~GHz.

\subsection{Interacciones hadr'onicas}

\begin{figure}
\begin{center}
\includegraphics[angle=270, width=0.45\textwidth]{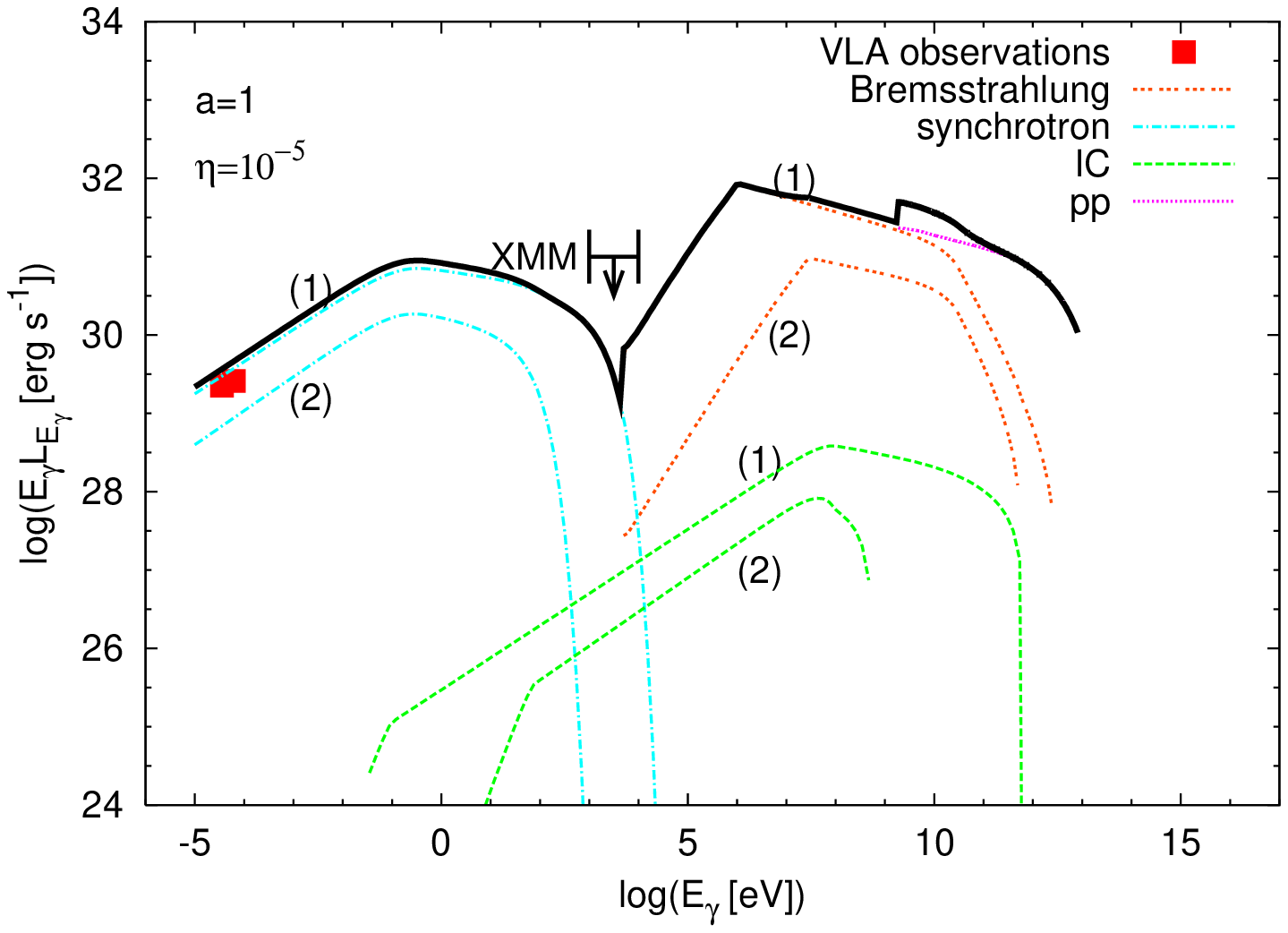}
\includegraphics[angle=270, width=0.45\textwidth]{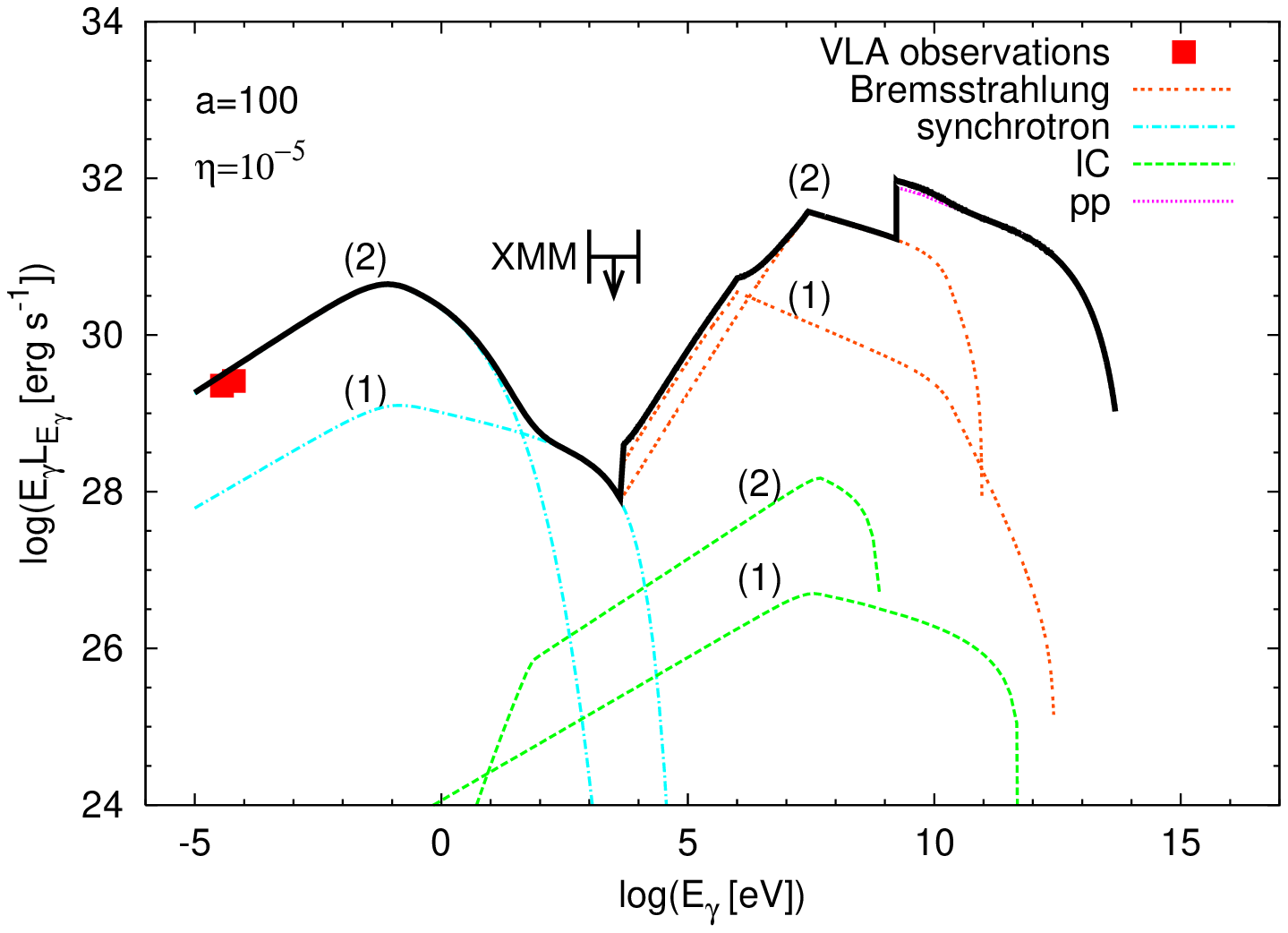}
\caption{Distribuciones espectrales de energ'ia para los casos 
con $a = 1$ (izquierda) y 100 (derecha), 
en los cuales consideramos 
tanto una poblaci'on lept'onica como hadr'onica de part'iculas primarias
y la poblaci'on de pares $e^{\pm}$ producida. 
Con (1) y (2) indicamos la contribuci'on de los leptones primarios
y secundarios, respectivamente. Como mencionamos en la Figura~\ref{SED1}, 
la l'inea negra indica la SED total (suma de todas las contribuciones, 
lept'onicas y hadr'onica) y los 
cuadrados rojos indican la luminosidad observada con VLA. La sensibilidad de
\emph{XMM} es tambi'en graficada mostrando nuevamente que los l'imites 
observacionales en rayos~X tampoco se violan  en los casos con
$a = 1$ y 100 (Araudo et al. 2007).}
\label{SED2}
\end{center}
\end{figure}

En los casos con $a = 1$ y $a = 100$ la emisi'on de los 
protones es significativa, ya que la densidad $n_{\rm nm}$ de la nube molecular 
es alta. 
Calculando la emisividad $pp$ a trav'es de la f'ormula~(\ref{q_pi}) obtenemos 
las luminosidades espec'ificas que se muestran en la Figura~\ref{SED2},
adem'as de las lept'onicas (sincrotr'on, IC y Bremsstrahlung relativista).
En el caso con $a = 1$ la luminosidad emitida por $pp$ es similar 
a la emitida por Bremsstrahlung relativista ($\sim 10^{32}$~erg~s$^{-1}$)
pero la primera se extiende hasta energ'ias m'as altas, $\sim 1$~TeV.
Los leptones primarios dominan la emisi'on por sincrotr'on y por 
Bremsstrahlung relativista en el caso $a = 1$, mientras que en el caso con
$a = 100$ la contribuci'on m'as importante a la SED es producida por los 
pares secundarios. 
En ning'un caso la emisi'on por IC es significativa. 
B'asicamente la emisi'on para $E_{\rm ph} > 1$~GeV es debida al decaimiento
de los $\pi^0$, mientras que la emisi'on en rayos~X y en rayos gamma blandos
es producida por Bremsstrahlung relativista. 
Para los casos $a=1$ y $a =100$, las componentes espectrales debidas a 
la radiaci'on 
sincrotr'on y Bremsstrahlung relativista de los pares secundarios tienen la 
misma forma que la emisi'on de los primarios (para los mismos casos) pero
diferentes a bajas energ'ias debido a que la energ'ia m'inima de cada 
distribuci'on de part'iculas (electrones primarios y pares secundarios) 
es distinta. En el caso de los electrones
primarios la energ'ia m'inima considerada es $\sim 2 m_ec^2$ mientras que los
pares $e^{\pm}$ tienen una energ'ia m'inima determinada por la energ'ia   
umbral para la creaci'on de los piones cargados: $E_{\rm u} \sim 1.22$~GeV.

\section{Discusi'on}

Las fuentes de rayos gamma del plano gal'actico
son usualmente asociadas con regiones de formaci'on estelar 
(Romero et al. 1999). Sin embargo las fuentes detectadas por EGRET\footnote{El
instrumento EGRET (a bordo del sat'elite Compton) funcion'o en los años
1991-2000 y detectaba fotones con energ'ias 
30~MeV~$\lesssim E_{\rm ph} \lesssim 300$~GeV. Con los datos obtenidos se 
confeccion'o el primer cat'alogo de fuentes de rayos gamma que permiti'o
realizar estudios poblacionales.}
son mucho m'as luminosas que los YSOs de gran masa, de acuerdo a los c'alculos
realizados con nuestro modelo. Las fuentes EGRET no identificadas  
tendr'ian como contrapartidas a p'ulsares, remanentes de supernovas, 
estrellas tempranas
(Romero 2001) y MQs (Bosch-Ramon et al. 2005). 
La emisi'on producida por YSOs de gran masa est'a por debajo de la 
sensibilidad de 
EGRET, pero 'estos podr'ian ser detectados por \emph{Fermi}. 
En el rango de las VHE, los futuros arreglos de telescopios Cherenkov 
como HESS~II, MAGIC~II y CTA\footnote{Los instrumentos HESS~II y MAGIC~II
ser'ian los que reemplazar'ian a HESS y MAGIC, respectivamente.
Por otro lado, los instrumentos que formar'an parte del arreglo de 
telescopios Cherenkov CTA (por \emph{Cherenkov Telescope Array}) ya comenzaron
a contruirse, aunque el sitio donde estar'an montados a'un no esta definido.}
podr'ian detectar fuentes como IRAS~16547-4247. 
Ya que estos instrumentos deber'ian llegar a medir hasta el \emph{cut-off} de 
altas energ'ias, la medici'on de este valor dar'ia una valiosa 
informaci'on acerca de la eficiencia de aceleraci'on de part'iculas en los 
choques terminales de los \emph{jets} de YSOs de gran masa. 

En rayos X, la fuente deber'ia ser detectable por \emph{Chandra} y 
\emph{XMM-Newton}, probablemente como una fuente puntual a trav'es de una 
observaci'on profunda en la regi'on. En los datos de archivo
de \emph{XMM-Newton}, no se ve una emisi'on significativa por encima de
la radiaci'on de fondo en la regi'on donde la fuente IRAS est'a localizada. 
Esto nos proporciona  una cota m'axima de $\sim 10^{31}$~erg~s$^{-1}$ (a la
distancia $\sim 2.9$~kpc) en el rango de 1 a 10~keV
a las luminosidades que obtenemos con nuestros modelos te'oricos, como se 
muestra en las Figuras~\ref{SED1} y  \ref{SED2}.
Sin embargo, una observaci'on m'as profunda (m'as de 30~ks) deber'ia detectar
radiaci'on por encima de la emisi'on de fondo de esta regi'on.
Si las temperaturas del material chocado son suficientemente altas, la 
componente t'ermica deber'ia ser tambi'en detectable en rayos~X. 
En la banda ultravioleta (UV), esta emisi'on t'ermica podr'ia ser 
a'un mayor que la producida
por las part'iculas no t'ermicas, pero ser'ia dif'icil de 
detectar\footnote{Para una discusi'on m'as detallada sobre la emisi'on
t'ermica, como as'i tambi'en un an'alisis m'as preciso del impacto del
medio en la evoluci'on de las part'iculas y en la radiaci'on producida, 
ver Bosch-Ramon et al. (2010).}.

Un hecho relevante respecto de la factibilidad del escenario discutido en 
este cap'itulo es que
las luminosidades no t'ermicas son bastante m'as bajas que la luminosidad
cin'etica del \emph{jet} de la fuente IRAS~16547-4247. La p'erdida de masa 
del \emph{jet}
ha sido estimada en $\sim 10^{-5} M_{\odot}$~yr$^{-1}$ y con una velocidad
$v_{\rm j} \sim 1000$~km~s$^{-1}$ con lo que la luminosidad cin'etica del 
\emph{jet} resulta 
$\sim 10^{36}$~erg~s$^{-1}$, m'as de tres 'ordenes de magnitud mayor que la 
luminosidad no t'ermica predicha por nuestro modelo. 
Esto es, nuestro modelo predice niveles de emisi'on que respetan los l'imites
energ'eticos de la fuente. Por lo tanto, todav'ia ser'ia posible un 
incremento de la eficiencia radiativa,   
incrementando as'i las posibilidades de detectecci'on de este tipo de objetos.
Adem'as de la fuente IRAS~16547-4247, otras protoestrellas de gran masa
son tambi'en potenciales candidatos a fuentes de rayos gamma. Entre ellas
podemos mencionar HH~80-81 (Bosch-Ramon et al. 2010) y W3(OH) (Araudo et al.
2008a).

%% file: MQs_final.tex
\chapter{Microcuasares}
\label{MQs}

\section{Introducci\'on}

Las estrellas en general no se encuentran aisladas, sino que la mayor'ia
de ellas forman sistemas binarios o c'umulos. 
En los primeros, si ambas componentes tienen masas diferentes, la estrella 
m'as temprana evolucionar'a m'as r'apido que la m'as tard'ia 
y as'i las estrellas estar'an en diferentes estados evolutivos.
Si la masa de la estrella m'as
evolucionada que queda luego de la explosi'on de supernova es 
$3 M_{\odot} \gtrsim M_{\star}\gtrsim  1.4 M_{\odot}$, entonces 
el objeto remanente al final de la evoluci'on ser'a una estrella de 
neutrones, mientras que si $M_{\star} \gtrsim 3 M_{\odot}$ ser'a un 
agujero negro.

Consideremos un sistema binario formado por una estrella 
no degenerada, la cual puede estar en diferentes estados de su evoluci'on, y el
objeto colapsado, el cual puede ser un agujero negro o una estrella de 
neutrones.
Cuando la estrella compañera se convierte en gigante roja, las capas externas
de la atm'osfera de la misma llenan de material el l'obulo de Roche. 
La materia aqu'i dentro se escapa a trav'es del punto lagrangiano L1
siendo
acretada por el objeto compacto. La materia cae a
la superficie del objeto a trav'es de un disco de acreci'on debido a la
conservaci'on del momento angular.
La materia en el disco rota a una velocidad que
disminuye con la distancia $r_{\rm d}$ al objeto compacto. Anillos
a diferentes $r_{\rm d}$ rotan a velocidades distintas lo que produce 
disipaci'on de energ'ia y calentamiento 
por roce de anillos contiguos. La materia del disco se puede calentar 
hasta temperaturas suficientemente altas como para radiar t'ermicamente
en rayos~X.

Los sistemas binarios que emiten fuertemente en rayos~X
debido a la  acreci'on, son llamados binarias de rayos X
(XRB, por \emph{X-Ray Binaries}). Si la estrella primaria es de gran masa
(tipo espectral O 'o B, $M_{\star} > 8 M_{\odot}$) entonces el sistema se 
dice de gran masa (HMXB, por \emph{High Mass X-ray Binary}), 
mientras que si es vieja el sistema binario es una binaria de rayos~X de 
baja masa (LMXB, por \emph{Low Mass X-ray Binary}). 
Una subclase de las XRB 
son los MQs, los cuales presentan \emph{jets} extendidos en la banda de radio 
(Mirabel \& Rodr'iguez 1999). Estos \emph{jets} son flujos colimados de 
materia que se mueve
a una velocidad de conjunto relativista. La materia que forma los \emph{jets} 
puede ser materia arrancada del disco de acreci'on por las l'ineas de
campo magn'etico (Blandford \& Payne 1981)
o bien pares $e^{\pm}$ generados en la ergosfera del agujero negro 
(Blandford \& Znajek 1977) y lanzados en forma de chorros por procesos 
MHDs. El campo magn'etico juega un rol 
importante en la formaci'on y colimaci'on de los \emph{jets}.

\begin{figure}
\begin{center}
\includegraphics[angle=0, width=0.7\textwidth]{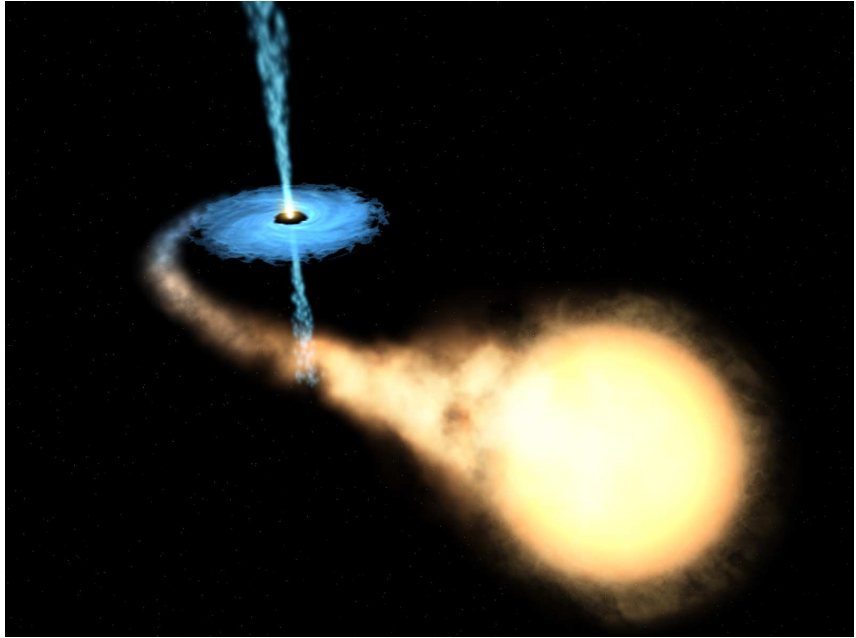}
\caption{Esquema de un MQ. Se muestra una estrella evolucionada
transfiriendo masa al objeto compacto a trav'es del disco de acreci'on.
\emph{Jets} bipolares emergen del objeto compacto.
(Cr'edito: ESA, NASA y Felix Mirabel.)}\label{mq}
\end{center}
\end{figure}

Se ha detectado emisi'on  no t'ermica proveniente de los \emph{jets} de los
MQs. Esta emisi'on abarca desde las frecuencias radio (Rib'o 2005) 
hasta los rayos~X (Corbel et al. 2002), aunque emisi'on de m'as alta
energ'ia (rayos gamma) tambi'en puede producirse 
(Kaufmann Bernad'o et al. 2002, Bosch-Ramon et al. 2006). 
Los telescopios MAGIC y \emph{AGILE} han detectado emisi'on transitoria
(\emph{flares}) en HE y VHE  asociada 
a los HMMQs Cygnus~X-3 (Tavani et al. 2009b; Abdo et al. 2009a) y  Cygnus~X-1 
(Albert et al. 2007; Sabatini et al. 2010), respectivamente. 
Eventos de emisi'on 
transitoria en rayos gamma tambi'en han sido observados en las 
HMXBs LS~5039 y LS~I+61~303 por los
telescopios HESS (Aharonian et al. 2005) y MAGIC (Albert et al. 2006),
respectivamente, aunque la naturaleza de las fuentes  no ha sido a'un 
confirmada en estos casos. (Una discusi'on sobre el tema puede encontrarse 
en Romero et al. 2007.) 
Por otro lado, el sat'elite \emph{Fermi} ha detectado 
emisi'on espor'adica proveniente de fuentes no identificadas del
plano gal'actico (ATel\footnote{La sigla ATel (por \emph{Astronomical 
Telegrams}) se utiliza para designar los reportes sobre descubrimientos
observacionales. La p\'agina de internet en la cual pueden hallarse estos 
reportes es http://www.astronomerstelegram.org/.}~1394, Abdo et al. 2009). 
Esta emisi'on altamente 
variable detectada en rayos gamma y generada en HMMQs podr'ia tener un origen 
similar. 
Por ejemplo, la interacci'on entre el \emph{jet} y el viento de la estrella
compañera podr'ia producir \emph{flares} en el rango de las HE y VHE.

La p'erdida de masa de las estrellas de gran masa  
forma un viento estelar que se propaga a velocidades supers'onicas. 
Observaciones en rayos~X de las l'ineas de emisi'on producidas en
el viento indican que 'este no tiene una estructura uniforme sino 
porosa, con \emph{clumps} (o grumos de materia) que se forman a una 
distancia del orden de un radio estelar ($R_{\star}$)
%$\sim R_{\star}$
de la superficie de la estrella (e.g. Puls et al. 2006, Owocki
\& Cohen 2006).  Sin embargo, las caracter'isticas de 
estos \emph{clumps} no se conocen ya que la resoluci'on
espacial de los telescopios a'un no es suficiente.
Por esto las propiedades
de los \emph{clumps}, como el tama~no, la densidad y el factor de llenado, 
no son bien conocidas y se estudian a trav'es de m'etodos indirectos como
el an'alisis de l'ineas espectrosc'opicas (Moffat 2008).

En este cap'itulo proponemos un modelo para explicar los \emph{flares} en 
rayos 
gamma detectados en algunas binarias de gran masa, basado en la interacci'on
de los \emph{jets} del HMMQ con las inhomogeneidades del viento de la estrella 
compañera. Los \emph{clumps} pueden eventualmente penetrar en el 
\emph{jet} y producir emisi'on
no t'ermica transitoria, generada al convertir parte de la energ'ia cin'etica 
del \emph{jet} en energ'ia de part'iculas relativistas, las cuales 
emitir'an radiaci'on sincrotr'on, IC y sufrir'an colisiones $pp$.

\section{Escenario}

\begin{figure}
\begin{center}
\includegraphics[angle=0, width=0.7\textwidth]{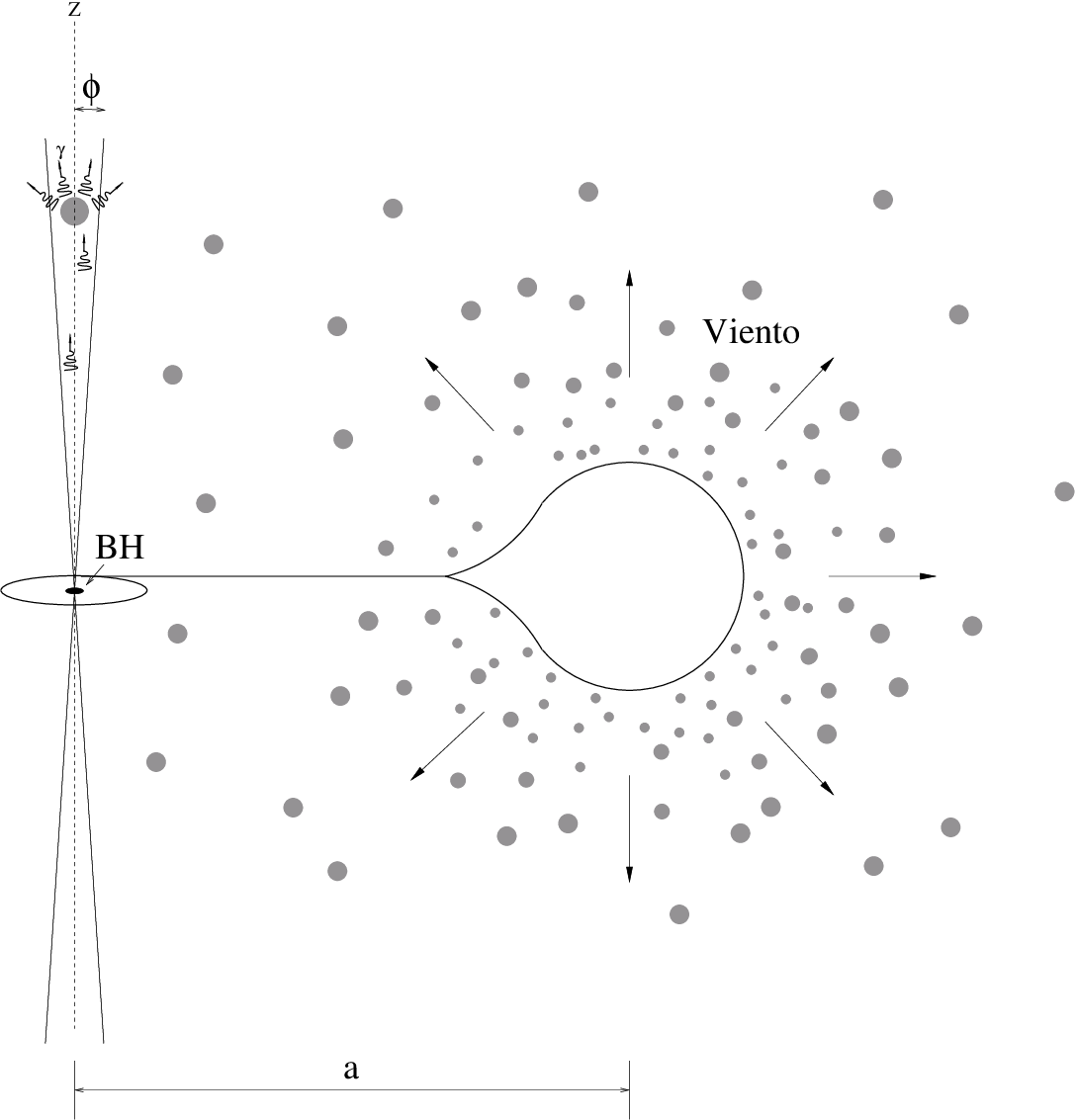}
\caption{Esquema de un MQ. El viento de la estrella
compañera presenta \emph{clumps}, algunos de los cuales pueden llegar
a penetrar en el \emph{jet} (Romero et al. 2008).}\label{mq-scenario}
\end{center}
\end{figure}

Para estudiar la interacci'on  entre un \emph{clump} del viento estelar
con uno de los \emph{jets} de un HMMQ adoptamos un escenario con 
caracter'isticas
similares a las del sistema binario Cygnus~X-1. 
Fijamos la separaci'on $a_{\rm mq}$ entre el objeto compacto y la 
estrella masiva 
en $3\times10^{12}$~cm ($\sim$ 0.2~UA). Para la luminosidad
y la temperatura de la estrella adoptamos los siguientes valores:
$L_{\star} = 10^{39}$~erg~s$^{-1}$ y $T_{\star} = 3\times10^4$~K,
respectivamente. 
Por otro lado suponemos que la p'erdida de masa de la estrella es
$\dot M_{\star} = 3\times10^{-6} M_{\odot}$~yr$^{-1}$ con una velocidad
terminal del viento $v_{\rm v} \sim 2.5\times 10^8$~cm~s$^{-1}$. 
Un esquema gr'afico de este escenario se muestra en la 
Figura~\ref{mq-scenario}.  

\subsubsection{Modelo para el \emph{clump}}

Debido a las incertezas en la 
determinaci'on de las caracter'isticas de 
los \emph{clumps}, suponemos que 'estos son esf'ericos y consideramos dos 
valores para el radio: $R_{\rm c}= 10^{10}$ y $10^{11}$~cm 
($\sim 3\times10^{-3}-3\times10^{-2}\,a_{\rm mq}$). A su vez adoptamos
una distribuci'on uniforme de densidad en el \emph{clump}, 
$n_{\rm c} = 10^{12}$~cm$^{-3}$, que corresponde a un factor de llenado 
$f_{\rm c} =\dot{M}_{\star}/4\pi\, a_{\rm mq}^2\, m_p\, v_{\rm c}\, n_{\rm c} 
\sim 0.005$,
donde $v_{\rm c}$ es la velocidad de los \emph{clumps} y consideramos que 
'estos se mueven a la velocidad del viento, esto es, $v_{\rm c} = v_{\rm v}$
(Owocki et al. 2009 y referencias en ese art'iculo).
La temperatura de los \emph{clumps}, $T_{\rm c}$, es fijada en $10^4$~K 
(Krti$\check{\rm c}$ka \& Kub\'at 2001), siendo moderadamente menor que
la temperatura superficial de la estrella primaria.

\subsubsection{Modelo para el \emph{jet}}

Para los \emph{jets}, adoptamos un modelo hidrodin'amico, es decir,
consideramos un \emph{jet} dominado din'amicamente por protones fr'ios
con una velocidad de conjunto moderadamente relativista: $v_{\rm j} = 0.3\,c$,
que corresponde a un factor de Lorentz $\Gamma_{\rm j} = 1.05$. 
Observaciones en radio muestran que los \emph{jets} de los MQs son 
extremadamente
colimados (Miller-Jones et al. 2006). En esta tesis hemos supuesto
que la relaci'on entre el radio y la altura es 
$R_{\rm j}(z_{\rm j}) = 0.1\,z_{\rm j}$,
lo que corresponde a un 'angulo de apertura $\phi = 6^{\circ}$.
La velocidad de expansi'on del \emph{jet} resulta 
$v_{\rm exp} = 0.1\;v_{\rm j}$, 
que para un \emph{jet} hidrodin'amico en expansi'on libre
implica un n'umero de Mach en la base 
de $\sim v_{\rm j}/v_{\rm exp} \sim 10$, con lo cual el \emph{jet} es 
supers'onico.
La luminosidad cin'etica del \emph{jet} es fijada en 
$L_{\rm j} = 3\times10^{36}\;\rm{erg\;s^{-1}}$, similar a la estimada para
Cygnus~X-1 (e.g. Gallo et al. 2005, Russell et al. 2007). Usando la 
ecuaci'on~(\ref{L_jet}) podemos estimar la densidad de part'iculas $n_{\rm j}$ 
del \emph{jet} en el sistema de referencia
del laboratorio. A la altura de la interacci'on \emph{jet-clump}, que hemos
fijado en $z_{\rm int} = a_{\rm mq}/2$, obtenemos 
$n_{\rm j} = 4.7\times10^7$~cm$^{-3}$.
De esta manera, el cociente entre la densidad del \emph{clump} y la del 
\emph{jet} resulta $\chi_{\rm mq} = 2.1\times10^4$. 
Finalmente mencionamos que
despreciamos la curvatura del \emph{jet} producida por la interacci'on 
con el viento estelar. En los \emph{jets} de HMMQs con luminosidades cin'eticas 
$> 10^{36}$~erg~s$^{-1}$, la geometr'ia de 'estos no deber'ia ser 
modificada considerablemente por el viento (Perucho \& Bosch-Ramon 2008), 
aunque este 
efecto puede ser importante en sistemas con \emph{jets} tipo HH 
(protoestelares)
interactuando con el viento de una estrella (Raga et al. 2009).

Los valores de los par'ametros del \emph{jet} y del \emph{clump} que suponemos 
o estimamos en este cap'itulo est'an listados  en la Tabla~\ref{parameters}.

\begin{table}[]
\begin{center}
\caption{Par'ametros adoptados y estimados para el escenario en el cual 
desarrollamos el estudio presentado en este cap'itulo.}
\label{parameters}
\begin{tabular}{lll}
\hline 
\hline
Par'ametro [unidades] & \emph{Clump} & \emph{Jet} \\
\hline 
Radio [cm] & $R_{\rm c} = 10^{10}-10^{11}$ & $R_{\rm j} = 1.5\times10^{11}$ \\  
Velocidad [$\rm cm\;s^{-1}$] & $v_{\rm c} = 2.5\times10^8$ & $v_{\rm j} = 10^{10}$ \\  
Densidad [$\rm cm^{-3}$] & $n_{\rm c} = 10^{12}$ & $n_{\rm j} =4.7\times10^7$ \\  
\hline
{} & Sistema binario & {} \\  
\hline  
Tamaño del sistema  [cm] & $a_{\rm mq} = 3\times10^{12}$ &{}\\
Luminosidad de la estrella [erg~s$^{-1}$] & $L_{\star} = 10^{39}$ & {}\\
Temperatura de la estrella [K] & $T_{\star} = 3\times10^4$ & {}\\
Tasa de p'erdida de masa [M$_{\odot}$ yr$^{-1}$] & $\dot M_{\star} =3\times10^{-6}$ &{}\\
Velocidad del viento [$\rm cm\;s^{-1}$] & $v_{\rm v} =2.5\times10^8$ & {} \\
\hline
\end{tabular}
\end{center}
\end{table}

\subsection{Interacci'on \emph{jet-clump}}
\label{jet-clump-times}

Los \emph{clumps} se forman en la regi'on de aceleraci'on del viento de la 
estrella, aproximadamente a una distancia $\sim R_{\star}$ de la superficie de 
la misma (Puls et al. 2006).
Algunos de estos \emph{clumps}  pueden llegar hasta el \emph{jet} y penetrar
en 'el, debido al gran contraste de densidades $\chi_{\rm mq}$ 
($\equiv n_{\rm c}/n_{\rm j}$).

Habiendo especificado las caracter'isticas que adoptamos para los 
\emph{clumps} y los \emph{jets} de nuestro modelo, nos concentramos ahora en la 
interacci'on de uno de estos
\emph{clumps} con uno de los \emph{jets} del HMMQ.
De acuerdo a los par'ametros considerados, las escalas de tiempo de
los procesos din'amicos descriptos en la Secci'on~\ref{pert-locales}
toman los valores que exponemos a continuaci'on y que resumimos
en la Tabla~\ref{Table_timescales_clump}.

El tiempo de penetraci'on del \emph{clump} en el \emph{jet} es:
\begin{equation} 
t_{\rm c} \sim 80\,\left(\frac{R_{\rm c}}{10^{10}\,{\rm cm}}\right)
\left(\frac{v_{\rm c}}{2.5\times10^8\,{\rm cm\,s^{-1}}}\right)^{-1}\,{\rm s},
\end{equation}
mientras que el tiempo de cruce a una altura
$z_{\rm int} = a_{\rm mq}/2 = 1.5\times10^{12}$~cm es 
\begin{equation} 
t_{\rm j} \sim 1.2\times10^3\,
\left(\frac{z_{\rm int}}{1.5\times10^{12}\,{\rm cm}}\right)\,{\rm s}.
\end{equation}

El \emph{bow shock}  en el \emph{jet} alcanza el 
estado estacionario en un tiempo muy corto.
Para los par'ametros de este sistema y considerando que
$v_{\rm j1} \sim v_{\rm j}/4$ y que $Z = 0.2\,R_{\rm c}$ (van Dicke \& Gordon,
1959) tenemos que 
\begin{equation} 
t_{\rm bs} \sim 0.8\,\left(\frac{R_{\rm c}}{10^{10}\,{\rm cm}}\right)
\left(\frac{v_{\rm j}}{10^{10}\,{\rm cm\,s^{-1}}}\right)^{-1}\,{\rm s}.
\end{equation}

El tiempo caracater'istico de este tipo de interacciones est'a dado, como
mencionamos en el Cap'itulo~\ref{cap2}, por aquel en el cual el choque que se 
propaga en el \emph{clump} lo recorre completamente. Esta escala 
temporal est'a caracterizada de la siguiente manera en un HMMQ:  
\begin{eqnarray} 
t_{\rm cc} &\sim& 5\times10^3\,\left(\frac{R_{\rm c}}{10^{10}\,{\rm cm}}\right)
\left(\frac{n_{\rm c}}{10^{12}\,{\rm cm^{-3}}}\right)^{1/2}
\left(\frac{z_{\rm int}}{1.5\times10^{12}\,{\rm cm}}\right) \nonumber\\
{}&{}& \times\left(\frac{\Gamma_{\rm j} -1}{0.06}\right)^{-1/2}
\left(\frac{L_{\rm j}}{3\times10^{36}\,{\rm erg\,s^{-1}}}\right)^{-1/2}
\left(\frac{v_{\rm j}}{10^{10}\,{\rm cm\,s^{-1}}}\right)^{-1/2}\,{\rm s}. 
\end{eqnarray}
La materia del \emph{jet} acelera al clump aplic'andole
una fuerza $\sim M_{\rm c}\, \vec g$, donde $M_{\rm c}$ es la masa del 
\emph{clump} y 
la aceleraci'on $g \sim 3\times10^4$ y $3\times10^3$~cm~s$^{-2}$, para 
$R_{\rm c} = 10^{10}$ y $10^{11}$~cm, respectivamente. 
Esta fuerza puede acelerar al \emph{clump} hasta la velocidad 
del \emph{jet} en un tiempo 
\begin{eqnarray} 
t_{\rm g} &\sim& 1.3\times10^6\,\left(\frac{R_{\rm c}}{10^{10}\,{\rm cm}}\right)
\left(\frac{n_{\rm c}}{10^{12}\,{\rm cm^{-3}}}\right)
\left(\frac{z_{\rm int}}{1.5\times10^{12}\,{\rm cm}}\right)^2 \nonumber\\
{}&{}& \times\left(\frac{\Gamma_{\rm j} -1}{0.06}\right)
\left(\frac{L_{\rm j}}{3\times10^{36}\,{\rm erg\,s^{-1}}}\right)^{-1}\,{\rm s}.
\end{eqnarray}
Sin embargo, antes de que el \emph{jet} acelere al \emph{clump} y 'este 
comience a ser arrastrado por el flujo, el \emph{clump} puede ser destruido por
las inestabilidades de Rayleigh-Taylor y Kelvin-Helmholtz, las cuales 
crecen en un tiempo $t_{\rm RT/KH} \sim t_{\rm cc} \ll t_{\rm g}$ hasta 
longitudes de escala $\sim R_{\rm c}$
(sin considerar el efecto estabilizador que puede llegar a tener el campo
magn'etico).

\begin{table}[]
\begin{center}
\caption{Valores obtenidos para las escalas de tiempo descriptas en la 
secci'on~\ref{jet-clump-times}, usando los  valores de los
par'ametros listados en la Tabla~\ref{parameters}.}
\label{Table_timescales_clump}
\begin{tabular}{c|cc}
\hline 
\hline
Escala de tiempo [s] & $R_{\rm c} = 10^{10}$~cm & $R_{\rm c} = 10^{11}$~cm  \\
\hline 
$t_{\rm c}$ & $80$ & $800$ \\ 
$t_{\rm j}$ & $1.2\times10^3$ & $1.2\times10^3$ \\ 
$t_{\rm cc}$ & $5\times10^3$ & $5\times10^4$ \\ 
$t_{\rm bs}$ & $0.8$ & $8$ \\
$t_{\rm g}$ & $7\times10^{10}$ & $7\times10^{11}$ \\
\hline
\end{tabular}
\end{center}
\end{table}

\begin{figure}
\begin{center}
\includegraphics[angle=0, width=0.5\textwidth]{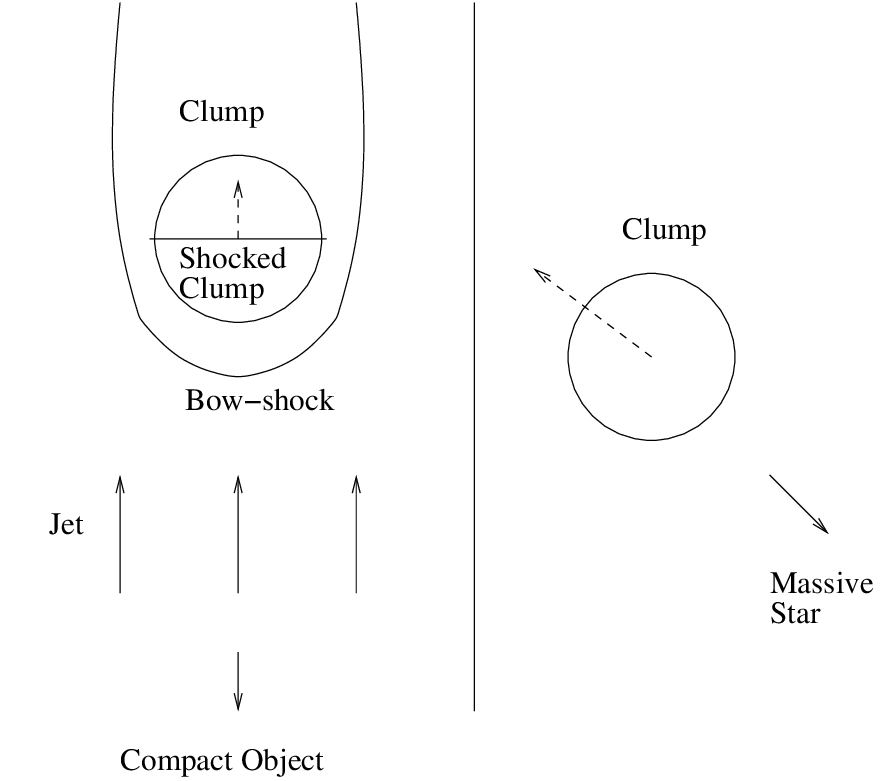}
\caption{Esquema de la interacci'on \emph{jet-clump} (Araudo et al. 2009).}
\label{jet-clump_int}
\end{center}
\end{figure}

A modo de resumen y
de acuerdo a las escalas de tiempo estimadas en los p'arrafos previos, el 
\emph{clump} puede penetrar completamente en el \emph{jet} si 
$t_{\rm c} < t_{\rm cc}$. En nuestros c'alculos radiativos no consideraremos 
los detalles de la penetraci'on del \emph{clump} en el \emph{jet}, 
sino que para un tiempo $> t_{\rm c}$ el primero se encuentra 
completamente dentro 
del 'ultimo (esto es, consideramos un sistema con simetr'ia cil'indrica 
como se muestra en la Figura~\ref{jet-clump_int}). El \emph{bow shock}
se forma r'apidamente en un tiempo $t_{\rm bs}$ mucho menor que
$t_{\rm c}$ y que $t_{\rm cc}$. 
Por 'ultimo, notamos que el \emph{clump} 
podr'ia no escapar del \emph{jet} ya que para los par'ametros considerados y el 
$z_{\rm int}$ que hemos fijado, $t_{\rm j} > t_{\rm RT/KH}$. 
Sin embargo, s'olo podemos cuantificar el tiempo
de las inestabilidades de una manera muy somera y no podemos afirmar que 
el \emph{clump} se destruir'a necesariamente antes de poder escapar del 
\emph{jet}. 
Simulaciones num'ericas muestran que las escalas de tiempo de las
inestabilidades pueden ser varias veces el tiempo de cruce del choque en el
\emph{clump}, es decir, $t_{\rm RT/KH} > t_{\rm cc}$ (Klein et al. 1994). 

Respecto de las propiedades de los choques y dadas las caracter'isticas
espec'ificas de este escenario (HMMQ), el choque en el \emph{clump} es fuerte,
radiativo y lento, mientras que el  \emph{bow shock} es tambi'en
fuerte pero adiab'atico y r'apido. Por estas razones, el material
chocado y calentado del \emph{clump} rad'ia una fracci'on significativa de 
la energ'ia que el choque le ha transferido (pero debido a que 
$\chi_{\rm mq} \gg 1$, esta energ'ia es baja comparada con la que 
transporta el \emph{jet}).

\subsection{Emisi'on t'ermica del \emph{clump}}
  
\begin{figure}
\begin{center}
\includegraphics[angle=270, width=0.49\textwidth]{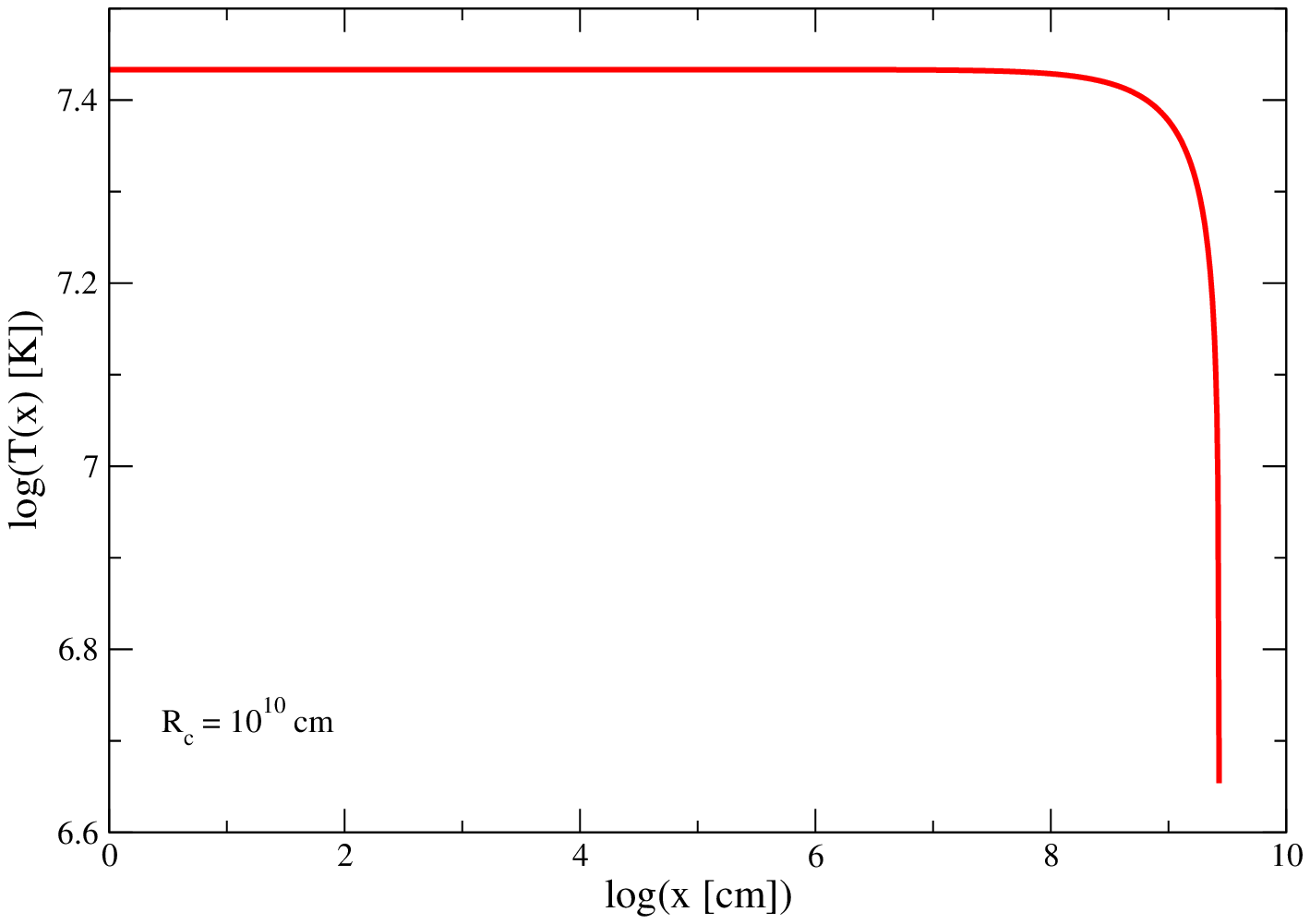}
\includegraphics[angle=270, width=0.49\textwidth]{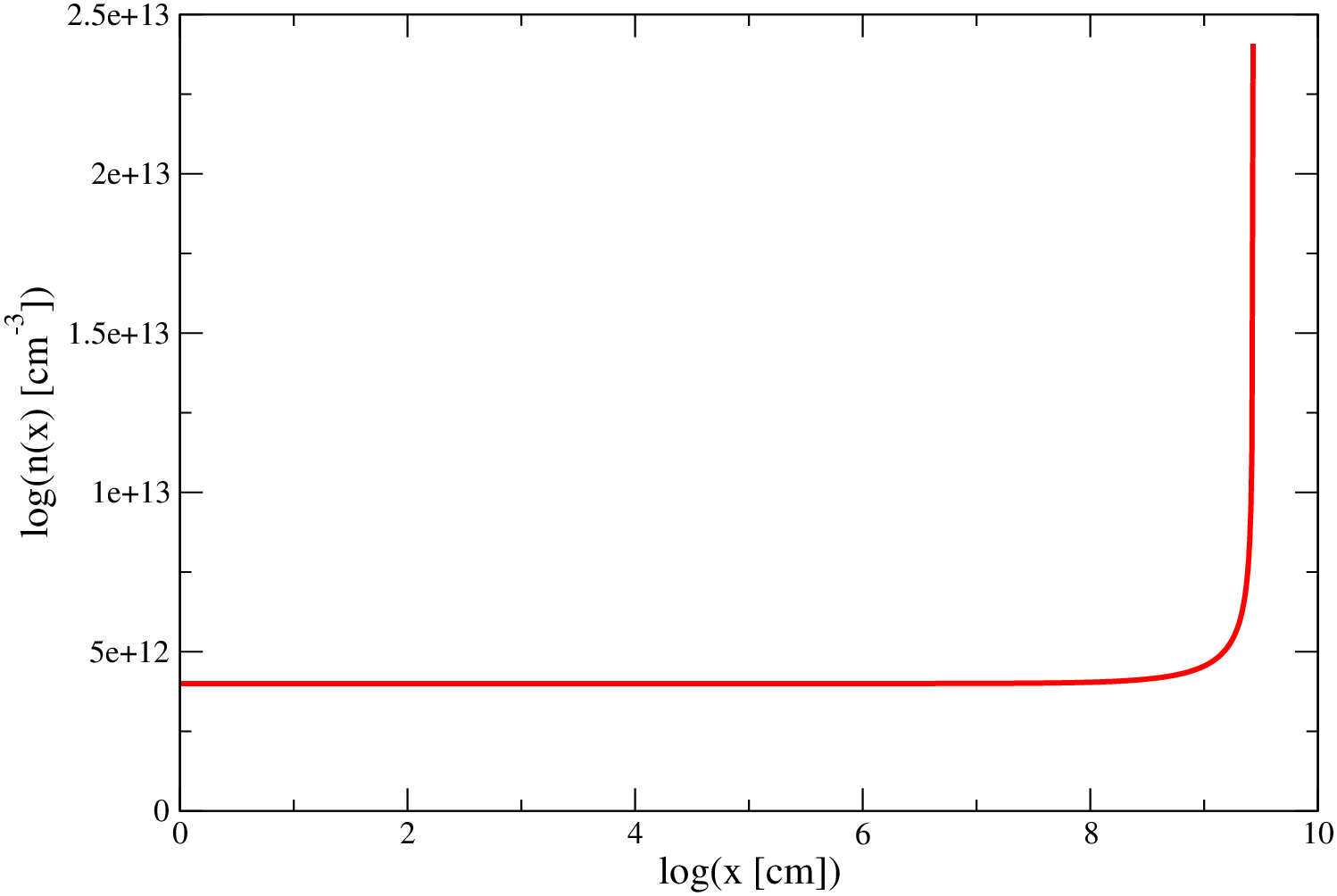}
\caption{Variaci'on de la temperatura (izquierda) y la densidad (derecha) 
en la regi'on chocada 
del \emph{clump} en funci'on de la distancia $x$ al choque.}
\label{Temp-clump}
\end{center}
\end{figure}

Para estimar la densidad $n_{\rm c1}(x)$ y la temperatura $T_{\rm c1}(x)$
del material del \emph{clump} chocado a una distancia $x$ del choque
usamos las relaciones~(\ref{n-rad}) y (\ref{T-rad}) considerando que
$\Lambda(T) = 7\times10^{-19} T^{-0.6}$~erg~cm$^{-3}$~s$^{-1}$
y que  la temperatura y la densidad en la zona adiab'atica de la regi'on chocada
del \emph{clump} toman los valores $8.5\times10^6$~K y 
$4\times10^{13}$~cm$^{-3}$, respectivamente. 
En la Figura~\ref{Temp-clump} se muestran los gr'aficos de $n_{\rm c1}(x)$ y
$T_{\rm c1}(x)$.
Luego, el tiempo de enfriamiento por radiaci'on t'ermica (de continuo y de 
l'ineas) resulta
\begin{equation} 
t_{\rm ter}= 3\times10^2\,\frac{T_{\rm c1}(x)^{1.6}}{n_{\rm c1}(x)}\,\,{\rm s}
\sim 10\,{\rm s}\,.
\label{T}
\end{equation}   
Siendo $t_{\rm ter} < t_{\rm cc}$ podemos decir que el choque en el 
\emph{clump} es
radiativo. Este tiempo de enfriamiento corresponde a una distancia 
$x_{\rm rad} \sim t_{\rm ter} v_{\rm cc} /4 \sim 2\times10^8$~cm, que es menor
que $2 R_{\rm c}$. La velocidad $v_{\rm cc}$
de propagaci'on del choque en el \emph{clump} es calculada a trav'es de la
ecuaci'on~(\ref{v-choque-obstaculo}) y resulta 
$v_{\rm cc} \sim 7\times10^7$~cm~s$^{-1}$.
A una distancia $> x_{\rm rad}$ la temperatura del material
chocado del \emph{clump} es muy baja  y la densidad crece hasta valores
$\sim 10^{14}$~cm$^{-3}$.

Aunque en este cap'itulo estamos interesados en la emisi'on de rayos gamma
producida por la interacci'on \emph{jet-clump}, hemos estimado por completitud
la radiaci'on libre-libre generada por el material chocado y calentado del
\emph{clump}. Considerando $n_{\rm c1}(x)$ y $T_{\rm c1}(x)$ estimamos la 
luminosidad por emisi'on libre-libre integrando la emisividad a lo largo 
del \emph{clump} chocado (Lang 1999).
Las luminosidades bolom'etricas obtenidas son $L_{\rm ter} \sim 5\times10^{30}$
y $5\times10^{32}$~erg~s$^{-1}$ para $R_{\rm c} = 10^{10}$ y $10^{11}$~cm, 
respectivamente, con un m'aximo alrededor de los rayos~X blandos ($\sim 1$~keV).
La luminosidad espec'ifica se muestra juntamente con la emisi'on no t'ermica
en la Figura~\ref{SEDs_MQs}. 

Contrariamente al choque en el \emph{clump}, el \emph{bow shock} es adiab'atico
y r'apido. Por esta raz'on es un lugar propicio para la aceleraci'on de
part'iculas hasta energ'ias relativistas, como veremos a continuaci'on.

\section{Poblaci'on de part'iculas relativistas}

En lo que sigue consideramos que las part'iculas relativistas son 
aceleradas en el \emph{bow shock} 'unicamente.
Los  electrones y protones acelerados en este choque 
son inyectados siguiendo una distribuci'on
\begin{equation} 
Q_{e_1,p} = K_{e_1,p}\, E_{e_1,p}^{-2}\, 
\exp\left(-\frac{E_{e_1,p}}{E_{e_1,p}^{\rm max}}\right).
\end{equation}
Determinamos la constante de normalizaci'on $K_{e_1,p}$ 
suponiendo que el 25~\% de la luminosidad del \emph{jet} inyectada
en el \emph{bow shock}, 
$L_{\rm bs} \sim (\sigma_{\rm c}/\sigma_{\rm j})\,L_{\rm j}$,
donde $\sigma_{\rm c} = \pi R_{\rm c}^2$ es la secci'on 
efectiva\footnote{Despreciamos
la regi'on donde el \emph{bow shock} se hace muy oblicuo y nos focalizamos 
en aquella en la cual el choque es m'as fuerte, es decir, en el frente del 
\emph{clump}.} del \emph{clump}, se convierte en potencia de inyecci'on de 
las part'iculas
relativistas. As'i, fijando que $L_{e,p} = 0.25\,L_{\rm bs}$ hallamos 
\begin{equation}
\label{K_bs_clump}
K_{e_1,p} = 0.25 \,\left(\frac{R_{\rm c}}{R_{\rm j}}\right)^2\,
\frac{L_{\rm j}}
{\ln(E_{e_1,p}^{\rm max}/E_{e_1,p}^{\rm min})}.
\end{equation}

\subsection{Aceleraci'on de part'iculas y p'erdidas radiativas}

Para calcular el tiempo de aceleraci'on $t_{\rm ac}$,
necesitamos conocer el valor del campo magn'etico $B_{\rm bs}$ en la regi'on
del \emph{bow shock}. Consideramos dos valores para esta magnitud.
En primer lugar, estimamos el $B_{\rm bs}$ resultante de imponer que la 
densidad de energ'ia magn'etica $u_{\rm B} = B_{\rm bs}^2/(8 \pi)$ es el 
10~\% de la densidad de energ'ia del material chocado del
\emph{jet}, $u_{\rm j1}$, cuya expresi'on es la siguiente: 
\begin{equation}
u_{\rm j1} = \frac{3}{2} P_{\rm cin} = \frac{9}{8}\, n_{\rm j} \,m_p\, v_{\rm j}^2,
\end{equation}
donde $P_{\rm cin} = \rho\,v^2$ es la presi'on cin'etica de un  medio con
densidad $\rho$ que se mueve con velocidad $v$. Fijando
\begin{equation}
\label{B_subequip_clump}
\frac{B_{\rm bs}^2}{8 \pi} = 0.1\, u_{\rm j1}
\end{equation}
obtenemos $B_{\rm bs} \sim 150$~G, lo que nos da un tiempo de aceleraci'on  
para choques perpendiculares $t_{\rm ac}^{\perp} \sim 10^{-2} E_{e_1,p}$~s 
(ver la ecuaci'on~(\ref{t_acc_perp})).
Por otro lado, hemos adoptado un valor mucho m'as bajo, $B_{\rm bs} \sim 1$~G,
para chequear el impacto de considerar un campo magn'etico mucho menos intenso
que 150~G.
Con $B_{\rm bs} \sim 1$~G, el tiempo de aceleraci'on resulta 
$t_{\rm ac}^{\perp} \sim 0.2 E_{e_1,p}$~s.

Adem'as de acelerarse, las part'iculas pueden escapar de la regi'on chocada del 
\emph{jet} ya sea por p'erdidas difusivas o convectivas. 
El tiempo de convecci'on por los costados del \emph{clump} 
puede estimarse de la siguiente manera:
\begin{equation}
t_{\rm conv} \sim \frac{R_{\rm c}}{v_{\rm bs}/4} \sim 
4\left(\frac{R_{\rm c}}{10^{10}\,{\rm cm}}\right)\,{\rm s}.
\end{equation}
Por otro lado, el tiempo de difusi'on desde el \emph{bow shock} hasta el
\emph{clump} considerando r'egimen de Bohm resulta
\begin{equation}
t_{\rm dif} = \frac{Z^2}{D_{\rm B}} \sim 0.2
\left(\frac{R_{\rm c}}{10^{10}\,{\rm cm}}\right)^2
\left(\frac{B_{\rm bs}}{1\,{\rm G}}\right)
\frac{1}{E_{e_1,p}}\,{\rm s}.
\end{equation}
Notamos entonces que las part'iculas m'as energ'eticas pueden difundir 
hasta el \emph{clump} 
antes de ser arrastradas por el material chocado del \emph{jet}. Tanto en 
el \emph{clump} como en la regi'on del  \emph{bow shock}, los electrones
y protones relativistas
pierden energ'ia a trav'es de los diferentes procesos
radiativos que  hemos descripto en el Cap'itulo~\ref{proc-rad} y 
cuyas f'ormulas aplicaremos a continuaci'on.

\subsubsection{P'erdidas lept'onicas}

En el caso con $B_{\rm bs} = 150$~G, la radiaci'on sincrotr'on es el mecanismo
m'as eficiente de enfriamiento de los electrones en la regi'on chocada
del \emph{jet}, con una escala de tiempo
\begin{equation}
t_{\rm sin} \sim \frac{1.8\times10^{-2}}{E_{e_1}}\,{\rm s}, 
\end{equation}
mientras que si $B_{\rm bs} = 1$~G, $t_{\rm sin} \sim 4\times10^2/E_{e_1}$~s. 
Para este 'ultimo valor de $B_{bs}$ el proceso radiativo
dominante resulta ser la dispersi'on IC. 

A la altura del \emph{jet} a la cual estamos considerando que ocurre la 
interacci'on \emph{jet-clump}, $z_{\rm int} = 1.5\times10^{12}$~cm, 
la densidad de energ'ia
de los fotones provenientes de la estrella compañera del HMMQ es
$u_{\rm ph\star} \sim 2.4\times10^{-2}$~erg~cm$^{-3}$ y la energ'ia promedio
de estos fotones es $E_{\rm ph\star} \sim 3 K_{\rm B} T_{\star} \sim 10$~eV.
Para $y \equiv E_{\rm ph\star} E_e/(m_e c^2)^2 > 1$, esto es, 
$E_e > 2.5\times10^{10}$~eV,
la interacci'on IC ocurre en el r'egimen de KN. El tiempo de enfriamiento 
tanto en el r'egimen de Th como en el de KN est'a dado por la expresi'on
(ver (\ref{t_ci})):
\begin{equation} 
t_{\rm IC} \sim  0.4\,
\frac{(1 + 8.3\,y)}{\ln(1+0.2\;y)}\frac{(1 + 1.3\,y^2)}{(1 + 0.5\,y +1.3\,y^2)}
~\rm{s}\,.
\end{equation}   
 
Por otro lado, las p'erdidas por Bremsstrahlung relativista no son relevantes
en ninguno de los dos casos (ni con $B_{\rm bs} = 150$~G ni con 
$B_{\rm bs} = 1$~G), 
ya que la densidad del \emph{jet}, $n_{\rm j}$, es muy baja a la
altura $z_{\rm int}$, con lo cual la densidad en la regi'on del 
\emph{bow shock}
tambi'en resulta pequeña ($n_{\rm bs} = 4 n_{\rm j} \sim 2\times10^8$~cm$^{-3}$). 
Con este valor de $n_{\rm bs}$
el tiempo de enfriamiento por Bremsstrahlung relativista resulta muy alto
(ver (\ref{t_Brem}))
\begin{equation} 
t_{\rm Brem} =  \frac{7\times10^{7}}{\ln\left(\frac{E_e}
{m_ec^2}\right) + 0.36}~\rm{s}.
\end{equation} 

\begin{figure}
\begin{center}
\includegraphics[angle=0, width=0.45\textwidth]{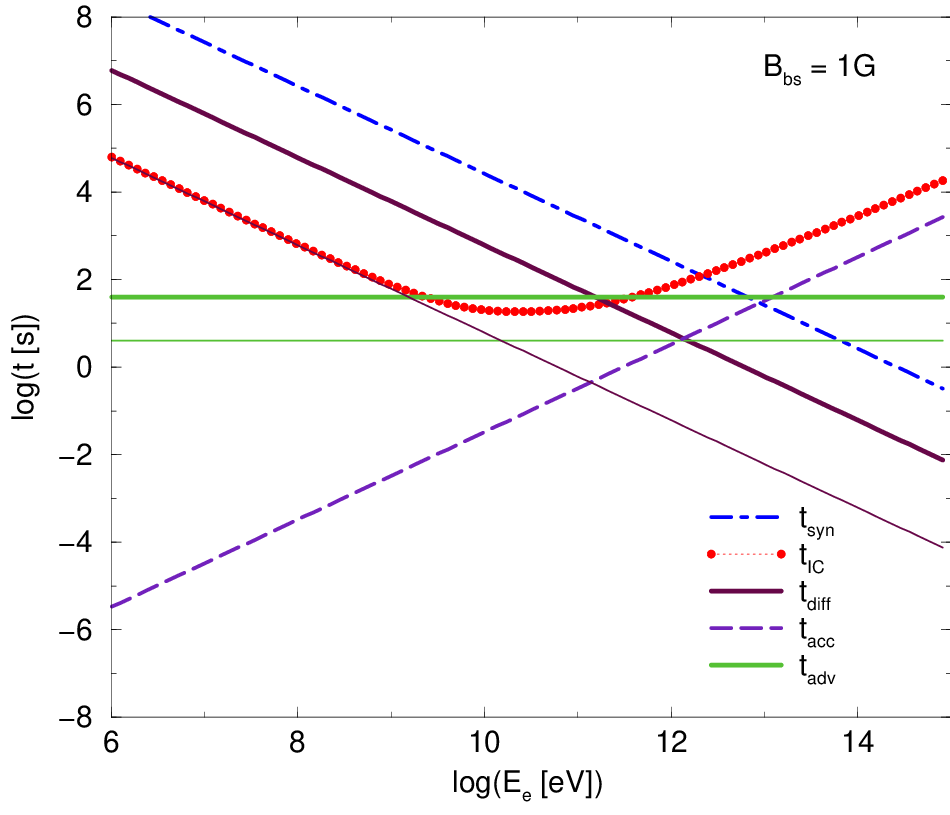}
\includegraphics[angle=0, width=0.45\textwidth]{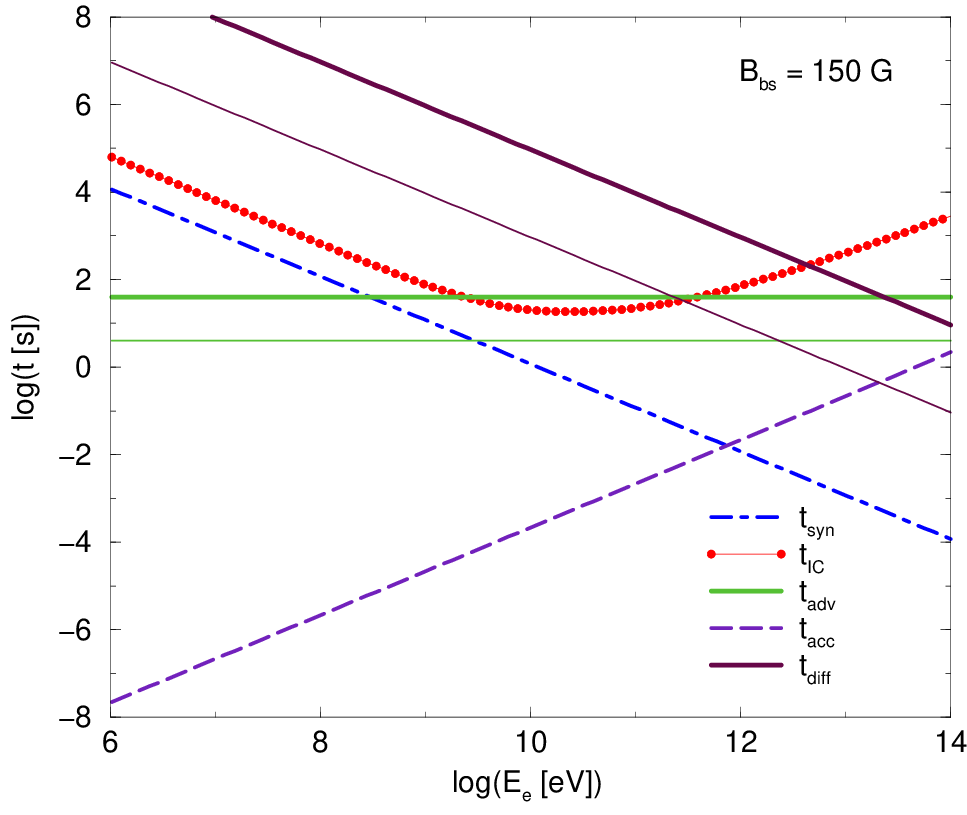}
\caption{Tiempos de aceleraci'on y de p'erdidas radiativas 
(sincrotr'on e IC) para electrones en la regi'on del \emph{bow-shock}
para los casos $B_{\rm bs} = 1$~G (izquierda) y $B_{\rm bs} = 150$~G (derecha). 
Los tiempos de convecci'on y difusi'on se
muestran para  $R_{\rm c} = 10^{10}$ (l'inea fina) y $10^{11}$~cm
(l'inea gruesa).}\label{fig_loss_1} 
\end{center}
\end{figure}

Teniendo en cuenta la ganancia por aceleraci'on, las p'erdidas radiativas y
los tiempos de escape calculamos la energ'ia m'axima que
pueden alcanzar los electrones acelerados en el \emph{bow shock}. 
En el caso con $B_{\rm bs} = 1$~G, la energ'ia m'axima est'a
determinada por las p'erdidas difusivas, mientras que en el caso
con $B_{\rm bs} = 150$~G es la radiaci'on sincrotr'on el mecanismo de 
enfriamiento dominante a 
altas energ'ias, como se muestra en la Figura~\ref{fig_loss_1}.
Como puede observarse en estos gr'aficos, la energ'ia de quiebre $E_{\rm q}$ 
del espectro
se obtiene igualando $t_{\rm conv} = t_{\rm dif}$, en el caso con
$B_{\rm bs} = 1$~G mientras que en el caso $B_{\rm bs} = 150$~G el quiebre ocurre
cuando $t_{\rm conv} = t_{\rm sin}$. Los valores de las energ'ias m'aximas
y de quiebre obtenidos en cada caso se listan en la Tabla~\ref{Table_energies}.

\begin{table}[]
\begin{center}
\caption{Energ'ias m'aximas alcanzadas por las part'iculas aceleradas 
en el \emph{bow shock}. Los  valores listados han sido calculados teniendo 
en cuenta los diferentes valores 
de $R_{\rm c}$ y $B_{\rm bs}$ considerados en este cap'itulo.}
\label{Table_energies}
\begin{tabular}{c|cccc}
\hline 
\hline
$R_{\rm c}$ [cm] & $10^{10}$ & $10^{10}$ & $10^{11}$ & $10^{11}$ \\
$B_{\rm bs}$ [G] & $1$ & $150$ & $1$ & $150$ \\ 
\hline 
$E_{e_1}^{\rm max}$ [eV] & $1.5\times 10^{11}$ & $8\times 10^{11}$ & 
$1.5\times10^{12}$ & $8\times 10^{11}$  \\  
$E_p^{\rm max}$ [eV]& $6\times 10^{11}$ & $9\times 10^{13}$ & 
$6\times10^{12}$ & $9\times 10^{14}$  \\   
$E_{\rm q}$ [eV]& $3.1\times 10^{10}$ & $2.8\times 10^{9}$ & 
$3.1\times10^{11}$ & $2.8\times 10^{8}$  \\ 
\hline
\end{tabular}
\end{center}
\end{table}

\begin{figure}
\begin{center}
\includegraphics[angle=270, width=0.49\textwidth]{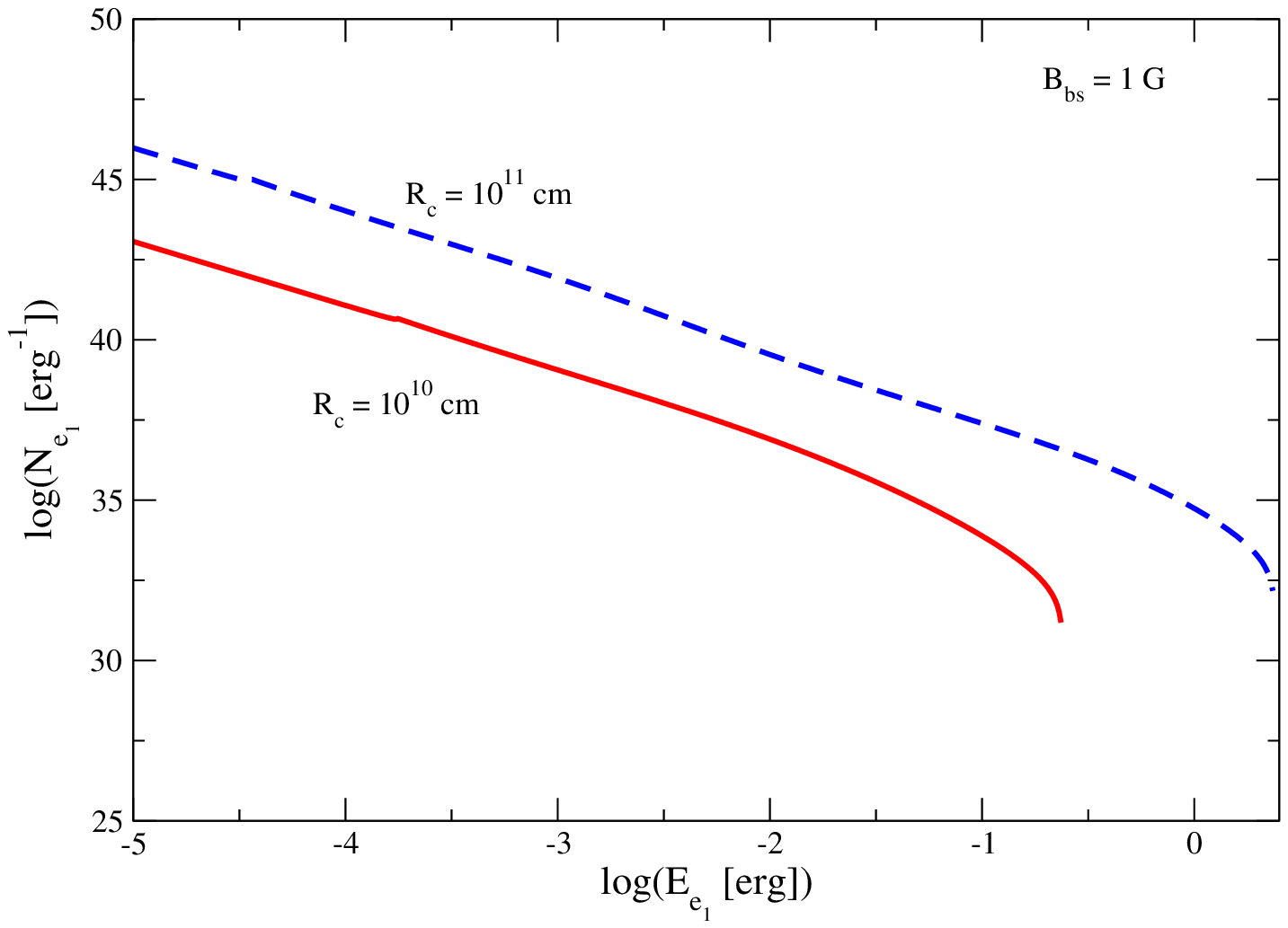}
\includegraphics[angle=270, width=0.49\textwidth]{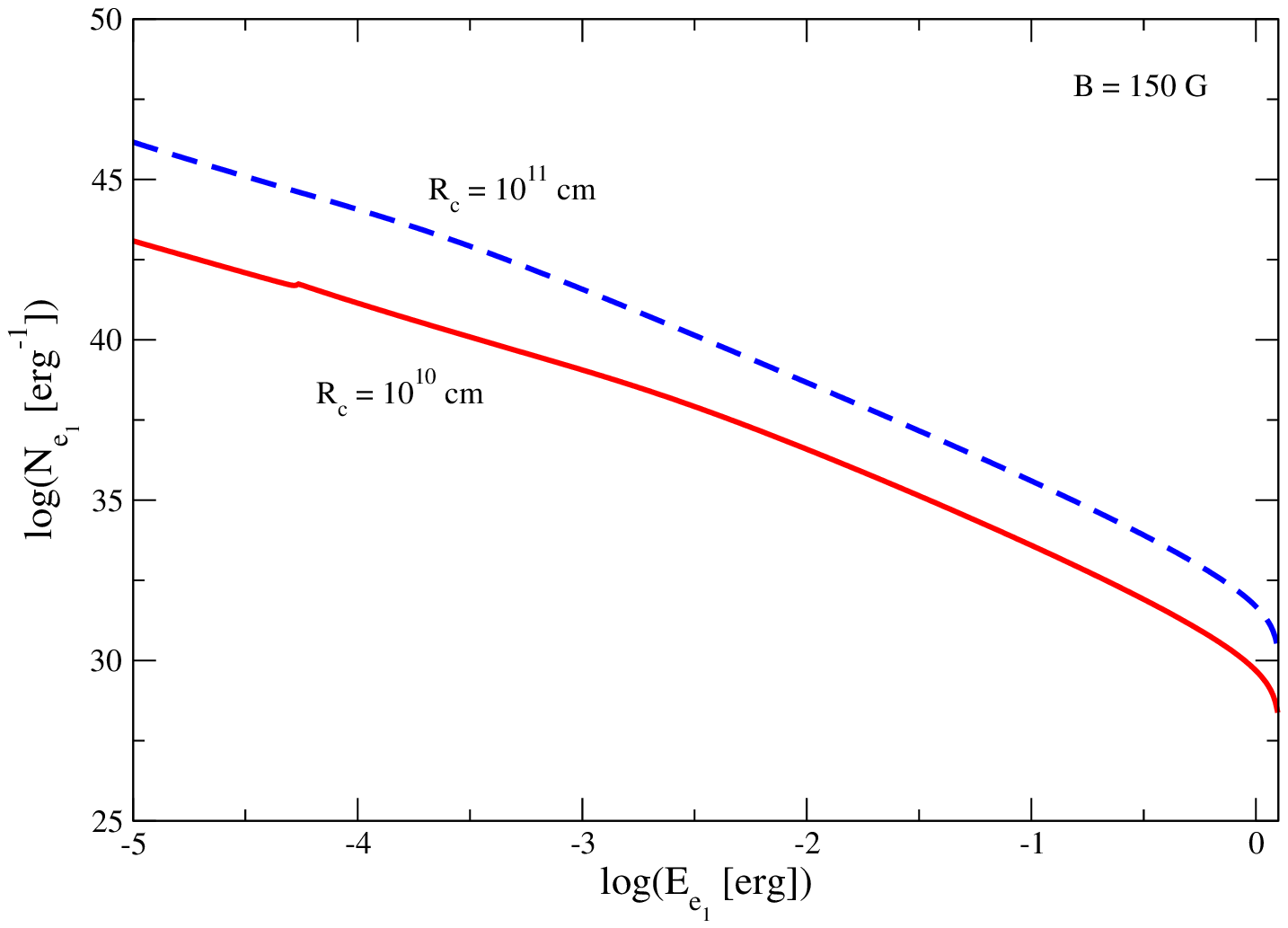}
\caption{Distribuci'on de energ'ia de los electrones acelerados en el 
\emph{bow shock} para los
casos $B_{\rm bs} = 1$ (izquierda) y 150~G (derecha).}
\label{time_evol_elec_MQs} 
\end{center}
\end{figure}

Teniendo en cuenta las p'erdidas radiativas que sufren los 
electrones relativistas en la regi'on del \emph{bow shock}, calculamos la
distribuci'on espectral de energ'ia de 'estos, $N_{e_1}(E_{e_1})$, a trav'es 
de la ecuaci'on~(\ref{kinetic}) y teniendo en cuenta la constante de 
normalizaci'on del 
espectro de inyecci'on dada por la ecuaci'on~(\ref{K_bs_clump}). 
Consideramos un tiempo de duraci'on de la interacci'on 
$\tau_{\rm vida} > t_{\rm cc}$ y obtenemos un
espectro en estado estacionario ya que los tiempos de p'erdidas
radiativas (sincrotr'on e IC) y de escape son menores que $t_{\rm cc}$. 
Los resultados obtenidos se muestran en la 
Figura~\ref{time_evol_elec_MQs}. 

Finalmente, los electrones m'as energ'eticos ($E_{e_1} > 0.3 E_{e_1}^{\rm max}$)
se podr'ian difundir hasta el \emph{clump}. Esto ocurre si 
$t_{\rm dif} < t_{\rm conv}$ y $t_{\rm dif} < t_{\rm rad}$, donde la distancia
que deben difundirse es el grosor del \emph{bow shock}. Si los electrones
llegan al \emph{clump}, probablemente rad'ien all'i toda su energ'ia.

\subsubsection{P'erdidas hadr'onicas}

De la misma manera que el Bremsstrahlung relativista es un canal de 
enfriamiento muy lento para los electrones en la regi'on chocada del \emph{jet},
el $pp$, con un ritmo de enfriamiento similar al Bremsstrahlung relativista,
es tambi'en un proceso poco eficiente para los protones. En la
regi'on del \emph{bow shock} $t_{pp}$ resulta $\sim 10^7$~s, ya que la densidad
del \emph{jet} chocado es $n_{\rm bs} \sim 2\times10^8$~cm$^{-3}$. 
La energ'ia m'axima de los protones relativistas acelerados en el
\emph{bow shock} queda  determinada por el 
%l'imite de Hillas, es 
%decir, imponiendo que $r_{\rm g} = Z$, con lo cual tenemos que 
%$E_p \leq Z q B_{bs}$ y la energi'a m'axima resulta
%
tiempo de difusi'on, dando
\begin{equation}
E_p^{\rm max} = 9\times10^{11} \left(\frac{R_{\rm c}}{10^{10}\,{\rm cm}}\right)  
\left(\frac{B_{\rm bs}}{1\,{\rm G}}\right)\,{\rm eV}. 
\end{equation}

Los protones con energ'ia $E_p > 0.025\,E_p^{\rm max}$  
pueden difundir hasta el clump, ya que para estas energ'ias 
$t_{\rm dif} < t_{\rm conv}$ en la direcci'on perpendicular al choque. 
Una vez all'i, estos protones pueden interactuar
con el material chocado del \emph{clump} y emitir rayos gamma en
un tiempo $t_{pp} \sim 500$~s. 
Para que los 
protones puedan estar confinados en el \emph{clump}, el campo magn'etico de
'este debe ser  $B_{\rm c} > 7.5\times10^{-3}\,B_{\rm bs}$, es decir,
$B_{\rm c} > 7.5\times10^{-3}$ 'o $1.125$~G, para el caso con 
$B_{\rm bs} = 1$ y 150~G,
respectivamente. Luego, los protones menos energ'eticos de aqu'ellos 
que lleguen al clump podr'an ser confinados para valores de $B_{\rm c}$ 
razonables. Pero este l'imite de confinamiento es muy laxo, y en realidad 
el campo $B_{\rm c}$
necesario para confinar a los protones podr'ia ser mucho mayor. 

Aqu'i consideraremos
el caso m'as conservativo para calcular la distribuci'on de energ'ia de 
los protones en el \emph{clump}. Es decir, teniendo en cuenta que 'estos  
permanecen en el \emph{clump} s'olo el tiempo que tardan en cruzarlo a 
una velocidad cercana a $c$, esto es, 
$t_{\rm cruce} \sim R_{\rm c}/c \sim 0.3$ 'o 3~s si 
$R_{\rm c} = 10^{10}$ 'o $10^{11}$~cm, respectivamente.
Siendo $t_{\rm cruce} < t_{pp}$,
la distribuci'on de energ'ia de los protones en el \emph{clump} resulta
\begin{equation}
\label{Np-jet-clump}
N_p(E_p) = \frac{R_{\rm c}}{c}\,Q_p(E_p),
\end{equation}
donde $0.025\,E_p^{\rm max} < E_p < E_p^{\rm max}$.

\section{Distribuci'ones espectrales de energ'ia}

\subsection{Emisi'on asociada al \emph{bow shock}}

Como mostramos en la secci'on de p'erdidas lept'onicas, los mecanismos m'as 
eficientes de p'erdidas radiativas son el sincrotr'on y la dispersi'on IC. 
Sin embargo, debido a que la densidad de energ'ia de la emisi'on 
sincrotr'on es  
menor que la correspondiente al campo magn'etico ($u_{\rm B}$) o a campos de
radiaci'on externos, la emisi'on
por SSC no ser'a importante y por lo tanto no la tenemos en cuenta en 
nuestros c'alculos. S'olo consideramos
los procesos sincrotr'on e IC externo.

\begin{figure}
\begin{center}
\includegraphics[angle=0, width=0.45\textwidth]{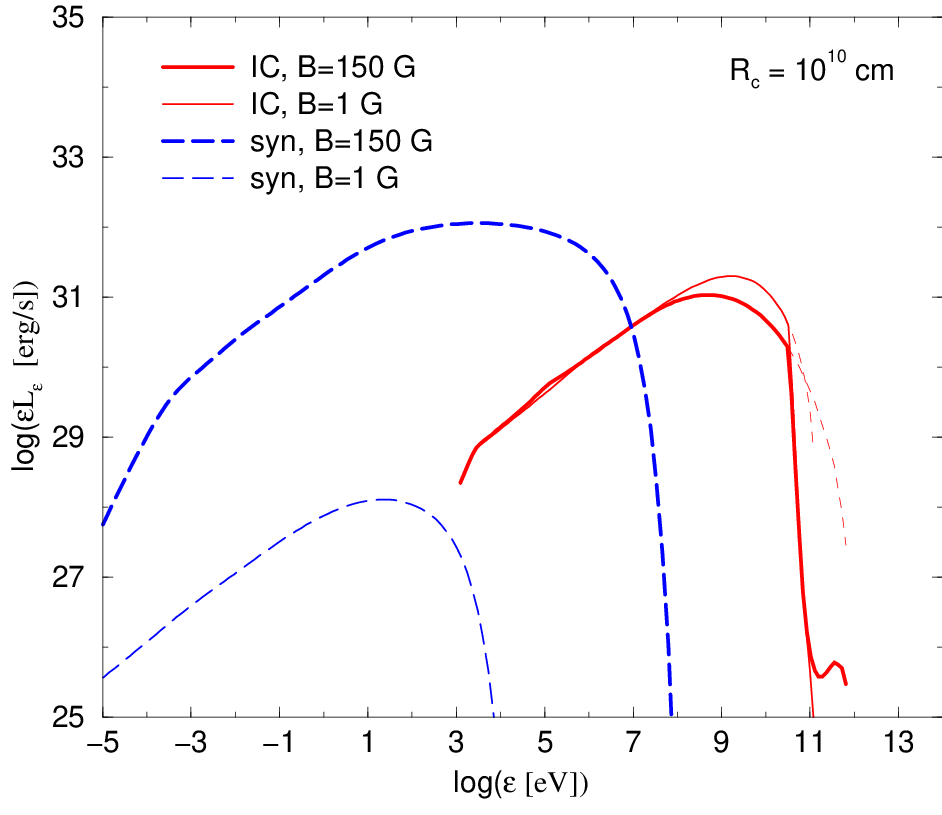}
\includegraphics[angle=0, width=0.45\textwidth]{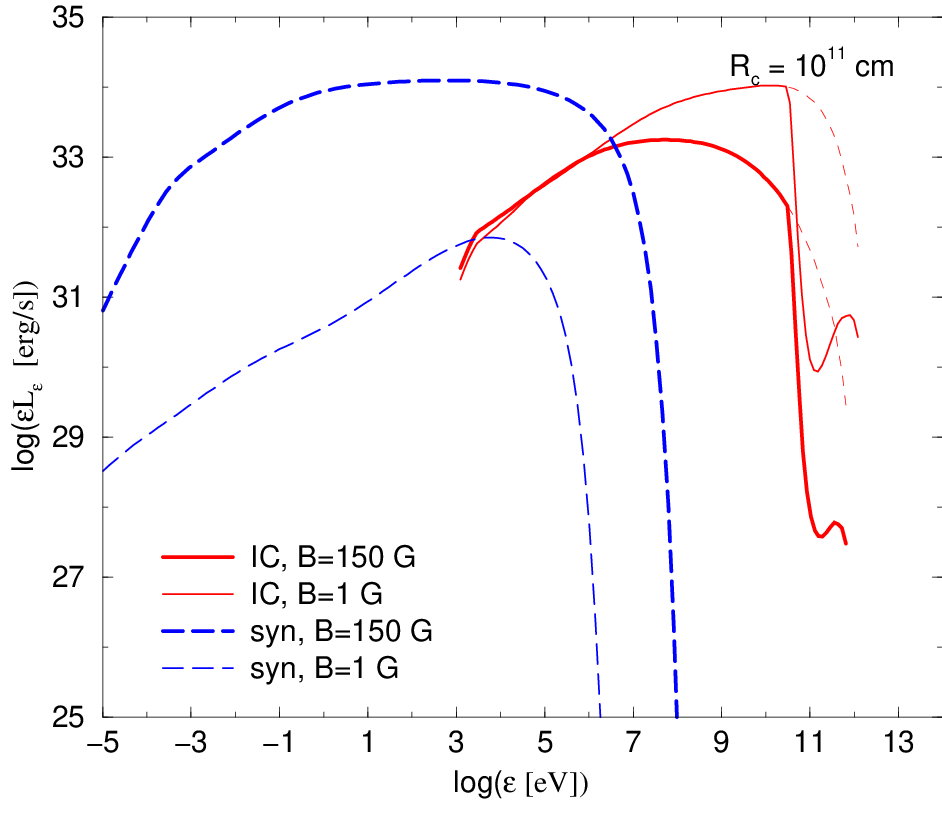}
\caption{Radiaci'on sincrotr'on (l'inea segmentada azul) y IC (l'inea roja) 
producida en la regi'on del \emph{bow-shock}, para
$R_{\rm c} = 10^{10}$ (izquierda) y $R_{\rm c} = 10^{10}$~cm (derecha), 
con $B_{\rm bs}=1$ (l'inea fina) y 150~G 
(l'inea gruesa). Los efectos de la absorci'on gamma-gamma 
son mostrados (l'inea gruesa), juntamente con el espectro no absorbido 
(l'inea punteada fina).}
\label{Synch-IC_1}
\end{center}
\end{figure}

Calculamos $E_{\rm ph}L_{\rm sin}(E_{\rm ph})$ y 
$E_{\rm ph}L_{\rm IC}(E_{\rm ph})$ para los diferentes valores de $B_{\rm bs}$ y
$R_{\rm c}$ considerados en este cap'itulo y los resultados se muestran en la 
Figura~\ref{Synch-IC_1}. 
Como puede verse en los gr'aficos, la componente debida a la radiaci'on 
sincrotr'on
es m'as luminosa que la correspondiente al IC en los casos con 
$B_{\rm bs} = 150$~G, alcanzando luminosidades bolom'etricas
$L_{\rm sin} \sim 10^{33}$ y $2\times10^{35}$~erg~s$^{-1}$ 
para $R_{\rm c} = 10^{10}$ y $10^{11}$~cm, respectivamente. 
Por el contrario, 
para $B_{\rm bs} = 1$~G, el proceso radiativo dominante es la interacci'on
IC, con luminosidades  bolom'etricas 
$L_{\rm IC}\sim 2\times10^{32}$ y $10^{35}$~erg~s$^{-1}$ para
$R_{\rm c} = 10^{10}$ y $10^{11}$~cm, respectivamente.
Las energ'ias m'aximas alcanzadas por los fotones emitidos  son 
$E_{\rm ph}^{\rm max} \sim 1$~TeV. Notamos que tanto el espectro debido a la
radiaci'on sincrotr'on como a la dispersi'on IC en el caso con 
$B_{\rm bs} = 150$~G se quiebra de manera clara por efecto
de las p'erdidas por convecci'on de las part'iculas de la regi'on del 
\emph{bow shock}. Adem'as, el efecto de las p'erdidas IC en el r'egimen de KN
endurecen el espectro de los electrones en el caso con 
$R_{\rm c} =10^{11}$~cm y $B_{\rm bs} = 1$~G,
lo cual se ve reflejado en el espectro
de los fotones producidos por los mecanismos sincrotr'on e IC.
Sin embargo, notamos que 
los electrones relativistas aunque son arrastrados de 
la regi'on del
\emph{bow shock} antes de que emitan significativamente en frecuencias radio, 
tambi'en pueden
emitir en esta banda de energ'ias en otras regiones del 
\emph{jet}\footnote{El c'alculo de esta emisi'on
escapa a los intereses de nuestro estudio. Por esta raz'on, no hemos tenido en
cuenta los efectos de la autoabsorci'on sincrotr'on en el espectro en 
frecuencias radio y no
haremos predicciones en este rango de energ'ias.}.

En los c'alculos de las SEDs hemos tenido en cuenta la absorci'on 
(gamma-gamma) por creaci'on de pares $e^{\pm}$ en el campo de los fotones 
producidos por la estrella. Debido a que no nos focalizamos en la 
geometr'ia del sistema HMMQ/observador, hemos supuesto que el campo
de fotones semilla  es isotr'opico (tambi'en hemos
despreciado los efectos angulares en el c'alculo de la interacci'on IC).
Como puede verse en la Figura~\ref{Synch-IC_1}, la absorci'on gamma-gamma
reduce los niveles de emisi'on varios 'ordenes de magnitud 
a energ'ias de cientos de GeV.
Solo en algunos casos, con geometr'ias espec'ificas en la interacci'on
gamma-gamma, la atenuaci'on puede ser despreciable (Khangulyan et al. 2008).

\subsection{Emisi'on asociada al \emph{clump}}

Las part'iculas m'as energ'eticas aceleradas en el \emph{bow shock} 
pueden difundir hasta el \emph{clump} y radiar all'i. En el caso de los 
protones, aquellos con $E_p > 0.025E_p^{\rm max}$ pueden llegar 
al \emph{clump}.
Por otro lado, en el caso $B_{\rm bs}= 1$~G, los electrones con 
$E_{e_1} > 0.3 E_{e_1}^{\rm max}$ son los que pueden difundir una distancia $Z$ y
perder toda su energ'ia en el \emph{clump}. 

\begin{figure*}
\centering 
\includegraphics[angle=0,width=0.45\textwidth]{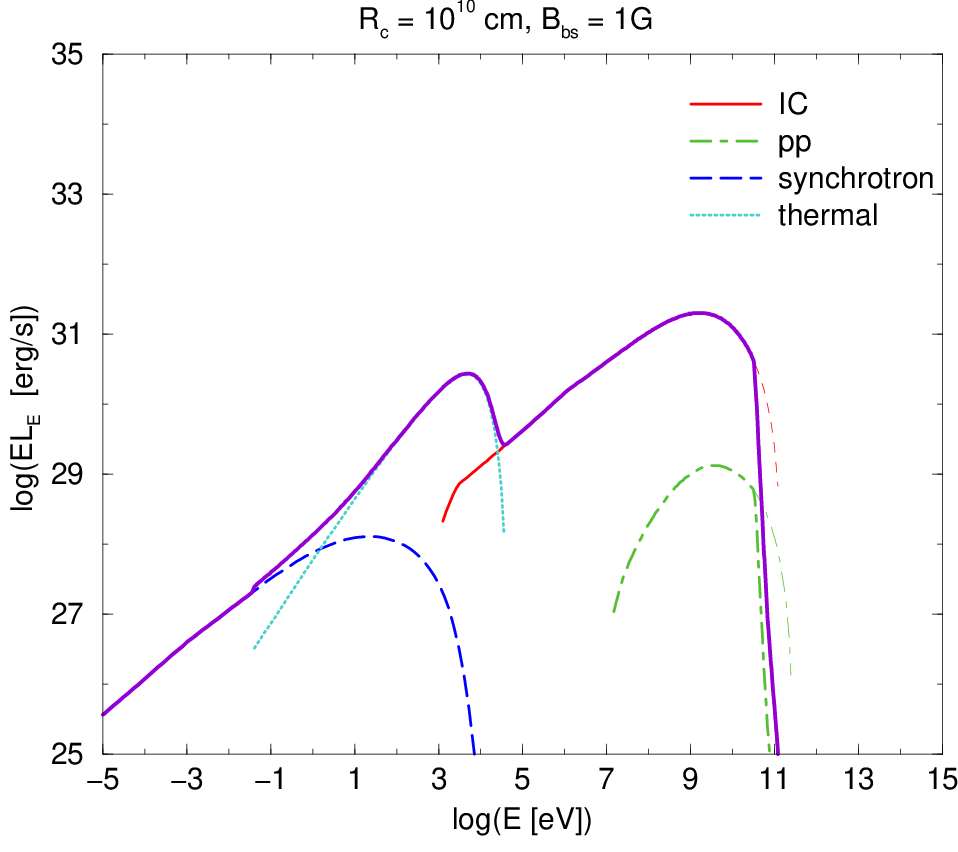}\qquad
\includegraphics[angle=0,width=0.45\textwidth]{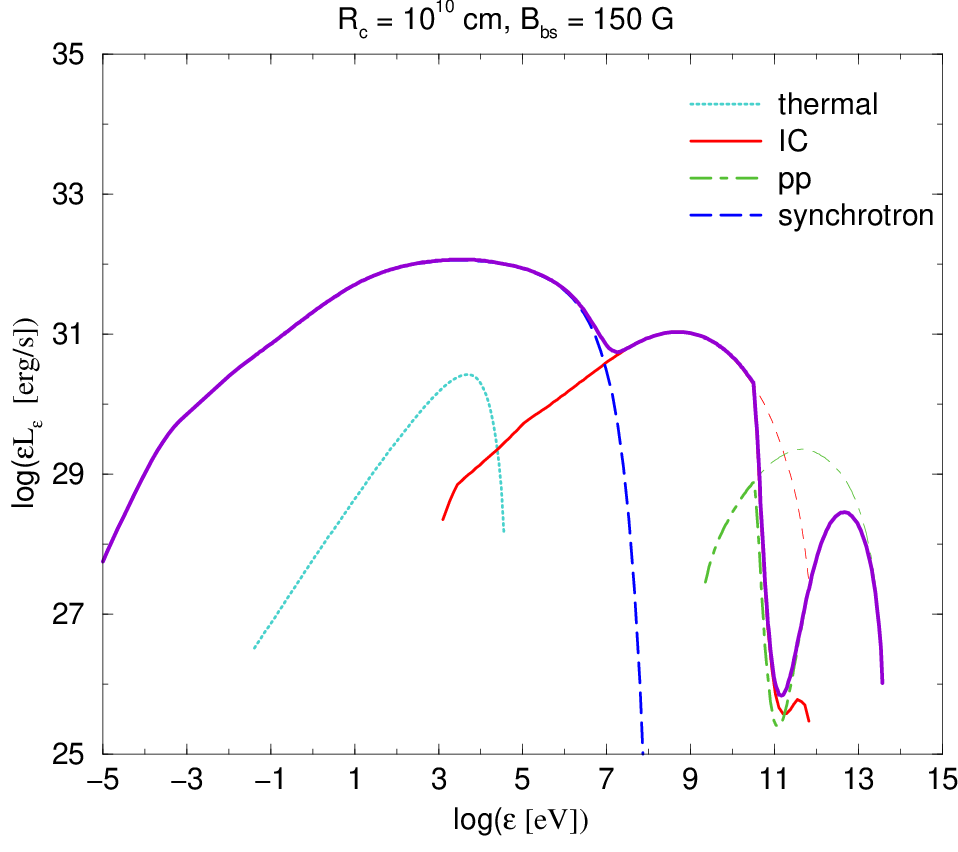}\\[10pt]
\includegraphics[angle=0, width=0.45\textwidth]{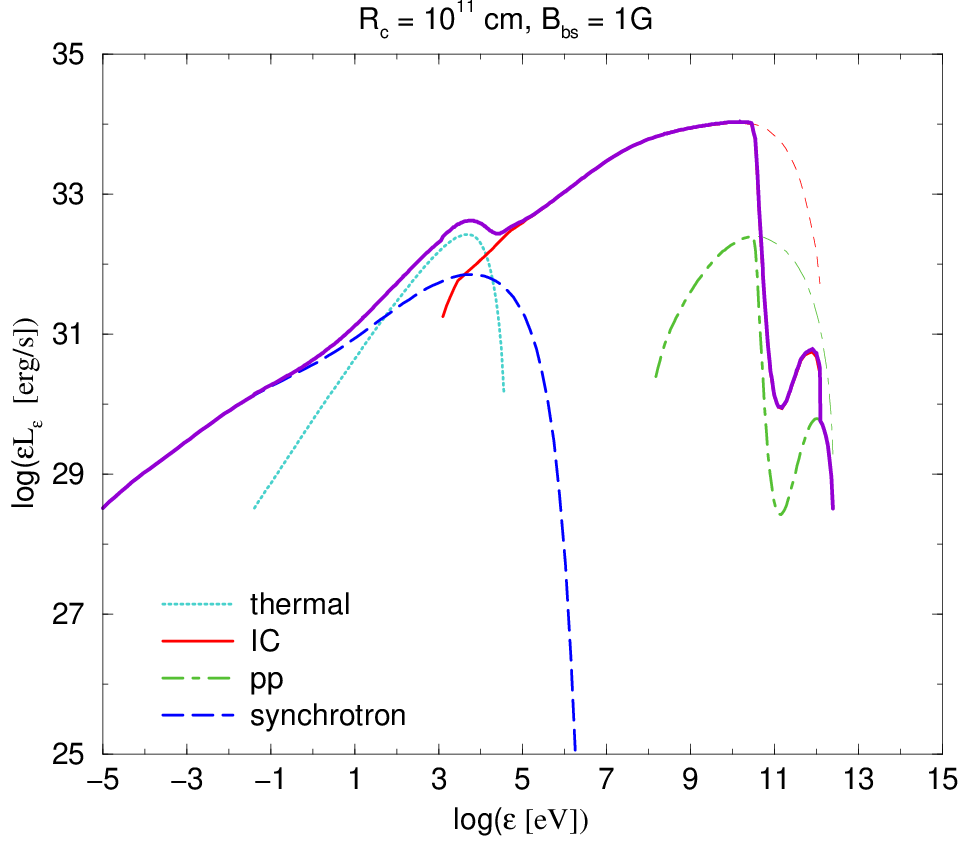}\qquad
\includegraphics[angle=0, width=0.45\textwidth]{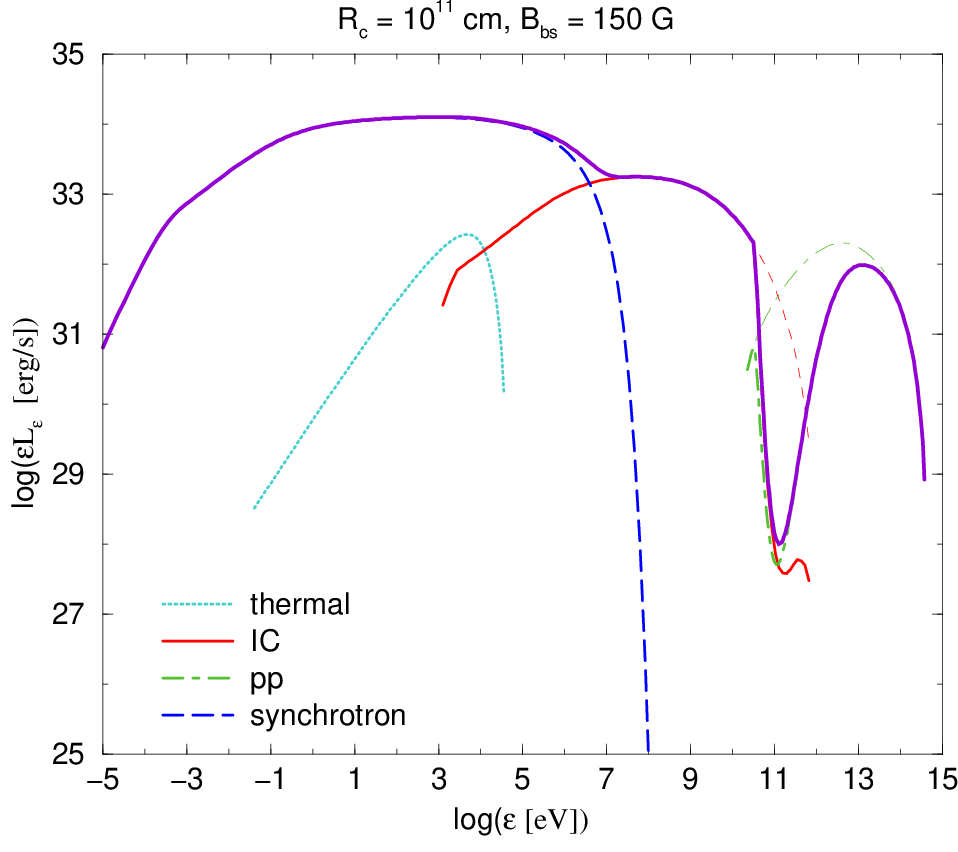}\\
\caption{SEDs de la emisi'on sincrotr'on, IC, $pp$ y t'ermica para 
diferentes valores de $B_{\rm bs}$ y $R_{\rm c}$;
las curvas de la componente absorbida y no absorbida (l'ineas finas) 
de la radiaci'on IC y $pp$ son graficadas.}
\label{SEDs_MQs}
\end{figure*}

A la emisi'on $pp$ de los protones que difunden hasta el \emph{clump} la 
calculamos usando las ecuaciones (\ref{q_pi}) y (\ref{q_pp}), 
y considerando que $N_p(E_p)$ est'a 
determinada por la f'ormula~(\ref{Np-jet-clump}). 
La emisi'on por $pp$ alcanza una luminosidad
$\sim 10^{32}$~erg~s$^{-1}$ a una energ'ia $E_{\rm ph} \sim 50$~GeV, 
como se muestra en la 
Figura~\ref{SEDs_MQs}, junto con las
contribuciones lept'onicas de la regi'on del \emph{bow shock}. Adem'as de 
rayos gamma, en las interacciones $pp$ tambi'en se producen
pares $e^{\pm}$ y neutrinos de muy altas energ'ias, siendo la luminosidad
de estos 'ultimos  $\sim L_{pp}$ (Aharonian et al. 2006; 
Reynoso \& Romero 2009). 
Los pares $e^{\pm}$ rad'ian casi toda su
energ'ia en el \emph{clump} por los procesos sincrotr'on, IC y Bremsstrahlung
relativista, pero la contribuci'on de los electrones primarios tanto en la
regi'on del \emph{bow shock} como en el \emph{clump} es mayor que la de los 
pares (Bosch-Ramon et al 2005; Orellana et al. 2007).

\begin{figure}
\begin{center}
\includegraphics[angle=0, width=0.45\textwidth]{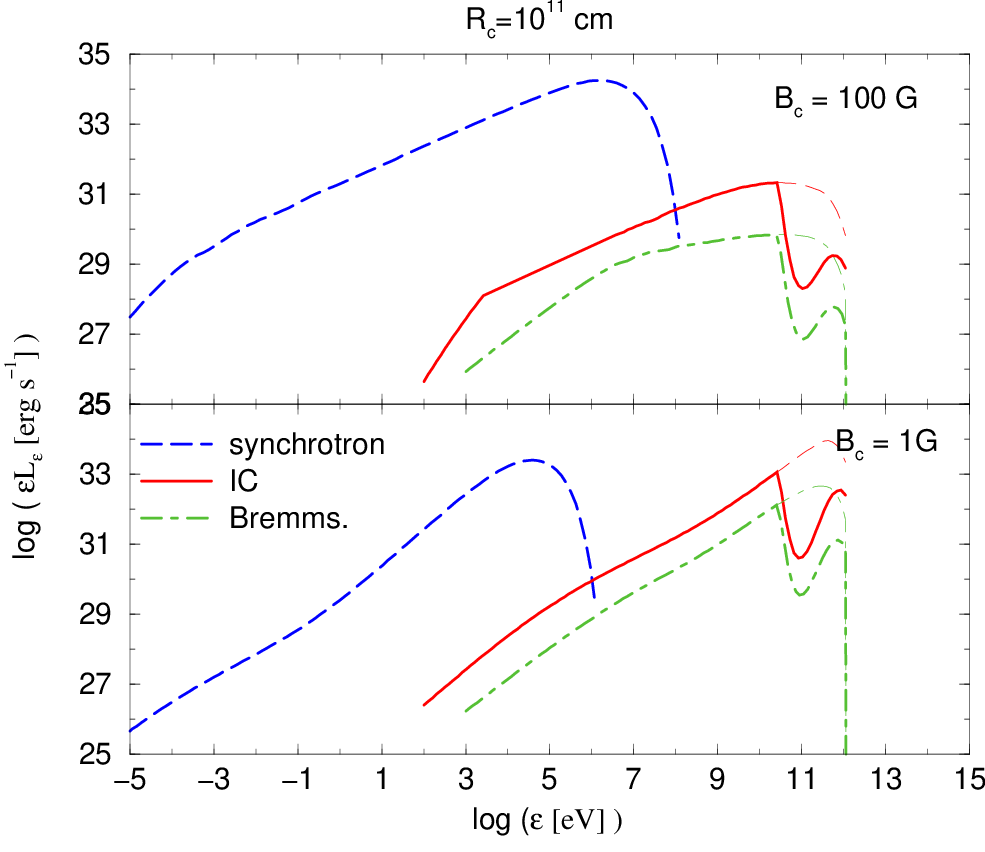}
\caption{Emisi'on no t'ermica del \emph{clump}, para el caso con 
$R_{\rm c} = 10^{11}$~cm y
para los dos valores de $B_{\rm c}$ asumidos. Las curvas correspondientes a la 
emisi'on por IC y Bremsstrahlung relativista se muestran absorbidas
(l'ineas gruesas) y no absorbidas (l'ineas finas).}
\label{electrones-clump}
\end{center}
\end{figure}

Para estimar la radiaci'on producida en el \emph{clump} por los electrones 
primarios acelerados 
en el  \emph{bow shock} y que difunden hasta all'i, suponemos 
dos valores
para el campo magn'etico del clump: $B_{\rm c} = 1$ y 100~G, siendo el 'ultimo 
similar al valor de equipartici'on entre la densidad de energ'ia magn'etica 
y t'ermica. %CHEQUEAR!!!
A las luminosidades espec'ificas de cada proceso (sincrotr'on, IC y 
Bremsstrahlung relativista) las calculamos con las f'ormulas dadas en el 
Cap'itulo~\ref{proc-rad}. Como se muestra en la Figura~\ref{electrones-clump}, 
la componente sincrotr'on domina a la debida a las interacciones IC en el 
caso con 
$B_{\rm c} = 100$~G, alcanzando una luminosidad $\sim 10^{34}$~erg~s$^{-1}$
($R_{\rm c} = 10^{11}$~cm) a $E_{\rm ph} \gtrsim 1$~MeV. Por otro lado, para
$B_{\rm c} = 1$~G la componente IC absorbida (por gamma-gamma) alcanza una 
luminosidad 
similar a la sincrotr'onica y es $\sim 10^{33}$~erg~s$^{-1}$. La emisi'on por
Bremsstrahlung relativista es despreciable en ambos casos.
Las SEDs para el caso $R_{\rm c} = 10^{10}$~cm, y para ambos valores de 
$B_{\rm c}$, son morfol'ogicamente similares a las
correspondientes mostradas en la Figura~\ref{electrones-clump}, pero las 
luminosidades alcanzadas son aproximadamente dos 'ordenes de magnitud 
menores, ya que 'estas son $\propto \sigma_c$.

Notamos finalmente que la emisi'on no t'ermica del \emph{clump} es 
similar, aunque levemente menos intensa, a la producida en la regi'on
del \emph{bow shock}, siendo el espectro m'as duro. Esto es debido a que
las part'iculas inyectadas en el \emph{clump} tienen energ'ias 
$> 0.3 E_{e_1}^{\rm max}$.
Por claridad, no graficamos la 
emisi'on lept'onica del \emph{clump} jutamente con las dem'as componentes 
(lept'onicas
de la regi'on del \emph{bow shock} y hadr'onica del \emph{clump}).

\section{Interacciones simult'aneas}

Hasta aqu'i solo hemos considerado la interacci'on de un \emph{clump} con el 
\emph{jet} a la altura $z_{\rm int} = a_{\rm mq}/2$. Sin embargo, muchos 
\emph{clumps} pueden
estar simult'aneamente interactuando con el \emph{jet} a diferentes $z_{\rm j}$
(Owocki et al. 2009).

\begin{figure}
\begin{center}
\includegraphics[angle=270, width=0.49\textwidth]{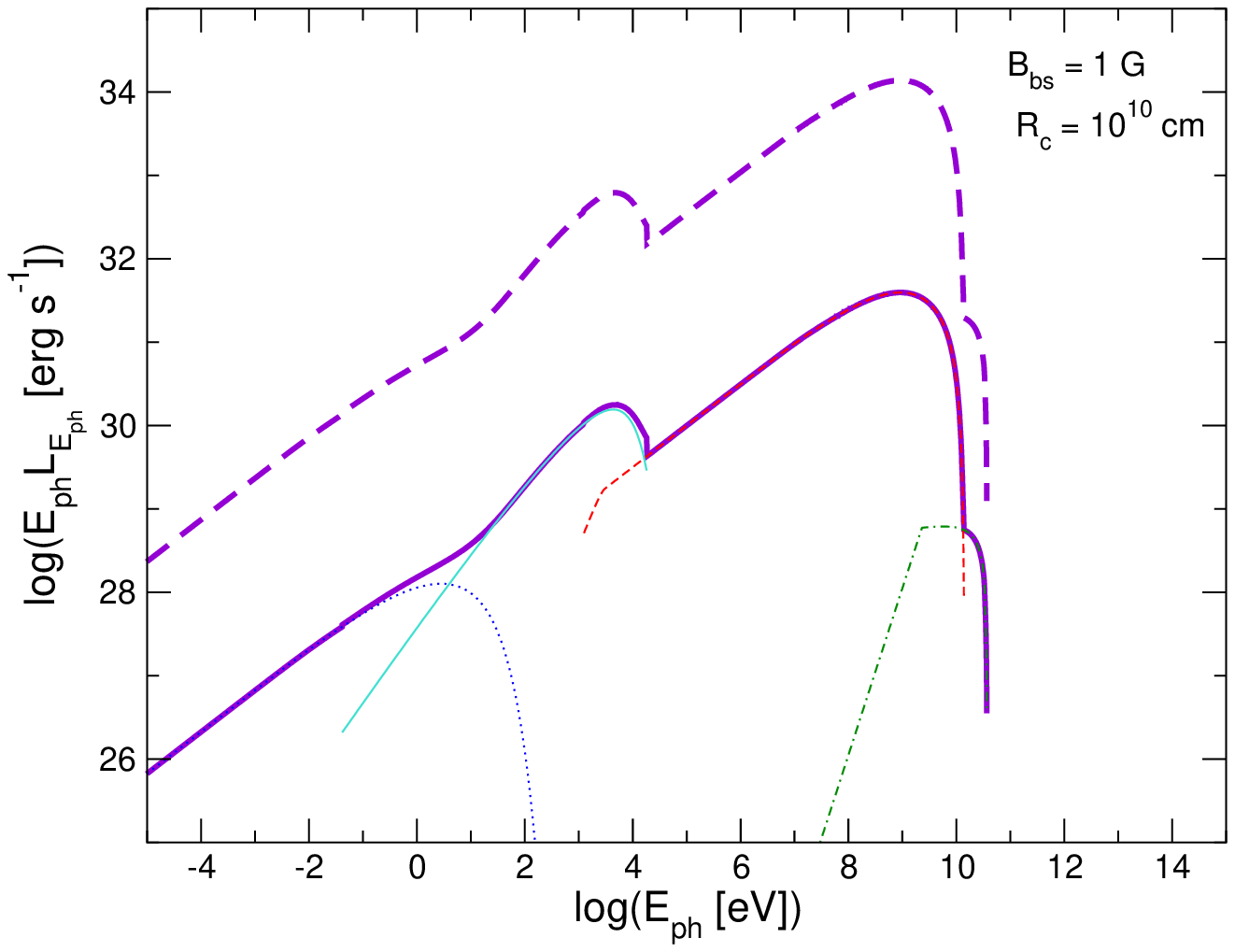}
\includegraphics[angle=270, width=0.49\textwidth]{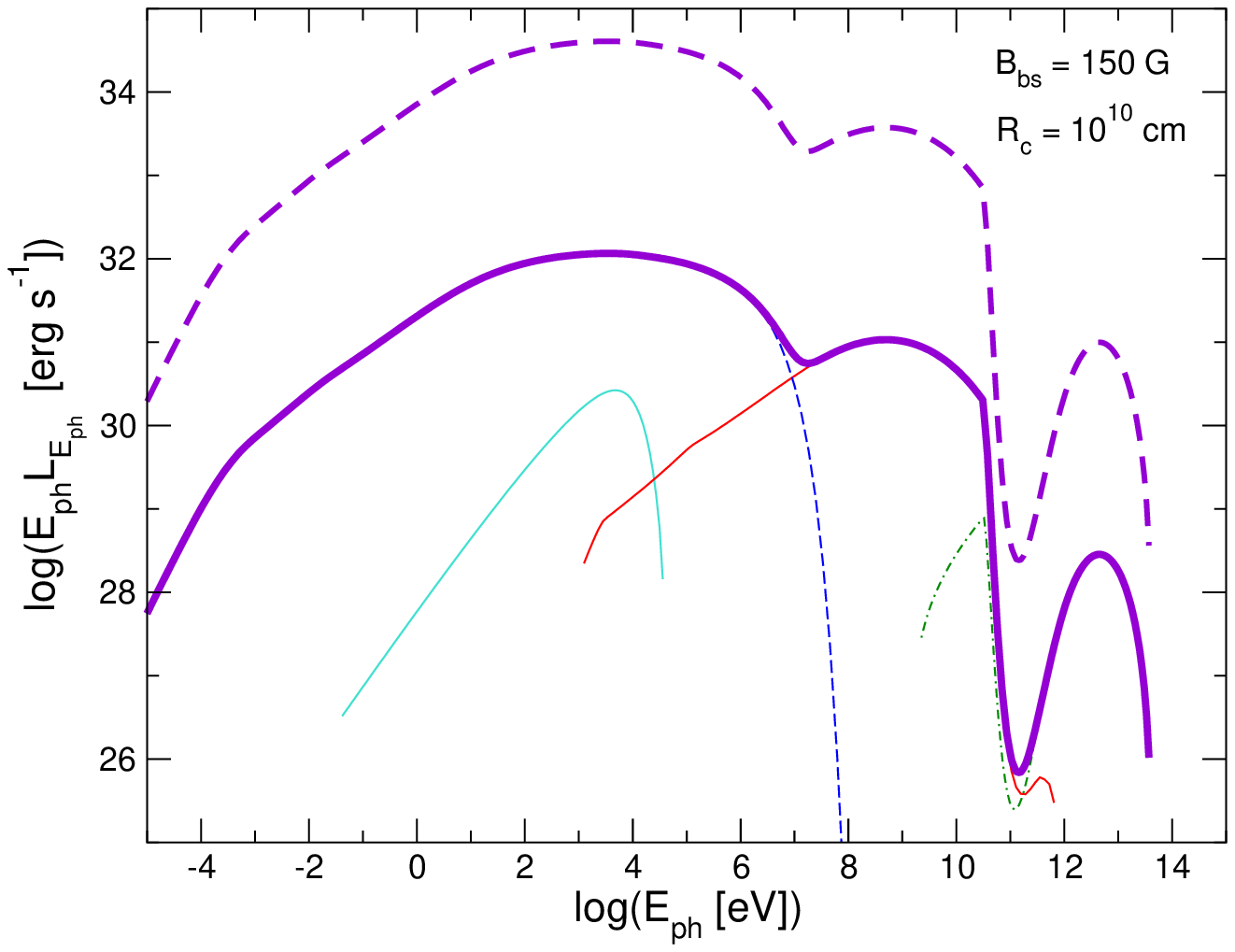}
\caption{Distribuciones espectrales de energ'ia de la emisi'on sincrotr'on, 
IC, $pp$ y t'ermica para el caso con \emph{clumps} de $R_{\rm c} = 10^{10}$~cm 
y $B_{\rm bs} = 1$ (derecha) y 150~G (derecha).
Hemos tenido en cuenta la contribuci'on de todos los \emph{clumps} que 
simult'aneamente se encuentran dentro del \emph{jet}.}
\label{SEDs-clumps}
\end{center}
\end{figure}

La altura m'inima $z_{\rm int}^{\rm min}$, a la cual los \emph{clumps} pueden 
penetrar completamente dentro 
del \emph{jet} sin ser destruidos en el proceso, es aquella para la cual 
$t_{\rm c} < t_{\rm cc}$. Esto determina un 
$z_{\rm int}^{\rm min} \sim 4\times10^{11}$~cm.
De esta manera, el modelo presentado en este cap'itulo para la interacci'on
\emph{jet-clump} es v'alido s'olo para $z_{\rm int} > 4\times10^{11}$~cm 
(y para $R_{\rm c} < R_{\rm j}$). 
Considerando que el \emph{jet} presenta una geometr'ia c'onica, calculamos 
el n'umero de \emph{clumps} $N_{\rm c}$ que pueden estar simult'aneamente 
dentro del \emph{jet}.
 Para esto integramos desde $z = z_{\rm int}^{\rm min}$ hasta $z = a_{\rm mq}$ y 
asumimos que el factor de llenado de \emph{clumps} en el \emph{jet} es el 
mismo que en el viento, $f = 0.005$. 
As'i hallamos $N_{\rm c} \sim 350$ y 0.5 para $R_{\rm c} = 10^{10}$
y $10^{11}$~cm, respectivamente.   Como consecuencia de estos resultados, 
los \emph{flares} producidos por la interacci'on de un \emph{clump} con el 
\emph{jet} 
ser'an un fen'omeno espor'adico para $N_{\rm c}$ bajo 
($R_{\rm c} = 10^{11}$~cm), o ser'a una modulaci'on de la emisi'on continua 
de la fuente, para  $N_{\rm c}$ alto 
($R_{\rm c} = 10^{10}$~cm) (ver Owocki et al. 2009). En el 'ultimo caso, 
la SED resultante ser'a 
morfol'ogicamente similar a las mostradas en las secciones previas, pero
multiplicada por $N_{\rm c}$, como se muestra en la Figura~\ref{SEDs-clumps}.
Sin embargo, notamos que por un lado el \emph{jet} puede verse din'amicamente 
afectado si muchos \emph{clumps} est'an simult'aneamente dentro de 'el. 
Por el otro, a 
diferentes alturas de interacci'on, la luminosidad no t'ermica disponible 
para radiar es distinta (disminuyendo con $z_{\rm j}$) y por lo tanto la SED 
resultante no es un simple escaleo con $N_{\rm c}$
como el mostrado en la Figura~\ref{SEDs-clumps}.

\section{Discusi'on}

En este cap'itulo exploramos los procesos f'isicos m'as relevantes y la 
naturaleza de la radiaci'on producida por la interacci'on de un 
\emph{clump} del 
viento de la estrella primaria con el \emph{jet} de un HMMQ.
La interacci'on \emph{jet-clump} produce dos choques: uno en el \emph{jet} y 
otro en el \emph{clump}. El primero alcanza r'apidamente el estado estacionario
formando un  \emph{bow shock} en el \emph{jet} mientras que el segundo se 
propaga a trav'es
del \emph{clump} empujado por la presi'on del medio chocado, que se equilibra con 
la presi'on del material del \emph{jet} chocado. 
El \emph{bow shock} es adiab'atico y
r'apido, y  part'iculas cargadas pueden acelerarse hasta energ'ias muy altas 
all'i mediante el mecanismo de Fermi. En la regi'on del \emph{bow shock} los 
electrones relativistas se enfr'ian eficientemente por radiaci'on sincrotr'on e
IC. Por otro lado, el choque en el \emph{clump} es lento y radiativo, no siendo
eficiente para acelerar part'iculas. Sin embargo, la emisi'on t'ermica del 
material 
chocado del \emph{clump} podr'ia ser significativa, como as'i tambi'en 
podr'ian serlo 
las interacciones $pp$ entre los protones relativistas acelerados en el 
\emph{bow shock} que difunden hasta el \emph{clump} y el material chocado
de 'este. Si los electrones acelerados en el
\emph{bow shock} llegan hasta el \emph{clump}, 'estos pueden a su vez 
radiar en el
\emph{clump} eficientemente a trav'es de los procesos sincrotr'on e IC.
Las SEDs de las componentes radiativas mencionadas anteriormente han sido
calculadas en el contexto de un HMMQ con par'ametros similares a los del
sistema Cygnus~X-1 y los resultados se muestran en las Figuras~\ref{SEDs_MQs} y 
\ref{electrones-clump}.

En rayos~X, la emisi'on es producida por radiaci'on sincrotr'on (en el 
\emph{bow shock} y en el \emph{clump}) y t'ermica (en el \emph{clump}).
En los casos con $B_{\rm bs}\sim 1$~G, la
emisi'on t'ermica alcanza luminosidades $L_{\rm ter}\sim 10^{32}$~erg~s$^{-1}$
siendo as'i mayor que la sincrotr'on emitida en la regi'on del 
\emph{bow-shock}, pero no mayor que la emitida en el \emph{clump} si 
$R_{\rm c}=10^{11}$~cm. 
La emisi'on sincrotr'on de la regi'on del \emph{bow-shock} 
es dominante en rayos~X para $B_{\rm bs}=150$~G, alcanzando
$L_{\rm sin} \sim 10^{35}$~erg~s$^{-1}$. En una fuente como
Cygnus~X-1, estos niveles de emisi'on en rayos~X 
ser'ian  superados por la radiaci'on del disco de acreci'on.
Sin embargo, en el caso de fuentes poco luminosas en rayos~X,
como LS~5039 y LS I +61 303 (Bosch-Ramon et al. 2007, 
Paredes et al. 2007), los rayos~X producidos via el proceso sincrotr'on 
durante la interacci'on \emph{jet-clump} deber'ian ser detectables, y a'un 
la componente t'ermica deber'ia ser  detectable bajo determinadas condiciones
(\emph{clumps} grandes con densidades relativamente bajas).

Las dispersiones IC en la regi'on del \emph{bow-shock} y en el
\emph{clump} producen rayos gamma hasta VHE,
dominando la SED en los  casos con campos magn'eticos
relativamente bajos ($B_{\rm bs}=1$~G). En nuestros
c'alculos, la luminosidad m'as alta alcanzada es $L_{\rm IC}\sim
10^{35}$~erg~s$^{-1}$ para $R_{\rm c}=10^{11}$~cm, aunque
la absorci'on gamma-gamma puede reducir  sustancialmente la emisi'on 
por encima de los 100~GeV (Romero et al 2010). Las interacciones $pp$ 
en el \emph{clump} pueden tambi'en producir 
rayos gamma a energ'ias tan altas como $\sim
10^{14}$~eV ($B_{\rm bs}=150$~G). La luminosidad m'axima obtenida por
$pp$ es sin embargo modesta, $L_{pp} \sim 10^{32}$~erg~s$^{-1}$
para $R_{\rm c}=10^{11}$~cm, aunque \emph{clumps} m'as densos y/o m'as 
grandes, y \emph{jets}
m'as poderosos producir'ian cantidades detectables de fotones 
fuera del rango (0.1-10~TeV) donde la absorci'on gamma-gamma es importante. 
Recordamos que una geometr'ia espec'ifica del sistema binario/observador
mas un  emisor de altas energ'ias lejos del objeto compacto puede dar
una atenuaci'on de los rayos gamma mucho menor
(e.g. Khangulyan et al. 2008). 

Como consecuencia de las caracter'isticas de la interacci'on, la emisi'on
esperada es transitoria (tipo \emph{flare}). La duraci'on de esta emisi'on
est'a relacionada con la permanencia del \emph{clump} dentro del \emph{jet}, 
lo cual 
depende fuertemente de las inestabilidades de RT y KH, las cuales pueden 
destruir el \emph{clump}. Debido a que el \emph{clump} puede ser 
acelerado dentro del \emph{jet}, el
tiempo de vida del primero es de algunas veces el tiempo caracter'istico 
$t_{\rm cc}$. Si el \emph{clump} no ha sido destruido, 'este puede 
eventualmente salir 
del \emph{jet} luego de haber sido chocado y calentado.
Dadas las escalas de tiempo din'amicas de la interacci'on,
el evento tendr'a una duraci'on de entre unos pocos minutos y algunas horas.
 
 Los \emph{flares} producidos en las interacciones \emph{jet-clump} pueden 
tener asociados
componentes espectrales a bajas (radiaci'on sincrotr'on y emisi'on t'ermica) 
y altas energ'ias (interacciones IC y $pp$), las cuales no tienen que estar 
correlacionadas con la actividad de acreci'on del disco. 
El nivel total de emisi'on, la importancia relativa de las diferentes 
componentes de la SED y la duraci'on de los \emph{flares} pueden proveer 
de informaci'on  
sobre la potencia del \emph{jet}, como as'i tambi'en del tamaño y la 
densidad de los \emph{clumps} y el valor del campo magn'etico en la regi'on
 donde ocurre 
la interacci'on (Romero et al. 2007). Por lo tanto, adem'as de las propiedades
del \emph{jet} mismo, las caracter'isticas de los \emph{clumps} pueden 
ser testeadas 
a trav'es de observaciones en HE y VHE (y probablemente tambi'en en frecuencias 
radio) de los \emph{flares} producidos en HMMQs, abriendo una nueva ventana del
espectro electromagn'etico para  
estudiar los vientos de las estrellas de gran masa.    

 Dependiendo del factor de llenado $f$ del viento (o de la densidad de los
\emph{clumps}) y de  $R_{\rm c}$, el n'umero de \emph{clumps} -$N_{\rm c}$- que 
simult'aneamente 
pueden estar dentro del \emph{jet} puede ser bajo ($< 1$) o alto, esto es, un
\emph{clump} eventualmente o muchos simult'aneamente. Luego, los \emph{flares} 
producidos por las interacciones de \emph{clumps} con los \emph{jets} en  
HMMQs pueden ser un fen'omeno espor'adico ($N_{\rm c}$ pequeño) o pueden 
aparecer como una modulaci'on
estacionaria en el espectro (\emph{flickering}) ($N_{\rm c}$ alto). 
Sin embargo, notamos que el \emph{jet} puede verse sustancialmente afectado si 
muchos \emph{clumps} est'an simult'aneamente dentro de 'el. Asumiendo que 
la ruptura del \emph{jet} tiene lugar para 
$\sigma_{\rm j} < N_{\rm c}\times\sigma_{\rm c}$, para los 
par'ametros del viento y del \emph{jet} adoptados en este cap'itulo, el 
\emph{jet} podr'ia ser destruido si
$R_{\rm c} < 10^{10}$~cm con el valor de $f$ adoptado. Sin embargo, 
c'alculos m'as detallados de la d'inamica 
de la interacci'on \emph{jet-clump} son requeridos para clarificar este hecho.  
En el pr'oximo cap'itulo, haremos un estudio un poco m'as minucioso de la 
interacci'on simult'anea de muchos obt'aculos con \emph{jets}, pero en el 
contexto de los AGNs.  

%% file: AGNs_final.tex
\chapter{N'ucleos de galaxias activas}
\label{AGNs}

\section{Introducci'on}

Los n'ucleos de las galaxias albergan agujeros
negros supermasivos (SMBHs, por \emph{Super Massive Black Holes}) 
los cuales abarcan un amplio rango de masas: 
$10^6 \lesssim M_{\rm smbh} \lesssim 10^{10} M_{\odot}$. 
El proceso de formaci'on de estos SMBHs es a 
trav'es de la captura de material del medio circundante, ya sean
nubes de gas, estrellas o  c'umulos de estrellas (Rees 1984). 
Dependiendo del momento angular del SMBH y de la materia, la acreci'on
ser'a esf'erica o no. 
Estos SMBHs pueden estar activos o no, dependiendo de la tasa de acreci'on
de materia. Podemos
decir que la actividad es una etapa en la vida de las galaxias que depende 
fuertemente de la cantidad de materia que haya en las cercan'ias del SMBH.
Si la acreci'on es suficiente como para que se forme un disco y 
consecuentemente 
los \emph{jets}, entonces se dice que tenemos una galaxia activa.

\begin{figure}
\begin{center}
\includegraphics[angle=0, width=0.5\textwidth]{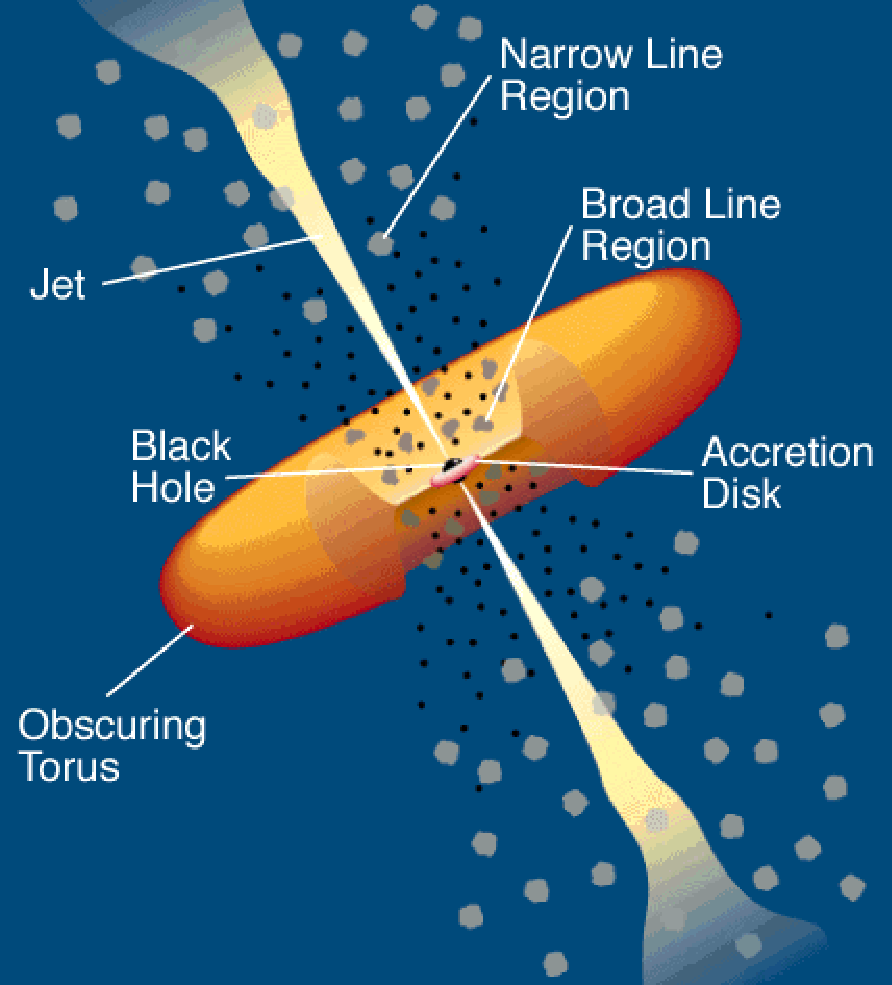}
\caption{Modelo est'andar de AGN.
Se muestran las componentes m'as importantes de acuerdo a este  modelo:
el SMBH, el disco de acreci'on, el toro, los \emph{jets} y las regiones 
donde se emiten las l'ineas  anchas  y delgadas (Urry \& Padovani 1995).}
\label{agn_mod-est}
\end{center}
\end{figure}

Actualmente, el modelo m'as aceptado para describir a los AGN es el
que se conoce como {\bf modelo est'andar}. Este considera que las galaxias
activas son un sistema compuesto por un SMBH rotante como objeto central, 
circundado por un disco de acreci'on el cual a su vez est'a rodeado por un
toro de gas y polvo. Adem'as el sistema se compone de dos \emph{jets} 
paralelos al eje de rotaci'on del SMBH y que se propagan en sentidos
opuestos. 
En la Figura~\ref{agn_mod-est} se muestran las principales componentes de
un AGN. 
De acuerdo al modelo est'andar, la actividad de los AGN es un fen'omeno
intr'insecamente anisotr'opico. Consecuentemente, la fenomenolog'ia
observada depender'a del 'angulo de inclinaci'on entre la l'inea
de la visual y alg'un eje de simetr'ia de la fuente, como por ejemplo
el eje de rotaci'on del SMBH. Cuando este 'angulo es cercano a 90$^{\circ}$
estamos observando o bien una galaxia de l'ineas delgadas (NLRG, por
\emph{Narrow Line Region Galaxy}) o bien una galaxia Seyfert del tipo II;
mientras que si dicho 'angulo es pr'acticamente nulo, tenemos un blazar. 
En cualquier caso intermedio  el objeto observado ser'a un cuasar 
radio-silencioso, una galaxia de l'ineas anchas (BLRG, por \emph{Broad
Line Region Galaxy}) o una galaxia Seyfert del tipo I.  
Por otro lado, las radiogalaxias son AGN que presentan
una fuerte emisi'on en frecuencias radio. Los \emph{jets} de estas fuentes 
no est'an alineados con la l'inea de la visual, es decir, no son blazares,
y se clasifican de acuerdo a la luminoisidad  de los mismos.
Las radiogalaxias Faranoff-Riley I (FR~I) son menos luminosas que las
FR~II (Fanaroff \& Riley, 1974) por lo cual los \emph{jets} de las
primeras pueden propagarse una 
distancia mayor que los \emph{jets} de las FR~II antes de ser frenados por el 
medio externo.

\begin{figure}
\begin{center}
\includegraphics[angle=0, width=0.5\textwidth]{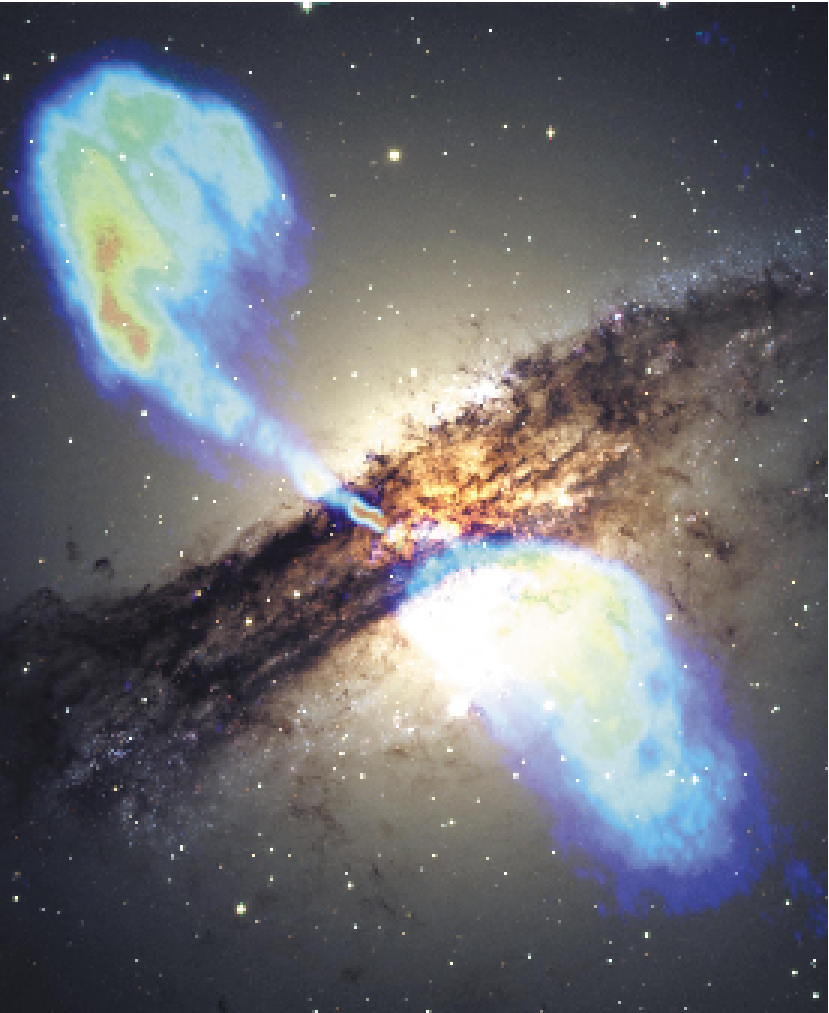}
\caption{Imagen de la radiogalaxia Centaurus A. La imagen muestra esta
galaxia activa en radio y en el 'optico. La emisi'on a bajas frecuencias 
corresponde a los \emph{jets} y a los l'obulos, y se piensa que es 
producida por el proceso sincrotr'on. La emisi'on 'optica corresponde a 
la radiaci'on t'ermica del gas y del polvo que 
se encuentran tapando la regi'on nuclear de la galaxia.
(Cr'edito: NASA.)}
\label{CenA-radio-optico}
\end{center}
\end{figure}

Una de las caracter'isticas principales de los AGN es que la 
emisi'on continua abarca casi todo el espectro electromagn'etico, desde
frecuencias radio hasta los rayos gamma.
Esta radiaci'on proviene b'asicamente del disco (t'ermica) y de los 
\emph{jets} (no t'ermica). En la Figura~\ref{CenA-radio-optico} se
muestra una imagen compuesta de la galaxia activa Centaurus~A (Cen~A) 
en radio y en 'optico.

Adem'as del espectro continuo, los AGN emiten l'ineas en 'optico y 
en UV. El mecanismo m'as aceptado para explicar la emisi'on de estas l'ineas 
es que el medio cerca del SMBH no es homog'eneo sino que tiene una estructura
porosa en forma de nubes o estrellas.    
La regi'on donde se forman las l'ineas de emisi'on m'as anchas
(BLR, por \emph{Broad Line Region}) est'a formada por nubes de material 
del disco o del medio circundante.
Este material est'a confinado por el medio externo caliente, cuya temperatura
es $\sim 10^8$~K, (Krolik et al. 1981)
o por campos magn'eticos (Rees 1987). La materia que forma estas nubes
puede ser ionizada  por los fotones emitidos en el disco de acreci'on
produciendo l'ineas de emisi'on. Luego estas l'ineas son ensanchadas debido a 
que las nubes se mueven en el pozo de potencial del SMBH con una velocidad
$v_{\rm n} > 1000$~km~s$^{-1}$. Otro modelo para la producci'on de las l'ineas
anchas detectadas en algunos AGN es aquel en el cual la BLR est'a compuesta 
por estrellas evolucionadas (gigantes rojas) cuyas crom'osferas son 
fotoionizadas (Penston 1988).

La BLR rodea al agujero negro y por lo tanto la interacci'on
de algunas de las nubes que la componen con la parte m'as interna de
los \emph{jets} es factible.
En este cap'itulo estudiamos la interacci'on de nubes de la BLR con la
base de los \emph{jets}, realizando un tratamiento similar al desarrollado 
en el cap'itulo anterior para los HMMQ.
 Los choques producidos por la penetraci'on de estas
nubes en los \emph{jets} pueden acelerar part'iculas hasta velocidades 
relativistas,
 las cuales  luego pueden enfriarse por diferentes procesos no t'ermicos 
produciedo niveles detectables de emisi'on en rayos gamma. La 
detecci'on de esta radiaci'on nos proveer'ia  informaci'on valiosa sobre las 
condiciones ambientales en las cercan'ias de la 
base de los \emph{jets} como as'i tambi'en de las propiedades de la BLR.

\section{Escenario}
\label{escenario}

\begin{figure}
\begin{center}
\includegraphics[angle=0, width=0.6\textwidth]{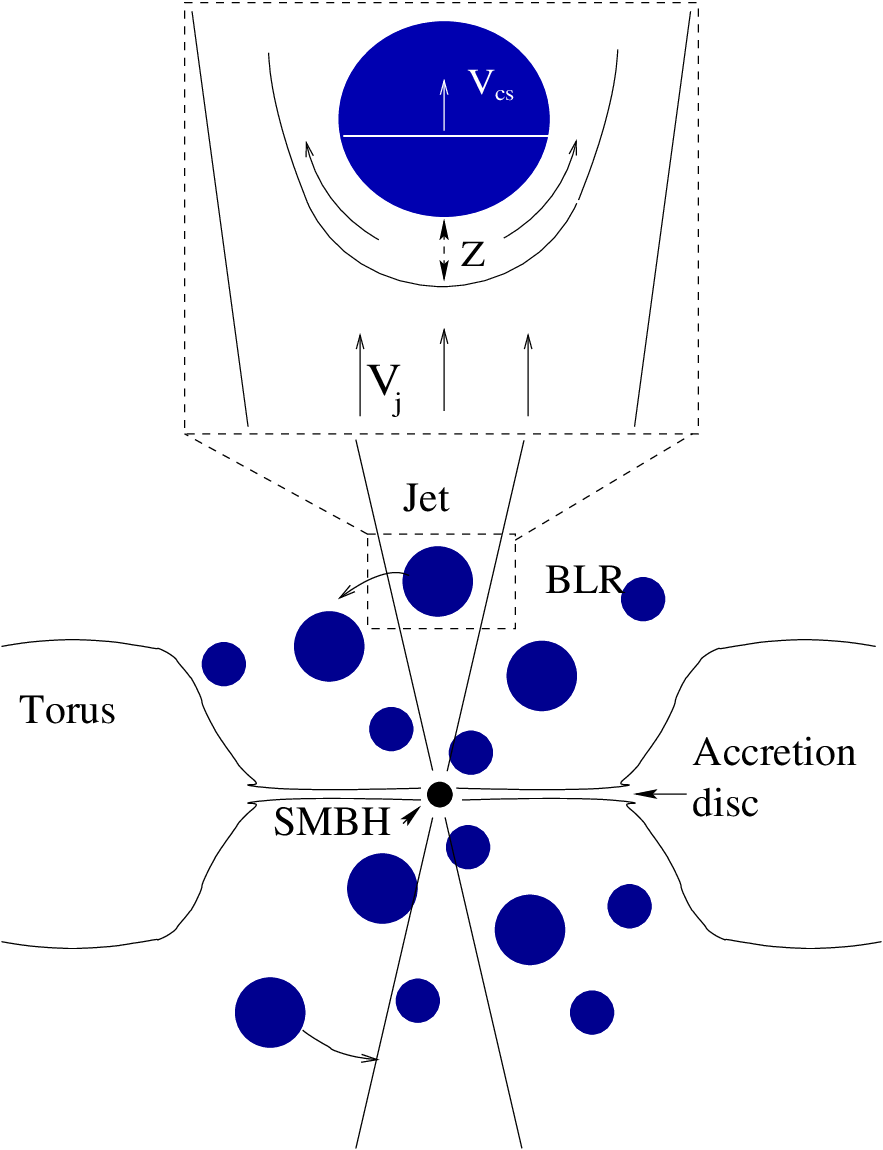}
\caption{Esquema, no a escala, del modelo unificado de los AGN  a las
escalas espaciales de la BLR. La interacci'on entre una nube de la BLR y uno
de los \emph{jets} se muestra en la parte superior de la figura.
El \emph{bow shock} en el \emph{jet} y el choque en la nube tambi'en son
esquematizados.}
\label{blr-sketch}
\end{center}
\end{figure}

Bajo ciertas relaciones entre la presi'on cin'etica del \emph{jet} y la 
densidad y el tamaño de las nubes, la penetraci'on de una nube en el 
\emph{jet} es factible.
Los detalles del proceso de penetraci'on en si mismo son 
complejos, y en esta tesis no tratamos esto en detalle aunque supondremos que
la penetraci'on ocurre si se satisfacen ciertas condiciones. Un 
esquema de la interacci'on se muestra en la Figura~\ref{blr-sketch}. 

A diferencia del tratamiento hecho en el cap'itulo anterior de la
interacci'on \emph{jet-clump}, donde consideramos un escenario similar a 
la fuente Cygnus~X-1, en este cap'itulo no planteamos un escenario definido,
asociado a un AGN espec'ifico, sino que hacemos un tratamiento m'as general 
de la interacci'on. Dejamos fijos a lo largo de todo el cap'itulo s'olo
aquellos par'ametros est'andares que, en principio, no var'ian sustancialmente
de una galaxia a otra.

\subsubsection{Modelo para las nubes}

Supondremos en este estudio nubes esf'ericas, cuyo radio fijamos en
$R_{\rm n}=10^{13}$~cm (Risaliti 2009), y con una densidad uniforme t'ipica 
$n_{\rm n} = 10^{10}$~cm$^{-3}$.
La velocidad de las nubes est'a fijada en
$v_{\rm n}=10^9$~cm~s$^{-1}$ (Peterson 2006).

\subsubsection{Modelo para los \emph{jets}}

Suponemos que los \emph{jets} son hidrodin'amicos y relativistas, con un
factor de Lorentz $\Gamma_{\rm j} = 10$, lo cual implica una velocidad 
$v_{\rm j} \sim c$.
A su vez fijamos el 'angulo de apertura  $\phi\approx 6^{\circ}$, 
es decir, la relaci'on entre el radio y la altura del jet resulta
$R_{\rm j}=\tan(\phi)\,z_{\rm j} \sim 0.1\,z_{\rm j}$. 
La densidad de los \emph{jets} a una altura $z_{\rm j}$ y en el SR del
laboratorio puede estimarse  a trav'es de la relaci'on (\ref{L_jet}), con la
cual obtenemos
\begin{equation}
n_{\rm j}  =   \frac{L_{\rm j}}{(\Gamma_{\rm j}-1)\,m_{p}\,c^3 \sigma_{\rm j}} 
\approx 8\times 10^{4}\left(\frac{L_{\rm j}}{10^{44}\,\rm{erg\, s^{-1}}}\right)
\left(\frac{\Gamma_{\rm j}-1}{9}\right)^{-1}
\left(\frac{z_{\rm j}}{10^{16}\,{\rm cm}}\right)^{-2} \,\rm{cm^{-3}},
\end{equation}
donde $L_{\rm j}$ es la luminosidad cin'etica de los jets y 
$\sigma_{\rm j} = \pi\,R_{\rm j}^2$.
En la Tabla~\ref{const_AGN} se listan los valores de los par'ametros de las
nubes y de los \emph{jets} que quedar'an fijos a lo largo de todo el cap'itulo. 

\begin{table}[]
\begin{center}
\caption{Valores asumidos en este trabajo para las nubes de la BLR 
y para los \emph{jets}.}
\label{const_AGN}
\begin{tabular}{ll} 
\hline
\hline
Descripci'on   & Valor \\  
\hline 
Tama~no de las nubes  & $R_{\rm n} = 10^{13}$~cm \\ 
Densidad de las nubes  & $n_{\rm n} = 10^{10}$~cm$^{-3}$ \\ 
Velocidad de las nubes  & $v_{\rm n} = 10^{9}$~cm~s$^{-1}$ \\ 
Temperatura  de las nubes & $T_{\rm n} = 2\times10^{4}$~K \\ 
Factor de Lorentz de los \emph{jets} & $\Gamma_{\rm j} = 10$ \\ 
Angulo de semi-apertura de los \emph{jets} & $\phi\approx 6^{\circ}$ \\ 
\hline
\end{tabular}
\end{center}
\end{table}

\subsection{Interacci'on \emph{jet}-nube}

El modelo de interacci'on de una nube de la BLR con uno de los \emph{jets} 
del AGN
es similar al descripto en el cap'itulo anterior pero en este escenario 
el \emph{jet} es relativista.

Una de las condiciones que deben satisfacerse para que la nube pueda entrar
entera en el \emph{jet} es que la presi'on cin'etica de 'este no la 
destruya en el
proceso de penetraci'on.  Esto significa que el tiempo de penetraci'on, 
\begin{equation} 
t_{\rm n}\sim \frac{2 R_{\rm n}}{v_{\rm n}} =
2\times 10^4\left(\frac{R_{\rm n}}{10^{13}\,\rm{cm}}\right)
\left(\frac{v_{\rm n}}{10^9\,\rm{cm\, s^{-1}}}\right)^{-1}\,{\rm s},
\end{equation} 
debe ser menor que el tiempo de vida de la nube dentro del \emph{jet}. 
Para estimar 
esta escala temporal debemos conocer la velocidad del choque en la 
nube, $v_{\rm cn}$, que se obtiene, como vimos en el Cap'itulo~\ref{cap2}, 
igualando 
las presiones cin'eticas del \emph{jet} y de la nube: 
$(\Gamma_{\rm j}-1)\,n_{\rm j}\,m_{p}\,c^2=n_{\rm n}\,m_{p}\,v_{\rm cn}^2$,
v'alido mientras $v_{\rm cn}\ll c\,$. As'i obtenemos 
\begin{equation} 
v_{\rm cn} \sim  \frac{c \,(\Gamma_{\rm j} -1)}{\chi_{\rm agn}^{1/2}}\sim
3\times10^8\left(\frac{n_{\rm n}}{10^{10}\,\rm{cm^{-3}}}\right)^{-1/2}
\left(\frac{z_{\rm j}}{10^{16}\,{\rm cm}}\right)^{-1}
\left(\frac{L_{\rm j}}{10^{44}\,\rm{erg\,
s^{-1}}}\right)^{1/2} \, \rm{\frac{cm}{s}}, 
\end{equation}  
donde $\chi_{\rm agn} = n_{\rm n}/n_{\rm j}(z_{\rm j})$. 
Luego, el tiempo de cruce del choque a trav'es de toda la nube resulta
\begin{equation} 
t_{\rm cn} \sim \frac{2R_{\rm n}}{v_{\rm cn}}\simeq
7\times10^4 \left(\frac{R_{\rm n}}{10^{13}\, \rm{cm}} \right)\,
\left(\frac{n_{\rm n}}{10^{10}\,\rm{cm^{-3}}}\right)^{1/2} 
\left(\frac{z_{\rm j}}{10^{16}\,{\rm cm}}\right)\left(\frac{L_{\rm j}}{10^{44}\,
\rm{erg\,s^{-1}}}\right)^{-1/2}\,{\rm s}\,.
\end{equation}
Para un tiempo tan corto como $\sim t_{\rm cn}$, la nube se comporta 
como un obst'aculo efectivo para el material del \emph{jet}. Fijando 
$t_{\rm n} \sim t_{\rm cn}$ nos permite obtener valores m'inimos para 
$\chi_{\rm agn}$ y $z_{\rm j}$, por debajo de los cuales la nube no llega a
penetrar efectivamente  en el \emph{jet}.

Debido a la interacci'on con el material del  \emph{jet}, inestabilidades
hidrodin'amicas afectan  a la nube. El  \emph{jet} ejerce una fuerza
en la nube a trav'es de la superficie de discontinuidad. La aceleraci'on 
aplicada a la nube puede estimarse a trav'es de la ecuaci'on~(\ref{g_general})
 y en el caso que estamos estudiando resulta 
\begin{equation}
\label{g_agn} 
g=
%\frac{P_{\rm j}\,\sigma_{\rm n}}{M_{\rm n}}\sim
%\frac{3}{4}\frac{c^2}{\chi_{\rm agn}\,R_{\rm n}}=
\frac{3}{2}\,\frac{v_{\rm cn}}{t_{\rm cn}}\, = 
\frac{3}{4}\,\frac{v_{\rm cn}^2}{R_{\rm n}}. 
\end{equation}
Dada la aceleraci'on $g$, las inestabilidades
de RT se desarrollar'an en la nube
con una escala de tiempo 
\begin{equation} 
t_{\rm RT} \sim \sqrt{\frac{l}{g}}=
\sqrt{\frac{4\,\chi_{\rm agn}\,l\,R_{\rm n}}{3\,c^2}}.
\end{equation} 
Para perturbaciones con longitudes de escala 
$l\sim R_{\rm n}$, que son aquellas asociadas a la fragmentaci'on significativa
de la nube, el tiempo de crecimiento de la inestabilidad resulta
$t_{\rm RT}\sim t_{\rm cn}$.
Por otro lado, las inestabilidades de KH tambi'en pueden crecer
suficientemente como para destruir la nube. 
Dada la alta velocidad relativa entre el \emph{jet} chocado y el material de la
nube, $v_{\rm rel}\sim v_{\rm j}$, se obtiene
\begin{equation} 
t_{\rm KH}\sim =\frac{\chi_{\rm agn}\,l}{c}\,.
\end{equation} 
%
%donde $g_{\rm rel}\sim c^2/\chi_{\rm agn}\,l$. 
Para $l\sim R_{\rm n}$, obtenemos
nuevamente que $t_{\rm KH} \sim t_{\rm cn}$.  
Notamos que, dada la ecuaci'on~(\ref{g_agn}), el tiempo necesario para 
acelerar la nube hasta
la velocidad del choque $v_{\rm cn}$ es $\sim t_{\rm cn}$, mientras que  
para acelerar la nube hasta  $v_{\rm j}$ el tiempo requerido es 
$\gg t_{\rm cn}$. Por esto, antes de que la nube comience a moverse 
conjuntamente con el \emph{jet}, probablemente ser'a fragmentada. 

Finalmente, hay dos escalas de tiempo adicionales que tambi'en son
relevantes en nuestro estudio: el tiempo de formaci'on del \emph{bow shock},
$t_{\rm bs}$, y el tiempo requerido para que la nube cruce el jet, $t_{\rm j}$.
Considerando las ecuaciones (\ref{t_j}) y (\ref{t_bs}), y que en el caso de 
un \emph{bow shock}
relativista 'este se separa del obst'aculo una distancia 
$Z \sim 0.3\,R_{\rm n}$ (obtenido considerando que en un 
plasma relativista
las part'iculas se escapan de la regi'on chocada del \emph{jet} a la velocidad 
del sonido $c/\sqrt{3}$), estas escalas de tiempo resultan 
\begin{equation} 
t_{\rm j}\sim \frac{2 R_{\rm j}}{v_{\rm n}} =
2\times10^6 \left(\frac{z_{\rm j}}{10^{16}\,{\rm cm}}\right)
\left(\frac{v_{\rm n}}{10^9\,\rm{cm\, s^{-1}}}\right)^{-1} \,{\rm s},
\end{equation}  
y
\begin{equation} 
t_{\rm bs} \sim \frac{Z}{c}=10^2
\left(\frac{R_{\rm n}}{10^{13}\, \rm{cm}} \right)\,{\rm s}\,.
\end{equation} 

\begin{figure}
\begin{center}
\includegraphics[angle=270, width=0.7\textwidth]{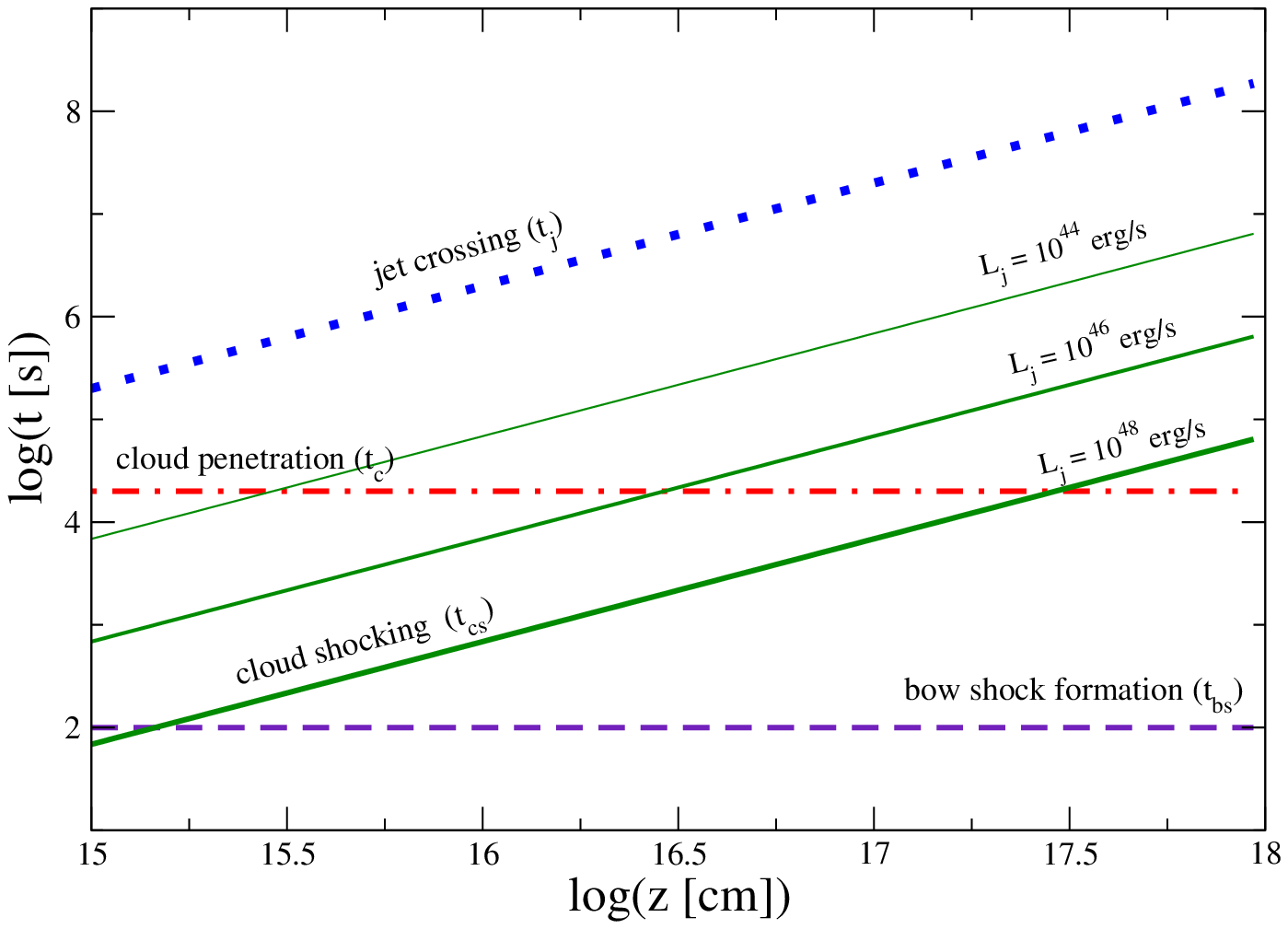}
\caption{Los tiempos de cruce de la nube a trav'es del \emph{jet} 
(l'inea punteada azul), de penetraci'on de la nube (l'inea punteada-rayada 
roja), de formaci'on de \emph{bow-shock} (l'inea rayada violeta) y el tiempo
en el cual el choque en la nube la recorre entera 
(l'ineas verdes) son graficados. Todos ellos han sido calculados usando los
valores dados en la Tabla~\ref{const_AGN}. El tiempo $t_{\rm cn}$ es 
graficado para $L_{\rm j}=10^{44}$, $10^{46}$ y $10^{48}$~erg~s$^{-1}$
(Araudo et al. 2010).}
\label{timescales}
\end{center}
\end{figure}

A modo de resumen del estudio anterior sobre la din'amica de la interacci'on 
\emph{jet}-nube, graficamos en la Figura~\ref{timescales}  $t_{\rm cn}$ (para 
diferentes valores de $L_{\rm j}$), $t_{\rm j}$, $t_{\rm n}$ y $t_{\rm bs}$ 
en funci'on de $z_{\rm j}$. Como se muestra en la figura, para algunos 
valores de $z_{\rm j}$ y $L_{\rm j}$
la nube podr'ia ser destruida por el \emph{jet} antes de entrar 
completamente, es
decir, $t_{\rm cn} < t_{\rm n}$. Esto nos provee de una condici'on 
para determinar 
la altura del \emph{jet} a la cual la nube puede penetrar entera dentro de 'el. 
Notamos tambi'en que, en general, $t_{\rm bs}$ es mucho m'as corto que las 
dem'as escalas de tiempo.

\subsubsection{Altura de la interacci'on}

La nube puede penetrar completamente en el \emph{jet} si el tiempo de vida 
de la nube despu'es del impacto con el \emph{jet} es m'as largo que el 
tiempo de penetraci'on.
Adem'as, la condici'on de que la presi'on lateral del \emph{jet} sea 
$<n_{\rm n}\,m_{p}\,v_{\rm n}^2$ se satisface autom'aticamente.
Esto determina la altura de interacci'on m'inima, $z_{\rm int}^{\rm min}$,
que no permite la destrucci'on de la nube antes de la penetraci'on completa.
Por otro lado, la interacci'on no puede ocurrir m'as abajo que la regi'on
de formaci'on del \emph{jet}, que se da en $z_{\rm 0}\sim 100\,R_{\rm g}\approx 
1.5\times 10^{15}\,(M_{\rm smbh}/10^8\,M_{\odot})$~cm (Junor et al. 1999),
donde $R_{\rm g}$ es el radio gravitacional del agujero negro.
Finalmente, para que la interacci'on entre nubes de la BLR y los 
\emph{jets} pueda occurrir, a una altura que llamaremos $z_{\rm int}$, 
el tamaño de la BLR debe ser $R_{\rm blr} > z_{\rm int} > z_{\rm 0}$.

El tiempo de vida de la nube depende del tiempo de fragmentaci'on (dado por el
crecimiento de las inestabilidades), que est'a
fuertemente relacionado con $t_{\rm cn}$. El valor de $z_{\rm int}$ puede luego
ser estimado fijando $t_{\rm n} < t_{\rm cn}$, ya que queremos que la
nube entre completamente en el \emph{jet} antes de ser sustancialmente 
distorcionada por el impacto con 'este. Una vez chocada, la nube puede 
sufrir expansi'on 
lateral y calentamiento por conducci'on lo cual har'ia que las inestabilidades
crezcan m'as r'apidamente. Sin embargo no hemos tenido en cuenta estos
dos 'ultimos efectos en nuestro estudio y
determinamos $z_{\rm int}$ imponiendo que $t_{\rm cn} = 2\,t_{\rm n}$, 
con lo cual 
\begin{equation}
z_{\rm int}   \approx  5\times10^{15}
\left(\frac{v_{\rm n}}{10^9\,\rm{cm\,s^{-1}}}\right)^{-1}
\left(\frac{n_{\rm n}}{10^{10}\,\rm{cm^{-3}}}\right)^{-1/2}
%{} & {\times} &
\left(\frac{L_{\rm j}}{10^{44}\,\rm{erg\, s^{-1}}}\right)^{1/2} \,\rm{cm}.
\end{equation}  
Notamos que la potencia disponible en el \emph{bow shock} es
$L_{\rm bs}\sim (\sigma_{\rm n}/\sigma_{\rm j})\,L_{\rm j}\propto z_{\rm j}^{-2}$, 
donde $\sigma_{\rm n} = \pi\,R_{\rm n}^2$,
por lo cual la interacci'on m'as luminosa tendr'a lugar en 
$z_{\rm j} \sim z_{\rm int}$.
Ahora debemos verificar si la elecci'on de $z_{\rm int}$ que hemos hecho
cumple con los
requisitos geom'etricos $z_0 < z_{\rm int} < R_{\rm blr}$.

El tamaño de la BLR puede estimarse a trav'es de las relaciones emp'iricas
obtenidas para galaxias FR~II, las cuales poseen una BLR bien determinada
por las observaciones.
Los ajustes que se obtienen a partir de las observaciones de numerosas 
galaxias son en general del tipo 
$R_{\rm blr} \propto L_{\rm blr}^{\alpha_{\rm blr}}$, donde $L_{\rm blr}$ es la 
luminosidad de la BLR y $\alpha_{\rm blr} \sim 0.5 - 0.7$ 
(Peterson et al. 2005; Bentz et al. 2006). 
En este cap'itulo usamos las siguientes relaciones:
\begin{equation}
\label{Rblr}
R_{\rm blr}\sim 6\times 10^{16} 
\left(\frac{L_{\rm blr}}{10^{44}\,\rm{erg\,s^{-1}}}\right)^{0.7}\,\rm{cm},
\end{equation}
y
\begin{equation}
\label{Rblr_07}
R_{\rm blr}\sim 2.5\times 10^{16} 
\left(\frac{L_{\rm blr}}{10^{44}\,\rm{erg\,s^{-1}}}\right)^{0.55}\,\rm{cm},
\end{equation}
obtenidas por Kaspi y colaboradores (2005, 2007).
En galaxias FR~I, en las cuales la detecci'on de la BLR es muy imprecisa
o ni siquiera se llega a detectar, las relaciones~(\ref{Rblr}) y 
(\ref{Rblr_07}) deben tomarse con cautela. 

\begin{figure}
\begin{center}
\includegraphics[angle=270, width=0.7\textwidth]{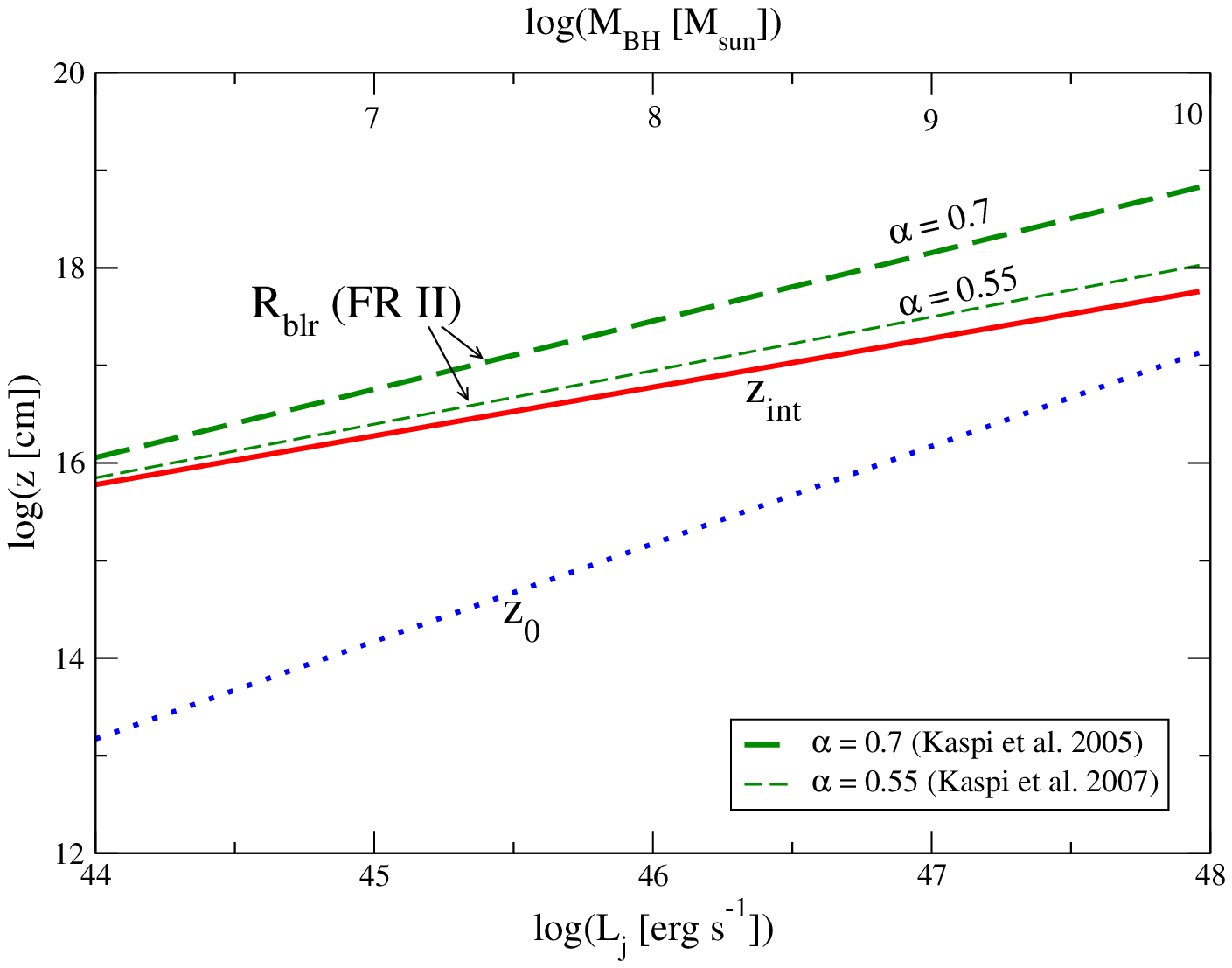}
\caption{La altura de interacci'on, $z_{\rm int}$ (l'inea s'olida roja), 
y el tamaño de la BLR, $R_{\rm blr}$ (l'inea rayada verde), se grafican para
diferentes valores de $L_{\rm j}$ (eje horizontal inferior). 
Derivamos $R_{\rm blr}(L_{\rm j})$ fijando que $L_{\rm blr} = 0.1\,L_{\rm j}$ y 
graficamos $R_{\rm blr}$ usando las ecuaciones~(\ref{Rblr}) y (\ref{Rblr_07}).
En la misma figura, la altura de la base del \emph{jet}, $z_0$ 
(l'inea punteada azul),
es graficada en funci'on de $M_{\rm smbh}$ (eje horizontal superior).}
\label{zint-rblr}
\end{center}
\end{figure}

En la Figura~\ref{zint-rblr} mostramos como var'ian $z_{\rm int}$ y 
$R_{\rm blr}$ con $L_{\rm j}$, asumiendo que $L_{\rm blr}$ es un 10\% de la 
luminosidad del disco, $L_{\rm d}$, y esta 'ultima es tomada igual a $L_{\rm j}$.
Como puede verse en la
figura, para valores razonables de los par'ametros, la condici'on 
$z_{\rm int} < R_{\rm blr}$
se satisface en un amplio rango de valores de $L_{\rm j}$. En la misma figura
tambi'en mostramos la relaci'on entre  $z_0$ y
$M_{\rm smbh}$, de la cual se deduce que para
$M_{\rm smbh} > 10^9\,M_{\odot}$
el \emph{jet} podr'ia no estar (completamente) formado a las escalas de 
la BLR para valores bajos de  $L_{\rm j}$.

\section{Poblaci'on de part'iculas relativistas}
\label{poblacion_nt}

En el \emph{bow shock} y en el choque en la nube se pueden acelerar part'iculas
a trav'es de un mecanismo de aceleraci'on difusiva como el descripto en el
 Cap'itulo~\ref{cap2} (Secci'on~\ref{particle_accel}). 
Sin embargo, debido a que el \emph{bow shock} es m'as fuerte que el 
choque en la nube ($v_{\rm bs} \gg v_{\rm cn}$), la aceleraci'on de 
part'iculas ser'a m'as eficiente en el primero.
Adem'as, la luminosidad del choque en la nube es menor que la luminosidad
del  \emph{bow shock} por un factor $\sim 1/(2 \chi_{\rm agn}^{1/2})$, con lo
cual la energ'ia disponible para entregar a las part'iculas aceleradas es 
menor tambi'en. Por estas razones,
s'olo consideraremos la aceleraci'on de electrones y protones en el 
\emph{bow shock} y 
en esta secci'on describiremos someramente la inyecci'on y evoluci'on de 
estas part'iculas no t'ermicas. 

La luminosidad inyectada por el \emph{jet} en el \emph{bow shock} localizado 
en $z_{\rm int}$ es
$L_{\rm bs} = (\sigma_{\rm n}/\sigma_{\rm j}) L_{\rm j}$. Una fracci'on 
$\eta_{\rm nt}$ de esta cantidad ser'a transferida a las part'iculas que se 
aceleran en este choque, con lo cual la luminosidad de 'estas es 
\begin{equation}
\label{L_nt_agn}
L_{e,p} = \eta_{\rm nt} L_{\rm bs} \approx 4\times10^{39} 
\left(\frac{\eta_{\rm nt}}{0.1}\right)
\left(\frac{R_{\rm n}}{10^{13}\rm{cm}}\right)^2
\left(\frac{L_{\rm j}}{10^{44}\,\rm{erg\,s^{-1}}}\right)\,\rm{\frac{erg}{s}}.
\end{equation}

La distribuci'on de energ'ia de la poblaci'on de part'iculas relativistas
inyectadas en la regi'on del  \emph{bow shock}  es una ley de potencias 
con un \emph{cut-off} exponencial: 
\begin{equation}
\label{Q_ep_agn}
Q_{e,p} = K_{e,p}\, E_{e,p}^{-2.2} \,\exp^{-E_{e,p}/E_{e,p}^{\rm max}}. 
\end{equation}
A la 
constante $K_{e,p}$ la determinamos a trav'es de la ecuaci'on~(\ref{L_nt_agn})
y considerando que 
\begin{equation}
\label{L_ep_agn}
L_{e,p} = \int_{E_{e,p}^{\rm min}}^{E_{e,p}^{\rm max}} Q_{e,p}\, E_{e,p}\,
{\rm d}E_{e,p},
\end{equation}
donde fijamos $E_{e,p}^{\rm min} = 2\,m_{e,p}c^2$ pero $E_{e,p}^{\rm max}$ es 
desconocida a'un. Sin embargo, igualando las ecuaciones (\ref{Q_ep_agn}) y 
(\ref{L_ep_agn}) obtenemos
\begin{equation}
\label{K_nt_agn}
K_{e,p} = \eta_{\rm nt}\, L_{\rm bs}\, 
\left(\frac{0.2}{{E_{e,p}^{\rm min}}^{-0.2} - {E_{e,p}^{\rm max}}^{-0.2}}\right)
\sim \eta_{\rm nt}\, L_{\rm bs}\,\frac{0.2}{{E_{e,p}^{\rm min}}^{-0.2}},
\end{equation}
si $E_{e,p}^{\rm max} \gg E_{e,p}^{\rm min}$.

\subsection{Aceleraci'on de part'iculas y p'erdidas radiativas}

El campo magn'etico $B_{\rm bs}$ en el acelerador/emisor 
puede determinarse relacionando las 
densidades de energ'ia magn'etica y no t'ermica de la siguiente manera:
$u_{\rm B} = \eta_{\rm B} u_{\rm e,p}$,
donde la densidad de energ'ia de las part'iculas relativistas es 
$u_{e,p} = L_{e,p}/(\sigma_{\rm n}\,c)$.
De manera tal que la radiaci'on  IC sea importante en rayos gamma,
$\eta_{\rm B} \ll 1$ es requerido. En este contexto, si $\eta_{\rm B} = 0.01$,
$B_{\rm bs}$ puede se parametrizado de la siquiente manera:
\begin{equation}
\label{Bbs_agn}
B_{\rm bs}\approx 10 \left(\frac{\eta_{\rm B}}{0.01}\right)
\left(\frac{v_{\rm n}}{10^{9}\, \rm{cm\,s^{-1}}}\right)^2
\left(\frac{n_{\rm n}}{10^{10}\, \rm{cm^{-3}}}\right)\,{\rm G}\,.
\end{equation}

Debido a que el  \emph{bow shock} es relativista y el tratamiento de estos
choques es complejo, adoptamos la prescripci'on 
$\dot E_{e,p}^{\rm ac} \sim 0.1\, e\, B_{\rm bs}\, c$~erg~s$^{-1}$ 
(de Jager et al. 1996) para la tasa de aceleraci'on, 
con la cual el tiempo de aceleraci'on est'a dado por la relaci'on
\begin{equation} 
t_{\rm ac} \sim 0.7 \,E_{e,p}
\left(\frac{B_{\rm bs}}{10\, \rm{G}}\right)^{-1}\,{\rm s}\,. 
\end{equation}

Las part'iculas est'an afectadas por diferentes tipos de p'erdidas que 
compensan las ganancias por aceleraci'on. Las p'erdidas por escape de las
part'iculas contemplan la convecci'on del material chocado, 
\begin{equation} 
t_{\rm conv}\sim \frac{3\,R_{\rm n}}{c} = 10^3\,
\left(\frac{R_{\rm n}}{10^{13}\, \rm{cm}}\right)\,{\rm s}\,, 
\end{equation}
y la difusi'on, la cual para que las part'iculas puedan difundir una
distancia $Z = 0.3 R_{\rm n}$ el tiempo requerido es
\begin{equation} 
t_{\rm dif}\sim 5\times10^{6}
\left(\frac{R_{\rm n}}{10^{13}\, \rm{cm}}\right)^2
\left(\frac{B_{\rm bs}}{10\, \rm{G}}\right)\,\frac{1}{E_{e,p}}\,{\rm s}\,. 
\end{equation}

\subsubsection{P'erdidas lept'onicas}

Los electrones en la regi'on chocada del \emph{jet} rad'ian 
principalmente por los procesos
Bremsstrahlung relativista, radiaci'on sincrotr'on y dispersiones IC.
En el 'ultimo caso los fotones semilla pueden ser externos (EC), 
producidos en el disco o en la BLR, o generados en la misma fuente
%la regi'on del \emph{bow shock}, 
por radiaci'on sincrotr'on (SSC).
Dada la baja densidad del \emph{jet} en 
$z_{\rm int}$ ($n_{\rm j} \sim 3.2\times10^5$~cm$^{-3}$)
las p'erdidas por Bremsstrahlung relativista ser'an despreciables, con
una escala de tiempo $t_{\rm Brem} \sim 10^{10}$~s. 
Considerando el valor de $B_{\rm bs}$ dado en la ecuaci'on~(\ref{Bbs_agn}), 
las p'erdidas por radiaci'on sincrotr'on resultan:
\begin{equation}
t_{\rm sin} \sim 4.1 \left(\frac{\eta_{\rm B}}{0.01}\right)^{-2} \frac{1}{E_e}\,
{\rm s}.
\end{equation}

Para estimar la relevancia de  las interacciones IC
(EC o SSC), necesitamos conocer la densidad $n_{\rm ph}$ de fotones semilla
para as'i luego calcular $t_{\rm EC}$ y $t_{\rm SSC}$  a trav'es de las 
f'ormulas dadas en el Cap'itulo~\ref{proc-rad}. A continuaci'on 
describimos como calculamos  $n_{\rm ph}$ de cada campo de fotones semilla
considerado en este cap'itulo. 
\begin{itemize} 
\item {\bf SSC:}
Estimando la luminosidad de la radiaci'on sincrotr'on,
$L_{\rm sin} \sim \int N_e(E_e) \,P_{\rm sin}(E_e, E_{\rm ph})\,{\rm d}E_e$,
donde $\mathcal{P}(E_e, E_{\rm ph})$ est'a dada por la 
f'ormula~(\ref{P_1e_exac}) y
considerando que  la distribuci'on $N_e$ se encuentra en  el estado
estacionario, podemos hallar la densidad de energ'ia de los fotones 
sincrotr'on: $u_{\rm sin} \sim L_{\rm sin}/(\sigma_{\rm n} \,c)$.
Como la distribuci'on de estos fotones es una ley de 
potencia, la densidad de fotones es
$n_{\rm ph}^{\rm sin} ={\rm d}U_{\rm sin}/{\rm d}E_0^{\rm sin}$.
Para calcular $t_{\rm SSC}$ desarrollamos un c'odigo
num'erico que describiremos someramente m'as abajo. 
\item {\bf EC (disco):}
La densidad de energ'ia $u_{\rm d}$ de los fotones emitidos por el disco 
puede estimarse de la siguiente manera:
$u_{\rm d} \sim L_{\rm d}/(4 \pi z_{\rm int}^2 c)$. 
De acuerdo al modelo de disco delgado de 
Shakura \& Sunyaev (1973), la temperatura $T_{\rm d}$ del material acretado 
decrece con la distancia $r_{\rm d}$ al SMBH de acuerdo a 
$T_{\rm d}(r_{\rm d}) = T_{\rm in} (r_{\rm d}/r_{\rm in})^{-3/4}$, donde 
$r_{\rm in} = 3 r_{\rm g} = 4.5\times10^5 (M_{\rm smbh}/M_{\odot})$~cm y
$T_{\rm in} = T_{\rm d}(r_{\rm in})$. Debido a que
la mayor intensidad de radiaci'on se produce en las partes m'as internas 
del disco, esto es, en $r_{\rm d} \sim r_{\rm in}$, podemos considerar que 
$T_{\rm d} \sim T_{\rm in}$ y luego 
$L_{\rm d} \sim 4\pi\, \sigma_{SB}\, T_{\rm in}^4 \,r_{\rm in}^2$, donde 
$\sigma_{\rm SB} = 5.67\times10^{-5}$~gr~K$^{-4}$~s$^{-2}$. 
Despejando de aqu'i la temperatura, 
la energ'ia de los fotones emitidos por el disco
resulta $E_0^{\rm d} \sim 2.4\times10^2 (L_{\rm d}/10^{44})^{1/4} (M_{\rm smbh}/10^6M_{\odot})^{-1/2}$~eV y los electrones
con $E_e > (L_{\rm d}/10^{44})^{-1/4} (M_{\rm smbh}/10^6M_{\odot})^{1/2}$~GeV
interact'uan con los fotones del disco en el r'egimen de KN.

\item  {\bf EC (BLR):}
La densidad de energ'ia de los fotones emitidos por la BLR es calculada 
mediante $u_{\rm blr} \sim L_{\rm blr}/(\pi R_{\rm blr}^2 c) \sim 0.08 (L_{\rm blr}/10^{44})^{-0.4}$, 
donde hemos considerado la relaci'on~(\ref{Rblr}).
Para nubes con temperaturas $T_{\rm n} \sim 2\times10^4$~K, la energ'ia de los
fotones resulta $E_0^{\rm blr} \sim 5.3$~eV y la interacci'on en
el r'egimen de KN se dar'a para dispersiones con electrones cuya energ'ia sea
$E_e > 48$~GeV.
\end{itemize}

\begin{figure}
\begin{center}
\includegraphics[angle=270, width=0.49\textwidth]{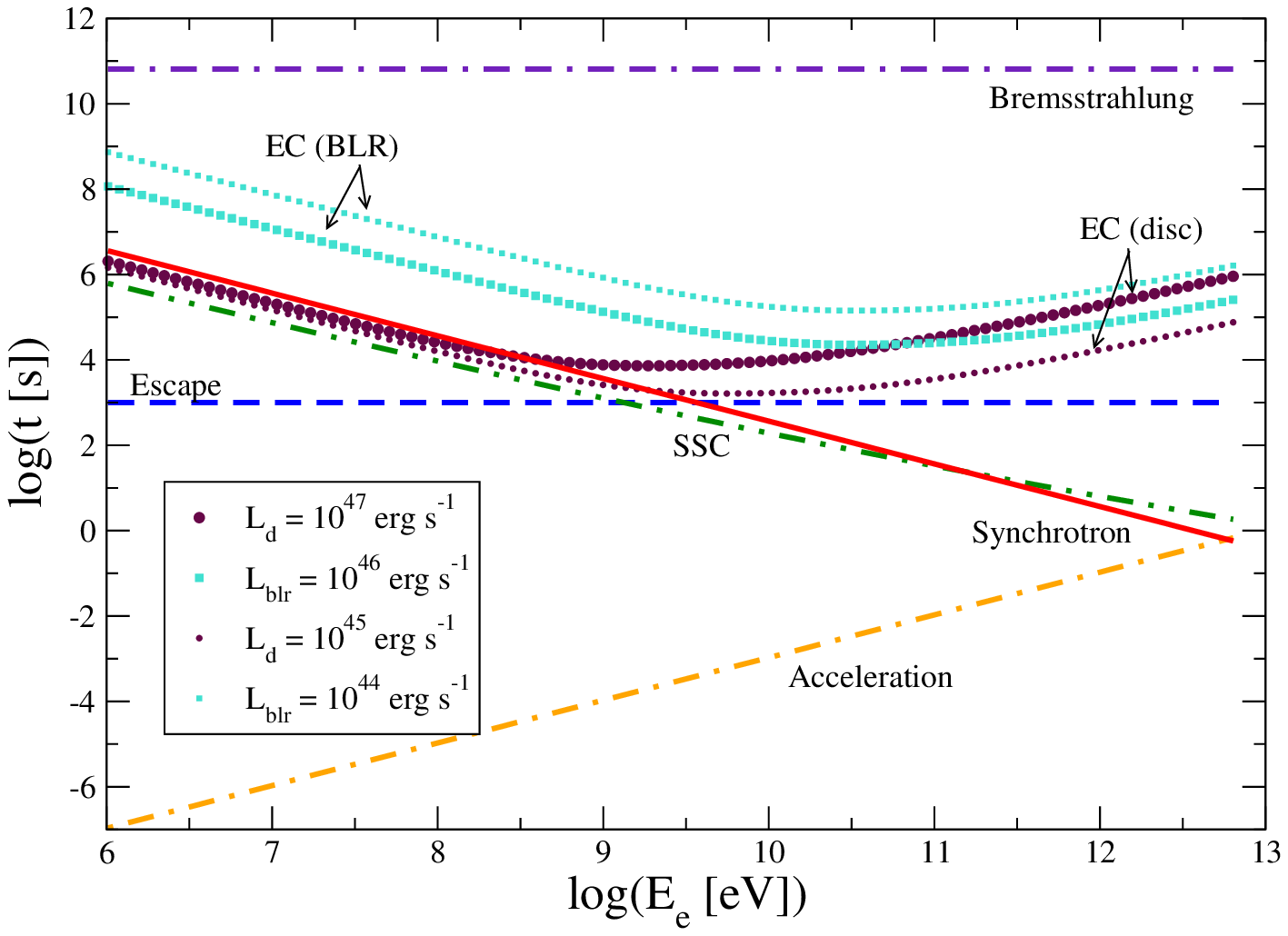}
\includegraphics[angle=270, width=0.49\textwidth]{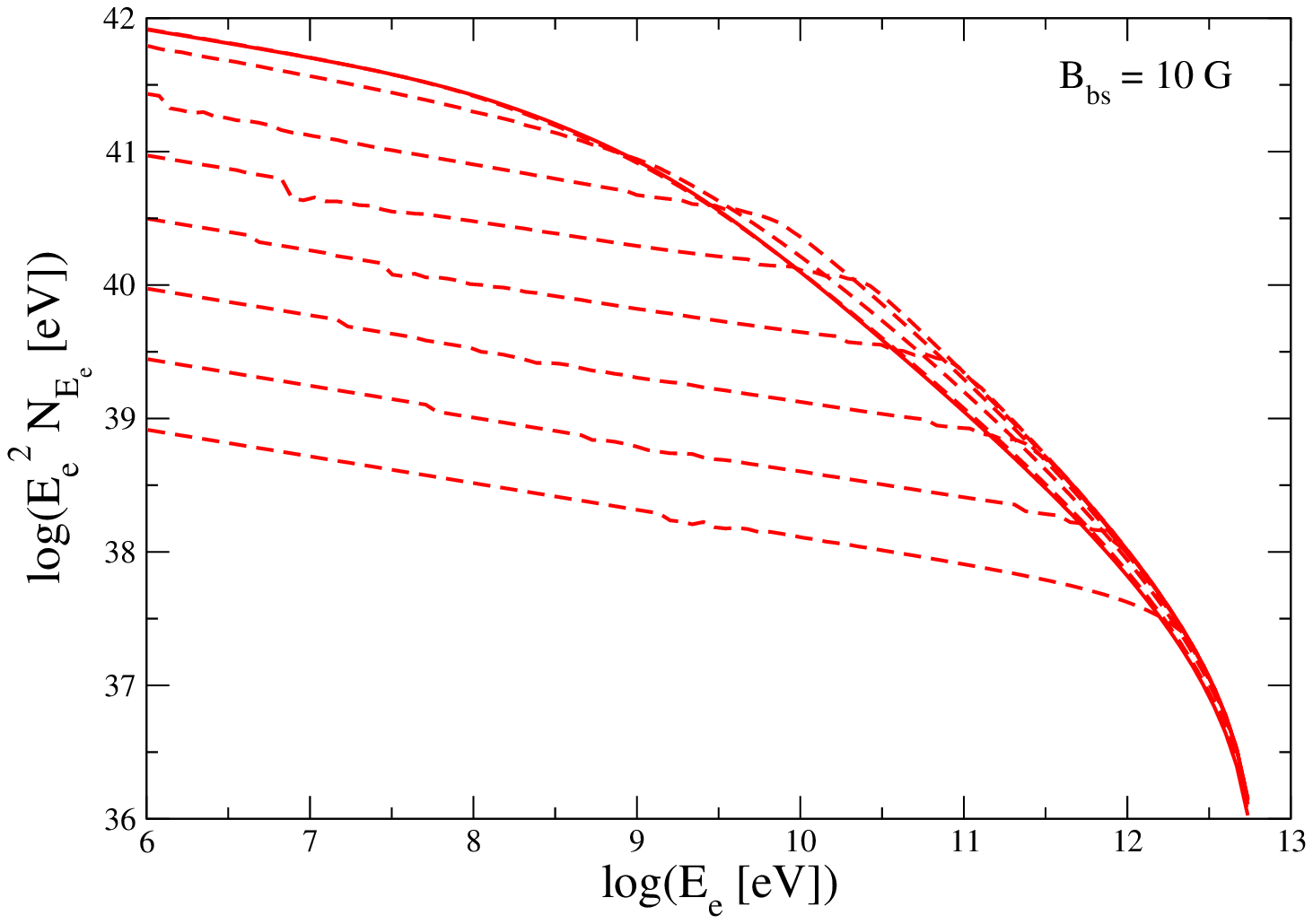}
\caption{Izquierda: Las ganancias por aceleraci'on (l'inea punteada-rayada 
anaranjada), 
convecci'on (l'inea rayada azul) y los tiempos de enfriamiento radiativos
son graficados.
El tiempo SSC (l'inea verde de rayas y puntos) graficado corresponde al 
calculado
con la distribuci'on de energ'ia estacionaria de los electrones.
El tiempo de enfriamiento por EC con los fotones de la BLR
(cuadrados turquesa) y del disco (l'inea punteada marr'on) son graficados
para fuentes d'ebiles
(BLR: $10^{44}$~erg~s$^{-1}$; disco: $10^{45}$~erg~s$^{-1}$) y 
brillantes (BLR: $10^{46}$~erg~s$^{-1}$; disco: $10^{47}$~erg~s$^{-1}$). 
Las p'erdidas por radiaci'on sincrotr'on (l'inea roja) y Bremsstrahlung 
relativista (l'inea violeta) son tambi'en graficadas.
Hemos fijado $\eta_{\rm B} = 0.01$.
Derecha: Evoluci'on temporal de la distribuci'on de los electrones 
relativistas  en la regi'on del \emph{bow shock}, $N_e$. Las l'ineas de rayas
indican el espectro a diferentes tiempos $t_i$, mientras que la l'inea llena
indica el espectro una vez que el estado estacionario ha sido alcanzado.
Como puede observarse, el quiebre en el espectro estacionario se produce 
en $E_e \sim 2\times10^9$~eV, como se deduce de la figura de la izquierda.}
\label{losses_elec_agn}
\end{center}
\end{figure}

En la Figura~\ref{losses_elec_agn} (izquierda) se muestran las escalas 
de tiempo de las
p'erdidas radiativas, junto con el tiempo de aceleraci'on y de convecci'on
para un \emph{bow shock} ubicado en  $z_{\rm int}$. Para 
$\eta_{\rm B}$ adoptamos un valor igual a 0.01 y las p'erdidas por SSC 
son graficadas para el estado estacionario de $N_{E_e}(E_e)$.
Notamos que tanto $B_{\rm bs}$
como $L_{e,p}$ y $z_{\rm int}$ son constantes para diferentes valores de
$L_{\rm j}$ y fijadas las fracciones $\eta_{\rm B}$ y $\eta_{\rm nt}$.
S'olo $L_{\rm blr}$ y $L_{\rm d}$ se espera que varien con $L_{\rm j}$. 
As'i, mientras los campos de fotones externos no sean relevantes,
la energ'ia m'axima que pueden alcanzar  los electrones en
$z_{\rm int}$ no cambia para diferentes potencias del \emph{jet}.

Los procesos radiativos dominantes a las energ'ias m'as altas de la poblaci'on
de electrones son el sincrotr'on, el EC y el SSC.
El Bremsstrahlung relativista es despreciable para todas las energ'ias y
el escape por convecci'on es relevante en la parte de m'as bajas energ'ias.
Esto produce un quiebre en el espectro de los electrones, cuando las
p'erdidas por sincrotr'on, EC o SSC se igualan a las de convecci'on.
La transici'on del r'egimen de Th al de KN se ve claramente en las curvas
de enfriamiento por EC, pero es menos notorio en el caso del SSC debido a
la naturaleza no t'ermica de los fotones semilla (sincrotr'on).
La energ'ia m'axima de los electrones es de $\sim 40$~TeV 
(para $\eta_{\rm B} = 0.01$) y est'a determinada
por la igualdad $t_{\rm ac} = t_{\rm sin}$.

A primer orden, la evoluci'on de la distribuci'on de los electrones 
relativistas en la regi'on del \emph{bow shock} puede ser calculada 
asumiendo condiciones homog'eneas
en el acelerador/emisor y los mecanismos de aceleraci'on y escape 
mencionados anteriormente. Debido a que el SSC es un
canal de enfriamiento importante, el c'alculo de 
$N_e(E_e,t)$ debe realizarse num'ericamente. Para esto desarrollamos un c'odigo
en el cual para un tiempo de observaci'on $t$, dividimos el per'iodo [0-$t$] 
en intervalos $\Delta t$. En cada intervalo,  $N_e$ se calcula teniendo 
en cuenta las p'erdidas por SSC con los fotones 
sincrotr'on radiados en el intervalo anterior. As'i se contin'ua hasta que
se alcanza el estado estacionario. 
La duraci'on de cada intervalo debe ser menor que la duraci'on de los 
intervalos previos de modo de calcular apropiadamente el crecimiento de 
$u_{\rm sin}$ en el emisor.
En la Figura~\ref{losses_elec_agn} (derecha) se muestra el resultado 
obtenido considerando $\tau_{\rm vida} > t_{\rm cn}$. El estado estacionario
se alcanza cuando $t = t_{\rm conv}$.

\subsubsection{P'erdidas hadr'onicas}

Debido a la baja densidad del \emph{jet} en $z_{\rm int}$, los protones no se 
enfr'ian eficientemente en la regi'on del \emph{bow shock} por interacciones
$pp$. Considerando que $n_{\rm j}(z_{\rm int}) \sim 3.2\times10^5$~cm$^{-3}$, el 
tiempo de enfriamiento por $pp$ resulta muy largo, 
$t_{pp}^{\rm bs} \sim 2\times10^9$~s. 
La energ'ia m'axima de los protones queda entonces determinada por el tiempo que
tardan 'estos en difundir una distancia $Z = 0.3 R_{\rm n}$ desde el 
\emph{bow shock} hasta la nube, lo cual nos da una energ'ia m'axima
\begin{equation}
E_{p}^{\rm max} \sim 0.1\,e\,B_{\rm bs}\,R_{\rm n}=
5\times 10^3\,\left(\frac{B_{\rm bs}}{10\,{\rm G}}\right)\,\left(\frac{R_{\rm
n}}{10^{13}\,{\rm cm}}\right)\,{\rm TeV}\,.  
\end{equation}
Los protones con energ'ias $E_p > 0.4\,E_{p}^{\rm max}$, es decir, 
$t_{\rm conv} > t_{\rm dif}$, podr'an difundir hasta la nube antes de ser 
convectados por el material chocado del \emph{jet}. 

En la nube, el campo magn'etico necesario para confinar a estos protones
tan energ'eticos es $B_{\rm n}^{\rm conf} \geq 0.7 (B_{\rm bs}/10\,{\rm G})$.
Si los protones estuviesen confinados en la nube, entonces radiar'ian una
fracci'on significativa de su
energ'ia por interacciones $pp$ con el material chocado de la nube en un 
tiempo  $t_{pp} \sim 2\times10^{15}/(4 n_{\rm n}) \sim 5\times10^4$~s. 
Sin embargo, como no conocemos el valor del campo magn'etico de 
las nubes de la BLR, adoptamos que las part'iculas solo permanecen en la 
nube el tiempo que tardan en cruzarla a una velocidad $\sim c$, con lo cual
el tiempo de cruce es  $t_{\rm cruce} \sim R_{\rm n}/c \sim 700$~s $< t_{pp}$. 
Luego, la distribuci'on energ'etica de los protones relativistas en la nube
ser'a
\begin{equation}
N_p(E_p) \sim \frac{R_{\rm n}}{c}\, Q_p(E_p),
\end{equation}
donde $Q_p = K_p E_p^{-2.2} \exp(-E_p/E_p^{\rm max})$
y $K_p$ est'a dada por la ecuaci'on~(\ref{K_nt_agn}).

%_________________________________________________________________________

\section{Distribuciones espectrales de energ'ia}

Una vez que las distribuciones de energ'ia de las part'iculas relativistas
($N_e(E_e)$ y $N_p(E_p)$)
han sido calculadas (en el estado estacionario), estamos en condiciones de 
calcular las SEDs de la radiaci'on no t'ermica. 

\subsection{Emisi'on asociada al  \emph{bow shock}}

En el \emph{bow shock} los electrones pierden energ'ia basicamente por
radiaci'on sincrotr'on, SSC y EC (con los fotones del disco y de la BLR), 
aunque este 'ultimo no es el proceso dominante si $B_{\rm bs} \sim 10$~G.
La autoabsorci'on sincrotr'on
es tenida en cuenta, pero solo afecta a la parte menos energ'etica de los 
fotones sincrotr'on. En rayos gamma, la absorci'on (gamma-gamma) por 
creaci'on de pares $e^{\pm}$ producida
por la emisi'on del disco de acreci'on y de la BLR tambi'en debe ser 
considerada, pero la absorci'on interna debida a la radiaci'on sincrotr'on
es despreciable y no la tenemos en cuenta.  Dadas las energ'ias t'ipicas
de los fotones emitidos por el disco y la BLR, $\sim 1$~keV y $\sim 10$~eV,
respectivamente, los rayos gamma con energ'ias por encima de 1 y 100~GeV 
pueden ser fuertemente afectados por la absorci'on  gamma-gamma.
Por otro lado, en la mayor'ia de los casos los fotones con energ'ias
$< 1$~GeV van a escapar del denso campo de fotones emitidos por
el disco.  

\begin{figure}
\begin{center}
\includegraphics[angle=270, width=0.6\textwidth]{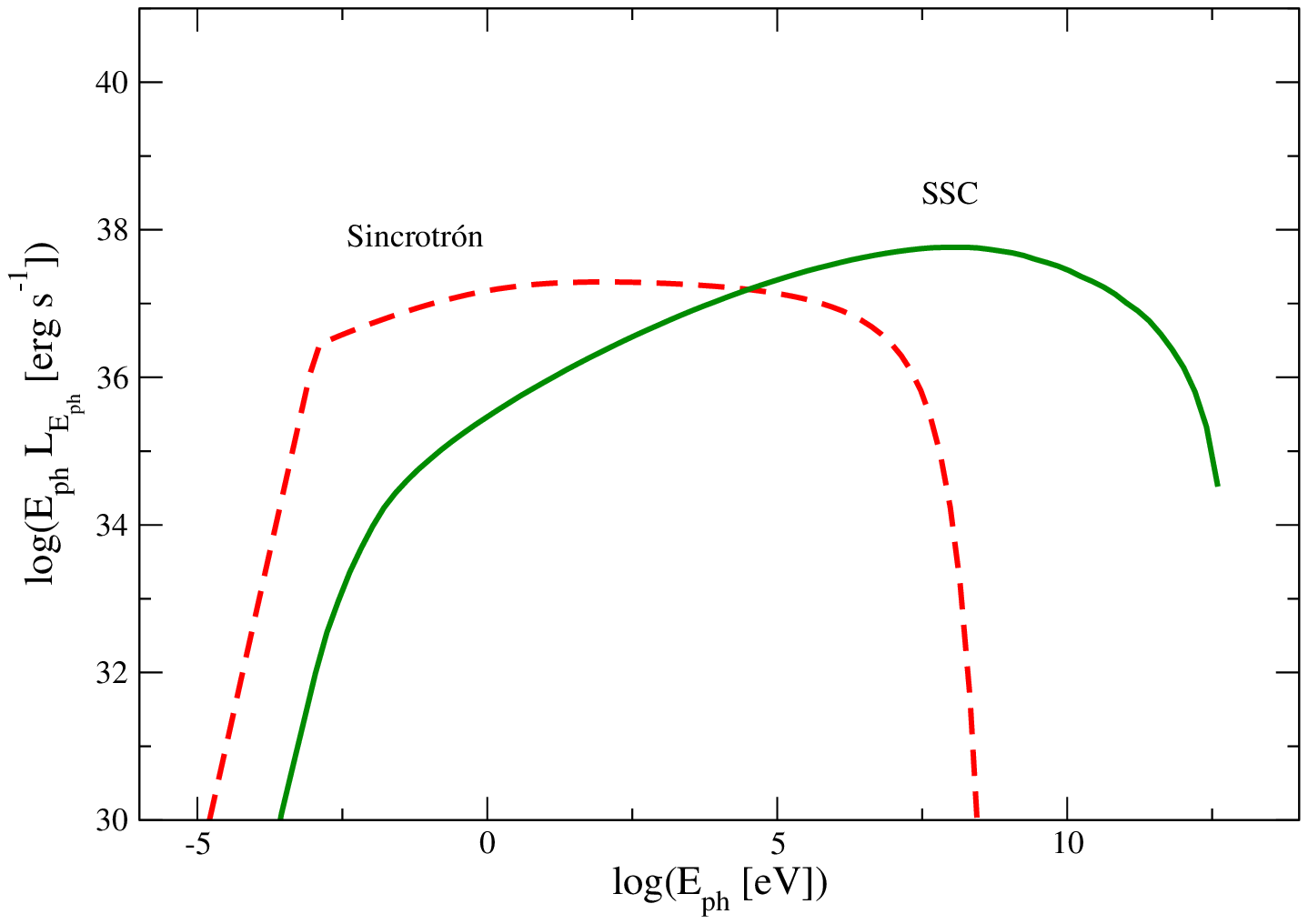}
\caption{Distribuci'on espectral de energ'ia calculada en el caso general 
considerando los procesos
sincrotr'on y SSC. Puede observarse el efecto de la
autoabsorci'on sincrotr'on, a $E_{\rm ph} < 10^{-3}$~eV.}
\label{SED_g}
\end{center}
\end{figure}

En la Figura~\ref{SED_g} se muestra  la SED de los procesos sinctrotr'on y
SSC en la regi'on del \emph{bow shock}. Debido a que este c'alculo fue hecho 
para un caso general, sin especificar las luminosidades ni del jet, ni de la
BLR, ni del disco, la absorci'on gamma-gamma no fue tenida en cuenta, pero
s'i la autoabsorci'on sinctrotr'on. Las luminosidades bolom'etricas 
alcanzadas por ambos
procesos de emisi'on son $L_{\rm sin} \sim 4\times10^{38}$~erg~s$^{-1}$ y 
$L_{\rm SSC} \sim 8\times10^{38}$~erg~s$^{-1}$.

\subsection{Emisi'on asociada a la nube}

Aunque los protones no se enfr'ien eficientemente en la regi'on del 
\emph{bow shock}, s'i pueden hacerlo en la nube aquellos que con 
$E_p > 0.4 E_p^{\rm max}$  llegan hasta all'i. Sin embargo, debido a que 
estos protones son muy  energ'eticos y el campo magn'etico necesario 
para confinarlos en la nube es muy alto, estos protones cruzar'an la nube a la
velocidad de la luz. Con lo cual, solo la pequeña fracci'on 
$\sim 0.17\,t_{\rm cruce}/t_{pp} \sim 10^{-3}$ de la energ'ia 
promedio por prot'on ser'a radiada por interacciones $pp$ en la nube y
la luminosidad bolom'etrica emitida resulta 
$L_{pp} \sim 2\times10^{36}$~erg~s$^{-1}$.
Debido a que no hemos considerado
confinamiento, este valor es un l'imite inferior para la emisi'on
por $pp$ de la nube.

%__________________________________________________________________________

\section{Interacciones m'ultiples}
\label{Many_Clouds_AGN}

\begin{figure}
\begin{center}
\includegraphics[angle=0, width=0.45\textwidth]{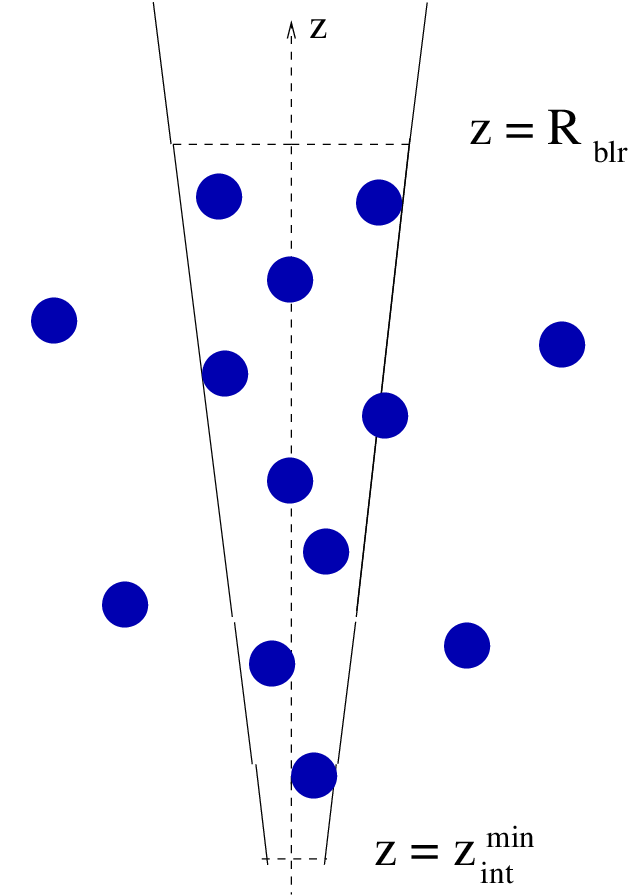}
\caption{Esquema de la interacci'on simult'anea de muchas nubes con el 
\emph{jet}.}
\label{clouds_jet}
\end{center}
\end{figure}

El tamaño de la BLR determina la regi'on en la cual pueden ocurrir
interacciones \emph{jet}-nube. 
Debido a que hay muchas nubes en la BLR, es l'icito pensar que algunas de ellas
pueden interactuar simult'aneamente con el \emph{jet} a diferentes alturas 
$z_{\rm int}^{\rm min} < z_{\rm j} < R_{\rm blr}$ y cada una de las cuales produce 
una cierta cantidad de 
radiaci'on no t'ermica. Esta luminosidad total puede ser mucho mayor que la
producida por la interacci'on de una 'unica nube con el \emph{jet}, que es
$\sim L_{e,p}$. El n'umero de nubes dentro de ambos \emph{jets}, 
$N_{\rm n}^{\rm j}$, 
puede ser calculado
a trav'es del volumen de cada \emph{jet} ($V_{\rm j}$) y de cada nube 
($V_{\rm n}$), resultando
\begin{equation}
\label{N_clouds}
N_{\rm n}^{\rm j} = 2\, f\, \frac{V_{\rm j}}{V_{\rm n}}\sim
9\left(\frac{L_{\rm j}}{10^{44}\,\rm{erg\,s^{-1}}}\right)^{2}
\left(\frac{R_{\rm n}}{10^{13}\,\rm{cm}}\right)^{-3},
\end{equation}
donde el factor 2 es debido a la presencia de dos \emph{jets} y 
$f \sim 10^{-6}$ es el
factor de llenado de nubes en la BLR (Dietrich et al. 1999).
Este c'alculo de $N_{\rm n}^{\rm j}$ es correcto si no tenemos en cuenta que
las nubes pueden ser destru'idas dentro del \emph{jet} y los fragmentos 
eventualmente
dilu'idos. Por ejemplo, Klein y colaboradores (1994) estimaron que el tiempo
de vida de una nube chocada es varias veces $t_{\rm cn}$, y Shin y 
colaboradores 
(2008) encontraron que a'un un campo magn'etico d'ebil en la nube puede 
incrementar significativamente el tiempo de vida de 'estas. Finalmente,
a'un en el caso de que las nubes se fragmenten, \emph{bow shocks} fuertes
pueden formarse alrededor de cada fragmento antes de que 'estos sean acelerados
hasta una velocidad $\sim v_{\rm j}$. Todas estas consideraciones hacen que el 
n'umero real de nubes que simult'aneamente est'an interactuando con el 
\emph{jet} 
sea dif'icil de estimar, pero este n'umero debe estar entre 
$(t_{\rm cn}/t_{\rm j})\,N_{\rm n}^{\rm j}$ y $N_{\rm n}^{\rm j}$.
La presencia de muchas nubes dentro del \emph{jet}, no solo en 
$z_{\rm int}$ sino 
tambi'en a $z_{\rm j}$ m'as altos, implica que la luminosidad no t'ermica total 
disponible en la intersecci'on entre el \emph{jet} y la BLR es 
\begin{equation}
\label{Lrad_tot}
L_{e,p}^{\rm tot} \sim 2\,\int^{R_{\rm blr}} 
\frac{{\rm d}N_{\rm n}^{\rm j}}{{\rm d}z} L_{\rm nt}(z)\, {\rm d}z
\sim 1.7\times10^{40} \left(\frac{\eta_{\rm nt}}{0.1}\right) 
\left(\frac{R_{\rm n}}{10^{13}}\right)^{-1}
\left(\frac{L_{\rm j}}{10^{44}\,\rm{erg s^{-1}}}\right)^{1.7},
\end{equation} 
donde ${\rm d}N_{\rm n}^{\rm j}$ es el n'umero de nubes
localizadas en un volumen del \emph{jet}  
${\rm d}V_{\rm j}=\pi\,(0.1z)^2\,{\rm d}z_{\rm j}$. 
En ambas ecuaciones~(\ref{N_clouds}) y (\ref{Lrad_tot}), 
$L_{\rm blr}$ ha sido fijada en $0.1\,L_{\rm j}$, aproximadamente como
en las galaxias FR~II, y $R_{\rm blr}$ ha sido derivada usando la
ecuaci'on~(\ref{Rblr}). 

\begin{figure}
\begin{center}
\includegraphics[angle=270, width=0.7\textwidth]{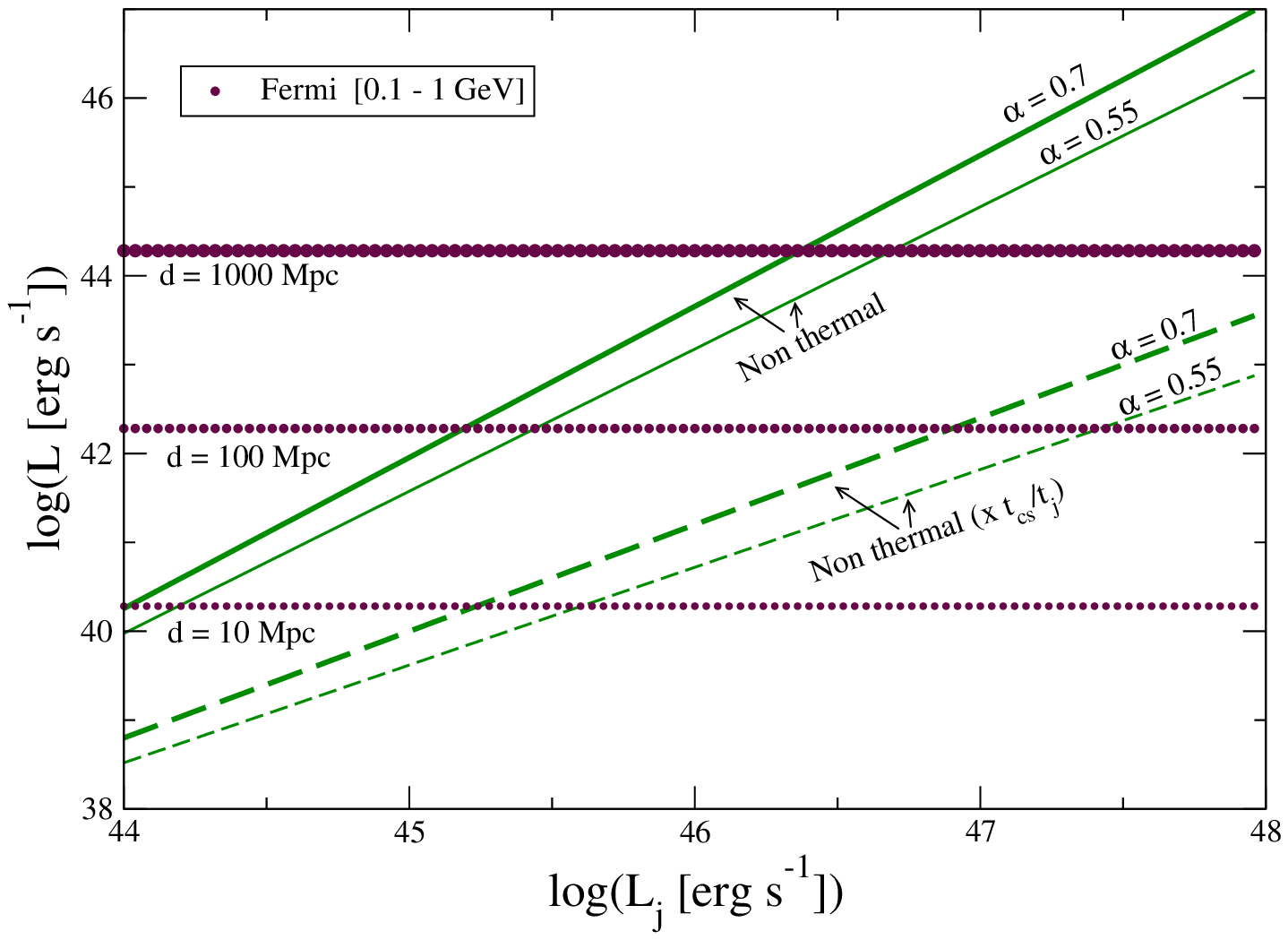}
\caption{L'imites superiores para la luminosidad de rayos gamma producida por 
$N_{\rm n}^{\rm j}$ nubes dentro del \emph{jet} como una funci'on de
$L_{\rm j}$ en fuentes FR~II. Dos casos son graficados, uno asumiendo que 
las nubes cruzan el \emph{jet} sin ser fragmentadas (l'ineas s'olidas verdes), y
otro en en el cual las nubes son destru'idas en un tiempo $\sim t_{\rm cn}$
(l'ineas rayadas verdes). 
Las l'ineas gruesas (s'olidas y rayadas) y delgadas (s'olidas y rayadas) 
corresponden a  
$R_{\rm blr} \propto L_{\rm blr}^{0.7}$ (Kaspi et al. 2007)  y
$R_{\rm blr} \propto L_{\rm blr}^{0.55}$ (Kaspi et al. 2005), respectivamente.
Adem'as, los niveles de sensibilidad de {\it Fermi} en el rango
0.1--1~GeV (l'ineas punteadas marrones) son graficadas para tres  
distancias diferentes $d=10$, 100 y 1000~Mpc (Araudo et al. 2010).}
\label{L_tot}
\end{center}
\end{figure}

En la Figura~\ref{L_tot} mostramos estimaciones de la luminosidad en rayos
gamma predicha para el caso de la interacci'on simult'anea de muchas nubes 
con el 
\emph{jet}. Para esto hemos seguido un procedimiento muy simple, asumiendo que
toda la luminosidad no t'ermica es radiada en forma de rayos gamma. Este puede 
ser el caso si los tiempos de escape y de enfriamiento por radiaci'on 
sincrotr'on son m'as largos que el tiempo de enfriamiento por IC (EC + SSC)
para los electrones m'as energ'eticos. Dado que se conoce muy poco de la
BLR en las galaxias FR~I, no consideramos este tipo de fuentes en la
figura.   

%_____________________________________________________________________

\section{Aplicaciones} 

En las pr'oximas dos subsecciones presentamos los resultados obtenidos de
aplicar el modelo expuesto en las secciones~\ref{escenario} y
\ref{poblacion_nt} a dos
fuentes caracter'isticas: Cen~A (FR~I, 'unica interacci'on) y
3C~273 (FR~II, m'ultiples interacciones).

%--------------------------------------------------------------------
\subsection{Galaxias FR~I: Cen A}
\label{FRI}

Cen~A es la galaxia activa m'as cercana, ubicada a una distacia 
$d\approx 3.7$~Mpc (Israel 1998). 
Esta fuente ha sido clasificada como una radio-galaxia FR~I y como un
objeto Seyfert~2 en el 'optico.
La masa del agujero negro es $M_{\rm smbh} \approx 6\times10^7$~$M_{\odot}$ 
(Marconi et al. 2000). 
El 'angulo entre los \emph{jets} y la l'inea de la visual es grande, 
$>50^\circ$ (Tingay et al. 1998), 
de tal manera que la radiaci'on producida en el \emph{jet} no est'a 
significativamente corrida en frecuencia por efecto Doppler.
Los \emph{jets} de Cen~A son frenados a una distancia $\sim$ kpc,
formando dos radio l'obulos gigantes que se extienden $\sim 10^{\circ}$ en
el cielo del hemisferio sur. En el 'optico, la zona del n'ucleo de Cen~A 
est'a oscurecida
por una regi'on  densa  de gas y polvo, formada probablemente en una
colici'on reciente  con otra galaxia
(Thomson 1992,  Mirabel et al. 1999). A energ'ias m'as altas, los
sat'elites {\it Chandra} y
{\it XMM-Newton} detectaron emisi'on continua en rayos~X 
proveniente de la regi'on nuclear, con una
luminosidad $\sim 5\times10^{41}$~erg~s$^{-1}$
entre 2--7~keV (Evans et al. 2004). Estos rayos~X pueden haber sido 
 producidos en el disco de acreci'on y en el \emph{jet} m'as interno, 
aunque su origen es a'un desconocido.
En HE, Cen~A ha sido detectada por encima de las $200$~MeV por
\emph{Fermi}, con una luminosidad bolom'etrica de 
$\sim 4\times10^{40}$~erg~s$^{-1}$  (Abdo et al. 2009b, y por encima de
 $\sim 200$~GeV por HESS, con una luminosidad bolom'etrica de 
$\approx 3\times10^{39}$~erg~s$^{-1}$ (Aharonian et al. 2009). En ambos casos,
la emisi'on en HE es asociada a la regi'on nuclear.
Cen~A ha sido propuesta como una fuente de rayos c'osmicos de alta energ'ia
por Romero y colaboradores (1996). 

Si bien no se han detectado a'un l'ineas de emisi'on que indiquen la presencia 
de una BLR en Cen~A (Alexander et al. 1999), esto puede ser una consecuencia de
que la regi'on nuclear de esta fuente est'a oscurecida en el 'optico por 
la ``estela de polvo''. Es por esto que  puede
haber nubes circundando el SMBH de Cen A (Wang et al.  1986, Risaliti et
al. 2002)  pero, debido a que el disco de acreci'on de este AGN es d'ebil, 
no se espera un nivel alto de fotoionizaci'on de estas posibles nubes
y por lo tanto el proceso ser'a ineficiente para producir l'ineas. Bajo 
esta hip'otesis podemos 
considerar que existe una poblaci'on de nubes oscuras en la regi'on nuclear de 
Cen~A y aplicar el modelo desarrollado a esta fuente, considerando 
interacciones EC solamente con los fotones emitidos por el material acretado.

\begin{figure}
\begin{center}
\includegraphics[angle=270, width=0.49\textwidth]{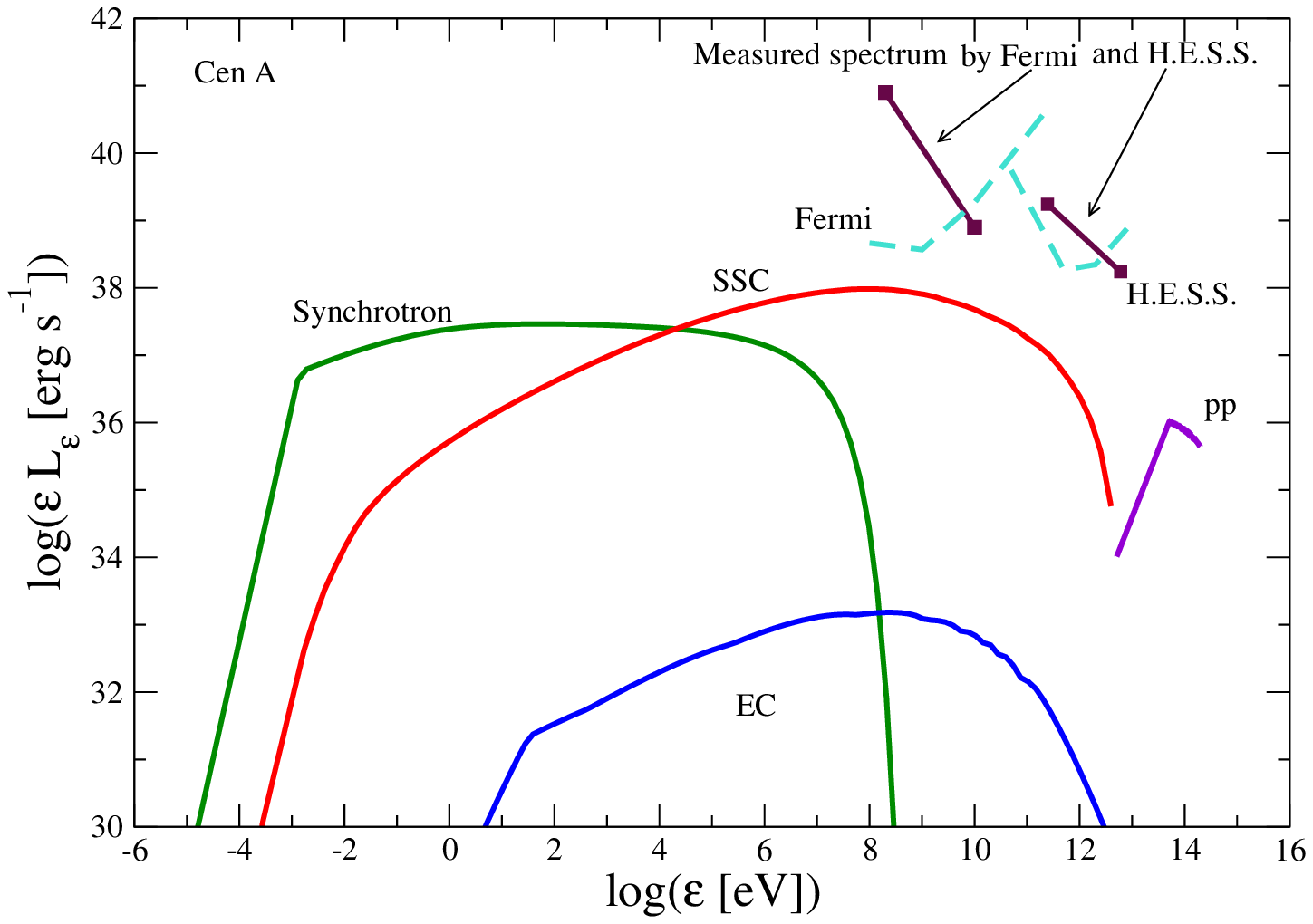}
\includegraphics[angle=270, width=0.49\textwidth]{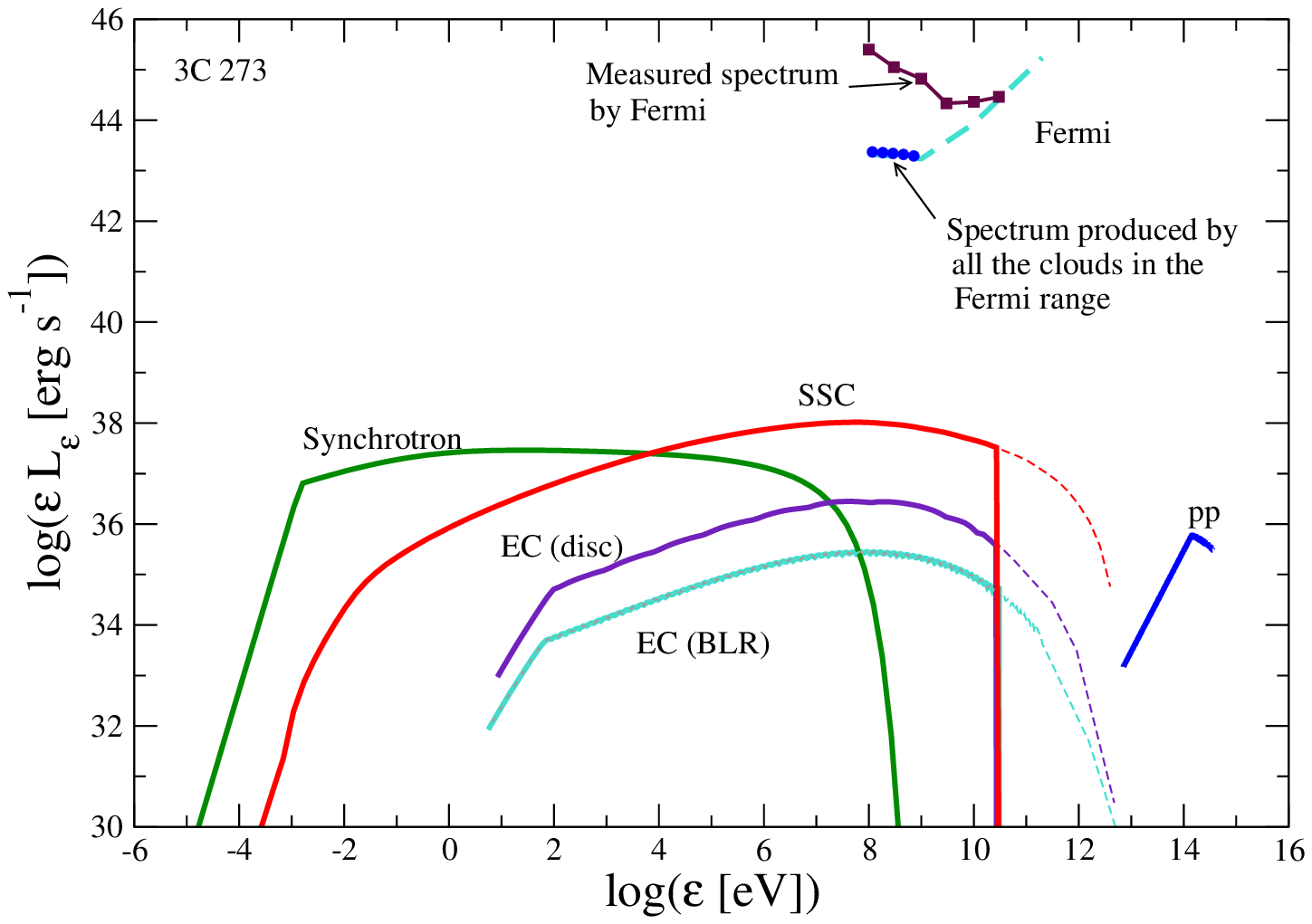}
\caption{Distribuciones espectrales de energ'ia calculadas para las fuentes 
Cen~A (izquierda) y 3C~273 (derecha).
En ambos casos se han graficado las luminosidades espec'ificas producidas
por los procesos sincrotr'on, EC, SSC y $pp$ obtenidos de la interacci'on de
una nube con uno de los \emph{jets}. Tambi'en mostramos la emisi'on
detectada por \emph{Fermi} y HESS y el espectro emitido considerando 
muchas nubes simult'aneamente interactuando con ambos \emph{jets}  en el
caso de 3C~273.}
\label{SEDs_AGN}
\end{center}
\end{figure}

Adoptando $L_{\rm j}=10^{44}$~erg~s$^{-1}$ para la luminosidad del \emph{jet}
de Cen~A y aquellos valores 
que han sido listados en la Tabla~\ref{const_AGN} para el resto de los
par'ametros, $z_{\rm int}$ resulta $\approx 5\times10^{15}$~cm. A
esta altura del \emph{jet}, la emisi'on producida por la interacci'on
con una nube  es calculada suponiendo que  
$\eta_{\rm B}=0.01$. La correspondiente SED se muestra en la 
Figura~\ref{SEDs_AGN} (izquierda). %Como mencionamos en la
%secci'on~\ref{acc}, 
La parte de m'as baja energ'ia del espectro sincrotr'on es
autoabsorbida a energ'ias menores que $\sim 10^{-4}$~eV. 
En rayos gamma, la absorci'on gamma-gamma se desprecia debido
a que la densidad de fotones ambiente es muy baja 
(e.g. Rieger \& Aharonian 2009, Araudo et al. 2009, 2010). 
En HE, la emisi'on por  SSC domina la SED, siendo la luminosidad
de este proceso a energ'ias mayores que 100~MeV $\sim
2\times10^{39}$~erg~s$^{-1}$, y por encima de 100~GeV aproximadamente 10 
veces menor. Estos valores son
un orden de magnitud menores que los correspondientes a los flujos
detectados por los telescopios \emph{Fermi} y HESS.
Notamos sin embargo que $L_{e,p}\propto R_{\rm n}^2$, y para
nubes apenas m'as grandes $L_{e,p}$ 
puede aumentar hasta niveles detectables. La penetraci'on de una
nube de tamaño $R_{\rm n} > 10^{13}$~cm en el \emph{jet} de Cen~A podr'ia 
producir un \emph{flare} de aproximadamente un d'ia de duraci'on.

%--------------------------------------------------------------------
\subsection{Galaxias FR~II: 3C 273}

\begin{table}[]
\begin{center}
\caption{Par'ametetros adoptados para  Cen~A y 3C~273.}
\label{applications}
\begin{tabular}{lcc}
\hline
\hline 
{}    & Cen A &  3C 273  \\ 
\hline 
Distancia [Mpc] & 3.7 &  $6.7\times10^2$\\
Masa del SMBH  [$M_{\odot}$] & $6\times10^7$& $7\times10^9$ \\
Angulo de inclinaci'on del \emph{jet} [$^\circ$] & $> 50$  & $\sim 15$ \\
Luminosidad del \emph{jet}  [erg~s$^{-1}$] & $10^{44}$ & $4\times10^{47}$\\  
Luminosidad del disco [erg~s$^{-1}$]&$5\times10^{41}$& $2\times10^{46}$ \\
Densidad de energ'ia de los fotones del disco [eV] & $\sim 5\times10^3$ & 54  \\
Luminosidad de la BLR  [erg~s$^{-1}$]& - & $4\times10^{45}$  \\
\hline
\end{tabular}
\end{center}
\end{table}

A la distancia $d=6.7\times10^2$~Mpc, 
3C~273 es el AGN activo en radio m'as potente (Courvoisier 1998) 
con un SMBH de masa $M_{\rm smbh} \sim 7\times10^9 M_{\odot}$
(Paltani \& T$\ddot{\rm u}$rler 2005). El 'angulo entre la direcci'on del
\emph{jet} con la l'inea de la visual es chico, $\approx 6^\circ$, 
lo que implica que 3C~273 es un blazar
(Jolley et al. 2009). El espectro completo de esta fuente es variable 
(e.g. Pian et al. 1999) con per'iodos que van desde años (en radio) hasta
unas pocas horas (en rayos gamma). En HE, 3C~273
fue el primer blazar detectedo en la banda MeV por el sat'elite COS-B,
y luego por  EGRET en la banda 0.1 - 10~GeV (Hartman et al. 1999). Esta 
fuente fue tambi'en detectada en energ'ias GeV por
\emph{Fermi} y \emph{AGILE}, pero a'un no ha sido detectada en TeV.
Dada la luminosidad del \emph{jet} de 3C~273, $L_{\rm j}\approx
4\times10^{47}$~erg~s$^{-1}$ (Kataoka et al. 2002), $z_{\rm int}$ resulta
 $\approx 3\times10^{17}$~cm. 
La luminosidad de la BLR de esta fuente es $L_{\rm blr} \approx
4\times10^{45}$~erg~s$^{-1}$ (Cao \& Jiang 1999), y su tamaño
$R_{\rm blr} \sim 7\times10^{17}$~cm (Ghissellini et al. 2010), 
lo que implica que las interacciones \emph{jet}-nube pueden ocurrir por ser 
$z_{\rm int} < R_{\rm blr}$.
La luminosidad del disco es alta,
$L_{\rm d}\approx 2\times 10^{46}$~erg~s$^{-1}$, con energ'ias t'ipicas de los
fotones de $\approx 54$~eV (Grandi \& Palumbo 2004).

La SED de la radiaci'on no t'ermica generada por las interacciones 
\emph{jet}-nube en 3C~273 es mostrada en la Figura~\ref{SEDs_AGN} (derecha). 
A la altura $z_{\rm int}$, los procesos radiativos m'as importantes son 
el sincrotr'on y el SSC. Las luminosidades bolom'etricas generadas 
por estos procesos en una interacci'on en
$z_{\rm int}$ son $L_{\rm sin} \sim 6\times10^{38}$~erg~s$^{-1}$ y
$L_{\rm SSC} \sim 2\times10^{39}$~erg~s$^{-1}$. 
Dados los fuertes
campos de radiaci'on del disco y de la BLR, la emisi'on por encima 
de $\sim 10$~GeV es absorbida por interacciones fot'on-fot'on,
y el m'aximo de la emisi'on ocurre alrededor de 0.1--1~GeV. 
Siendo que el n'umero estimado de nubes en la BLR de 3C~273 es $\sim 10^8$
(Dietrich et al. 1999), el factor de llenado que se deduce es
$f\sim 3\times10^{-7}$ y el n'umero de nubes en ambos \emph{jets} resulta
$\sim 2\times 10^3$ y $5\times 10^5$, para los valores m'inimo y m'aximo
dados en la Secci'on~\ref{Many_Clouds_AGN}. En el caso m'as optimista,
la luminosidad SSC podr'ia alcanzar valores
$\sim 2\times10^{44}$~erg~s$^{-1}$.  Este valor est'a por debajo de la
luminosidad detectada por \emph{Fermi} en HE, $\sim
3\times10^{46}$~erg~s$^{-1}$ en el estado estacionario y $\sim
1.7\times10^{47}$~erg~s$^{-1}$ en \emph{flares} (Soldi et
al. 2009). Sin embargo, la emisi'on detectada probablemente est'e muy 
amplificada por efecto
Doppler y esto tapa la emisi'on que no sufre tal amplificaci'on.
Sin embargo, para AGN que no son blazares (galaxias FR~II, como Cygnus~A), 
cuya
emisi'on no sufre amplificaci'on Doppler, la radiaci'on en GeV producida por
interacciones \emph{jet}-nube podr'ian ser detectadas. En este caso, 
dado que muchas nubes pueden interactuar simult'aneamente con el \emph{jet}, la
emisi'on ser'a estacionaria.

%_________________________________________________________________________

\section{Discusi'on}
\label{disc}

En este cap'itulo hemos estudiado la interacci'on de nubes de la 
BLR con la base de los \emph{jets} en los AGN. 
Considerando valores razonables para los par'ametros de las nubes y de los 
\emph{jets}, estimamos las escalas de tiempo de los procesos din'amicos m'as 
relevantes en el escenario analizado, concluyendo que las nubes de la BLR 
pueden entrar en el \emph{jet} s'olo a $z_{\rm j} \geq z_{\rm int}$. 
Para alturas menores que
este valor, el \emph{jet} es muy compacto y las presiones magn'etica y cin'etica
de 'este destruir'an a la nube antes de que 'esta entre completamente en el
\emph{jet}. Cuando la nube interact'ua significativamente con el \emph{jet}, choques
fuertes son generados con la subsecuente emisi'on en rayos gamma de las
part'iculas aceleradas en tales choques.    
 
Valores del campo magn'etico en la regi'on chocada del \emph{jet} ($B_{\rm bs}$)
menores que el correspondiente a la equipartici'on con la
energ'ia de las part'iculas no t'ermicas permiten una emisi'on significativa
de rayos gamma. Para valores m'as altos de $B_{\rm bs}$ (esto es, \emph{jets} 
dominados 
por el flujo de Poynting), el tratamiento desarrollado aqu'i no es v'alido.
En tal caso,  $z_{\rm int}$ puede a'un ser definido si en vez de considerar 
la presi'on cin'etica del \emph{jet}, consideramos la magn'etica. 
Si una nube entra al \emph{jet}, la aceleraci'on de part'iculas en el 
\emph{bow shock}
puede darse por reconeci'on magn'etica, por ejemplo. El estudio de este caso
requiere de un tratamiento completamente diferente al expuesto en este 
cap'itulo. En general, para valores de   $B_{\rm bs}$ mayores que el de 
equipartici'on la emisi'on de rayos gamma por IC es suprimida en favor de la
radiaci'on sincrotr'on al menos que la disipaci'on magn'etica reduzca
la intensidad del campo magn'etico lo sufiente como para que el canal
de p'erdidas IC sea el dominante. 

En fuentes cercanas, como por ejemplo Cen~A, la interacci'on de nubes 
grandes con los \emph{jets} se deber'ia detectar como un evento espor'adico,
aunque el n'umero de estas nubes m'as grandes y el 
\emph{duty cycle}\footnote{Se define el \emph{duty cycle} de una fuente
como la fracci'on del tiempo de observaci'on durante la cual se producen 
variaciones significativas de flujo.}
de los \emph{flares} son dif'iciles de estimar.
Dado que los campos de fotones externos son d'ebiles en estas galaxias,
fotones de VHE pueden escapar del emisor sin ser absorbidos por interacciones 
con fotones ambientales. Las interacciones \emph{jet}-nube en galaxias 
FR~I cercanas
ser'ian detectables tanto en HE como en VHE como emisi'on espor'adica
con escalas de tiempo de aproximadamente un d'ia. 
La detecci'on de esta emisi'on nos proveer'ia de informaci'on sobre
las condiciones del medio de la regi'on nuclear de los AGN 
(oscurecida en general en otras frecuencias por la absorci'on del polvo), 
como as'i tambi'en de la base de los \emph{jets}.  

En fuentes tipo FR~II, muchas nubes pueden interactuar simult'aneamente con
el \emph{jet}. El n'umero de 'estas -$N_{\rm n}^{\rm j}$- depende fuertemente 
del tiempo de vida de las nubes dentro del \emph{jet}, que es del orden de 
algunas veces  $t_{\rm cn}$.
Sin embargo, notamos que a'un despu'es de la fragmentaci'on de las nubes,
los \emph{bow shocks} asociados a cada fragmento pueden todav'ia formarse 
 y acelerar part'iculas eficientemente si los fragmentos se mueven 
lentamente, a una velocidad $\ll v_{\rm j}$.
Debido a que las galaxias FR~II tienen tasas de acreci'on altas, la radiaci'on
por encima de 1~GeV producida cerca de la base del \emph{jet} puede ser 
sustancialmente atenuada por los densos campos de fotones externos 
(del disco y de la BLR) aunque los rayos gamma de energ'ias $< 1$~GeV
no deber'ian ser afectados significativamente.  Ya que la emisi'on producida en
las interacciones \emph{jet}-nube es isotr'opica, 'esta puede ser tapada por la 
emisi'on producida en los \emph{jets} de los blazares y amplificada por efecto 
Doppler. No obstante, los 
\emph{jets} de las galaxias  FR~II cercanas no muestran un 
incremento muy importante del flujo y estas 
fuentes pueden emitir rayos gammas por interacciones \emph{jet}-nube. 
En el contexto de la unificaci'on de los AGN  (Urry \& Padovani 1995),
el n'umero de AGN que no son blazares (intensos en radio) debe ser mucho 
mayor al de blazares para el mismo valor de  $L_{\rm j}$.
Como se muestra en la Figura~\ref{L_tot}, fuentes cercanas e intensas
podr'ian ser detectables por observaciones suficientemente profundas del
sat'elite {\it Fermi}. Luego de unos pocos años de exposici'on, 
señales significativas de estas fuentes podr'ian ser encontradas, 
aportando esto una evidencia de que los \emph{jets} est'an dominados por la 
materia a las alturas de las interacciones con las nubes, adem'as
de informaci'on sobre las caracter'isticas de la BLR.

%% file: clusters_final.tex
\chapter{C'umulos de galaxias}
\label{clusters}

\section{Introducci'on}

Las galaxias no est'an aisladas en el Universo sino que se agrupan de a cientos
en regiones de un tamaño $\sim 1-10$~Mpc, donde forman parte de la 
estructura filamentosa del Universo. 
Para explicar la formaci'on de estas estructuras a gran escala, 
el modelo actualmente aceptado contempla la existencia de materia oscura fr'ia
(CDM, por \emph{Cold Dark Matter}). Bajo este paradigma, las galaxias se 
forman por la fusi'on de estructuras menores que dan origen a sistemas de 
mayor escala, como as'i 
tambi'en por la acreci'on de material en forma de gas. 
Durante el proceso de acreci'on y fusi'on,
la materia se acumula en una red c'osmica con sobredensidades moderadas
en forma de filamentos, en cuyas intersecciones se ubican las galaxias y los
c'umulos de galaxias.

El proceso de acreci'on contempla la ca'ida de gas difuso proveniente del
espacio circundante al c'umulo. Este mecanismo es inestable y anisotr'opico, ya 
que ocurre a trav'es de flujos de material que se propagan a lo largo de los 
filamentos que convergen en el c'umulo produciendo fuertes choques de 
acreci'on al impactar con el medio intra c'umulo (ICM, por 
\emph{Intra Cluster Medium}). Por otro lado, la 
fusi'on de dos estructuras
virializadas puede contemplar la colisi'on entre dos galaxias, como se cree que
ha ocurrido 
en Cen A, como as'i tambi'en la fusi'on de dos c'umulos de galaxias. 
Tanto los procesos de acreci'on como los de fusi'on son mucho m'as
intensos en los c'umulos de galaxias que en las galaxias individuales.
Esto hace que se produzcan fuertes choques (de acreci'on y de fusi'on), los 
cuales calientan el
ICM y aceleran part'iculas cargadas hasta velocidades relativistas. 
Tanto la emisi'on del gas caliente como de las part'iculas no t'ermicas 
caracterizan a los c'umulos de
galaxias en diferentes bandas del espectro electromagn'etico.

En rayos~X, los c'umulos de galaxias son los objetos m'as brillantes 
del cielo alcanzando una luminosidad 
$L_{\rm X}\sim 10^{43}-10^{45}$~erg~s$^{-1}$. 
Esta emisi'on es t'ermica,  producida por el calentamiento del 
gas intra c'umulo que alcanza temperaturas $\sim 10^8$~K.  
Sin embargo, la detecci'on de estas estructuras en frecuencias radio como 
as'i tambi'en en el extremo UV ($0.07-0.4$~keV) y en los rayos~X duros 
($20-80$~keV)
indica la presencia de actividad no t'ermica en estos sitemas.
La emisi'on en frecuencias radio detectada en los c'umulos 
de galaxias es originada en la interacci'on de electrones relativistas con 
el campo magn'etico del c'umulo  ($\sim \mu{\rm G}$).
Por otro lado, la radiaci'on en el extremo UV debe ser producida o bien por
la componente t'ermica m'as fr'ia o bien por interacciones IC con los 
fotones del fondo c'osmico de radiaci'on 
(CMB, por \emph{Cosmic Microwave Background})
con la misma poblaci'on de electrones que produce la emisi'on sincrotr'on 
en radio. La radiaci'on sincrotr'on tambi'en podr'ia explicar la emisi'on 
de los rayos~X duros, aunque esto es todav'ia tema de debate.

\begin{figure}
\begin{center}
\includegraphics[angle=0, width=0.99\textwidth]{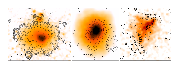}
\caption{Los c'umulos de galaxias Abell~2163 (izquierda), RX~J1347.5-1145 
(medio) y ZwCl~2341.1 (derecha) observados en rayos~X 
(amarillo-anaranjado-marr'on) y en 
frecuencias radio (contornos de l'ineas llenas) son mostrados. 
Uno de los halos m'as 
luminosos que se han detectado es aquel asociado al c'umulo A2163.
En el centro de la fuente RX~J1347.5-1145, la m'as luminosa en rayos~X, se 
ha detectado un halo pequeño, mientras que remanentes dobles han sido 
detectados en los bordes del c'umulo ZwCl~2341.1 (Ferrari 2010).}
\label{halos-remanentes}
\end{center}
\end{figure}

Respecto de la emisi'on en radio, aunque su naturaleza no t'ermica est'a 
bien establecida por las observaciones, 'esta puede ser debida a dos 
tipos de fuentes diferentes:
halos o remanentes (\emph{relics}). Los primeros 
est'an localizados en los centros de los c'umulos 
y la emisi'on en radio de 'estos no est'a polarizada y se superpone a la 
emisi'on t'ermica en rayos~X. Los 'ultimos se encuentran en la periferia 
de los c'umulos
y son regiones de emisi'on polarizada y con 
una estrucura m'as irregular. El modelo actualmente aceptado para la formaci'on
 de estos halos y remanentes es que  se producen
por choques de acreci'on y de fusi'on, respectivamente.  
En la Figura~\ref{halos-remanentes} se muestran estos dos tipos de fuentes.

Los remanentes han sido observados en muchos c'umulos, aunque los 
m'as poderosos y extendidos han sido detectados en
fuentes con halos centrales, como el c'umulo Coma (Giovannini et al. 1991) y 
las fuentes Abell~2163 (Feretti et al. 2001),
Abell~2255 (Feretti et al. 1997), Abell~2256 (Rottgering et al. 1994) y 
Abell~2744  (Govoni et al. 2001). 
S'olo muy pocos c'umulos presentan dos remanentes (opuestos el uno al otro),
siendo los ejemplos m'as prominentes aquellos encontrados en los c'umulos
 Abell~3667 (Rottgering et al. 1997) y Abell~3376 (Bagchi et al. 2006).

Dado que hay una poblaci'on de part'iculas relativistas que emiten la 
radiaci'on sincrotr'on y quizas la emisi'on de rayos~X duros por IC, 
estas mismas part'iculas pueden producir tambi'en 
emisi'on de m'as alta energ'ia. 
La radiaci'on gamma ser'ia  generada por interacciones $pp$ con el ICM 
si protones son acelerados 
(ver Völk et al. 1996), o por dispersiones IC entre los fotones del CMB 
y los electrones relativistas (Atoyan \& Völk, 2000).
Las simulaciones cosmol'ogicas hidrodin'amicas son 'utiles para 
predecir los niveles de emisi'on de rayos gamma en objetos 
virializados a gran escala (Keshet et al. 2003; Pfrommer et al. 2007, 2008).  
Si embargo, los resultados de estos estudios no predicen una 
contribuci'on significativa al fondo difuso de rayos gamma (Berrington \&
Dermer 2003).
Estos resultados te'oricos son consistentes con el hecho de que 
no se hayan detectado c'umulos de galaxias en HE y VHE con los telescopios
de rayos gamma que actualmente ent'an funcionando.
S'olo han sido reportadas evidencias marginales de emisi'on de altas energ'ias 
del c'umulo Abell~1758 
dentro de la caja de error de la fuente 3~EG~J1337+5029 (Fegan et al. 2005). 
Los c'umulos de galaxias Perseus y Abell 2029 fueron observados por Perkins y
colaboradores (2006) con el telescopio Cherenkov Whipple, aunque 
no encontraron evidencias de fuentes de rayos gamma puntuales ni 
extendidas en el rango de los TeV.
Por otro lado, observaciones recientes de los c'umulos Coma y Abell 496
llavadas a cabo con HESS (Domainko et al 2007) no detectaron emisi'on 
significativa en rayos gamma a partir de observaciones con tiempos de 
exposici'on moderados, de $\sim 10-20$ horas.

 Todas las estimaciones hechas al momento de la posible emisi'on
de rayos gamma de los c'umulos de galaxias han sido realizadas
para fuentes que no son reales sino simuladas
num'ericamente. Por esto, en el presente cap'itulo 
 estudiaremos los procesos no t'ermicos 
que tienen  lugar en los bordes del c'umulo Abell 3376, en donde se
han detectado en frecuencias radio dos remanentes gigantes y sim'etricos. 
En lo que sigue,
analizaremos tanto el contenido de part'iculas relativistas en los remanentes
de este c'umulo como as'i tambi'en la emisi'on no t'ermica producida por 
las part'iculas aceleradas, concentrandon'os en el an'alisis del espectro 
de rayos gamma.

\section{El c'umulo Abell 3376}

\begin{figure}
\begin{center}
\includegraphics[angle=270, width=0.7\textwidth]
{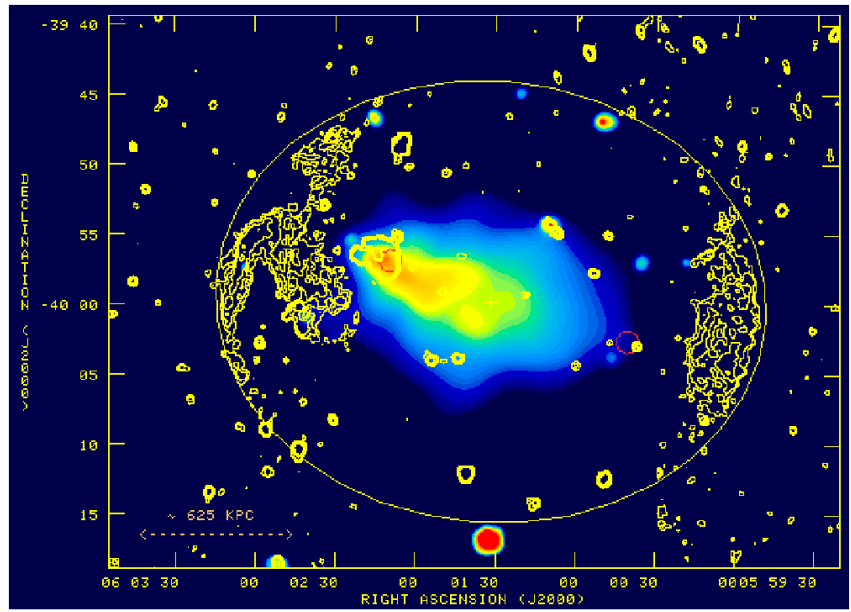}
\caption{Composici'on de la emisi'on en frecuencias radio 
(contornos amarillos) y en rayos~X (estructura central) del
c'umulo Abell~3376. Con amarillo se dibuja una elipse dentro de la
cual est'an la emisi'on de rayos X (en el centro) y los remanentes (en el borde)
(Bagchi et al. 2006).}
\label{Bagchi_figs_X}
\end{center}
\end{figure}

El c'umulo de galaxias Abell~3376 est'a ubicado a una distancia 
$d \sim 187$~Mpc, con un corrimiento al rojo (\emph{redshift})
$z \sim 0.046$ si consideramos una cosmolog'ia est'andar de materia oscura 
fr'ia $\Lambda$CDM ($\Omega_{\rm m}$=0.3, $\Omega_{\Lambda}$=0.7, 
${\rm H}_{\rm o}= 100\, h^{-1} \, {\rm km} \, {\rm s}^{-1} \, {\rm Mpc}^{-1}$,
con $h=0.7$).
La masa del c'umulo ha sido estimada aplicando el teorema del virial 
a las galaxias miembro del mismo. Asumiendo que la distribuci'on de 
masa de la fuente se comporta de la misma manera que  la distribuci'on de masa 
de las galaxias miembro (Girardi et al. 1998), obtenemos una masa virial
$M_{\rm A3376} \sim 3.64\times 10^{14}\;h^{-1}\,M_{\odot}$.
El correspondiente radio virial resulta $R_{\rm A3376} \sim 1.4 \, h^{-1}$~Mpc.

El c'umulo Abell~3376 ha sido detectado por los sat'elites
\emph{ROSAT} y \emph{XMM}-Newton en rayos X 
siendo esto una evidencia de la actividad de fusi'on de subc'umulos.
La luminosidad en  X de esta fuente es                                 
$L_{\rm X}(0.1-2.4\;\rm{keV}) \simeq 2.48\times 10^{44}\;\rm{erg\;s^{-1}}$,
producida t'ermicamente por el ICM caliente, que tiene una temperatura promedio
$T_{\rm X}\approx 5.8\times 10^7$~K ($\approx 5$~keV).

\begin{figure}
\begin{center}
\includegraphics[angle=0, width=0.47\textwidth]{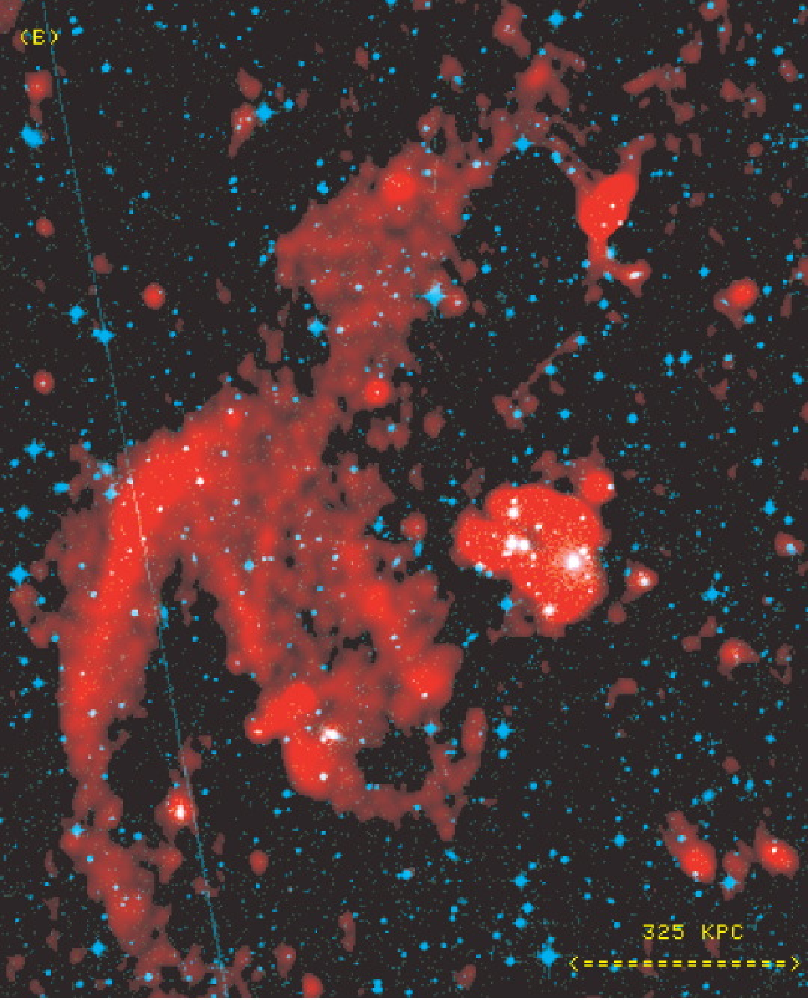}
\includegraphics[angle=0, width=0.35\textwidth]{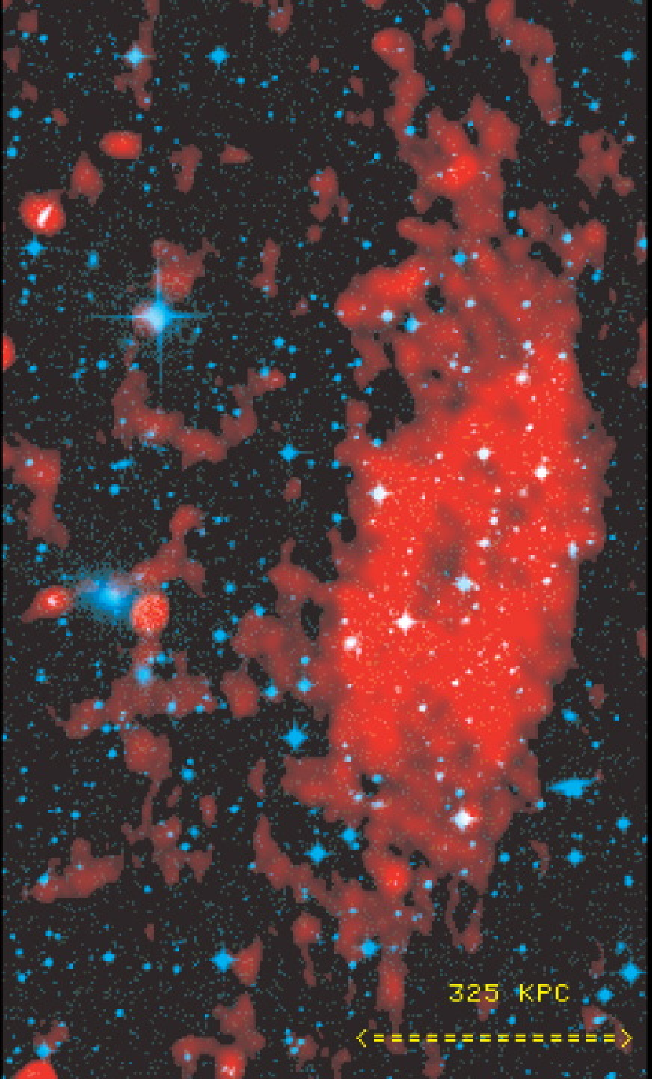}
\caption{Mapas compuestos en frecuencias radio y 'opticas de los remanentes 
este (izquierda) y oeste (derecha) del c'umulo Abell~3376. En rojo se 
muestra la emisi'on en radio y en azul
la radiaci'on detectada en 'optico (Bagchi et al. 2006).}
\label{Bagchi_figs_opt}
\end{center}
\end{figure}

Observaciones en radio llevadas a cabo por Bagchi y colaboradores (2006)
con el arreglo de telescopios VLA  muestran la presencia de dos estructuras 
gigantes en el borde del c'umulo, a una distancia $\sim 1$~Mpc del centro. 
El flujo detectado en radio a $\nu = 1.4$~GHz es
$S_{\nu} = 302$~mJy, que corresponde a una luminosidad de
$2.1\times 10^{40} \,{\rm erg}\,{\rm s}^{-1}$.
Estas estructuras tienen las caracter'isticas t'ipicas de los radio remanentes
(\emph{relics}) (Giovannini \& Ferreti 2004).
Ellas se ajustan bien con una elipse proyectada en el plano del cielo 
de semiejes menor y mayor  $\sim 1.6$ y $\sim 2$~Mpc, respectivamente, como
se muestra en la Figura~\ref{Bagchi_figs_X}. 
Adoptando una profundidad de la l'inea de la visual en la elipse de
$\sim 270$~kpc (Bagchi et al. 2006), el volumen del elipsoide 
tridimensional resulta %$V_{\rm A3376} 
$\sim 0.45\,{\rm Mpc}^3$. 
De la figura mencionada anteriormente inferimos que s'olo el 20~\% de este 
volumen corresponde a los remanentes en radio, es decir, 
$V_{\rm r} \sim 0.09\,{\rm Mpc}^3$. En la Figura~\ref{Bagchi_figs_opt}
se muestra con m'as detalle la estructura y tamaño de cada uno de
los remanentes. 

Algunos par'ametros del c'umulo que son relevantes para los prop'ositos 
de nuestro estudio, como la densidad  del gas en los remanentes,
$n_{\rm r}$, y la velocidad del choque, $v_{\rm ch}$, no son provistos por 
las observaciones y para hallarlos nos valemos de simulaciones num'ericas.
Considerando una simulaci'on hidrodin'amica de {\em N}-cuerpos/SPH 
(\emph{Smoothed Particle Hydrodynamics})
para un c'umulo cuya masa virial es
$M_{\rm vir} \sim 1.3 \times 10^{15} \, h^{-1} \,M_\odot$ ($\sim M_{\rm A3376}$), 
del an'alisis de los resultados de esta simulaci'on obtenemos que
$n_{\rm r} = 2\times 10^{-5}\,{\rm cm}^{-3}$  y  
$v_{\rm ch} = 1000\,{\rm km\;s^{-1}}$.
Teniendo en cuenta que la temperatura del ICM, donde el choque se est'a 
propagando, es $T_{\rm ICM} \sim 0.1 T_{\rm r}$ (Hoeft et al. 2004) el 
n'umero de Mach resulta $M \sim 4.2$ (Gabici \& Blasi 2003), donde
$T_{\rm r}$ es la temperatura del ICM en la regi'on de los remanentes. 
En el resto del cap'itulo, 
expresaremos los resultados num'ericos adoptando $h = 0.7$. En la 
Tabla~\ref{Param_cluster} listamos los valores de los par'ametros (observados 
y estimados) de la fuente Abell 3376.

\begin{table}[h]
\begin{center}
\begin{tabular}{ll}
\hline
\hline
Par'ametro & valor \\
\hline
Corrimiento al rojo & $z =0.046$\\
Distancia & $d = 187$~Mpc \\
Masa & $M_{\rm A3376} \sim 5.2\times10^{14}\,M_{\odot}$\\
Tamaño & $R_{\rm A3376} \sim 2$~Mpc\\
Volumen de los remanentes &$V_{\rm r} \sim 0.09\,{\rm Mpc}^3$\\
Densidad & $n_{\rm r} = 2\times 10^{-5}\,{\rm cm}^{-3}$  \\
Velocidad de los choques &$v_{\rm ch} = 1000\,{\rm km\;s^{-1}}$\\
Luminosidad en X& $L_{\rm X} = 2.48\times 10^{44}\;\rm{erg\;s^{-1}}$ \\
Luminosidad en radio & $2.1\times 10^{40} \,{\rm erg}\,{\rm s}^{-1}$\\
Temperatura del gas & $T_{\rm X}\approx 5.8\times 10^7$~K\\
\hline
\end{tabular}
\caption{Par'ametros del c'umulo de galaxias Abell~3376.
Los valores est'an dados considerando $h = 0.7$.}\label{Param_cluster}
\end{center}
\end{table}

\begin{figure}
\begin{center}
\includegraphics[angle=0, width=0.39\textwidth]
{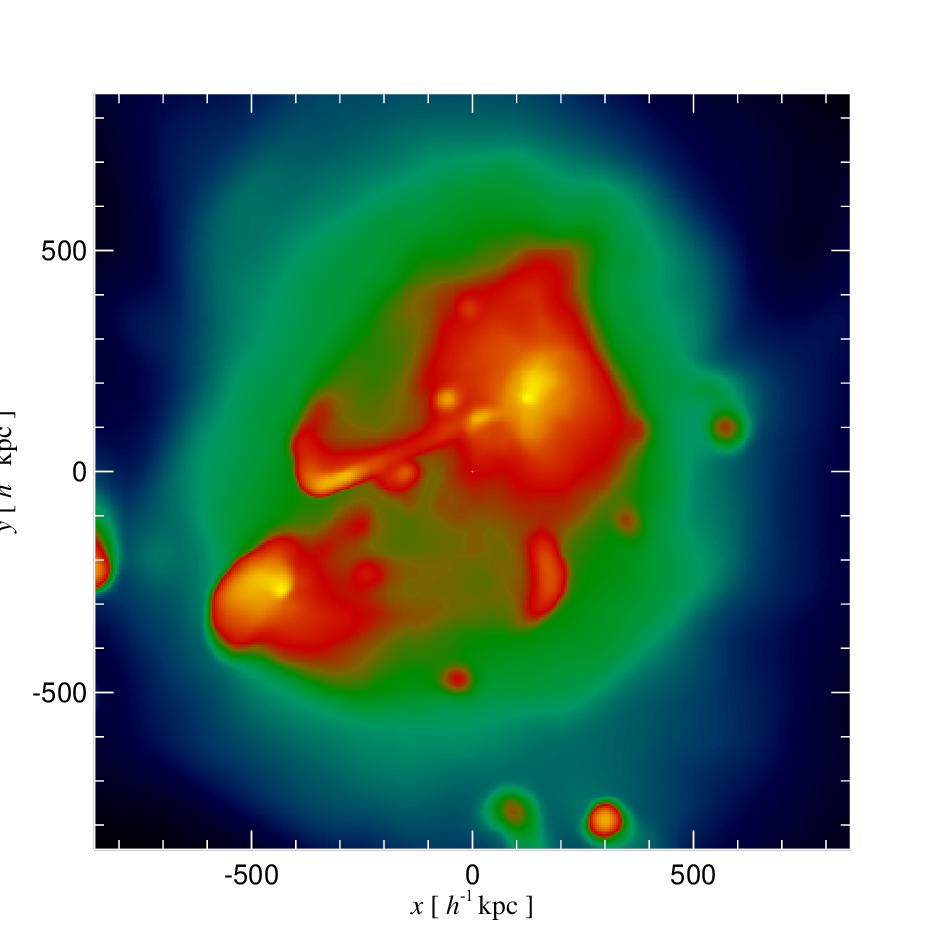}
\includegraphics[angle=0, width=0.09\textwidth]{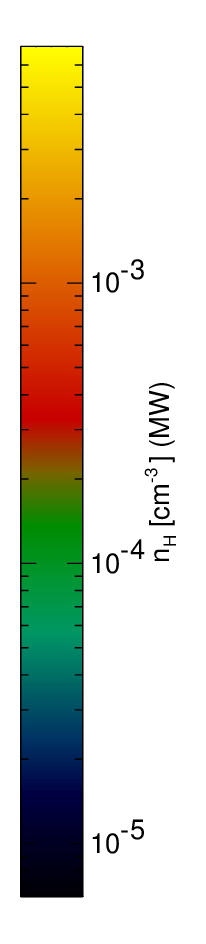}
\includegraphics[angle=0, width=0.39\textwidth]
{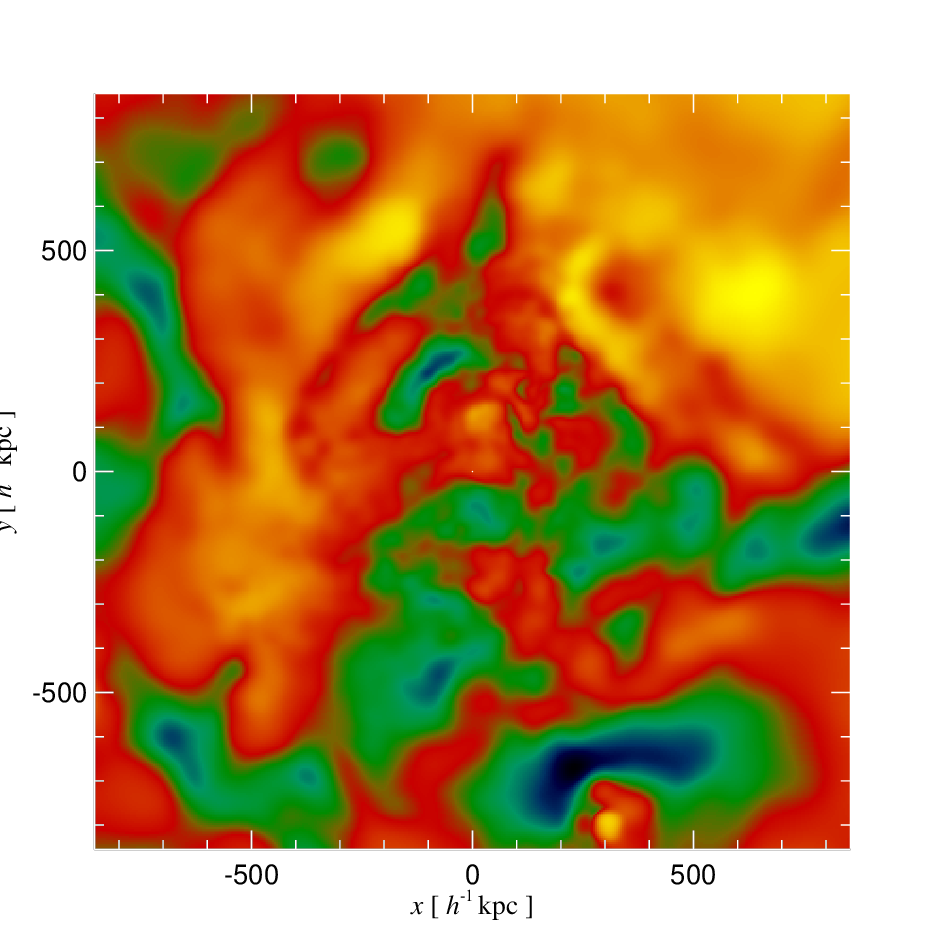}
\includegraphics[angle=0, width=0.09\textwidth]{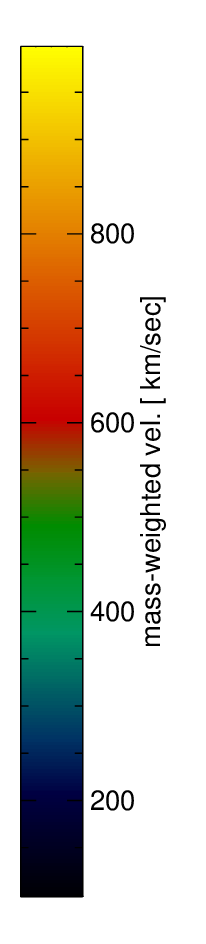}
\caption{Mapas de densidad (izquierda) y de velocidad (derecha) de una 
simulaci'on de un 
c'umulo de galaxias cuya masa virial es 
$\sim 1.3 \times 10^{15} \, h^{-1} \,M_\odot$. Del an'alisis de
estas figuras podemos obtener una estimaci'on de los valores de 
la densidad y de la velocidad de los choques
a una distancia $\sim 1$~Mpc del centro (donde se localizan los remanentes en 
la fuente Abell~3376). 
Hallamos $n_{\rm r} \sim 2\times 10^{-5}\,{\rm cm}^{-3}$ y 
$v_{\rm ch} \sim 1000\,{\rm km\;s^{-1}}$.}  
\label{simulaciones}
\end{center}
\end{figure}

\section{Poblaci'on de part'iculas relativistas}

Las ondas de choque que se generan (por acreci'on y fusi'on) durante la 
formaci'on y evoluci'on de los 
c'umulos de galaxias son la fuente principal de la termalizaci'on del ICM
y de la aceleraci'on de part'iculas (Pfrommer et al. 2006). 
La actividad de las radiogalaxias inmersas 
en los c'umulos tambi'en contribuye a la poblaci'on de part'iculas 
relativistas, inyectando un plasma f'osil que se detecta como cavidades
en los mapas de brillo superficial de rayos~X (Churazov et al. 2000).
En el c'umulo Abell 3376 se han detectado dos radiogalaxias. 
Una de ellas es la fuente MRC 0600-399, asociada con el segundo miembro 
m'as luminoso del c'umulo, y la otra radiofuente es originada posiblemente por
una galaxia el'iptica. 
Estas dos radiogalaxias se encuentran dentro de la regi'on central del 
c'umulo, donde la emisi'on t'ermica es detectada. Por esto, consideramos que 
las estructuras  que se detectan en frecuencias radio en el borde del 
c'umulo no est'an asociadas a estas radiogalaxias (que son fuentes 
puntuales).   

La morfolog'ia de la emisi'on en frecuencias radio y en rayos~X observada 
en el c'umulo
Abell~3376 sugiere que 'este est'a sufriendo una fusi'on con otro c'umulo
de masa similar.
Como muestran los c'alculos  num'ericos  que simulan la fusi'on de dos
c'umulos de galaxias (Hoeft et al. 2004), las ondas de
choque se propagan en ambas direcciones a lo largo de la l'inea
que conecta los centros de ambos c'umulos
con los remanentes casi exclusivamente observados  en la localizaci'on
de los frentes de choque. En los casos en los cuales s'olo se detecta un
'unico remanente, probablemente la fusi'on haya ocurrido entre dos c'umulos de
masas muy diferentes y s'olo el choque que se propaga a trav'es del cumulo m'as 
masivo perdura en el tiempo y se detecta.   
El tiempo de vida de estas estructuras suele estimarse como 
$\tau_{\rm vida} \sim R_{\rm r}/v_{\rm ch}$, donde $R_{\rm r}$ es la distancia 
recorrida por los choques. En el caso de la fuente que estamos estudiando, 
$R_{\rm r} \sim 1$~Mpc y $v_{\rm ch} \sim 1000$~km~s$^{-1}$, con lo cual
$\tau_{\rm vida}$ resulta $\sim 1$~Gyr. Notamos que estos choques no llegan a un
estado estacionario ya que son el producto de un evento puntual en el tiempo:
una fusi'on. 
La energ'ia que llevan los choques de fusi'on se 
disipa a medida que 'estos se propagan por el ICM.

\begin{figure}
\begin{center}
\includegraphics[angle=0, width=0.4\textwidth]{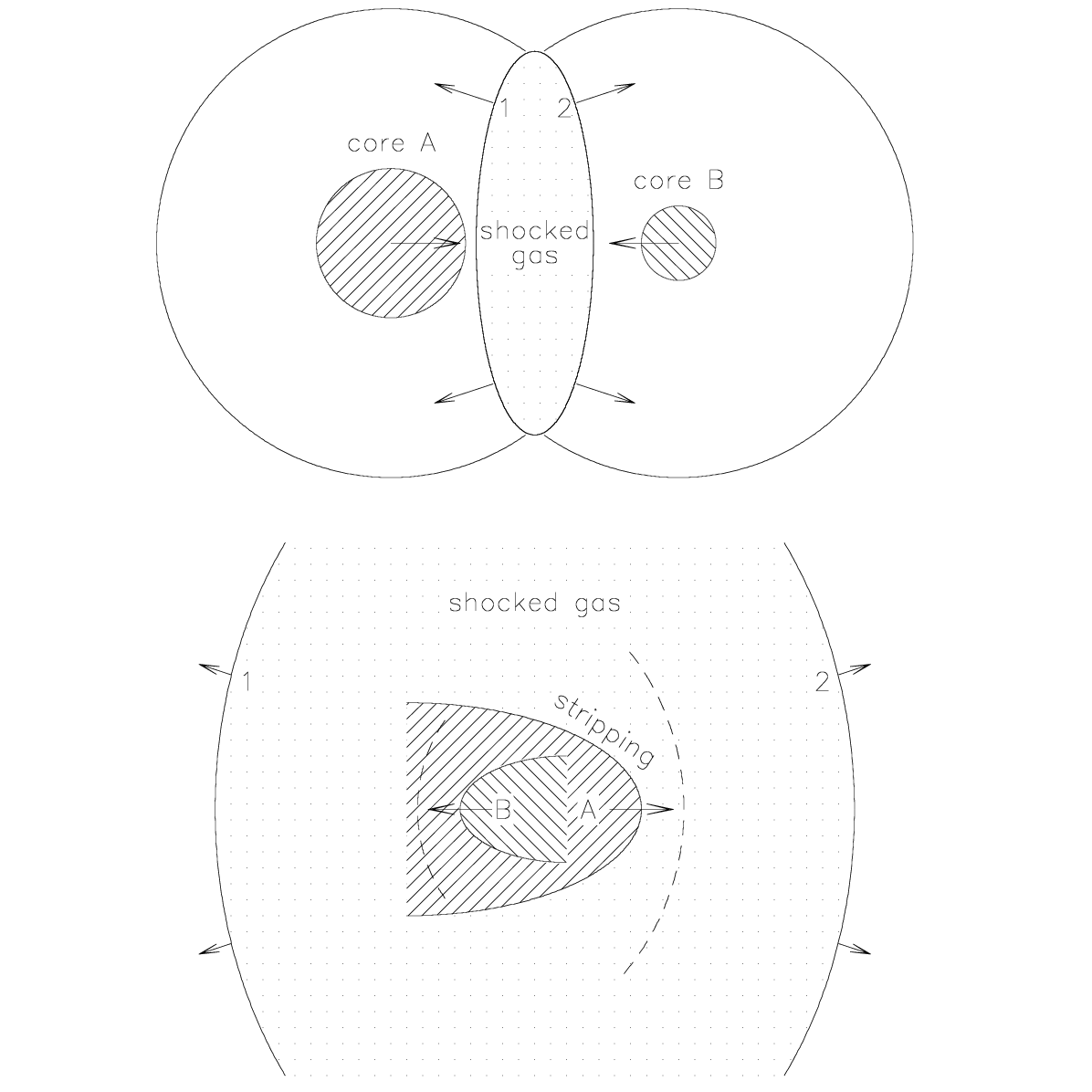}
\caption{Esquema de la fusi'on de dos c'umulos (\emph{cores}). En la figura de 
arriba se muestra el sistema antes de la fusi'on y en la de abajo despu'es.
Como puede observarse en el 'ultimo gr'afico, luego de la colisi'on queda un
n'ucleo compuesto por la fusi'on de ambos c'umulos y dos choques que se 
propagan en direcciones opuestas y hacia afuera del sistema 
(Markevitch et al. 2000).}  
\label{esquema_merger_shocks}
\end{center}
\end{figure}

Teniendo en cuenta las evidencias observacionales y los resultados num'ericos,
suponemos que el contenido de part'iculas relativistas en los remanentes
es debido a la aceleraci'on en los choques de fusi'on, despreciando
la posible contribuci'on de las radiogalaxias y de los choques de acreci'on. 
En las siguientes subsecciones describiremos los procesos de 
aceleraci'on y enfriamiento tanto de electrones (de los cuales tenemos 
evidencia observacional) como de protones, que determinan las 
distribuciones de energ'ia de estas part'iculas y su subsiguiente evoluci'on.

\subsection{Aceleraci'on de part'iculas y p'erdidas radiativas}

La emisi'on difusa en radio producida en la localizaci'on de los
remanentes es evidencia de la presencia de electrones relativistas en esa 
regi'on. 
Por otro lado, aunque no tengamos evidencia observacional de la presencia
de protones en los remanentes, estas part'iculas pueden ser aceleradas 
tambi'en de la misma manera que los electrones.
Para estimar el tiempo en el cual estas part'iculas se aceleran hasta una 
energ'ia dada necesitamos conocer la magnitud del campo magn'etico $B_{\rm r}$
en los remanentes/acelerador.

Para estimar $B_{\rm r}$ suponemos que la energ'ia se reparte de la misma manera
tanto en el campo magn'etico, $u_{\rm B}$, como en las part'iculas 
relativistas, $u_{\rm nt}$: 
\begin{equation}
\label{equipartition_clusters}
\frac{B_{\rm r}^2}{8\pi} = u_{\rm nt} = u_{e_1} + u_{p} + u_{e_2},  
\end{equation}
donde $u_{e_1}$ y $u_{p}$ son las densidades de energ'ia de los electrones
y protones primarios, respectivamente, y  $u_{e_2}$ la densidad de energ'ia de 
los pares secundarios. De la misma manera que en el Cap'itulo~\ref{yso},
para estimar $u_{\rm B}$ a partir de 
(\ref{equipartition_clusters}) necesitamos conocer las distribuciones de 
part'iculas, es decir, las constantes de normalizaci'on $k_i^{\prime}$, para 
$i= e_1, p, e_2$, y los 'indices $p_i^{\prime}$ de las potencias 
($n_i(E_i) = k_i^{\prime}\, E_i^{-p_i^{\prime}}$).
Para esto procedemos de la misma manera que se ha expuesto en  la 
Secci'on~\ref{Accel_part_yso}, fijando $u_p = a\, u_{e_1}$ y
considerando tres valores para el par'ametro libre $a$: 0, 1 y 100;
adem'as de tener en cuenta que $u_{e_2} = f_{e^{\pm}} u_p$. 

Para que la producci'on de pares secundarios sea eficiente, es necesario
un medio con una densidad alta. 
Sin embargo, esta condici'on no se satisface en las regiones perif'ericas 
de los c'umulos, las cuales est'an caracterizadas por densidades muy bajas 
del gas
($\sim  10^{-5}\,{\rm cm}^{-3}$).  Por esto, la contribuci'on de $u_{e_2}$ en
la ecuaci'on~(\ref{equipartition_clusters}) es despreciable y no la
tenemos en cuenta en los c'alculos de $B_{\rm r}$.
Luego, aplicando la restricci'on de que el flujo observado en radio
de la fuente Abell 3376 es $S (\nu =1.4\,{\rm GHz}) = 302$~mJy,
podemos estimar el valor de $B_{\rm r}$ y de las constantes $k_{i}^{\prime}$
asumiendo que  $p_{e_1}^{\prime}= p_p^{\prime} = 2.1$.
Los resultados obtenidos se muestran en la Tabla~\ref{table1_clusters}.
Considerando estos valores obtenemos que $p_{e_2}= 2.08$ y el cociente 
entre las densidades
de energ'ia de los pares $e^{\pm}$ y de los protones es muy chico,
$u_{e_2}/u_p \sim 10^{-5}$,
lo que justifica nuestra suposici'on previa sobre la contribuci'on de 
los pares secundarios al flujo detectado en frecuencias radio.
  
\begin{table}
\begin{center}
\caption{Campo magn'etico, $B_{\rm r}$, y constantes de normalizaci'on 
de la distribuci'on de electrones  primarios, $k_{e_1}^{\prime}$, 
y protones, $k_p^{\prime}$, acelerados en los remanentes para los
tres casos considerados y caracterizados por el par'ametro $a$.}
\begin{tabular}{lccc}
\hline
\hline
$a$ & $B_{\rm r}$ & $k_{e_1}^{\prime}$ & $k_p^{\prime}$ \\
{}  & [G] & [erg$^{p_{e_1} - 1}$ cm$^{-3}$] & [erg$^{p_p - 1}$ cm$^{-3}$]\\
\hline
0    & $9\times 10^{-7}$  &  $1.4\times 10^{-15}$ & -\\
1    & $1.1\times 10^{-6}$  &  $1.1\times 10^{-15}$ & $8.7\times 10^{-16}$\\
100  & $3.4\times 10^{-6}$  &  $3\times 10^{-16}$ & $2.5\times 10^{-14}$\\
\hline
\end{tabular}
\medskip
\label{table1_clusters}
\end{center}
\end{table}

Asumiendo que la velocidad del choque es $v_{\rm ch} = 1000$~km~s$^{-1}$,
el tiempo de aceleraci'on resulta
\begin{equation}
t_{\rm{ac}} \sim  0.5 \,\left(\frac{B_{\rm r}}{1\,\mu{\rm G}}\right)^{-1}\, 
\left(\frac{E_e}{1\,{\rm erg}}\right)\,\rm{s}\,.
\label{eqBremss}
\end{equation}   
Las p'erdidas por escape no son relevantes ya que las escalas espaciales que
deben recorrer las part'iculas para escapar de los remanentes son muy grandes.
Siendo el ancho del remanente $\sim 2 l_{\rm r} \sim 0.6$~Mpc 
(ver la Figura~\ref{Bagchi_figs_opt}) tenemos que el tiempo
de convecci'on a una velocidad $\sim v_{\rm ch}/4$ es muy largo,
\begin{equation}
t_{\rm{conv}} \sim  \frac{2\,l_{\rm r}}{v_{\rm ch}/4} \sim 7\times10^{16}\,{\rm s}.
\label{t_adv_clusters}
\end{equation}   
Para estimar el tiempo de difusi'on consideramos que la escala espacial m'inima
que deben recorrer las part'iculas para difundir y escapar del remanente es
$l_{\rm r} \sim 0.3$~Mpc y hallamos que
\begin{equation}
t_{\rm{dif}} \sim  2\times10^{22}\,\left(\frac{B_{\rm r}}{1\,\mu{\rm G}}\right) 
\left(\frac{E_i}{1\,{\rm erg}}\right)^{-1}\,{\rm s}.
\label{t_dif_clusters}
\end{equation}

\subsubsection{P'erdidas lept'onicas}
\label{perd_lept_clusters}

Las p'erdidas radiativas m'as importantes que sufrir'an los electrones en los
remanentes de fusiones son por radiaci'on sincrotr'on, por Bremsstrahlung 
relativista y por dispersiones IC. Las primeras est'an caracterizadas
por un tiempo de enfriamiento (ver (\ref{t_sin}))
\begin{equation}
t_{\rm sin} \sim 4.1\times10^{14}\,
\left(\frac{B_{\rm r}}{\mu{\rm G}}\right)^{-2}
\left(\frac{E_{e_1, p}}{1\,{\rm erg}}\right)^{-1}\,\rm{s}\,.
\end{equation}

Para las p'erdidas por IC consideramos el campo de fotones provistos por el CMB
como as'i tambi'en los rayos~X  emitidos t'ermicamente por el ICM caliente. 
En el primer caso, la densidad de energ'ia del CMB es
$u_{\rm CMB} = 1.2\times 10^{-13}\,{\rm erg}\,{\rm cm}^{-3}$
al \emph{redshift} $z$ correspondiente al c'umulo Abell 3376.
En el 'ultimo caso
$u_{\rm X}=L_{\rm X}\,(4\,\pi\,R_{\rm X}^2\,c)^{-1}\,(1+z)^4$,
donde  $R_{\rm X}$ es el radio de la regi'on emisora de los fotones X y es
 $\sim 0.15-0.3 R_{\rm vir}$ (Balestra 2007), con lo cual en la fuente 
que estamos estudiando resulta $R_{\rm X} \approx 0.3$ Mpc, dando
$u_{\rm X} \approx 1.3\times 10^{-15}\,{\rm erg}\,{\rm cm}^{-3}$.
Debido a que $u_{\rm X} \ll u_{\rm CMB}$ s'olo consideraremos los fotones del CMB
para estimar las p'erdidas por IC. 
Teniendo en cuenta que la energ'ia caracter'istica de los fotones del CMB es 
$E_{\rm ph}^{\rm CMB} \sim 1.9\times10^{-4}$~eV, 
s'olo aquellos electrones con energ'ias mayores que  $1.6\times10^{15}$~eV
van a interactuar en el r'egimen de KN.
El tiempo de enfriamiento por IC con los fotones del CMB resulta
\begin{equation} 
%\label{t_ci}
t_{\rm IC} =  15.25\,
\frac{(1 + 8.3\,y)}{\ln(1+0.2\;y)}\frac{(1 + 1.3\,y^2)}{(1 + 0.5\,y + 1.3\,y^2)}
~\rm{s}\,,
\end{equation}
donde $y \equiv E_{\rm ph}^{\rm CMB} E_e/(5.1\times10^5\,\rm{eV})^2$.

\begin{figure}
\begin{center}
\includegraphics[angle=270, width=0.6\textwidth]{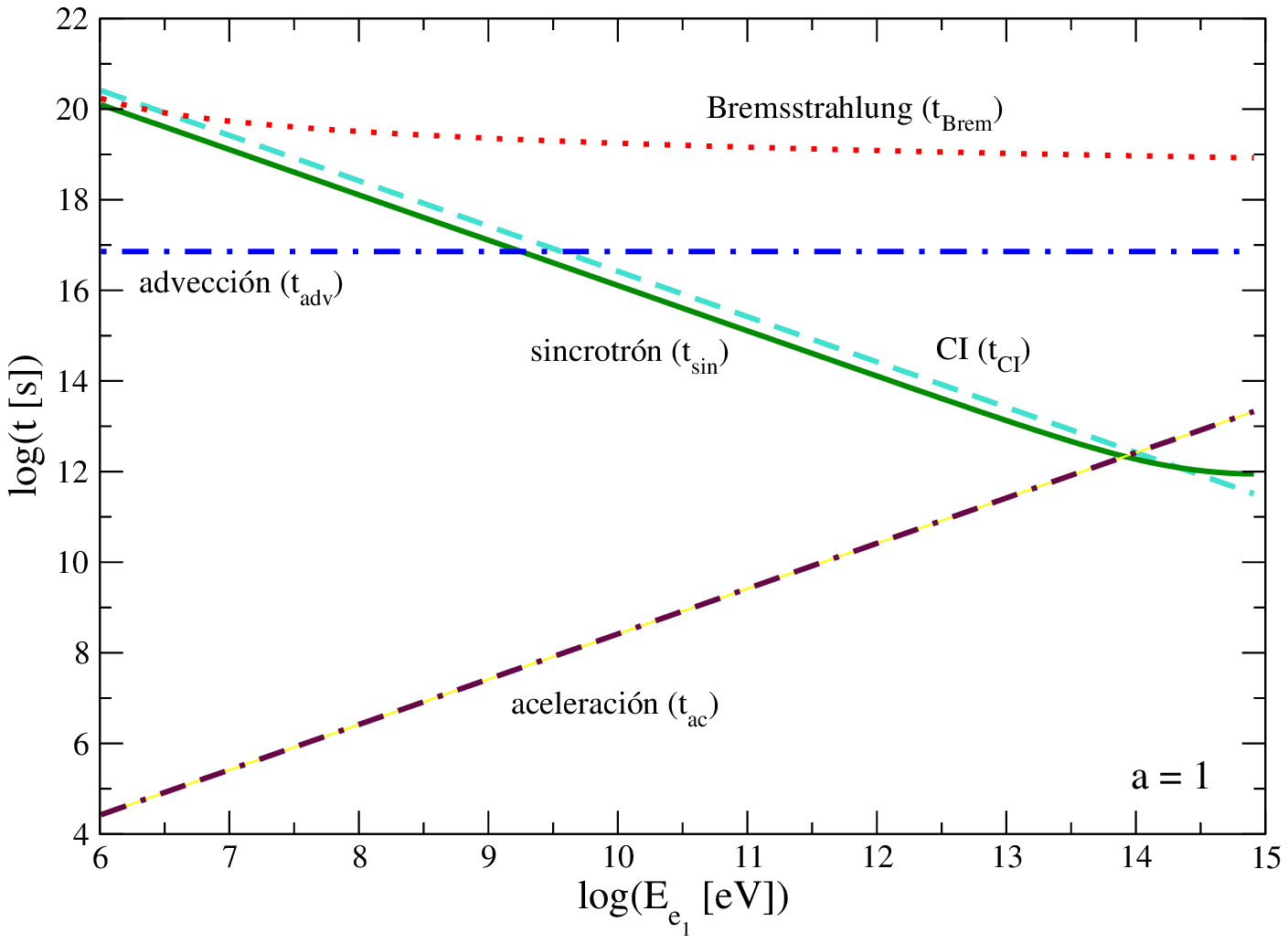}
\caption{Tiempos de aceleraci'on y de enfriamiento 
para los electrones primarios.
Esta figura corresponde al caso con $a =1$. Al resto de los casos
les corresponden gr'aficos similares, ya que la 'unica diferencia est'a en 
el valor de $B_{\rm r}$ (que afecta a $t_{\rm ac}$, $t_{\rm sin}$ y $t_{\rm dif}$)
 y 'este no cambia demasiado ($B_{\rm r} \sim 10^{-6}$~G para los 3 valores de
$a$ considerados).}  
\label{losses_cluster}
\end{center}
\end{figure}

Finalmente, considerando que la densidad del medio en la localizaci'on de 
los remanentes es $n_{\rm r} = 2\times 10^{-5}\,{\rm cm}^{-3}$, el tiempo 
de enfriamiento por Bremsstrahlung relativista es muy grande, 
$t_{\rm{Brem}} \sim  2\times10^{22}$~s, con lo cual este proceso no ser'a 
relevante.
En la Figura~\ref{losses_cluster} se muestran los tiempos de enfriamiento
calculados anteriormente juntamente con los tiempos de aceleraci'on y de escape
(convecci'on) para el caso $a = 1$.
Como puede observarse, la energ'ia m'axima de los electrones primarios queda 
determinada por la igualdad $t_{\rm ac} = t_{\rm IC}$ y da un valor
$E_{e_1}^{\rm max} \sim 9\times10^{13}$~eV para diferentes valores de $a$, como
se muetra en la segunda columna de la Tabla~\ref{Emax_clusters}. 
Para los electrones con energ'ias $\sim E_{e_1}^{\rm max}$, encontramos
que el radio de giro de estas part'iculas es $r_{\rm g}\sim 0.1$~pc 
($\ll l_{\rm r}$), con lo cual est'an contenidas dentro de la
regi'on de aceleraci'on.

\begin{figure}
\begin{center}
\includegraphics[angle=0, width=0.45\textwidth]{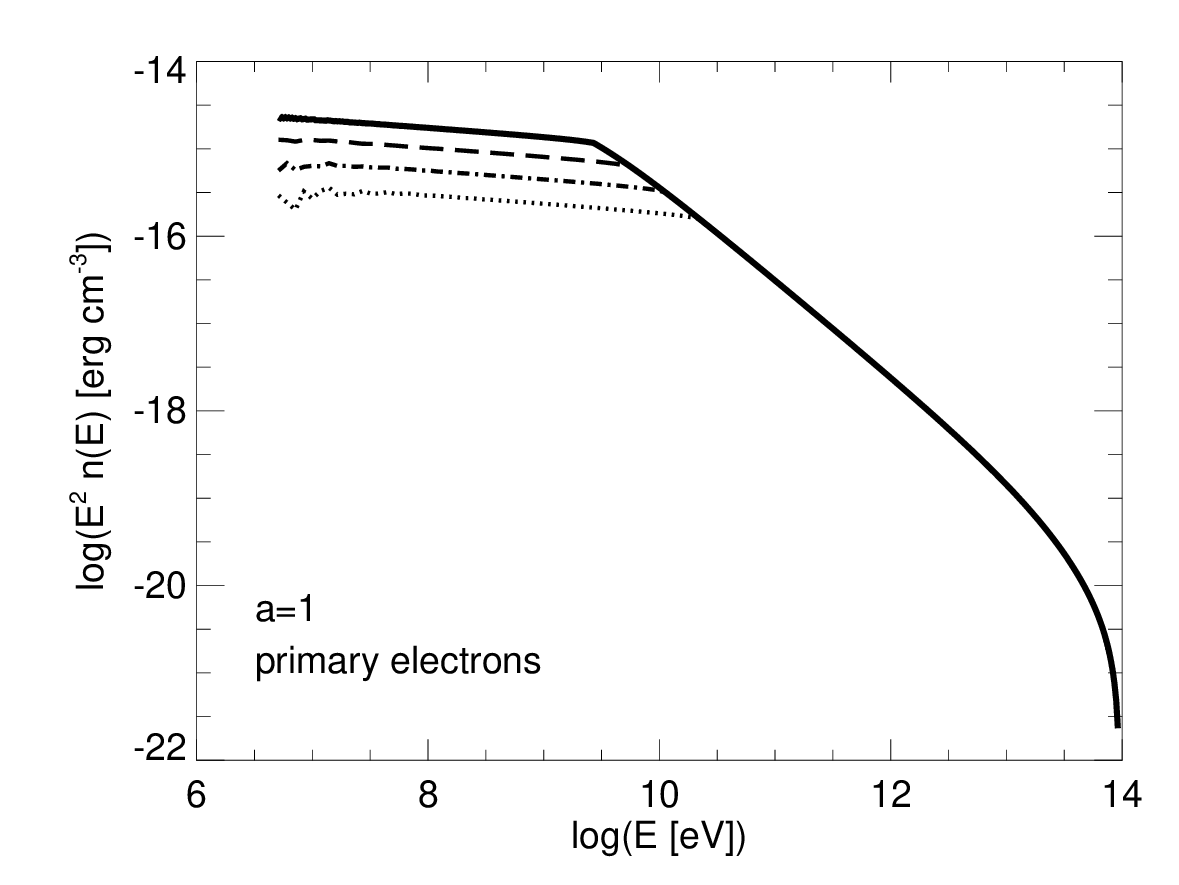}
\includegraphics[angle=0, width=0.45\textwidth]{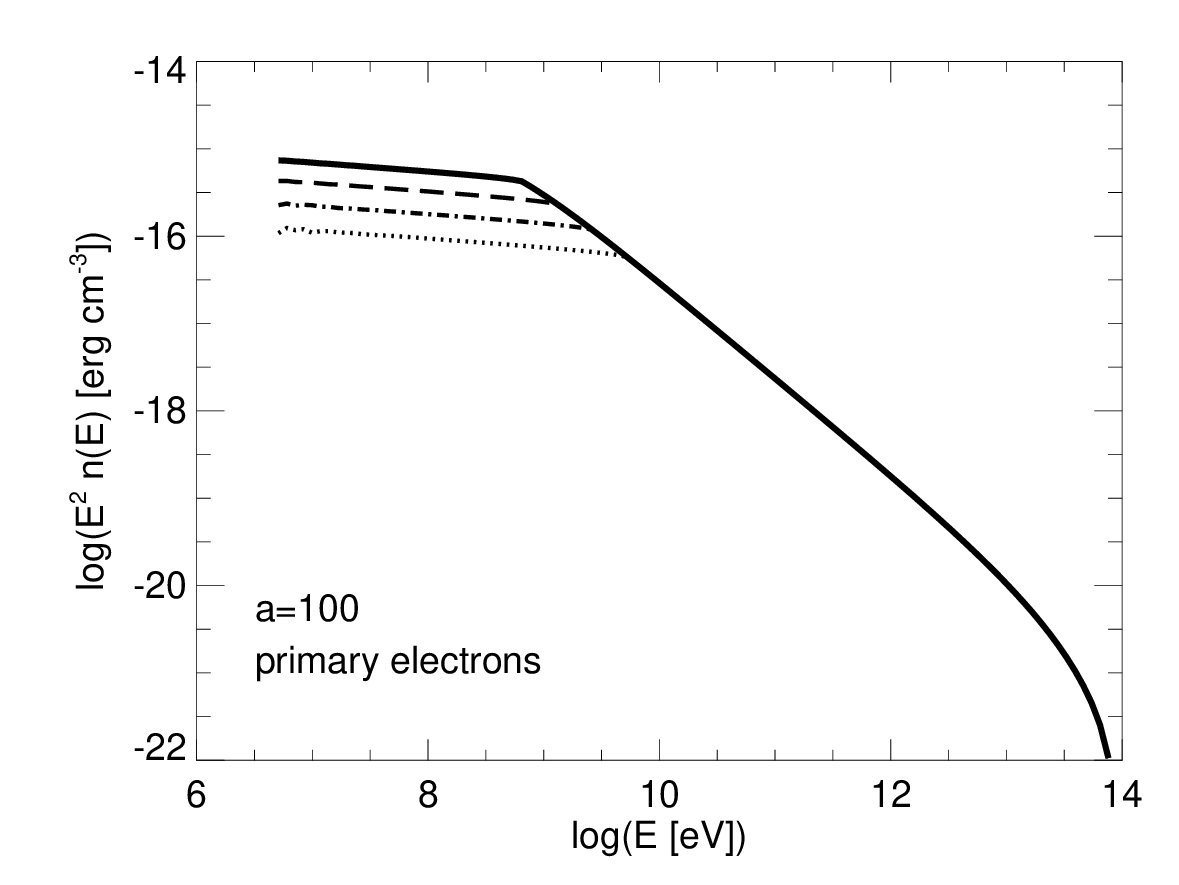}
\caption{Distribuciones de energ'ia ($\times E_{e_1}^2$) de los 
electrones primarios
para los casos $a = 1$ (derecha) y 100 (izquierda). Se muestran los estados 
evolutivos a diferentes tiempos de vida de los remanentes
$t_{\rm r}$: $0.125$~Gyr (l'inea punteada), 
$0.25$~Gyr (l'inea de puntos y rayas), $0.5$~Gyr
(l'inea rayada) y $1$~Gyr (l'inea llena). El estado presente se alcanza
para edades $\sim \tau_{\rm vida}$.}
\label{evol_prim_clusters}
\end{center}
\end{figure}

Para estimar el estado evolutivo de la poblaci'on de electrones relativistas
acelerados e inyectados en los remanentes, consideramos la 
ecuaci'on~(\ref{N_solution}) 
con un tiempo de vida de la fuente 
$\tau_{\rm vida}\sim 1$~Gyr. 
Los resultados se muestran en la Figura~\ref{evol_prim_clusters} para los
casos $a = 1$ y 100 y para diferentes tiempos de inyecci'on $t$. 
Como se muestra en estos gr'aficos, los espectros presentan
un quiebre en $E_{\rm q} \sim 5$ y 1~GeV, para los casos  
$a = 1$ y 100, respectivamente. Este quiebre se debe a que para 
$E_{e_1} < E_{\rm q}$, las p'erdidas de energ'ia por escape son m'as eficientes
 que el enfriamiento radiativo.   
Finalmente, dado que $t_{\rm conv} \sim \tau_{\rm vida}$, podemos considerar 
que el espectro de los electrones relativistas inyectados en los remanentes
alcanza  el estado presente despu'es de un tiempo
$\sim \tau_{\rm vida}$.

\begin{table}[]
\begin{center}
\caption{Las energ'ias m'aximas obtenidas para los electrones primarios, 
$E_{e_1}^{\rm max}$, y protones, $E_p^{\rm max}$, acelerados en los remanentes 
de los choques de fusi'on del c'umulo Abell 3376. 
Las energias m'aximas $E_{e_2}^{\rm max}$ alcanzadas por los pares $e^{\pm}$
son tambi'en presentadas.}\label{Emax_clusters}
\begin{tabular}{cccc}
\hline
\hline
a  & $E_{e_1}^{\rm{max}}$ & $E_{p}^{\rm{max}}$ & $E_{e_2}^{\rm{max}}$ \\
{} & [eV]                 & [eV]             & [eV]                 \\
\hline
$0$   & $9\times10^{13}$ & -                  & -                   \\
$1$   & $9.3\times10^{13}$ & $5.0\times10^{17}$ & $4\times10^{16}$ \\ 
$100$ & $8.8\times10^{13}$ & $1.3\times10^{18}$ & $10^{17}$ \\
\hline
\end{tabular}
\label{table2}
\end{center}
\end{table}

\subsubsection{P'erdidas  hadr'onicas}

El tiempo de enfriamiento debido a las colisiones inel'asticas $pp$
resulta muy largo,
$t_{pp} \sim 10^{20}$~s, debido a las bajas densidades del medio en los
bordes del c'umulo. Por esto, la producci'on de pares $e^{\pm}$ no ser'a
muy eficiente en los remanentes.
Considerando las p'erdidas por $pp$ la energ'ia m'axima que se obtiene 
es $\sim 2\times10^{21}$ y $5\times10^{21}$~eV para los casos
con $a = 1$ y 100, respectivamente.
Sin embargo, para energ'ias $E_p > 5\times10^{18}$~eV la
producci'on de pares por la interacci'on prot'on-fot'on del CMB es relevante,
y para energ'ias m'as all'a de $\sim 5\times10^{19}$~eV las p'erdidas por 
producci'on de piones son las dominantes (Berezinsky \& Grigorieva, 1988; 
Kelner et al. 2008).
Sin embargo, no es necesario tener en cuenta estos procesos ya que la energ'ia
m'axima de los protones queda determinada por restricciones adicionales:
el tamaño del acelerador y el tiempo de vida de la fuente.

Los valores de las energ'ias m'aximas hallados tanto para los electrones como
para los protones son v'alidos siempre y cuando 'estos permanezcan
dentro de la regi'on de aceleraci'on cuyo tamaño definimos como $l_{\rm a}$.
Por esto, las part'iculas deben satisfacer el requerimiento  
$r_{\rm g} < l_{\rm a}$. Asumiendo que las part'iculas se aceleran en los choques
trazados por los remanentes observados en frecuencias radio, adoptamos 
para el tamaño del acelerador  
$l_{\rm a} \simeq l_{\rm r}\sim 0.3$~Mpc.
%; donde este valor es obtenido de 
%una examinaci'on somera del mapa en radio del c'umulo proyectado en el plano
%del cielo.
Debido a que el valor de la energ'ia m'axima de los protones obtenido en el 
p'arrafo anterior a partir de las p'erdidas radiativas
corresponde a $r_{\rm g} > l_{\rm r}$, determinamos la energ'ia m'axima
imponiendo que  $r_{\rm g} = l_{\rm r}$, con lo cual obtenemos 
$E_p^{\rm{max}} \sim 10^{20}$  y $2.6\times 10^{20}$~eV, para $a=1$ y $100$, 
respectivamente. 
Sin embargo, el tiempo requerido para que los protones alcancen esta energ'ia
mediante el mecanismo de aceleraci'on de Fermi es mayor que el tiempo de vida 
de los remanentes, $\tau_{\rm vida} \sim 1$~Gyr. As'i, la verdadera 
energ'ia m'axima hasta la cual pueden acelerarse los protones en los 
remanentes  se obtiene
igualando $t_{\rm acc}$ y $\tau_{\rm vida}$, dando 
$E_p^{\rm{max}}\sim 5\times 10^{17}$ y $1.3\times 10^{18}$~eV para
$a=1$ y $100$, respectivamente, como se muestra en la Tabla~\ref{table2}.

La distribuci'on de energ'ia de los protones acelerados  en los remanentes 
queda determinada por el tiempo de vida de los 'ultimos 
$\tau_{\rm vida} \sim 1$~Gyr y debido a que
$\tau_{\rm vida} \sim t_{\rm conv}$ podemos considerar que $n_p(E_p)$ 
alcanza el estado 
presente despu'es de un tiempo $\sim \tau_{\rm vida}$. Por esto, 
\begin{equation}
n_p(E_p) \sim q_p(E_p) \,\tau_{\rm vida}.
\end{equation}

\begin{figure}
\begin{center}
\includegraphics[angle=0, width=0.45\textwidth]{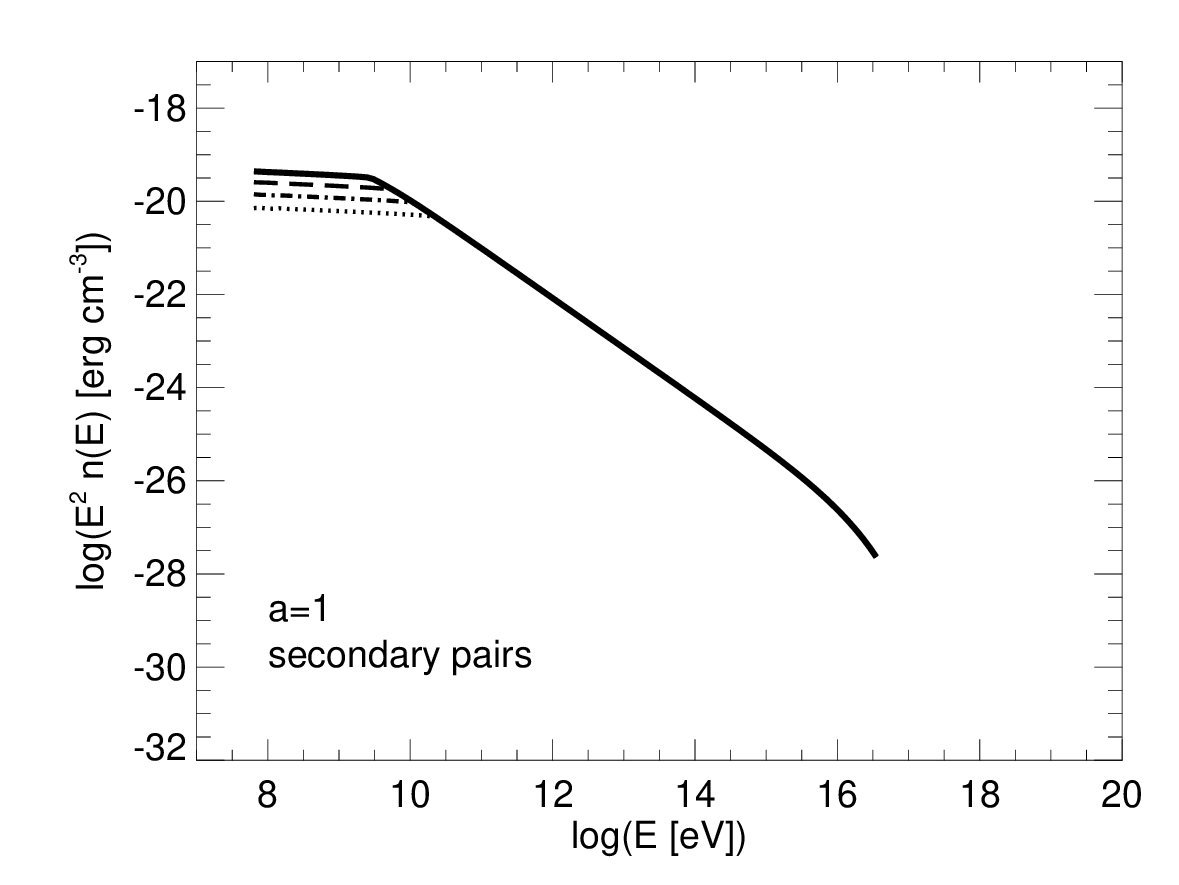}
\includegraphics[angle=0, width=0.45\textwidth]{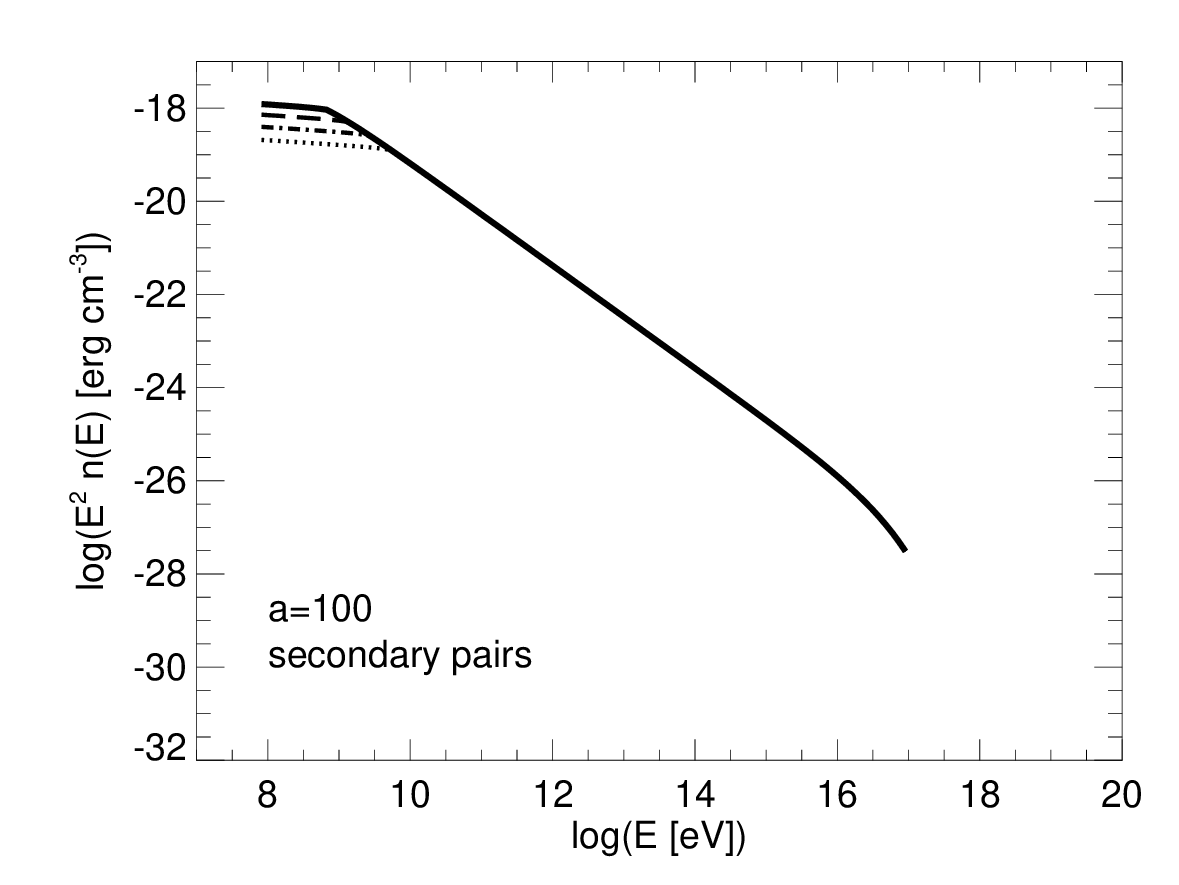}
\caption{Distribuciones de energ'ia ($\times E_{e_2}^2$) de los pares 
secundarios
para los casos $a = 1$ (derecha) y 100 (izquierda). Se muestran los estados 
evolutivos a diferentes tiempos de vida de los remanentes
$t_{\rm r}$: $0.125$~Gyr (l'inea punteada), 
$0.25$~Gyr (l'inea de puntos y rayas), $0.5$~Gyr
(l'inea rayada) y $1$~Gyr (l'inea llena). El tiempo m'aximo de evoluci'on es
$\sim \tau_{\rm vida}$.}  
\label{evol_sec_clusters}
\end{center}
\end{figure}

Respecto de la producci'on de pares $e^{\pm}$, si bien la contribuci'on
de 'estos a la SED ser'a despreciable, la calcularemos por completitud. 
Para esto necesitamos conocer como es la distribuci'on energ'etica de estas
part'iculas. 
La evoluci'on del  espectro de inyecci'on de los leptones secundarios 
en los remanentes queda determinada, al igual que en el caso
de los electrones primarios, por las p'erdidas radiativas (IC y sincrotr'on) 
para $E_{e_2} > E_{\rm q}$ y por el tiempo de escape para $E_{e_2} < E_{\rm q}$,
donde $E_{\rm q} \sim 5$ y 1~GeV, para $a = 1$ y 100, respectivamente.
El espectro alcanza el estado de inter'es, como se muestra en la 
Figura~\ref{evol_sec_clusters}, en un tiempo $\sim \tau_{\rm vida}$.

\section{Distribuciones espectrales de energ'ia}

Conociendo las distribuciones de energ'ia $n_i$,
podemos calcular la emisi'on que producen las poblaciones de
part'iculas relativistas. 
Las  luminosidades producidas en los remanentes est'an dadas por la 
ecuaci'on
\begin{equation}
E_{\rm ph}L_{\rm ph} = E_{\rm ph}^{2}\; j_{\rm ph}(E_{\rm ph})\; 
V_{\rm{r}},
\label{luminosity_clusters}
\end{equation}
donde $V_{\rm{r}} \sim 0.1\,\rm{Mpc^3}$ es el volumen de la regi'on emisora,
es decir, de los remanentes.

\subsection{Interacciones lept'onicas}

Las emisividades diferenciales $j_{\rm ph}(E_{\rm ph})$ producidas por
los leptones por radiaci'on sincrotr'on, dispersiones IC y Bremsstrahlung
relativista son calculadas usando las f'ormulas dadas en el 
Cap'itulo~\ref{proc-rad}. Las luminosidades espec'ificas son halladas 
a trav'es de
la ecuaci'on~(\ref{luminosity_clusters}).
En los casos $a=1$ y 100, en los cuales se asume que existe una poblaci'on 
de protones relativistas, las contribuciones de los pares secundarios a las 
SEDs tambi'en se estiman por completitud.

\begin{figure}
\begin{center}
\includegraphics[angle=270, width=0.5\textwidth]{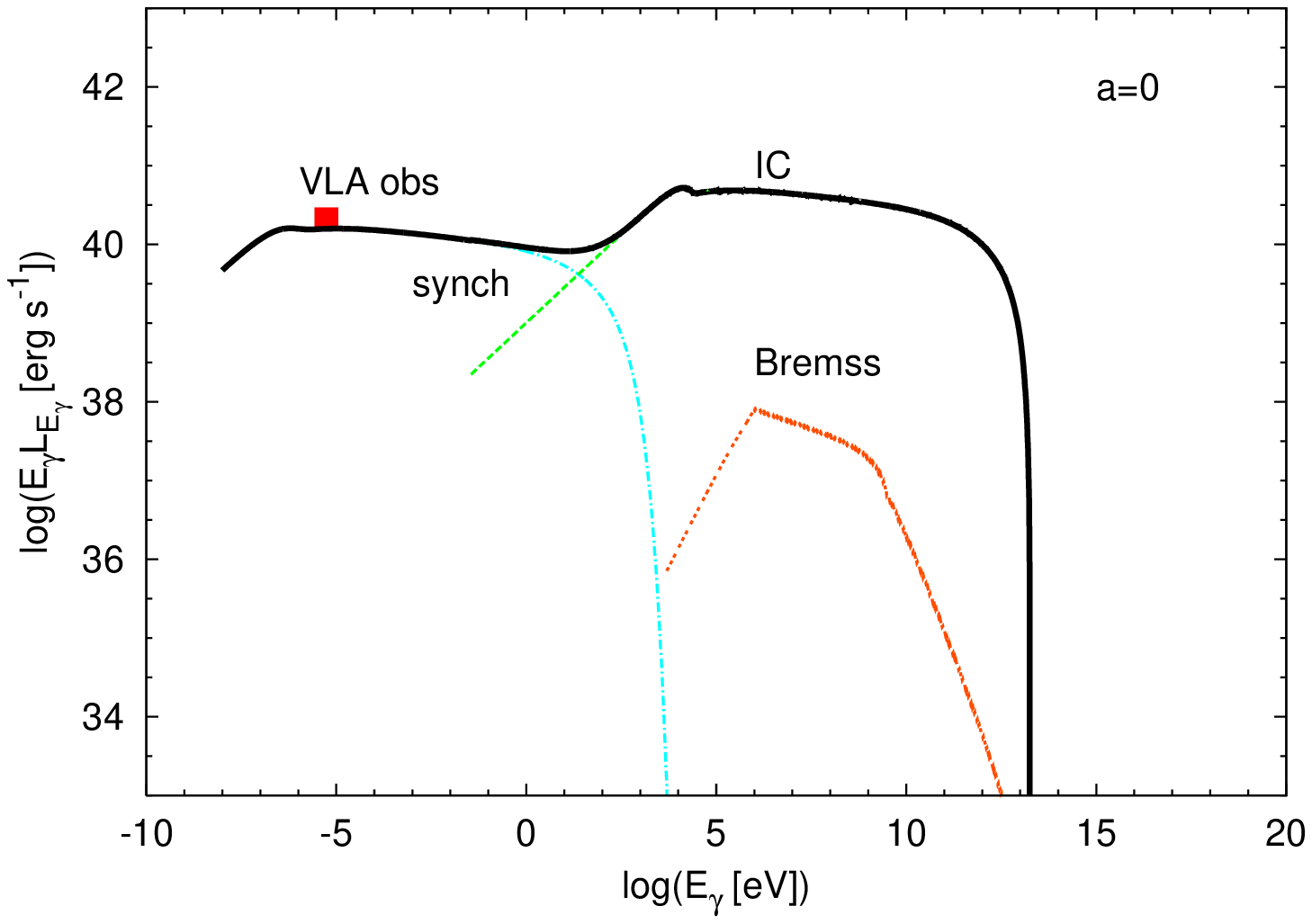}
\caption{Distribuci'on espectral de energ'ia para el caso puramente 
lept'onico, $a = 0$.
La l'inea negra representa la suma de las tres contribuciones (sincrotr'on,
IC y Bremsstrahlung relativista) a la SED total. 
La luminosidad observada con VLA a la frecuencia $\nu = 1.4$~GHz 
 del c'umulo Abel~3376 se representa con un cuadrado rojo.}
\label{SED0_clusters}
\end{center}
\end{figure}

Los resultados de las luminosidades espec'ificas producidas en el caso 
donde s'olo se aceleran electrones, caracterizado por  el par'ametro $a = 0$,
se muestran en la Figura~\ref{SED0_clusters}. Podemos ver que
las interacciones IC son el proceso radiativo m'as importante, con
una luminosidad 
$L_{\rm{IC}} \sim  9.1\times10^{41}\,{\rm erg}\,{\rm s}^{-1}$, a 
energ'ias $E_{\rm ph} \geq 0.1$~MeV y con un \emph{cut-off} en $\sim 10$~TeV.
La luminosidad producida por Bremsstrahlung relativista es despreciable,
en concordancia con los resultados mostrados en la Figura~\ref{losses_cluster}.

\subsection{Interacciones hadr'onicas}

\begin{figure}
\begin{center}
\includegraphics[angle=270, width=0.45\textwidth]{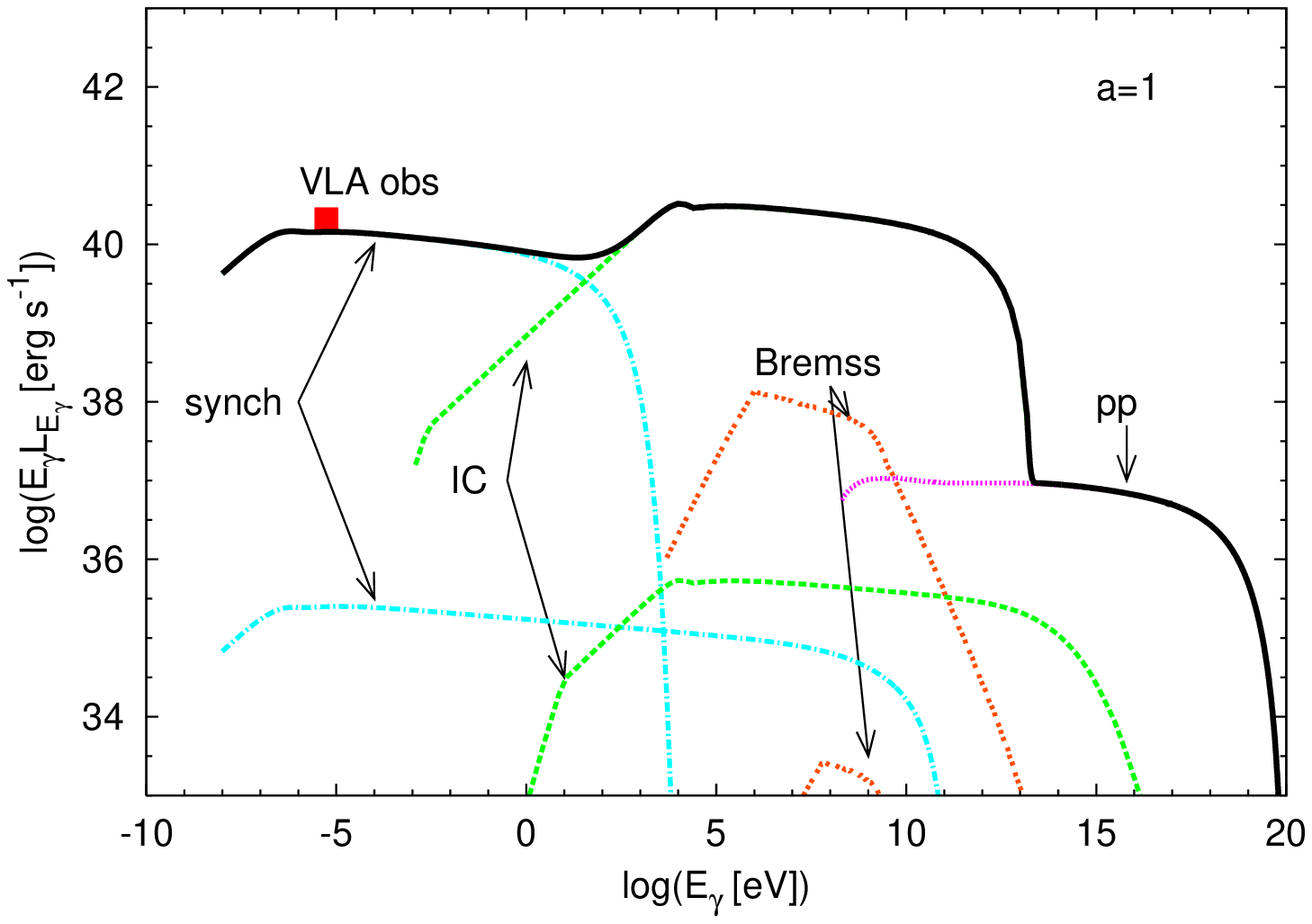}
\includegraphics[angle=270, width=0.45\textwidth]{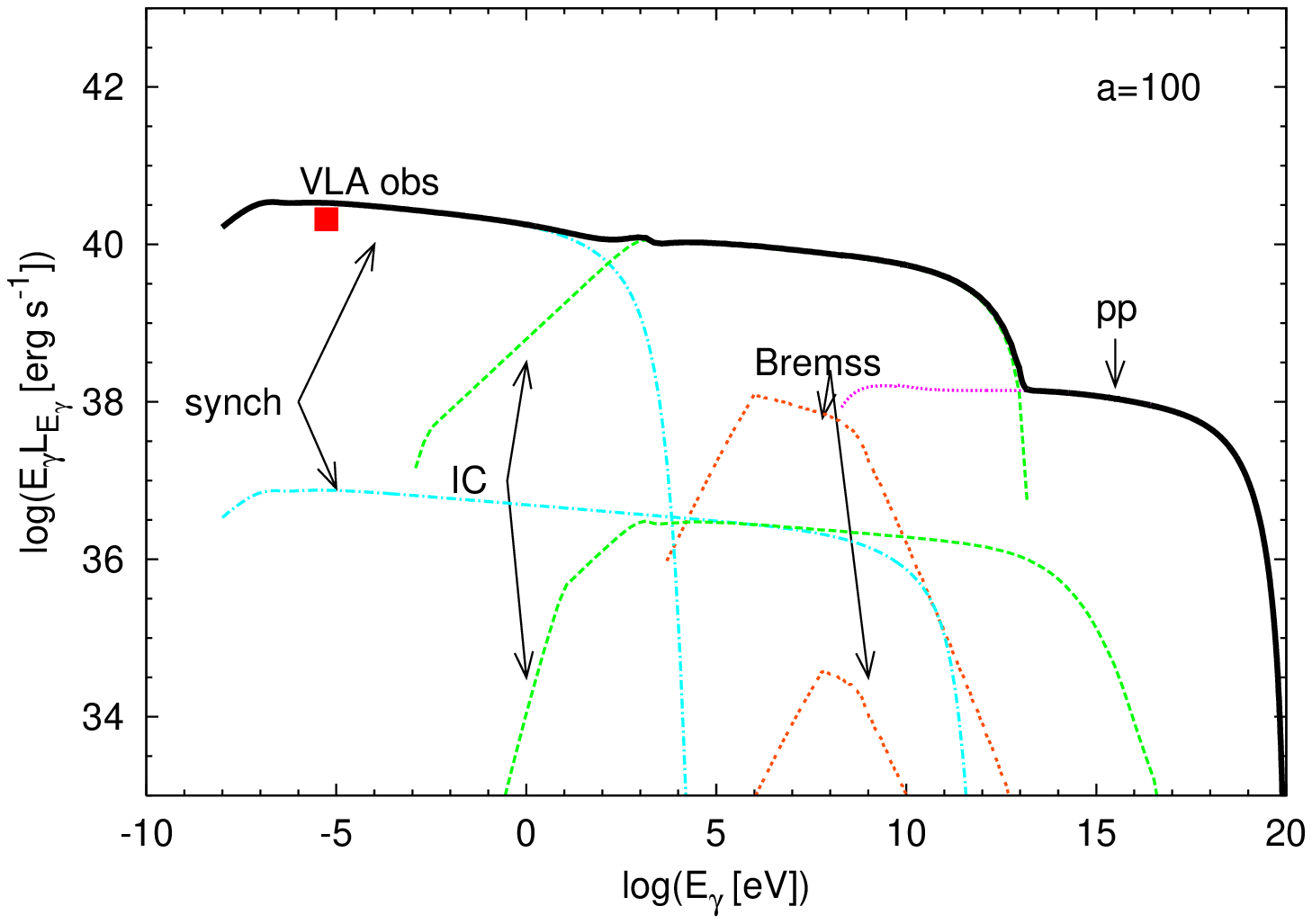}
\caption{Distribuciones espectrales de energ'ia para los casos con 
$a = 1$ (izquierda) y 100 (derecha), 
en los cuales consideramos 
tanto una poblaci'on lept'onica como hadr'onica de part'iculas primarias
y la poblaci'on de pares $e^{\pm}$ producida. 
La emisi'on producida por los pares secundarios es menor que la 
correspondiente a los electrones primarios.
Como mencionamos en la Figura~\ref{SED0_clusters}, 
la l'inea negra indica la SED total (suma de todas las contribuciones, 
lept'onicas y hadr'onica) y el
cuadrado rojo indica la luminosidad observada con VLA.} 
\label{SED2_cluster}
\end{center}
\end{figure}

La emisividad correspondiente a las interacciones $pp$ se calcula considerando 
las ecuaciones~(\ref{q_pi}) y (\ref{q_pp}),
y luego la luminosidad espec'ifica se calcula a trav'es de la
ecuaci'on~(\ref{luminosity_clusters}).
Los resultados de nuestros c'alculos 
se muestran en la Figura~\ref{SED2_cluster}. 

En el caso $a =1$, de la misma manera que como ocurre en el caso $a =0$,
la SED est'a dominada por las interacciones IC, con una luminosidad
$L_{\rm IC} \sim 7.4\times 10^{41}$~erg~s$^{-1}$. La emisi'on
a energ'ias mayores que $\sim 1$~GeV es producida por el decaimiento de
piones neutros, alcanzando una luminosidad
$L_{pp} \sim 1.6\times 10^{38}$~erg~s$^{-1}$ con un \emph{cut-off} en
$E_{\rm ph} \sim 10^{17}$~eV. Sin embargo, su contribuci'on a la SED
es evidente para energ'ias $> 10$~TeV, como se muestra en la 
Figura~\ref{SED2_cluster} (izquierda).

En el caso correspondiente al valor $a = 100$, tenemos que $u_p > u_{e_1}$, pero
esto no implica un flujo significativo de fotones
producidos en las interacciones $pp$ ya que la densidad de part'iculas blanco
es constante y muy baja en la localizaci'on de los remanentes.
Por esto, el espectro de rayos gamma producidos por el decaimiento de
los $\pi^0$ no domina la SED, como puede observarse en la
Figura~\ref{SED2_cluster} (derecha).
La luminosidad $L_{pp}$ es $\sim 4.2\times10^{39} \, {\rm erg}\,{\rm s}^{-1}$, 
siendo apenas mayor que la correspondiente al caso con $a=1$. 
Por otro lado, contrariamente a lo que ocurre en los casos con
$a=0$ y $a=1$, la emisi'on por IC es menor que la radiaci'on sincrotr'on,
con luminosidades
$L_{\rm IC} \sim 7.1\times10^{40}\,{\rm erg}\,{\rm s}^{-1}$ y 
$L_{\rm sin} \sim 3.8\times10^{41}\,{\rm erg}\,{\rm s}^{-1}$, respectivamente.
El hecho de que la radiaci'on por dispersiones IC sea reducida en el caso
con  $a=100$ puede explicarse como una consecuencia de la equipartici'on
asumida entre la densidad de energ'ia magn'etica y de las part'iculas
relativistas. Como puede verse en la 
Tabla~\ref{table1_clusters}, el campo magn'etico se hace m'as intenso
cuando $a$ crece. Esto reduce la cantidad de energ'ia de los electrones
que se rad'ia por interacciones IC en comparaci'on con la que se pierde a 
trav'es del proceso sincrotr'on.
Adem'as, la densidad de energ'ia en electrones relativistas tambi'en se reduce 
para poder ajustar el espectro  con el flujo observado con un valor 
de $B_{\rm r}$ mayor. Esto 'ultimo hace que $L_{\rm IC}$ sea menor 
en t'erminos absolutos en el caso $a = 100$.

Respecto de la emisi'on producida por los pares $e^{\pm}$, encontramos
que sus contribuciones a la SED por los diferentes procesos expuestos en la
secci'on~\ref{perd_lept_clusters} son mayores en el caso con $a=100$ que con 
$a =1$, pero a'un son
irrelevantes. Esto es consistente con la baja eficiencia de las 
interacciones $pp$ que tienen lugar en la periferia del c'umulo.

\section{Discusi'on}

En este cap'itulo hemos presentado el estudio realizado sobre el c'umulo de
galaxias Abell~3376. Esta fuente presenta fuertes evidencias de una actividad 
de fusi'on, caracterizada por grandes estructuras detectadas en frecuencias 
radio en la periferia del c'umulo y que 
han sido asociadas con remanentes (Bagchi et al. 2006).
Hemos modelado la emisi'on producida en estos remanentes en un 
amplio rango de frecuencias, desde radio hasta rayos gamma, considerando
la contribuci'on de diferentes procesos radiativos y diferentes relaciones entre
las densidades de energ'ia de los protones y los electrones relativistas.

Las p'erdidas lept'onicas m'as importantes que afectan la distribuci'on
de part'iculas son la radiaci'on sincrotr'on y las dispersiones IC con 
fotones del CMB. Los protones relativistas se enfr'ian por interacciones
$pp$ con el ICM, produciendo rayos gamma e inyectando
una poblaci'on de pares $e^{\pm}$ los cuales se enfr'ian por los mismos procesos
que los electrones primarios. 
Los par'ametros involucrados en estos procesos no t'ermicos, como el campo 
magn'etico y las distribuciones de part'iculas relativistas, se han estimado
a trav'es del flujo observado en radio de la fuente y de suponer 
equipartici'on de la energ'ia magn'etica y no t'ermica. Adem'as, las 
observaciones 
en rayos~X nos permiten extraer informaci'on sobre
el volumen de los remanentes. Por otro lado, los resultados de simulaciones 
num'ericas nos proveen de valores t'ipicos para la densidad del medio en los
bordes del c'umulo, como as'i tambi'en de la velocidad de los choques de 
fusi'on.

Las SEDs mostradas en las Figuras~\ref{SED0_clusters} y \ref{SED2_cluster} 
se han calculado considerando la contribuci'on de diferentes procesos
radiativos no t'ermicos en un amplio rango de energ'ias, desde radio hasta
rayos gamma. En este cap'itulo hemos estudiado particularmente
la detectabilidad del c'umulo Abell~3376 en HE y VHE.
Encontramos que los procesos radiativos m'as importantes que contribuyen a la
emisi'on de rayos gamma son las interacciones IC y el decaimiento de $\pi^0$,
siendo la luminosidad de este 'ultimo mucho menor que la alcanzada por el
primero, a'un en el caso en el cual $u_p = 100\,u_{e_1}$. La luminosidad
por $pp$ obtenida en este caso es $L_{pp} \sim 4\times 10^{39}$~erg~s$^{-1}$. 
La luminosidad por dispersiones IC de
fotones del CMB es $L_{\rm IC} \sim 7.4\times 10^{41}$~erg~s$^{-1}$.  
As'i, los rayos gamma emitidos en el c'umulo  Abell~3376, y bajo las 
suposiciones de nuestro modelo, podr'ian ser detectables con \emph{Fermi} y
HESS, que operan en las bandas (aproximadas) 100~MeV-100~GeV y 
0.1-10~TeV, respectivamente. El futuro telescopio CTA 
podr'ia permitir incluso
la detecci'on de la componente hadr'onica a m'as altas energ'ias.

Finalmente, la proximidad del c'umulo  Abell~3376 y su posible 
contenido alto de
part'iculas relativistas hacen  esta fuente interesante para ser observada
tanto con instrumentos que actualmente est'an funcionando
 como as'i tambi'en con futuros telescopios Cherenckov ya planeados.
Notamos que hasta el d'ia de hoy no se han detectado c'umulos de galaxias 
en rayos gamma y que la fuente Abell~3376 es uno de los mejores candidatos 
para ser observado en esta banda de energ'ia.

%% file: Conclusiones_final.tex
\chapter{Conclusiones y perspectivas}

A lo largo de esta tesis hemos estudiado los procesos no t'ermicos que tienen 
lugar en diferentes fuentes astrof'isicas, tanto 
gal'acticas (YSOs, MQs) como extragal'acticas (AGNs, c'umulos de galaxias).
A continuaci'on resumimos las investigaciones realizadas en cada tipo de fuente
y luego damos las conclusiones generales de la tesis.

\begin{itemize}

\item YSOs: Algunas  estrellas masivas en formaci'on presentan \emph{jets}
y cuando 'estos son frenados por el material de la nube
  molecular en la cual se hallan embebidos, 
se producen choques terminales fuertes  que pueden
  acelerar part'iculas hasta energ'ias relativistas.  Estas part'iculas
  emiten en frecuencias radio por mecanismo sincrotr'on y es as'i como
  se detectan l'obulos no-t'ermicos en las regiones terminales de los
  \emph{jets} que emanan de protoestrellas de gran masa.

Hemos modelado uno de los l'obulos que forman parte el sistema triple en 
radio asociado a la fuente
IRAS~16547-4247, la protoestrella m'as luminosa
detectada a la fecha ($L_{\star} \sim 6.2\times10^4 L_{\odot}$). 
Suponiendo equipartici'on entre las densidades de energ'ia magn'etica y de
las part'iculas relativistas y valiendon'os de los flujos observados
en frecuencias radio hemos estimado el campo magn'etico y las distribuciones
espectrales de energ'ia para diferentes relaciones entre 
las densidades de energ'ia de los protones y electrones acelerados en el
l'obulo. Luego, calculamos las SEDs
considerando diferentes procesos radiativos no t'ermicos.

Los procesos dominantes resultan ser Bremsstrahlung
relativista y las interacciones $pp$, ya que la densidad de la nube
molecular es muy
grande, $\sim 10^5$~cm$^{-3}$. Los niveles de emisi'on predichos son 
$\sim 10^{32}-10^{33}$~erg~s$^{-1}$ en altas y muy altas energ'ias, con lo 
cual la fuente ser'ia
detectable con el sat'elite \emph{Fermi} y con el futuro arreglo de 
telescopios Cherenkov CTA.

Las protoestrellas de gran masa, de ser detectadas en altas energ'ias, ser'ian
un nuevo tipo de fuentes de rayos gamma.
De esta manera, se
abre una nueva ventana del espectro electromagn'etico a trav'es de la cual se 
puede obtener informaci'on sobre como es el proceso de formaci'on de las 
estrellas tempranas. Este proceso a'un hoy no est'a claramente
establecido. Espec'ificamente, se podr'an estudiar con nuevas herramientas
las propiedades f'isicas (como la densidad y campo magn'etico) de los 
\emph{jets}, as'i como las del medio cincundante.

\item MQs: Existen evidencias observacionales de que los vientos de
  las estrellas de gran masa no son homog'eneos sino que tienen una
  estructura porosa.  En los HMMQs, algunos \emph{clumps} del viento de la
  estrella  compañera pueden llegar hasta los \emph{jets} generados por
el objeto compacto.
   Debido a la  interacci'on del \emph{clump} con el material del 
\emph{jet} se producen dos choques: uno en el \emph{jet} y otro en el
  \emph{clump}. En el primero se pueden acelerar part'iculas que luego
  radiar'an localmente por diferentes procesos no t'ermicos, y el segundo
  calienta el material del \emph{clump}, que rad'ia t'ermicamente. 
Las part'iculas aceleradas en el \emph{bow shock} se enfr'ian
  tanto en el jet como en el \emph{clump}, ya que las m'as energ'eticas se
difunden
  hasta all'i.  Hemos
  calculado los procesos din'amicos y radiativos m'as importantes,
  considerando diferentes valores para el tamaño
  de los \emph{clumps} y para el campo magn'etico en la
  regi'on del \emph{jet} donde ocurre la interacci'on ($\sim a_{\rm mq}/2$).

De acuerdo a un estudio de la din'amica de la interacci'on 
\emph{jet-clump}, hemos podido estimar la velocidad del choque en el 
\emph{clump} y 
predecimos que la nube no ser'a destru'ida antes de que las part'iculas 
aceleradas en el
\emph{bow shock} se enfr'ien significativamente. 
Considerando  diferentes valores para el tamaño de los \emph{clumps} 
($R_{\rm c} = 10^{10}$ y $10^{11}$~cm) y del campo magn'etico en la regi'on de
interacci'on ($B_{\rm bs} = 150$ y 1~G), hemos calculado las SEDs, prediciendo
 emisi'on 
significativa a lo largo de  todo el espectro electromagn'etico. 
Las luminosidades m'as
altas obtenidas  en  rayos gamma son las producidas por interacciones IC con 
los fotones emitidos por la estrella compañera, $L_{\rm IC}\sim
10^{35}$~erg~s$^{-1}$, en HE, y en VHE se han alcanzado luminosidades 
tan altas como  $L_{pp}\sim 10^{32}$~erg~s$^{-1}$ producidas por colisiones 
$pp$ en el \emph{clump}.

De acuerdo al factor de llenado de \emph{clumps} en el viento que hemos
supuesto ($f \sim 0.005$), el n'umero de \emph{clumps} que simult'aneamente 
pueden interactuar con el \emph{jet} es $\sim 350$ si  $R_{\rm c} = 10^{10}$~cm
y $\sim 0.5$ si $R_{\rm c} = 10^{11}$~cm. Luego,
la emisi'on ser'a estacionaria en el primer caso, con una luminosidad total 
$\sim$ 350 veces m'as alta que la de una interacci'on simple, 
con una escala de tiempo de los \emph{flares} de $\sim 1$~hora, y
recurrencia en el plazo de varias horas.
De esta manera, las interacciones \emph{jet-clump} son una posible
explicaci'on a la producci'on de los \emph{flares} observados en 
algunas binarias de rayos gamma.

Los  niveles de emisi'on alcanzados son detectables por
intrumentos como \emph{Fermi} en HE y los telescopios Cherenkov de nueva 
generaci'on en VHE. 
 Esta emisi'on, de ser detectada, nos proveer'ia importante informaci'on sobre
los \emph{jets} de los HMMQs, como as'i tambi'en de los vientos de las
estrellas de gran masa (e.g. Cygnus~X-1, Cygnus~X-3).

\item AGNs: Un estudio similar al realizado sobre los HMMQs ha sido desarrollado
en el escenario de los AGNs, considerando que nubes de la BLR pueden 
penetrar en los
  \emph{jets} que emanan de las cercan'ias del SMBH. Debido a la interacci'on 
de una nube con uno de los \emph{jets} se produce un \emph{bow shock} 
fuerte en el \emph{jet}, donde pueden acelerarse part'iculas eficientemente, 
y otro choque m'as d'ebil
  en la nube. Como en el caso de los HMMQs, las part'iculas aceleradas 
en el \emph{bow shock} se enfr'ian tanto en el \emph{jet} como en la nube, 
ya que las m'as energ'eticas se difunden hasta all'i.  

Los procesos din'amicos que se desarrollan permiten que
la nube entre y permanezca entera suficiente tiempo como para que las
part'iculas relativistas aceleradas en el \emph{bow shock} puedan radiar una
fracci'on significativa de su energ'ia.  
Las luminosidades obtenidas por la interacci'on de una sola nube con
el \emph{jet} son relativamente bajas, $\sim 10^{38}$~erg~s$^{-1}$, 
pero si muchas nubes
se encuentran simult'aneamente interactuando con ambos \emph{jets} entonces la
contribuci'on total es significativa y detectable en fuentes no alineadas
(e.g. radiogalaxias). 
Debido a que el emisor est'a practicamente quieto, la
radiaci'on no es amplificada por efecto Doppler como acurre en
la emisi'on producida en los \emph{jets} de los blazares. 
La interacci'on de nubes de la BLR
con \emph{jets} podr'ia ser un mecanismo que explique
la emisi'on observada en algunos AGNs.

Finalmente, hemos aplicado nuestro modelo a galaxias FR~I y FR~II. 
En las primeras, si bien no hay detecciones claras de la presencia de 
una BLR, es posible suponer una poblaci'on de nubes oscuras aunque 
no podemos estimar el n'umero de ellas. Por esto, consideramos aqu'i s'olo
la interacci'on de una nube con el \emph{jet}. Debido a que estas fuentes son 
cercanas (Cen~A, M~87) la interacci'on de una sola nube de tamaño 
$R_{\rm n} \sim 10^{14}$~cm podr'ia producir \emph{flares} detectables. 
Sin embargo, en las galaxias FR~II la cantidad de nubes es grande, 
y la interacci'on 
de muchas de ellas con ambos \emph{jets} del AGN puede producir 
emisi'on detectable 
si los \emph{jets} no est'an alineados con la l'inea de la visual.

\item C'umulos de galaxias: 
Los c'umulos de galaxias son candidatos a
  ser fuentes de rayos gamma, ya que hay evidencias de la existencia
  de part'iculas relativistas que emiten en frecuencias radio.
  En particular, en esta tesis hemos estudiado la
  producci'on de rayos gamma en los remanentes (\emph{relics}) detectados 
en el
  borde de la fuente cercana Abell~3376. Los remanentes son trazadores 
de choques de fusi'on. El c'umulo Abell~3376 presenta
  grandes estructuras  en frecuencias radio, y que han sido 
asociadas a
  remanentes de choques producidos por la fusi'on entre dos c'umulos de 
galaxias.  Estos remanentes
sugieren un rico contenido de part'iculas relativistas del ICM en estas 
estructuras. Suponemos que
  la aceleraci'on de part'iculas tiene lugar en los bordes de los
  c'umulos donde se detectan los remanentes radio. Estas part'iculas
  luego se enfr'ian por diferentes procesos no t'ermicos.  

Adem'as de los electrones primarios y protones acelerados en los
choques, consideramos tambi'en una poblaci'on de pares secundarios
generados por interacciones $pp$ de los protones relativistas con el
ICM. Los par'ametros involucrados en los procesos no t'ermicos han sido
estimados considerando los datos observacionales como as'i tambi'en
suposiciones como la de equipartici'on de la energ'ia magn'etica y no 
t'ermica. Por otro lado,  hemos
hecho uso de simulaciones num'ericas para obtener aquellos  par'ametros
del medio (velocidad del choque y densidad del ICM) no provistos por
las observaciones.

La emisi'on no t'ermica m'as intensa producida en los bordes de este
c'umulo es debida a las interacciones IC de electrones primarios 
con los fotones del CMB,
alcanzando luminosidades $\sim 10^{41}$~erg~s$^{-1}$ en HE y VHE. Si bien al
d'ia de hoy no se han detectado c'umulos de galaxias en rayos gamma, la
fuente Abell~3376 es un buen candidato para ser detectado por medio de 
exposiciones prolongadas de acuerdo
a los resultados de nuestro modelo y a su proximidad.

\end{itemize}

Las ondas de choque se producen en diferentes tipos 
de fuentes astrof'isicas y por diversos mecanismos. Los frentes 
de choque asociados a estas ondas tienen tamaños que van desde $\sim 10^{-2}$~pc
en los \emph{jets} de los YSOs hasta $\sim 1$~Mpc en los c'umulos de galaxias.
Sin embargo, las velocidades de estos choques no son proporcionales a su 
tamaño, ya que en YSOs y en c'umulos de 
galaxias $v_{\rm ch} \sim 1000$~km~s$^{-1}$ mientras que en los \emph{jets} 
relativistas de los MQs y AGNs $v_{\rm ch} \sim c$. 
En todos estos choques se pueden acelerar part'iculas cargadas hasta 
velocidades relativistas.
El tamaño de los frentes de choque influye en las p'erdidas de escape, 
que se tornan muy 
lentas si aqu'ellos son muy grandes. El tamaño tambi'en es relevante para las
densidades de energ'ia de los campos ambientales. 
Si las densidades de energ'ia magn'etica y de part'iculas 
(fotones y materia) son bajas, 
entonces las p'erdidas
radiativas tampoco son muy eficientes y las energ'ias m'aximas ser'an 
altas, $> 1$~PeV. 
Sin embargo, si las densidades de energ'ia de los campos ambientales 
son altas, es posible alcanzar energ'ias m'aximas de $\sim 1$-10~TeV
en el r'egimen de saturaci'on. 

La competencia entre los tiempos din'amicos y radiativos 
determina la eficiencia radiativa de la fuente. Si $t_{\rm rad} < t_{\rm din}$
en alg'un rango de energ'ias de la distribuci'on de part'iculas relativistas,
entonces las part'iculas con esas energ'ias radiar'an significativamente
y se dice que han alcanzado el r'egimen de saturaci'on. 
En todos los escenarios explorados en esta tesis, 
el r'egimen de saturaci'on es alcanzado por los electrones en alg'un rango
 de energ'ia y para alg'un conjunto de par'ametros considerado. Sin embargo,
s'olo en aquellos casos en los cuales las part'iculas m'as energ'eticas 
son las que saturan, la emisi'on de la fuente resulta intensa y
detectable. De ser confirmados los resultados hallados en cada tipo de 
fuente que hemos estudiado, se aportar'ian nuevos conocimientos en cada campo
de investigaci'on respectivo.  

En lo que refiere a las estrellas de gran masa, por un lado modelamos 
la emisi'on 
de la protoestrella asociada a la fuente IRAS~16547-4247 obteniendo 
flujos detectables en rayos gamma, lo cual s'olo hab'ia sido someramente
sugerido anteriormente por Henriksen y colaboradores (1991),
y siendo as'i estos objetos un nuevo tipo de posibles de fuentes de 
rayos gamma. 
Esto ha sido estad'isticamente comprobado a trav'es de la muy buena 
correlaci'on encontrada entre las fuentes detectadas por \emph{Fermi} y la
localizaci'on en el plano del cielo de protoestrellas de gran masa 
(Munar y colaboradores, comunicaci'on personal). 

Por otro lado, las propiedades de las inhomegeneidades de los vientos de 
las estrellas tempranas pueden conocerse a trav'es de observaciones en rayos 
gamma, ya que las interacciones de 'estas con los \emph{jets} de los 
MQs producir'ian efectos detectables. 
Para poder determinar la microestructura de la variabilidad es necesario
el uso de telescopios con gran resoluci'on temporal (gran sensibilidad). 
Un instrumento apropiado
para esto ser'a el arreglo de telescopios Cherenkov CTA.

Las interacciones de obst'aculos con \emph{jets} 
nos proveen un nuevo mecanismo de radiaci'on, ya que adem'as de acelerarse 
las part'iculas en los \emph{bow shocks} que se forman en los \emph{jets}, 
el obst'aculo sirve de blanco para las interacciones
de estas part'iculas relativistas. 
En el campo de las binarias de rayos gamma de alta masa, el proceso descripto
puede 
ocurrir si los \emph{clumps} del viento de la estrella compañera 
llegan hasta el \emph{jet} generado por el agujero negro y as'i podri'an 
producirse los \emph{flares} detectados en algunas de estas fuentes.
En el campo de las galaxias activas, las interacciones de nubes de la BLR
con la base de los \emph{jets} podr'ian explicar la emisi'on espor'adica 
observada en 
galaxias FR~II cercanas y la emisi'on estacionaria 
de los AGNs m'as lejanos que no son blazares. Como trabajo a futuro en este
campo, haremos una aplicaci'on del modelo  expuesto en el 
Cap'itulo~\ref{AGNs} a la fuente 
 3C~120, de la cual contamos con datos observacionales de 
alta resoluci'on espacial. De estos datos es posible extraer informaci'on
sobre la microestructura del \emph{jet}, que presenta inhomogeneidades de 
$\sim $~cm (G'omez et al. 2008).
Como una consecuencia
natural de los modelos desarrollados de interacciones de \emph{jets} con 
obst'aculos,
nos proponemos estudiar que sucede 
cuando  una estrella masiva interact'ua con el \emph{jet} de un AGN. 
Si bien aqu'i el choque en el \emph{jet} se producir'ia por la colisi'on 
entre el viento de la estrella
y el material del \emph{jet}, gran parte del modelo ya desarollado puede
aplicarse a este nuevo escenario.

En el t'opico de los c'umulos de galaxias ofrecemos un candidato
concreto para ser observado con los telescopios de rayos gamma. A trav'es de 
una modelizaci'on de la fuente Abell~3376, estimamos que el flujo emitido
por 'esta ser'ia detectable por los instrumentos actuales que observan en HE
como as'i tambi'en por los futuros telescopios ya planeados.
Sin embargo, el sat'elite \emph{Fermi} no ha detectado ning'un
c'umulo de galaxias en tres años de observaci'on.
Esto muestra que alguno de los valores de los par'ametros supuestos por 
los modelos actuales
est'a siendo sobreestimado. Debido a que el efecto m'as importante es el IC
sobre los fotones del CMB, y el valor de $u_{\rm cmb}$ est'a bien
determinado, es l'icito pensar que quiz'as la poblaci'on de part'iculas 
relativistas no es tan
importante. Esto podr'ia ocurrir si los choques de fusi'on no fuesen tan
 intensos como resulta de las simulaciones num'ericas y la velocidad de
'estos fuese menor que  1000~km~s$^{-1}$.
Como trabajo a futuro en este campo, nos proponemos hacer un an'alisis del
espacio de valores de los
par'ametros considerados en la modelizaci'on de la emisi'on de rayos
gamma de los c'umulos de galaxias. 
%De esta manera se podr'ia determinar 
%cuales par'ametros ser'ian m'as sencibles . 

Finalmente,
de ser detectadas las fuentes propuestas en esta tesis y de comprobarse
los resultados obtenidos al estudiar  las interacciones de \emph{jets} con 
obst'aculos, 
se incrementar'ian las clases de emisores de rayos gamma y se 
podr'ian explicar la emisiones espor'adicas producidas en
algunas fuentes, mediante mecanismos no explorados anteriormente.

%% file: referencias_final.tex
\chapter*{Referencias}
%\section*{Referencias}

\noindent
Abdo, A.A. et al. (Fermi Collaboration) 2009a, ApJS, 183, 46\\
Abdo A.~A. et al. (Fermi Collaboration) 2009b, ApJ, 707, 1310\\
Aharonian, F.A., Atoyan, A.M., 2000, A\&A, 362, 937 \\
Aharonian, F.A., et al., 2005, Science, 309, 746\\ 
Aharonian, F.A., et al., 2007, A\&A, 467, 1075\\ 
Aharonian, F.~A., Anchordoqui, L.~A., Khangulyan, D., 
\& Montaruli, T. 2006, J.Phys.Conf.Ser., 
\indent 39, 408\\ 
Aharonian, F.A. et al. (H.E.S.S. Collaboration) 2007, A\&A, 467, 1075 \\
Aharonian, F.A. et al. (H.E.S.S. Collaboration) 2009, ApJ, 695L, 40 \\
Albert, J., et al. 2006, Science, 312, 1771\\
Albert, J., et al., 2007, ApJ, 665, L51 \\
Alexander, D.M., Hough, J.H., Young, S., Bailey, J.A., Heisler, C.A., Lumsden S.L., \\
\indent Robinson, A., 1999, MNRAS 303, L17\\
Araudo A., Romero G.E., Bosch-Ramon V., Paredes J.M. 2007, A\&A, 476, 1289 \\
Araudo A., Romero G.E., Bosch-Ramon V., Paredes J.M., 2008a, IJMPD, 17, 1889\\
Araudo A., Cora S., Romero G.E. 2008b, MNRAS, 390, 323\\
Araudo, A. T., Bosch-Ramon, V., Romero, G. E. 2009, A\&A, 503, 673 \\
Araudo, A. T., Bosch-Ramon, V., Romero, G. E. 2010, A\&A (en prensa) 
[arXiv:1007.2199] \\
Atoyan A.M., V\"olk H.J., 2000, ApJ, 535, 45\\
Axford, W. I.; Leer, E.; Skadron, G., 1977, ICRC 11, 132 \\	
Bagchi J., Durret F., Neto G.B.L., Paul S., 2006, Sci, 314, 791\\
Balestra I., Tozzi P., Ettori S., Rosati P., Borgani S., Mainieri V., 
Norman C., Viola M., 
\indent 2007, A\&A, 462, 429 \\
Bell, A.R., 1978, MNRAS, 182, 147\\
Bentz, M.C.,  Peterson, B.M., Pogge, R.W., Vestergaard, M., 
Onken, C.A., 2006, ApJ, 644, \indent 133\\
Berezinsky V.S. \& Grigorieva S.I., 1988, A\&A, 199, 1\\
Berrington, R.C. \& Dermer C.D., 2003, ApJ, 594, 709\\
Blake, G.M., 1972, MNRAS 156, 67\\
Blandford, R.D. \& Znajek, R.L., 1977, MNRAS 179, 433 \\
Blandford, R.D. \& Payne, D.G., 1982, MNRAS 199, 883 \\
Blondin, J.M., Königl, A., Fryxell, B.A., ApJ, 1989, 337L\\
Blumenthal, G.R., Gould, R.J., 1970, Rev. Mod. Phys., 42, 237\\
Bonnell, I.A., Bate, M.R., Zinnecker, H., 1998, MNRAS, 298, 93\\
Bosch-Ramon, V., Aharonian, F.A., \& Paredes, J.P., 2005, 
A\&A, 432, 609\\
Bosch-Ramon, V., Romero, G.E., Paredes, J.M., 2005, A\&A, 429, 267\\
Bosch-Ramon, V., Romero, G.E., \& Paredes, J.P., 2006,
A\&A, 447, 263\\
Bosch-Ramon, V., 2006, Tesis doctoral: \emph{Broadband emission from 
high energy processes in} \indent \emph{microquasars}\\
Bosch-Ramon, V., Motch, C., Rib\'o, M., Lopes de Oliveira, R.,
Janot-Pacheco, E., \\
\indent Negueruela, I., Paredes, J.M., \& Martocchia, A., 2007, A\&A, 473, 545\\
Bosch-Ramon, V. \& Khangulyan, D., 2009, IJMPD, 18, 347\\
Bosch-Ramon, V., Romero, G.E., Araudo, A.T., Paredes, J.M., 2010, A\&A, 511, 8\\
Bowyer S., Korpela E.J., Lampton M., Jones T.W., 2004, ApJ, 605, 168\\
Brooks, K., Garay G.,  Mardones, D., Bronfman, L., 2003, ApJ, 594, L131\\ 
% \bibitem{} Drury, L.O'C., Aharonian, F.A., \& Voelk, H.J., 1994, A\&A, 287, 959
Cao, X., Jiang, D.R., 1999, MNRAS, 307, 802\\
Churazov E., Forman W., Jones C., B\"ohringer H., 2000, A\&A, 356, 788\\
Coppi, P., Blandford R., 1990, MNRAS, 245, 453\\
Corbel, S., Fender, R.P., Tzioumis, A.K., Tomsick, J.A., Tomsick, J.A., 
Orosz, J.A., Miller 
\indent J.M., Wijnands, R., Kaaret, P., 2002, Sci, 298, 196\\
Courvoisier, T.J.-L. 1998, A\&A Rv, 9, 1\\
de Jager, O.C., Harding, A.K., Michelson, P.F., Nel, H.I., Nolan, P.L., 
Sreekumar, P., \indent Thompson, D.J., 1996, ApJ, 457, 253\\
Dietrich, M.,  Wagner, S.J., Courvoisier, T.J.-L., 
Bock, H., North, P., 1999, A\&A, 351, 31\\
Dolag K., Vazza F., Brunetti G., Tormen G., 2005, MNRAS, 354, 753   \\
Domainko W., Benbow W., Hinton J.A., Martineau-Huynh O., de Naurois M., 
Nedbal \indent D., Pedaletti G., Rowell G., for the H. E. S. S. Collaboration
2007, 30th International
\indent Cosmic Ray Conference, Merida, Mexico, astro-ph/0708.1384v1\\
Drury, L.O.'C., 1983, RPPh, 46, 973 \\
En$\beta$lin T. A., Biermann P. L., 1998, A\&A, 330, 90\\
En$\rm{\beta}$lin T. A., Biermann P. L., Klein U., Kohle S. 1998,
A\&A, 332, 395 \\
En$\rm \beta$lin T. A., Gopal-Krishna, 2001, A\&A, 366, 26\\
Evans, D.A., Kraft, R.P., Worrall, D.M., Hardcastle, M.J., 
Jones, C., Forman, W.R., \\
\indent Murray, S.S., 2004, ApJ, 612, 786\\
Fanaroff, B.L.,Riley, J.M.: 1974, MNRAS 167, 31\\
Fegan S.J., Badran H.M., Bond I.H., Boyle P.J., Bradbury S.M., Buckley J.H., \\
\indent Carter-Lewis D.A., Catanese M., et al., 2005, ApJ, 624, 638\\
Feretti L., Givannini G., 1996. In R. Ekers, C. Fanti \^ L. Padrielli (eds.)
IAU Symp. 175, \indent 
Extragalactic Radio Sources. Kluwer Academic Publisher, p. 333\\ 
Feretti L., B\"ohringer H., Giovannini G., Neumann D., 1997, A\&A, 317, 432\\
Feretti L., Fusco-Femiano R., Giovannini G., Govoni F., 2001, A\&A, 373, 106\\
Feretti L., Burigana C., En$\rm\beta$lin T.A., 2004, New Astron. Rev., 48, 1137\\
Feretti L., Giovannini G., 2008, in  Plionis M., Lopez-Cruz O., Hughes D. (eds.)
\\
\indent Panchromatic View of Clusters of Galaxies and the Large-Scale 
Structure, Lecture 
\indent Notes Physics 740, Springer, Dordrecht, p. 143\\
Fermi, E., 1949, Physics Review 75, 1169\\
Ferrari, C., 2010 (en prensa) [arXiv:1005.3699]\\
Fragile, P.C., Murray, S.D., Anninos, P. \& van Breugel, W., 
2004, ApJ, 604, 74\\
Fusco-Femiano R., dal Fiume D., Feretti L., Giovannini G., Grandi P., Matt G.,
Molendi \indent S., Santangelo A., 1999, ApJ, 513, L21\\
Fusco-Femiano R., Orlandini M., Brunetti G., Feretti L., Giovannini G., Grandi P., Setti \indent G., 2004, ApJ, 602, L73\\
Gabici S., Blasi P., 2003, ApJ, 583, 695\\
Gallo, E., Fender, R., Kaiser, C., Russell, D., Morganti, R.,
Oosterloo, T., \& Heinz, S., \indent 2005, Natur, 436, 819\\
Garay, G.; Rodriguez, L.F.; Moran, J.M.; Churchwell, E., 1993, ApJ 418, 368\\
Garay, G., Brooks, K., Mardones, D., Norris, R.P., 2003, ApJ, 537, 739\\
Garay, G.; Mardones, D.; Bronfman, L.; Brooks, K.J.; Rodríguez, L.F.; 
Güsten, R.; \\
\indent Nyman, L.-A.; Franco-Hernández, R.; Moran, J.M.; A\&A, 463, 217\\
Gaisser, T.K., 1990, \emph{Cosmic Rays and Particle Physics}, Cambridge University Press, \\ \indent Cambridge\\
Ginzburg, V.L., Syrovatskii, S.I., 1964, \emph{The Origin of Cosmic Rays}, 
Pergamon Press, \indent New York\\
Giovannini G. Feretti L., Stanghellini C., 1991, A\&A, 252, 528\\
Giovannini G., Tordi M., Feretti L., 1999, New Astron., 4, 141\\
Giovanninii G., Ferreti L., 2004, JKAS, 37, 323\\
Girardi M., Giuricin G., Mardirossian F., Mezzetti M., Boschin W. 1998, ApJ, 505, 74\\
Ghisellini, G., Maraschi, L., Treves, A. 1985, A\&A, 146, 204\\	
Ghisellini, G., Tavecchio, F., Foschini, L., Ghirlanda, G., Maraschi, L., Celotti, A. 2010, \indent MNRAS, 402, 497\\
G'omez J.L., Marscher A.P., Jorstad S.G., Agudo I., Roca-Sogorb M.,
2008, ApJL, 681, \indent L69\\
Govoni F, Feretti L., Giovannini G., B\"ohringer H., Reiprich T.H., Murgia M. 2001, A\&A, \indent 376, 803 \\
Govoni F, Feretti L., 2004, Int. J. Mod. Phys. D, 13, 1549 \\
Grandi, P., Palumbo G.G.C., 2004, Sci, 306, 998\\
Hartman, R.C. et al., 1999, ApJS, 123, 79\\
Henriksen, R.N.; Mirabel, I.F.; Ptuskin, V.S.; 1991, A\&A, 248, 221\\
Hillas, A.M., 1984, ARA\&A, 22, 425\\
Hoeft M., Br\"ueggen M., Yepes G. 2004, MNRAS, 347, 389\\
Israel, F.P., 1998, A\&AR, 8, 237\\
Jaffe W.J., 1977, ApJ, 212, 1\\
Jokipii, J.R.; 1987, ApJ, 313, 842\\
Jolley, E.J., Kuncic, Z., Bicknell, G.V., Wagner, S., 2009, MNRAS,
400, 1521  \\
Junor, W., Biretta, J.A., Livio, M. 1999, Nature, 401, 891\\
Kaspi, S., Maoz, D., Netzer, H., Peterson, B.M., Vestergaard, M., Jannuzi, B.T., 2005, ApJ \indent 629, 61\\
Kaspi, S., Brandt, W.N., Maoz, D., Netzer, H., Schneider, D.P.,
 Shemmer, O., 2007, \indent ApJ 659, 997\\
Kataoka, J., Tanihata, C., Kawai, N., Takahara, F., 
Takahashi, T., Edwards, P.G., Makino, \indent F., 2002, MNRAS 336, 932\\
Khangulyan D., Hnatic S., Aharonian F., Bogovalov S., 2007, MNRAS, 380, 312 \\
Kaufman Bernad'o, M.M., Romero, G.E., Mirabel, I.F., 2002, 385, L10\\
Kelner, S.R., \& Aharonian, F.A., 2008, PhysRevD 78, 034013\\
Kelner, S.R., Aharonian, F.A., \& Vugayov, V.V., 2006, Phys.Rev.D 74, 034018\\
Keshet U., Waxman E., Loeb A., Springel V., Hernquist L., 2003, ApJ, 585, 128\\
Khangulyan, D., Hnatic, S., Aharonian F., Bogovalov S. 2007, MNRAS 380, 320\\
Khangulyan, D., Aharonian, F., Bosch-Ramon, V., 2008, MNRAS, 383, 467\\
Klein, R.I., McKee, C.F. \& Colella, P., 1994, ApJ, 420, 213\\
Krolik, J.H., McKee, C.F., \& Tarter, C.B., 1981, ApJ, 249, 422\\
Krti$\check{\rm c}$ka, J., \& Kub\'at, J., 2001, A\&A, 377, 175 \\
Landau, L.D. \& Lifshitz, E., 1951, \emph{The classical theory of fields},
Addison Wesleey Press, Cambridge\\ 
Landau, L.D. \& Lifshitz, E., 1959, \emph{Fluid Mechanics}, \\
Lang, K.R., 1999, Astrophysical Formulae, Springer, Berlin\\
Lieu R., Mittaz J.P.D., Bowyer S., Lockman F., Hwang C.-Y., Schmitt J.H.M.M.,
1996, ApJ, 458, L5\\
Lieu R., Axford W. I., Bonamente M., 1999, ApJ, 510, L25\\
Mannheim, K., Schlickeiser, R., 1994, A\&A, 286, 983\\ 
Marconi A., Schreider, E.J., Koekemoer, A., Capetti, A.,
Axon, D., Macchetto, D., \& Kaon, \indent N., 2000, ApJ, 528, 276\\
Markevitch, M. et al.; 2000, ApJ, 541, 542\\ 
Mart{\'\i}, J., Rodr{\'\i}guez, L.F., Reipurth, B., 1995, ApJ, 449, 184\\
Miller-Jones, J.C.A., Fender, R.P. \& Nakar, E., 2006, MNRAS, 367, 1432\\
Mirabel, I.F. \& Rodr{\'\i}guez, L.F., 1999, ARA\&A 37, 409 \\
Mirabel, I.F., Laurent, O., Sanders, D.B., Sauvage, M., 
Tagger, M., Charmandaris, V., \indent Vigroux, L., Gallais, P., Cesarsky, C., 
Block, D.L. 1999, A\&A 341, 667\\
Mittaz J.P.D., Lieu R., Lockman F.J., 1998, ApJ, 498, L17\\
Moffat, A.F.J., 2008, Proceedings of the conference ``Clumping 
in hot-star winds'' 
%held in Potsdam, Germany, June 2007. 
Eds.:  \indent Hamann, W.R., Feldmeier, A. \& Oskinova, L., 17\\
Myasnikov, A.V.; Zhekov, S.A.; Belov, N.A., 1998, MNRAS, 298, 1021\\ 
Orellana, M., Bordas, P., Bosch-Ramon, V., Romero, G.~E., 
\& Paredes, J.~M. 2007, A\&A, \indent 476, 9 \\
Orosz, J.A., Miller J.M., Wijnands, R., Kaaret, P., 2002, Sci, 298, 196\\
Owocki, S.P., \& Cohen D.H., 2006, ApJ, 648, 5650\\
Owocki, S.P., Romero G.E., Townsend, R. \& Araudo, A.T., 2009, ApJ 696, 690\\
Pacholczyk, A.G., 1970, \emph{Radio Astrophysics}, Freeman, San Francisco\\
Paltani, S., \& T$\ddot{\rm u}$rler, M., 2005, A\&A, 435, 811\\
Paredes, J.M., Rib\'o, M., Bosch-Ram\'on, V., et al., 2007, ApJ, 664, L39\\ 
Paredes, J.M. 2008, Int. Jour. Mod. Phys. D, 17, 1849 \\
Penston, M.V.,1988, MNRAS, 233, 601\\
Perkins J.S., Badran H.M., Blaylock G., Bradbury S.M., Cogan P., Chow Y.C.K., 
Cui W., \indent Daniel M.K., et al., 2006, ApJ, 644, 148 \\
Perucho, M., \& Bosch-Ramon, V, 2008, A\&A, 482, 917\\
Peterson, B.M.; Bentz, M.C.; Desroches, L.-B.; Filippenko, A.V.; Ho, L.C.; 
Kaspi, S.; Laor, \indent A.; Maoz, D.; Moran, E.C.; Pogge, R.W.; Quillen, A.C., 
2005, ApJ, 632, 799\\
Peterson, B.M., 2006, LNP 693, 77\\
Petrosian V., 2001, ApJ, 557, 560\\
Pfrommer C., Springel V., En$\beta$lin T. A., Jubelgas M., 2006, MNRAS, 367, 
113\\
%Pfrommer, C., En$\beta$lin, T.A., 2004, A\&A, 413, 17 \\
Pfrommer C., En$\beta$lin T.A., Springel V., Jubelgas M., Dolag K., 2007, 
MNRAS, 378, 385\\
Pfrommer C., En$\beta$lin T. A., Springel V., 2008, MNRAS, 385, 1211\\
Pian, E. et al., 1999, ApJ, 521, 112\\
Platzek, A.M., Apuntes de la materia {\it Introducci'on a la magnetohidrodin'amica}\\
Priest, E.R., 1982, \emph{Solar Magnetohydrodynamics}, Reidel, Dordrecht\\ 
Protheroe, R.J., 1999, 
%in: Acceleration and Interaction of Ultra High Energy Cosmic Rays in  
{\it Topics in cosmic-ray astrophysics}, eds. M. A. DuVernois
(Nova \\
\indent Science Publishing), p. 240  [astro-ph/9812055]\\
Puls, J., Markova, N., Scuderi, S., Stanghellini, C., 
Taranova, O.G., Burnley, A.W., Howarth \indent I.D., 2006, A\&A, 454, 625\\
Raga, A.C., Cant\'o, J., Rodr{\'\i}guez-Gonz\'alez, A. \& 
Esquivel, A., 2009, A\&A, 493, 115\\ 
Reimer O., Pohl M., Sreekumar P., Mattox J. R., 2003, ApJ, 588, 155\\
Rees, M., 1984, ARA\&A 22, 471\\
Rees, M.J., 1987, MNRAS, 228, 47\\
Reynoso, M. \& Romero, G.~E. 2009, A\&A, 493, 1\\
Rib\'o, M., 2005, ASPC, 340, 269\\
Rieger, F.M., Aharonian, F.A., 2009, A\&A\\
Risaliti, G., 2009 [arXiv:0912.2118]\\
Risaliti, G., Elvis, M., Nicastro, F., 2002, ApJ 571, 234\\
Rodr'iguez, L.F., Marti, J., Canto, J., Moran, J.M., Curiel, S., 1993,
RMxAA, 25, 23\\
Rodr'iguez L.F., Garay G., Brooks, K., Mardones, D., 2005, ApJ 626, 953\\
Romero, G.E., 1995, Ap\&SS 234, 49\\
Romero, G.E., Combi, J.A., Perez Bergliaffa S.E., Anchordoqui,
L.A., 1996, APh, 5, 279 \\
Romero, G.E., Benaglia, P., Torres, D.F., 1999, A\&A, 348, 868\\
Romero, G.~E. 2001, \emph{The Nature of Unindentified Galactic
High-Energy Gamma-Ray Sources}, \indent ed. A. Carraminana, O. Reimer, \& 
D. Thompson, Kluwer 
Academic Publishers, \\ \indent Dordrecht, 65\\
Romero, G.~E., Torres, D.~F., Kaufman Bernad\'o, M.~M., 
\& Mirabel, I. F., 2003, A\&A, \indent 410, 1 \\
Romero, G.~E. \& Orellana, M. 2005, A\&A, 439, 237\\
Romero, G.E.; Okazaki, A.T.; Orellana, M.; Owocki, S.P., 2007, A\&A, 474, 15\\
Romero, G. E., Owocki, S. P., Araudo, A. T., Townsend, R. H. D., 
\& Benaglia,
P., Actas \indent del congreso \emph{Clumping in Hot Star Winds}, 
W. R. Hamann, A. Feldmeier \& L. M. \\
\indent Oskinova  (eds.), Potsdam, Univ. Verl., 2008, p. 191 \\
Romero, G.E., del Valle, M.V., Orellana, M., 2010, A\&A, 518, 12\\ 
Romero, G.E., 2010, Apuntes de la materia \emph{Introducci'on a la 
Astrof'isica Relativista.} \\
\indent (http://www.iar.unlp.edu.ar/garra/AR/apunte.html)\\
R\"ottgering H.J.A., Snellen I., Miley G., de Jong J.P., Hanisch R.J., Perley R., 1994, ApJ, \indent 436, 654\\
R\"ottgering H.J.A., Wieringa M.H., Hunstead R.W., Ekers R.D., 1997, MNRAS, 
290, 577\\
Russell, D.M., Fender, R.P., Gallo, E., \& Kaiser, C.R.,
2007, MNRAS, 376, 1341 \\
Sabatini, S. et al., 2010, ApJ, 712, L10 \\ 
Shakura, N.I., Sunyaev, R.A., 1973, A\&A 24, 337\\
Shin, M.-S., Stone, J.~M., Snyder, G.~F. 2008, ApJ, 680, 336\\
Shu, F.H, Adams, F.C., Lizano, S. 1987, ARA\&A, 25, 23 \\
Soldi, S., Beckmann, V., T$\ddot{\rm u}$rler M., 2009 [arXiv:0912.2266v1]\\
Sturrock, P.A., 1971, ApJ 164, 529\\ 
Taub, A.H., 1948, PhRv 74, 328\\
Tavani, M. et al. 2009a (AGILE Collaboration), ApJ, 698, L142 \\
Tavani, M. et al. 2009b (AGILE Collaboration), Nature, 462, 620 \\
Thomson, R.C., 1992, MNRAS, 257, 689\\
Tingay S.j. et al., 1998, AJ, 115, 960\\
Urry, C. \& Padovani, P., 1995, PASP, 107, 803 \\
van Dike \& Gordon, H., 1959, NASA TR R-1\\
Vila, G., Aharonian F., 2009, en `\emph{Compact Objects and their Emission},
Eds: Romero, G.E., \indent Benaglia, P.\\
V\"olk H.J., Aharonian F.A., Breitschwerdt D., 1996, SSRv, 75, 279\\
Wang, B., Inoue, H., Koyama K., Tanaka, Y., 1986, PASJ, 38, 685\\
Zhekov, S.A. \& Palla, F., 2007, MNRAS, 382, 1124\\
Zinnecker, H.; Yorke, H.W., 2007, ARA\&A 45, 481\\

%Lista de siglas:

%A\&A: \emph{Astronomy \& Astropfysics}\\
%ASPC: 
%IJMPD: \emph{International Journal of Modern Physics D}\\
%ICRC: \emph{International Cosmic Ray Conference}\\

%% file: acronimos.tex
\chapter{Lista de acr'onimos}

 Debido al uso generalizado de las siglas inglesas en la jerga astron'omica 
referida al tema de esta tesis, se han utilizado a lo largo de este 
trabajo una gran cantidad de acr'onimos que se refieren a palabras en la 
lengua mencionada.
\bigskip

\noindent 
AGN: \emph{Active Galactic Nucleus}\\
ATCA: \emph{Australia Telescope Compact Array}\\
BH: \emph{Black Hole} \\
BLR: \emph{Broad Line Region}\\
%BLRG: \emph{broad line radio-galaxy} (radio-galaxia de l'ineas anchas)\\
CDM: \emph{Cold Dark Matter} \\
%DC: ciclo de actividad (\emph{duty cycle})\\
CMB: \emph{Cosmic Microwave Background} \\
CTA: \emph{Cherenkov Telescope Array} \\
EC: \emph{External Compton} \\
%EGRET \emph{External Compton}\\
FR: Faranoff-Rayleigh\\
FSRQ: \emph{flat spectrum radio-quasar} \\
%FWHM: ancho total a medio m'aximo (\emph{full width half maximum})\\
%HPQ: cu'asar altamente polarizado (\emph{highly polarized quasar})\\
GRB: \emph{Gamma-Ray Burst} \\
HE: \emph{High Energy}\\
HH: Herbig-Haro\\
HMMQ: \emph{High Mass Microquasar}\\
IC: \emph{Inverse Compton} \\
ICM:\emph{Intra Cluster Medium} \
IR: \emph{Infra Red}\\
KH: Kelvin-Helmholtz\\
%LBV: \emph{Luminous Blue Variable} ((estrellas) luminosas variables azules)\\
MHD: \emph{Magnetohydrodynamics} \\
MQ: \emph{Microquasar}\\
%NLR: regi'on de l'ineas angostas (\emph{Narrow Line Region})\\
%NLRG: radio-galaxia de l'ineas angostas (\emph{narrow line radio-galaxy})\\
RT: Rayleigh-Taylor\\
SED: \emph{Spectral Energy Distribution} \\
SMBH: \emph{Super-Massive Black Hole} \\
SPH: \emph{Smoothed Particle Hydrodynamics}\\
SSC: \emph{Synchrotron Self Compton} \\
VHE: \emph{Very High Energy} \\
VLA: \emph{Very Large Array}\\
WR: Wolf-Rayet\\
YSO: \emph{Young Stellar Object}\\
%ZAMS: Secuencia principal de edad cero (\emph{Zero Age Main Sequence})\\

%% file: PublicationList.tex
\chapter{Lista de publicaciones}

\begin{itemize}
\item Publicaciones en revistas internacionales con referato

\begin{enumerate}

\item  \emph{Gamma-ray emission from massive young stellar objects}\\
Anabella T. Araudo, Gustavo E. Romero, Valent{\'\i} Bosch-Ramon \& 
Josep M. Paredes\\
A\&A, 476, 1289-1295, 2007
\item  \emph{Gamma-ray emission from massive star forming regions}\\
Anabella T. Araudo, Gustavo E. Romero, Valent{\'\i} Bosch-Ramon \& 
Josep M. Paredes\\
%\emph{International Journal of Modern Physics D}
IJMPD,
%\footnote{International Journal of Modern Physics D}, 
17, 1889-1894, 2008

\item \emph{Non-thermal processes in the cluster of galaxies Abell 3376} \\
Anabella T. Araudo, Sof{\'\i}a A. Cora \& Gustavo E. Romero\\
MNRAS, 390, 323, 2008

\item \emph{High-energy emission from jet-clump interactions in microquasars}\\
Anabella T. Araudo, Valent{\'\i} Bosch-Ramon \& Gustavo E. Romero\\
A\&A, 503, 673-681, 2009

\item \emph{Gamma-Ray Variability from Wind Clumping in High-Mass X-Ray Binaries with Jets}\\
Stanley P. Owocki; Gustavo E. Romero; Richard H.D. Townsend; Anabella T. 
Araudo\\
ApJ, 696, 690-693, 2009 

\item \emph{Massive protostars as gamma-ray sources}\\
Valent{\'\i} Bosch-Ramon, Gustavo E. Romero,  Anabella T. Araudo\&
Josep M. Paredes\\
A\&A, 511, 1-10, 2010

\item \emph{High-Energy Emission from Jet-Cloud Interactions in AGNs}\\
Anabella T. Araudo, Valent{\'\i} Bosch-Ramon \& Gustavo E. Romero\\
IJMPD, 19, 931-936, 2010

\item \emph{Gamma rays from cloud penetration at the base of AGN jets}\\
Anabella T. Araudo, Valent{\'\i} Bosch-Ramon \& Gustavo E. Romero\\
A\&A, en prensa, 2010 [arXiv:1007.2199]

\end{enumerate}

\item Publicaciones en revistas nacionales con referato

\begin{enumerate}
\item  \emph{High-energy emission from Abell 3376}\\
Anabella T. Araudo, Sof{\'\i}a A. Cora \& Gustavo E. Romero\\
BAAA\footnote{Bolet'in de la Asociaci'on Argentina de Astronom'ia}, 50, 303-306, 2007

\item  \emph{Gamma-ray emission from jet-clump interactions}\\
Gustavo E. Romero, Anabella T. Araudo, Stanley P. Owocki \& Richard Townsend\\
BAAA,
%\emph{Bolet{\'\i}n de la Asociaci\'on Argentina de Astronom{\'\i}a}, 
50, 319-322, 2007

\item  \emph{Interactions of jets and clumpy stellar winds in high-mass 
microquasars}\\
Anabella T. Araudo, Valent{\'\i} Bosch-Ramon \& Gustavo E. Romero\\
BAAA, 51, 305-308, 2008

\item  \emph{Jet-cloud interactions in the BLR of Centaurus A}\\
Anabella T. Araudo, Valent{\'\i} Bosch-Ramon \& Gustavo E. Romero\\
BAAA, 52, 255-258, 2009

\end{enumerate}

\item Publicaciones en actas de congresos 

\begin{enumerate}

\item \emph{Gamma-ray emission from massive young stellar objects: 
the case of IRAS 16547-4247}\\
Anabella T. Araudo, Gustavo E. Romero, Valent{\'\i} Bosch-Ramon \& 
Josep M. Paredes\\
Revista Mexicana de Astron. Serie de Congresos, 33, 159, 2008
%RMxAC, 33, 159, 2008

\item  \emph{Using gamma-rays to probe the clumped structure of 
stellar winds}\\
Gustavo E. Romero, Stanley P. Owocki,  Anabella T. Araudo, Richard Townsend
\& Paula Benaglia\\
\emph{Clumping in hot-star winds},  Potsdam: Univ.-Verl., 191-194, 2008

\item \emph{Non-thermal emission from massive YSOs. Exploring the 
spectrum at high energies}\\
Anabella T. Araudo, Gustavo E. Romero, Valent{\'\i} Bosch-Ramon \& 
Josep M. Paredes\\
First La Plata International School: Compact Objects and their Emission,
I. Andruchow \& G.E. Romero, eds., (2008) [arXiv:0806.2306]

\item \emph{Gamma-radiation from the galaxy cluster Abell 3376}\\
Sof{\'\i}a A. Cora, Anabella T. Araudo \& Gustavo E. Romero\\
American Institute of Physics, Conference Proceedings (AIPC), 1085, 573-576, 2008

\item \emph{Jet-Cloud Interactions in AGNs}\\
Anabella T. Araudo, Valent{\'\i} Bosch-Ramon \& Gustavo E. Romero\\
2009. [arXiv:0908.0926]

\item \emph{High-energy flares from jet-clump interactions}\\
Anabella T. Araudo, Valent{\'\i} Bosch-Ramon \& Gustavo E. Romero\\
Astron. Soc. of the Pacific Conference Series, Eds.: Josep Martí, 
Pedro L. Luque-Escamilla and Jorge A. Combi, 422, 32-40, 2010

\item \emph{Gamma-rays from massive protostars}\\
Gustavo E. Romero, Anabella T. Araudo, Valent{\'\i} Bosch-Ramon \& Josep M. Paredes\\
Astron. Soc. of the Pacific Conference Series, Eds.: Josep Martí, 
Pedro L. Luque-Escamilla and Jorge A. Combi, 
422, 100-108, 2010

\item \emph{Gamma-Ray Variability from Stellar Wind Porosity in Microquasar 
Systems}\\
Stanley P. Owocki; Gustavo E. Romero; Richard H.D. Townsend; Anabella T. 
Araudo\\
Astron. Soc. of the Pacific Conference Series, Eds.: Josep Martí, 
Pedro L. Luque-Escamilla and Jorge A. Combi, 
422, 49-54, 2010

\end{enumerate}

%\item Publicaciones en actas de congresos sin referato

\item Publicaciones no relacionadas a la tesis

\begin{enumerate}

\item \emph{Extreme microvariability of blazars: fact and fiction}\\
Anabella T. Araudo, Sergio A. Cellone \& Gustavo E. Romero \\ 
BAAA, 48, 379, 2005

\item \emph{Multifrecuency variability of the blazar AO 0235+164}\\
\emph{The WEBT campaign in 2004-2005 and long-term SED analysis}\\
C.M. Raiteri, M. Villata, M. Kadler et al.\\
A\&A,  459, 731, 2006 

\item \emph{Extremely violent optical microvariability in blazars: fact or fiction?}\\
Sergio A. Cellone, Gustavo E. Romero \& Anabella T. Araudo\\
MNRAS, 374, 357, 2007

\item \emph{Detection of nonthermal emission from the bow shock of a 
massive runaway star}\\
Paula Benaglia, Gustavo E. Romero, Josep Mart'i, Cintia S. Peri \&
Anabella T. Araudo\\
A\&A Letters, en prensa (2010) [arXiv:1007.3279]

\end{enumerate}
\end{itemize}